\documentclass[twoside,12pt]{article}
\usepackage{epsfig}
\def\Journal#1#2#3#4{{#1} {#2} (#4) #3}
\def\NCA{{\em Nuovo Cimento} A}

\def\NPA{{\em Nucl. Phys.} A}

\def\PRO{{\em Prog. Theor. Phys.}}
\def\NPB{{\em Nucl. Phys.} B}

\def\PLB{{\em Phys. Lett.} B}

\def\PRL{\em Phys. Rev. Lett.}
\def\PREV{\em Phys. Rev.}
\def\PREP{\em Phys. Rep.}

\def\PRD{{\em Phys. Rev.} D}
\def\PRC{{\em Phys. Rev.} C}

\def\ZPC{{\em Z. Phys.} C}
\def\ZPA{{\em Z. Phys.} A}
\def\ANNP{\em Ann. Phys. (N.Y.)}
\def\RMP{{\em Rev. Mod. Phys.}}

\def\INT{{\em Int. J. Mod. Phys.} E}
\def\r{\vec r}

\def\JPG{{\em J. Phys.} G}
\def\EPJC{{\em Eur. Phys. J.} C}
\def\EPJA{{\em Eur. Phys. J.} A}
\def\PTPS{{\em Prog. Theor. Phys. Suppl.}}
\def\PPNP{{\em Prog. Part. Nucl. Phys.}}

\def\alt{\stackrel{<}{\sim}}
\def\agt{\stackrel{>}{\sim}}
\def\ubar{\bar{u}}
\def\dbar{\bar{d}}
\def\sbar{\bar{s}}

\def\qbar{\bar{q}}
\def\Qbar{\overline{Q}}
\def\Kbar{\overline{K}}
\def\Dbar{\overline{D}}
\def\Bbar{\overline{B}}

\def\vr{\vec{r}}

\def\JPsi{J/\Psi}
\def\bra#1{{\langle #1{\left| \right.}}}
\def\ket#1{{{\left.\right|} #1\rangle}}

\newcommand{\be}{\begin{equation}}
\newcommand{\ee}{\end{equation}}
\newcommand{\bge}{\begin{equation}}
\newcommand{\ene}{\end{equation}}
\newcommand{\bea}{\begin{eqnarray}}
\newcommand{\eea}{\end{eqnarray}}
\newcommand{\bg}{\begin{eqnarray}}
\newcommand{\en}{\end{eqnarray}}
\newcommand{\nn}{\nonumber}
\begin{document}

\title{ \vspace{1cm} Nucleon and hadron structure changes 
in the nuclear medium and impact on observables}
\author{K.\ Saito,$^{1}$ K.\ Tsushima,$^{2}$ A.W.\ Thomas$^3$\\ 
\\
$^1$Department of Physics, Faculty of Science and Technology,\\
Tokyo University of Science, Noda 278-8510, Japan\\
$^2$Physics Division, National Center for Theoretical Sciences, 
Taipei 10617, Taiwan\\
$^3$Thomas Jefferson National Accelerator Facility,\\ 
12000 Jefferson Ave., Newport News VA 23606 USA}
\maketitle
%
\vspace{-12cm}
\hfill JLAB-THY-05-326 \hspace*{2em}\\
\vspace{11cm}

\noindent
\hrulefill\\
{\small
\noindent{\bf Abstract}\\

We study the effect of hadron structure changes in a nuclear 
medium using the quark-meson coupling (QMC) model. The QMC model is based 
on a mean field description of non-overlapping nucleon (or baryon) bags bound 
by the self-consistent exchange of scalar and vector mesons 
in the isoscalar and 
isovector channels. The model is extended to investigate the properties of 
{}finite nuclei, in which, using the Born-Oppenheimer 
approximation to describe the 
interacting quark-meson system, one can derive the effective equation of motion for the nucleon 
(or baryon), 
as well as the self-consistent equations for the meson mean fields.  

In conventional nuclear physics, the Skyrme 
effective forces are very popular, 
but, there is no satisfactory interpretation of the parameters appearing 
in the Skyrme forces. Comparing a many-body Hamiltonian generated by 
the QMC model in the zero-range limit with that of the Skyrme effective 
forces, it is possible to obtain a remarkable agreement between 
the Skyrme force and the QMC effective interaction.  
One can also investigate the relationship between the QMC model and Quantum Hadrodynamics,  
by carrying out a re-definition of the scalar field in matter.  Furthermore, 
by using {\it naive dimensional analysis}, it is possible to see that the QMC model can provide 
remarkably {\it natural} 
coupling constants and hence the model itself is regarded as a 
{\it natural} effective field theory for nuclei. 

The model is first applied to nuclear matter, where the coupling constants are determined 
so as to produce the saturation condition at normal nuclear matter density. 
We find a new, simple scaling relation for the changes of hadron masses in 
a nuclear medium, 
which can be described in terms of the number of light quarks in a 
hadron and the value of the scalar mean-field in matter. 
Once the coupling constants are fixed, the model can be applied to various finite nuclei, 
including strange and exotic hypernuclei.  
In this article, we discuss in detail the properties of hypernuclei and meson-nucleus deeply bound 
states. 

It is also of great interest that the QMC model predicts 
a variation of the nucleon form 
factors in nuclear matter, which will affect certainly the 
analysis of electron scattering off nuclei, 
including polarization-transfer experiments. 
Recent experimental analysis of data taken at Jefferson Laboratory (JLab) 
and MAMI 
indeed seems to support such a variation of nucleon form factors 
in nuclei. The change of nucleon structure in a nuclear medium  
is also expected to modify nuclear 
structure functions (the nuclear EMC effect), which are measured by 
scattering with high-energy charged-leptons and/or neutrinos.
We study such possibilities, including consideration of the 
empirically observed, Bloom-Gilman (quark-hadron) 
duality.

We also study hadronic reactions in nuclear medium, namely, 
subthreshold kaon production in heavy ion collisions, 
$D$ and $\overline{D}$ meson production in antiproton-nucleus 
collisions, and $J/\Psi$ suppression.
In particular, the modification of the $D$ and $\overline{D}$ meson properties 
in nuclear medium derived by the QMC model, can lead a large 
$J/\Psi$ absorption cross section, 
which is required to explain the observed 
$J/\Psi$ suppression in microscopic hadronic comover 
dissociation scenario, without assuming the formation of a 
QGP phase.  

The present investigation indicates that the traditional nuclear/hadronic 
physics approach may have its limitations. It further 
suggests the need for the study 
of alternate approaches  which include subhadronic degrees of freedom, 
even at normal nuclear matter density. 

\vspace{1cm}

\noindent
{\it Keywords:} Quarks in nuclei, Quark-meson coupling model, 
Relativistic nuclear model, Properties of finite nuclei, A new scaling 
formula for hadron masses in matter, Nucleon form factors, 
Hyper and exotic nuclei, Meson-nucleus bound states, 
Nuclear structure functions, Hadron reactions in matter\\

\noindent
\hrulefill\\
\\
\noindent {\bf Contents}
\begin{enumerate}
\item{Introduction\dotfill\\}
\item{Foundation of the QMC Model\dotfill\\ 
2.1.\hspace{1em} Non-relativistic treatment\dotfill\\
\hspace*{3em}2.1.1. Classical consideration\dotfill\\
\hspace*{3em}2.1.2. Quantization at the quark level\dotfill\\
\hspace*{3em}2.1.3. Correction due to the center of mass motion\dotfill\\
\hspace*{3em}2.1.4. Correction due to $H_1$ and Thomas precession\dotfill\\
\hspace*{3em}2.1.5. Total Hamiltonian in the mean fields\dotfill\\
\hspace*{3em}2.1.6. Equations for the meson fields\dotfill\\
\hspace*{3em}2.1.7. Self-consistency procedure\dotfill\\
2.2.\hspace{1em} Relativistic treatment\dotfill\\
\hspace*{3em}2.2.1. Relativistic model\dotfill\\
\hspace*{3em}2.2.2. Properties of the in-medium nucleon mass\dotfill\\
\hspace*{3em}2.2.3. Effect of meson structure\dotfill\\
\hspace*{3em}2.2.4. Naturalness\dotfill\\
2.3.\hspace{1em} Relationship between the QMC model and conventional nuclear models\dotfill\\
2.4.\hspace{1em} Relationship between the QMC model 
and Quantum Hadrodynamics\dotfill\\
2.5.\hspace{1em} Modified quark-meson coupling model\dotfill\\
2.6.\hspace{1em} Variants of the QMC model\dotfill\\}
\item{Properties of nuclei\dotfill\\
3.1.\hspace{1em} Nuclear matter properties - saturation properties and incompressibility\dotfill\\
\hspace*{3em}3.1.1. A new scaling formula for hadron masses in matter\dotfill\\
\hspace*{3em}3.1.2. Variations of quark and gluon condensates in nuclear matter\dotfill\\
\hspace*{3em}3.1.3. Nuclear matter at finite temperature and neutron stars\dotfill\\
3.2.\hspace{1em} Finite nuclei - $^{16}$O, $^{40}$Ca, $^{48}$Ca, $^{90}$Zr, $^{208}$Pb \dotfill\\
3.3.\hspace*{1em} Strange, charm and bottom hadrons in nucleus\dotfill\\
\hspace*{3em}3.3.1. Strange, charm and bottom hadron properties 
in nuclear matter\dotfill\\
\hspace*{3em}3.3.2. Mean-field equations of motion for
strange, charm and bottom hypernuclei\dotfill\\
\hspace*{3em}3.3.3. Spin-orbit potential in the QMC model
- strange hypernuclei \dotfill\\
\hspace*{3em}3.3.4. Pauli blocking and channel coupling effects\dotfill\\
\hspace*{3em}3.3.5. Results for strange hypernuclei\dotfill\\
\hspace*{3em}3.3.6. Results for charm and bottom hypernuclei\dotfill\\
3.4.\hspace*{1em}Meson-nucleus bound states\dotfill\\}
\item{Effects of nucleon substructure on lepton-nucleus scatterings\dotfill\\
4.1.\hspace{1em} Nucleon form factors\dotfill\\
\hspace*{3em}4.1.1. Electromagnetic form factors\dotfill\\
\hspace*{3em}4.1.2. The QMC predictions and experimental results
(EM interactions)\dotfill\\
\hspace*{3em}4.1.3. Nucleon substructure effect on 
the longitudinal response functions\dotfill\\
\hspace*{3em}4.1.4. Axial form factors - neutrino-nucleus scattering\dotfill\\
4.2.\hspace{1em} Quark-hadron duality, the nuclear EMC effect
and nuclear structure functions\dotfill\\
\hspace*{3em}4.2.1. The nuclear EMC effect\dotfill\\
\hspace*{3em}4.2.2. Bloom-Gilman duality and the nuclear EMC effect
\dotfill\\
\hspace*{3em}4.2.3. Bloom-Gilman duality and the nuclear structure
functions\dotfill\\}
\item{Hadronic reactions in nuclear medium\dotfill\\
5.1.\hspace{1em} Subthreshold kaon production and in-medium effects\dotfill\\
5.2.\hspace{1em} $D$ and $\Dbar$ meson production in nuclei\dotfill\\
5.3.\hspace{1em} $\JPsi$ suppression\dotfill\\}
\item{Summary and outlook\dotfill\\}
\end{enumerate}

\section{Introduction 
\label{sec:Intro}}

It is very well recognized that the dynamics of electrons in a molecular system 
can be completely separated from the dynamics of the 
nucleons in the nuclear core, because 
the ratio of the typical energy scales of a molecular 
system ($\sim 10$ eV) to a nucleus ($\sim 10$ MeV) is as much 
as $10^{-6}$.  One is naturally led to ask whether the dynamics of nucleons in 
a nucleus can also be described independent of the underlying 
degrees of freedom, namely the quarks and gluons.  
However, in this case the ratio of these energy scales is at most 
of order $10^{-1}$. 

One of the most exciting topics in nuclear physics is thus to study how the 
hadron properties are modified by the nuclear environment and how 
such modifications affect the properties of nuclei. Since nucleons and 
mesons are 
made of quarks, anti-quarks and gluons, one expects their internal 
structure to change when placed inside nuclear matter 
or atomic nuclei~\cite{BR}. 
There is little doubt that, at sufficiently high nuclear 
density and/or temperature, 
quarks and gluons are the correct degrees of freedom. 
By contrast, the general success 
of conventional nuclear physics (with {\em effective} interactions) 
indicates that nucleons and mesons provide 
a good starting point for describing  
a nucleus at low energy. Therefore, a 
consistent nuclear theory describing the transition from 
nucleon and meson degrees of freedom to 
quarks and gluons is truly required to describe nuclei 
and nuclear matter over a wide range of 
density and temperature.  
Of course, theoretically, lattice QCD simulations may eventually give  
reliable information on the density and temperature dependence of hadron 
properties in nuclear medium.  
However, current simulations have mainly been performed  
for finite temperature systems with zero baryon density~\cite{LAT}, 
which is very far from what is needed for the description of a finite nucleus.  
As a first step in this direction, it is therefore of great interest 
necessary to build a 
nuclear model which incorporates the internal quark and gluon degrees 
of freedom of hadrons themselves. 

We know that explicit quark degrees of freedom 
are certainly necessary to understand 
deep-inelastic scattering (DIS) at momentum transfer of several GeV.  
In particular, the nuclear EMC effect~\cite{EMC} has suggested that it is 
vital to include some effects involving dynamics beyond the conventional 
nucleon-meson treatment of nuclear physics to explain the whole EMC 
effect~\cite{MILL-EMC,SMITH}. Furthermore, 
the search for evidence of some modification of nucleon properties 
in medium has recently been extended to the electromagnetic form factors 
of nucleon in polarized (${\vec e},e^\prime{\vec p}$) scattering experiments 
on $^{16}$O and $^4$He nuclei at MAMI and JLab~\cite{POLE}. 
These experiments observed 
the double ratio of proton-recoil polarization-transfer coefficients 
in the scattering off nuclei with respect to the 
elastic $^1$He(${\vec e},e^\prime{\vec p}$) reaction at momentum 
transfers of several 100 MeV to a few 
GeV.  It strongly hints at the need to include medium modifications 
of the proton electromagnetic form factors. 

Some early attempts to construct a model which bridges the gap 
between conventional nuclear physics 
and the quark-gluon picture were made in the mid 80's. 
Using a covariant nontopological soliton model of the nucleon, 
Celenza {\it et al.}~\cite{CEL} discussed the effect of medium modification of 
the nucleon structure on its size to understand the nuclear EMC effect. 
They introduced a coupling between quarks and a scalar 
field generated by the nuclear 
environment, and predicted that the size of nucleon in a nucleus becomes larger 
than that of free nucleon. The increase in radius, of about 
15\% in an iron nucleus (Fe), was able  
to reproduce the observed nuclear structure functions. 
Stimulated by Celenza {\it et al.}~\cite{CEL}, 
Wong investigated the effect of nuclear 
mean fields on quarks in a nucleon using the MIT bag model~\cite{WON}.  
He pointed out that an attractive scalar potential makes quarks more 
relativistic by decreasing their masses in nuclear medium.  J{\"a}ndel and 
Peters also studied the change of the in-medium nucleon 
in size within the framework of the Friedberg-Lee nontopological-soliton 
bag model~\cite{JAN}. They reported that 
the swelling of the nucleon is about 30\% at normal nuclear matter density. 

In 1988, Guichon~\cite{GUI-1} developed a novel model for nuclear matter, 
in which quarks in non-overlapping nucleon bags interact 
{\em self-consistently} with (structureless) 
isoscalar-scalar ($\sigma$) and isoscalar-vector 
($\omega$) mesons in the mean-field approximation (MFA). 
The mesons couple not to point-like nucleons but directly to confined quarks. 
He discussed a novel, possible mechanism for nuclear matter saturation in terms of quark degrees of freedom.  
(It should be noted here that Frederico {\it et al.}~\cite{FRE} 
studied a similar model for nuclear 
matter by using a harmonic oscillator potential for the 
quark confinement and discussed the dependence of the 
$\sigma$-nucleon coupling constant on nuclear density.)  After that work, 
the model was studied further by Fleck {\it et al.}~\cite{FLE} and  
refined and extended by  
Saito and Thomas~\cite{SAI-1} in the 90's -- 
the model is called the quark-meson coupling (QMC) model. 
(We also note that in the early 90's Banerjee~\cite{BAN} and Naar and 
Birse~\cite{NAA} studied nucleon 
properties in nuclear matter using the color-dielectric model, 
and that Mishra {\it et al.}~\cite{MIS}  
investigated a nonlocal $\sigma$-$\omega$ model in the 
relativistic Hartree approximation, 
including short distance vertex form factors.) 

The QMC model may be viewed as an extension of 
Quantum Hadrodynamics (QHD)~\cite{QHD}. 
In studies of infinite nuclear matter, as a result of the quark-$\sigma$ 
scalar coupling, the internal structure of the nucleon is modified with respect to the free case. In 
particular, an attractive force associated with the $\sigma$ 
meson exchange decreases the quark mass in matter, 
which leads to an enhancement of the small component of the confined quark 
wave function.  Because of this enhancement, 
the quark scalar density in a nucleon, which is itself the source of 
the $\sigma$ field, is reduced in matter 
compared with that in the free case.  
This can provide a new mechanism for the saturation of nuclear matter,  
where the quark structure of the nucleon plays a vital role~\cite{GUI-1}.  
The model can give a satisfactory description of the bulk properties of 
(symmetric) nuclear matter.  
Of particular interest is the fact that the decrease of the quark scalar 
density, depending on the density of nuclear matter, 
can provide a lower value of the incompressibility of 
nuclear matter than those obtained in approaches based on point-like 
nucleons such as QHD. This is a significant 
improvement on QHD at the same level of sophistication. 
A remarkable feature of this picture is that specific details 
of the model are mostly irrelevant: 
for example, the results do not depend on whether or not one uses  
the MIT bag model, all that is important is the confined 
quarks are treated relativistically. It also does not matter whether or not 
the sources of the repulsive and attractive forces are generated by 
the $\omega$ and $\sigma$, respectively, all that is needed is 
that they have respectively Lorentz-vector and scalar character. 

The original version of the QMC model consists of the 
nucleon, $\sigma$ and $\omega$ mesons. 
It is easy to include the (isovector-vector) $\rho$ meson, which is vital to produce the correct symmetry 
energy, in addition to the usual ingredients~\cite{SAI-1}.  It is also possible to incorporate the (isovector-scalar) 
$\delta$ meson into the model.  
Such a model allows us to study the effect of charge symmetry breaking 
in nuclei~\cite{SAI-4}.  In those cases, all the mesons are assumed to be structureless. This version is sometimes 
called the QMC-I model. However, it is true that the mesons are also built of quarks and anti-quarks and that 
they may change their character in matter.  
To incorporate the effect of meson structure, one can suppose that the vector mesons are 
again described by a relativistic quark model (like a bag) with {\it common} scalar and vector mean-fields in matter. 
In this case, the vector meson mass in matter will also depend on 
the scalar mean-field. The $\sigma$ meson itself 
is, however, not readily represented by a simple quark model, 
because it couples strongly to the pseudoscalar 
($2\pi$) channel and thus a direct treatment of chiral symmetry 
in medium is important.  If, however, we choose 
an appropriate parametrization of the $\sigma$ mass in matter, 
it is possible to construct a model in 
which the structure effects of both the nucleon and the mesons are 
included~\cite{SAI-3}.  This is called the QMC-II model. 
QMC-I predicts a non-linear coupling between the nucleon 
and the $\sigma$ meson and linear couplings of the 
vector mesons to the nucleon, while QMC-II yields many interaction terms 
because of the internal structure 
of the nucleon and mesons~\cite{SAI-3}.  Extending this idea, 
one can find a novel, scaling mass formula for 
various hadrons in matter (see section~\ref{subsubsec:scaling}).

In general, an effective field theory at low energy will contain an infinite number of interaction terms, which 
incorporate the compositeness of hadrons~\cite{compkt}.  It is then expected to involve numerous couplings which may be 
nonrenormalizable. Manohar and Georgi~\cite{MAN} have proposed a systematic 
way to manage to organize such complicated, effective field theories, 
which is called {\it naive dimensional analysis} (NDA). 
NDA gives rules for assigning a coefficient of the appropriate size to any 
interaction term in an effective Lagrangian. After extracting the 
dimensional factors and some appropriate counting factors using NDA, 
the remaining dimensionless coefficients 
are all assumed to be of order unity.  This is the so-called {\it naturalness} assumption. If naturalness 
is valid, we can control the effective Lagrangian, at least 
at the tree level. As seen in Ref.~\cite{SAI-5}, 
QMC-I provides remarkably {\it natural} coupling constants, while 
in QMC-II the coupling constants are 
almost all {\it natural}.  Therefore, the QMC model itself can be regarded as a {\it natural} effective field 
theory for nuclei. 

One can investigate the relationship between QMC and QHD~\cite{MUL,SAI-2}. 
Carrying out a re-definition of the scalar field in matter, 
it is possible to transform the QMC picture 
into a QHD-type model with a non-linear scalar potential 
(if we ignore the change in the internal quark wave function of the bound  
nucleon). Some potentials generated from the QMC model can be compared 
with those used in relativistic mean-field 
(RMF) models.  The QMC model always predicts positive coefficients for 
the cubic and quartic terms in the non-linear 
scalar potential~\cite{SAI-3,SAI-2}, while in the RMF 
model those coefficients are parameterized phenomenologically 
and thus sometimes become negative.  From the point of 
view of field theory, the coefficient of 
quartic term should be positive. 

It is of interest to compare the QMC model with sophisticated 
versions of RMF models: for example, 
the recent version of QHD which was proposed by Furnstahl, 
Serot, Tang and Walecka~\cite{FUR-1}.  
This model is constructed in terms of nucleons, pions and 
the low-lying non-Goldstone bosons, and  
chiral symmetry is realized nonlinearly with a light 
scalar ($\sigma$) meson included as a chiral singlet 
to describe the mid-range nucleon-nucleon attraction. 
This has a total of 16 coupling constants and they 
are almost {\it natural}.  
In this case, the coupling constants were determined so as to fit 
measured ground-state observables of several nuclei by solving the model 
equations for the nuclei simultaneously and minimizing the difference 
between the measured and calculated quantities using a 
nonlinear least-squares adjustment algorithm.  
Therefore, the coupling constants were fixed entirely 
phenomenologically. Such a phenomenological determination 
of the coupling constants is also typical of the ordinary RMF approach.
In contrast, the QMC model has basically three coupling constants, which are 
determined to fit the saturation properties of nuclear matter, 
whereas the other coupling constants are 
automatically generated through a model for the structure of the hadrons. 
Therefore, the physical meaning of the coupling constants is quite clear.  
In the QMC model, the meson masses ($\omega$, $\rho$, etc.) decrease 
in matter. However, in some RMF models, since the coupling constants 
were chosen phenomenologically, particular  
parameter sets lead to an increase of the effective meson masses 
in matter, which seems unlikely from the highlight 
of recent discussions~\cite{QNP2002,SUZ}. 
It should be noted that the chiral symmetry and its restoration 
have been recently discussed in the QMC model 
and that the pion cloud effect is studied in detail~\cite{DEL}.  

Furthermore, it is very remarkable that the QMC model can provide 
a many-body effective Hamiltonian in non-relativistic 
approximation~\cite{GUI-2}. It naturally leads to the appearance of 
many-body forces. In the conventional nuclear physics, 
the Skyrme effective forces are 
very popular and often used to estimate various properties of finite nuclei. 
However, there is no satisfactory interpretation of the parameters appearing 
in the Skyrme forces. Comparing the many-body Hamiltonian generated by 
the QMC model in the zero-range limit with that of the Skyrme effective 
forces, one can obtain a remarkable agreement between 
the Skyrme force and the effective interaction corresponding 
to the QMC model. It allows us to recognize that, indeed,  
the response of nucleon internal structure to the nuclear medium 
does play a vital role in nuclear structure. 

There have been a lot of interesting applications to the properties 
of infinite nuclear matter and finite nuclei. 
It is possible to extend the QMC model of infinite nuclear matter 
to finite nuclei if we assume 
that the meson fields do not vary rapidly across the interior of 
the nucleon in a nucleus and that the nucleon is not moving too rapidly 
-- i.e., the Born-Oppenheimer  
approximation~\cite{GUI-3,BLU}.  
It is, for example, possible to study the properties of 
strange, charm, and bottom hypernuclei 
(see section~\ref{subsec:scbA}), as well as the   
bound states of various mesons and nuclei   
(see section~\ref{subsec:mesonA}).
In this way we can investigate the properties of flavored hadrons in  
medium, as well as the partial restoration of chiral symmetry  
for heavy hadrons in a nuclear medium, where light quarks play 
crucial roles in those hadrons.
We should emphasize again that, we do not have to determine various 
coupling constants for many hadrons in the QMC model.
They are automatically determined at the quark level and this  
enables the QMC model to handle various hadrons in a systematic 
and unified manner based on the quark model.
This strengthens the predictive power of the model enormously, 
once it has been calibrated using  
successfully using the available experimental data. 

Furthermore, the QMC model predicts the variation of the nucleon form 
factors in nuclear matter (see section~\ref{subsec:formf}), 
which will certainly effect the 
analysis of electron scattering off nuclei~\cite{POLE,BOD}, 
including the polarization-transfer 
coefficients discussed at the top of this section. 
Recent experimental analysis~\cite{POLE} of results from Mainz and 
Jefferson Laboratory (JLab)
do indeed seem to support such  a variation of nucleon form factors 
in nuclei. The modification of nucleon structure in the nuclear medium  
is also expected to modify nuclear 
structure functions~\cite{Fernando}  
(the nuclear EMC effect), which are measured by 
scattering with high-energy charged-leptons and/or neutrinos.
We study such possibilities in section~\ref{subsec:duality} 
based on empirically observed, Bloom-Gilman (quark-hadron) 
duality~\cite{BG}, which bridges between the low energy nucleon form factors 
with high energy deep-inelastic scattering nuclear structure functions.

The discrepancy between the calculated binding energy differences
of mirror nuclei and those measured is a long-standing problem 
in nuclear physics, known as the
Okamoto-Nolen-Schiffer (ONS) anomaly~\cite{ONSREV}. 
It is recognized that the ONS anomaly may not be understood by only 
the traditional approach, and conventional 
nuclear contributions to the anomaly are thought 
to be at the few percent level of the experimental findings. 
The effects of charge symmetry breaking (CSB) in the 
nuclear force, especially $\rho$-$\omega$ mixing, seemed 
to reduce this discrepancy~\cite{BLU-2}.  
However, the importance of the momentum dependence 
of the mixing amplitude has been pointed out using 
several different models~\cite{MOM,MORI}. It suggests 
that the $\rho$-$\omega$ mixing amplitude at space-like 
momenta is quite different from the value at the $\omega$ pole and 
implies that the nucleon-nucleon (NN) potential 
given by the off-shell amplitude of the $\rho$-$\omega$ mixing is very far from the ones used in the successful 
phenomenology. Thus, it may be difficult to explain the ONS anomaly by only the meson-mixing potentials. 
When the mass difference between $u$ and $d$ quarks is taken into account,  
the nucleon substructure may produce the difference between the meson couplings to proton and neutron, that leads to 
CSB in the NN interaction.  It may be possible to extend the QMC model to study such 
CSB in nuclei~\cite{SAI-4,TSU-1}. The change of the form factors due to the substructure effect of the nucleon 
may also give a previously unknown correction to the extraction of the 
Cabibbo-Kobayashi-Maskawa (CKM) matrix element, $V_{ud}$, from super-allowed Fermi $\beta$-decay~\cite{CKM,SAI-6}. 
To draw more definite conclusions, there is, however, 
a need for further investigation. 

In a recent helicity analysis of subthreshold $\rho^0$ production on $^2$H, $^3$He and $^{12}$C at low photo-production 
energies, the results are indicative of a large 
longitudinal $\rho^0$ polarization and this signature is used to 
extract in-medium $\rho^0_L$ invariant mass distributions for 
all three nuclei~\cite{HUB}. 
Then, the $^2$H and $^3$He data distributions support the role of $N^*$(1520) excitation in shaping the in-medium 
$\rho^0_L$ invariant mass distribution, while the $^{12}$C 
distributions are consistent with quasi-free $\rho^0_L$ 
production and the data support an in-medium modification 
of the $\rho^0_L$ invariant mass distribution. 
The in-medium mass of the $\rho$ meson predicted by the QMC 
model is consistent with the observed data~\cite{HUB,SAI-7}. 
Two interesting experiments concerning the modification of the $\omega$ 
mass in nuclear matter have also been performed very recently at 
the ELSA tagged photon facility (the CBELSA/TAPS 
collaboration)~\cite{elsa} and KEK~\cite{kek}. 
Using the Crystal Barrel/TAPS experiment at ELSA, the in-medium 
modification of $\omega$ meson was studied via 
the reaction $\gamma + A \to 
\omega + X \to \pi^0 \gamma + X^\prime$, and results obtained for 
Nb were compared to a reference measurement on 
a liquid hydrogen target. 
While for recoiling, long-lived mesons ($\pi^0$, $\eta$ and $\eta^\prime$), 
which decay outside of the nucleus, 
a difference in the lineshape for the two data samples was 
not observed, they found a significant enhancement toward 
lower masses for omega mesons with low momenta produced on the Nb 
target.  For momenta less than $500$ MeV/c, they 
have concluded that the in-medium $\omega$ mass is 
about $722$ MeV at $0.6 \rho_0$ (where $\rho_0$ is the normal nuclear 
matter density), which is just the value predicted 
by the QMC model~\cite{SAI-3} (see also 
section~\ref{subsubsec:scaling}). 
At KEK, the invariant mass spectra of $e^+ e^-$ 
pairs produced in 12-GeV proton-induced nuclear reactions were measured
using the KEK Proton-Synchrotron. On the low-mass side of 
the $\omega$-meson peak, a significant enhancement over the 
known hadronic sources has been observed. The mass spectra, 
including the excess, are well reproduced by a model 
that takes into account the density dependence of the vector 
meson mass modification. 

Furthermore, in recent experimental work in relativistic heavy-ion 
collisions, the $J/\psi$ suppression is thought to be a 
promising candidate as a signal of a quark-gluon plasma (QGP) and the 
experimental data show an anomalous result~\cite{RHI}. 
The modification of the $D$ and $\overline{D}$ meson properties 
in nuclear medium, derived within the QMC model, can lead to a large 
$J/\Psi$ absorption cross section~\cite{jpsi}, 
which is required to explain the observed 
$J/\Psi$ suppression in the microscopic, hadronic comover 
dissociation scenario, without assuming any formation of a 
QGP phase (see section~\ref{subsec:jpsi}).  
(In QMC, the mass modification of $J/\Psi$ in nuclear medium is expected to 
be very moderate.)

These facts seem to indicate that the traditional nuclear/hadronic 
physics approach may have its limitations and suggest a need for the study 
of alternate approaches including subhadronic degrees of freedom. 
Exciting a single quark in the nucleon costs 
about 400 MeV. This is not significantly different from the energy 
required to excite a $\Delta$.  It is also the 
same order of magnitude as the scalar and vector potentials required in QHD.  
Furthermore, the $u$ and $d$ quarks are very light compared to 
nucleon and mesons and should be able to respond faster 
to their environment. We know of {\it no} physical argument why this 
response should be ignored. It is, therefore, very important to consider 
and answer the following question: {\it Do quarks plays an important 
role in nuclei and nuclear matter ?}  
Our aim in this review article is to show how 
subnucleonic (subhadronic) degrees of freedom do indeed appear 
in nuclear physics. 

This paper is organized as follows. We first review the foundation of 
the QMC model in section~\ref{sec:foundation}, 
in which the relationship between the QMC model and widespread 
nuclear models is also discussed. It is shown how the 
nucleon mass is modified by the nuclear environment.  
Several modified versions of the QMC model are also reviewed. 
In section~\ref{sec:properties}, various 
properties of infinite nuclear matter and 
finite nuclei are summarized.  We propose a new 
scaling formula for hadron masses in a nuclear medium. 
In particular, we study in detail the properties of strange, 
charm and bottom hypernuclei, which may be observed in forthcoming 
experiments.  A new type of the spin-orbit force as well as the 
effects of Pauli blocking and channel coupling are also discussed. 
We review the effects of nucleon substructure on lepton-nucleus scattering 
in section~\ref{sec:leptonA}. It is very interesting to study nucleon 
form factors in matter and nuclear structure functions because 
lepton-nucleus scattering can directly probe the quark 
substructure of in-medium nucleon. We compare several experimental results with the QMC predictions. In section~\ref{sec:reactions}, we summarize 
important implications predicted by the QMC model for hadronic reactions. 
In the last section, we give a summary and an outlook regarding future work. 

\section{Foundation of the QMC Model \label{sec:foundation}}
\subsection{\it Non-relativisitic treatment 
\label{subsec:nonrelativistic}}
\subsubsection{\it Classical consideration  
\label{subsubsec:classical}}

Following Guichon {\it et al.,} we first treat 
a nucleon in a nucleus classically~\cite{GUI-3}.  
We generally denote a coordinate in the nuclear rest frame (NRF) 
as $(t, \vec r)$, while we define an instantaneous rest frame for a nucleon at each time $t$ 
(IRF), which is denoted with primes $(t^\prime, {\vec r}^{\,\prime})$: 
\begin{eqnarray}
r_L &=& {\vec r}\cdot {\vec v} = r_L^\prime \cosh \xi + t^\prime \sinh \xi , \nonumber \\
{\vec r}_\perp &=& {\vec r}_\perp^{\,\prime} , \label{irf1} \\
t &=& t^\prime \cosh \xi + r_L^\prime \sinh \xi . \nonumber 
\end{eqnarray}
In the NRF, the nucleon follows a classical trajectory, ${\vec R}(t)$, and the instantaneous velocity of the 
nucleon is given by ${\vec v} = d{\vec R}/dt$. In Eq.~(\ref{irf1}), $r_L$ and ${\vec r}_\perp$ are respectively the 
components parallel and transverse to the velocity and $\xi$ is the rapidity defined by 
$\tanh \xi = \mid {\vec v}(t)\mid$.  

Let us suppose that quarks in the nucleon have {\em enough} time to adjust to the local fields in which the nucleon is 
moving~\cite{BLU}.  It is exact if the fields are constant, because the motion of the nucleon has no acceleration. 
We shall examine this approximation for a typical nuclear environment. 
Assume that at $t=0$ the nucleon is at ${\vec R}_0$.  After a short time $t (\ll 1)$, the coordinate of the 
nucleon can be described as
\begin{equation}
{\vec R}(t) = {\vec R}_0 + {\vec v}_0 t + \frac{1}{2} t^2 {\vec \alpha}_0 , \label{Rt}
\end{equation}
where the velocity and acceleration at $t=0$ are respectively denoted 
by ${\vec v}_0$ and ${\vec \alpha}_0 = {\vec F}/M_N = - {\vec \nabla} V/M_N$ 
(with $M_N$ the free nucleon mass and $V$ potential). 
If we take a typical nuclear potential, $V$, as the Woods-Saxon form 
\begin{equation}
V(r) = \frac{V_0}{1+\exp\left[\frac{r-R_A}{a}\right]} , \label{wspot}
\end{equation}
with depth $V_0 \sim -50$ MeV, $a \sim 0.5$ fm (surface thickness is about 2 fm) and radius $R_A \sim 1.2 A^{1/3}$, 
the maximum acceleration occurs at the nuclear surface, and it is  
\begin{equation}
{\vec \alpha}_{max} = \frac{V_0}{4aM_N}{\hat R} , \label{accmax}
\end{equation}
where ${\hat R}$ is a unit vector in the radial direction.  
Therefore, in the IRF, since the coordinates 
$(t, {\vec R}_\perp, R_L)$ are transformed as Eq.~(\ref{irf1}), 
one finds $R_L^\prime = R_L \cosh \xi - 
t \sinh \xi$. Assuming $\mid {\vec v}_0\mid \ll 1$ and 
combining the transverse part, we get 
\begin{equation}
{\vec R}^{\,\prime}(t) \simeq {\vec R}_0^{\,\prime} +  \frac{V_0}{8aM_N} t^2 {\hat R} , \label{vecRpr1}
\end{equation}
which gives 
\begin{equation}
R^\prime - R_0^\prime \equiv ({\vec R}^{\,\prime}(t) - {\vec R}_0^{\,\prime})\cdot{\hat R} = 
\frac{V_0}{8aM_N} t^2 . \label{vecRpr}
\end{equation}
In the worst case, the departure from the initial 
position ${\vec R}_0$, relative to 
the typical size of the nucleon itself ($R_N \sim 1$ fm) is then expressed as 
\begin{equation}
\left| \frac{R^{\prime}(t)- R_0^{\prime}}{R_N} \right| \sim \frac{|V_0|}{8aM_NR_N} t^2 \sim  \frac{t^2}{75}, 
\label{relsize}
\end{equation}
where $t$ is in fm. Thus, as long as the time taken for 
the quark motion to change is less than $\sim 9$ 
fm, the nucleon position in the IRF can be considered as {\em unchanged}. Since the typical time for an adjustment 
in the motion of the quarks is given by the inverse of the typical excitation energy, which is of order 
0.5 fm (corresponding to the $\Delta$ excitation), this is quite safe. 

We begin by constructing an appropriate Lagrangian density in the IRF. 
As a nucleon model we adopt 
the static, spherical MIT bag~\cite{MITBAG}: 
\begin{equation}
{\cal L}_0 = {\overline \psi_q}^{\prime} 
(i\gamma\cdot\partial -m_q) \psi_q^\prime - BV_B , \ \ 
\mbox{for} \mid {\vec u}^{\,\prime} \mid \leq R_B , 
\label{L0}
\end{equation}
with $B$ the bag constant, $V_B$ the bag volume, $R_B$ the bag radius, 
$m_q$ the quark mass and ${\vec u}^{\,\prime}$ the position 
of the quark from the center of the bag in the IRF -- we denote 
as $u^\prime$ the 4-vector: 
$u^\prime = (t^\prime, {\vec u}^{\,\prime}) = (t^\prime, {\vec r}^{\,\prime} - {\vec R}^{\,\prime})$.  
The quark field in the IRF is expressed by $\psi_q^\prime(t^\prime, {\vec u}^{\,\prime})$, which satisfies the 
bag boundary condition at the surface $\mid {\vec u}^{\,\prime} \mid = R_B$: 
$(1+i{\vec \gamma}\cdot {\hat u}^{\,\prime})\psi_q^\prime =0$.  

We incorporate the interaction of the quark with the scalar ($\sigma$) and vector ($\omega$) mean fields, which 
are generated by the surrounding nucleons.  In the NRF, they are functions of 
position: $\sigma({\vec r})$ and $\omega^\mu =(\omega({\vec r}), {\vec 0})$ in MFA.  
Thus, in the IRF, these fields can be expressed by Lorentz transformation
\begin{eqnarray}
\sigma_{I}(t^\prime, {\vec u}^{\,\prime}) &=& \sigma({\vec r}) , \nonumber \\
\omega_I(t^\prime, {\vec u}^{\,\prime}) &=& \omega({\vec r}) \cosh \xi , \label{mf1} \\
{\vec \omega}_I^{\,\prime}(t^\prime, {\vec u}^{\,\prime}) &=&  - \omega({\vec r}) {\hat v} \sinh \xi , \nonumber 
\end{eqnarray}
where the subscript $I$ stands for the IRF.  Thus, the interaction can be written as 
\begin{equation}
{\cal L}_I = g_\sigma^q {\overline \psi_q}^\prime \psi_q^\prime(u^\prime) \sigma_I(u^\prime) - 
g_\omega^q {\overline \psi_q}^\prime \gamma_\mu \psi_q^\prime(u^\prime) \omega_I^\mu(u^\prime) , \ \ 
\mbox{for} \mid {\vec u}^{\,\prime} \mid \leq R_B , 
\label{LI}
\end{equation}
where $g_\sigma^q$ ($g_\omega^q$) is the quark-$\sigma$ ($\omega$) coupling constant. It is also possible to 
include the effect of $\rho$ meson field (see section~\ref{subsubsec:totalH}). 

Now let us construct a Hamiltonian in the IRF. Suppose that at time $t^\prime$ the bag is located at 
${\vec R}^{\,\prime}$ in the IRF, while, in the NRF, it is located at ${\vec R}$ at time $T$.  
Using Eq.~(\ref{irf1}), for 
an arbitrary position ${\vec r}^{\,\prime}$ in the bag (${\vec r}^{\,\prime} = {\vec u}^{\,\prime} 
+ {\vec R}^{\,\prime}$) at the same time $t^\prime$, we find 
\begin{eqnarray}
r_L &=& R_L^\prime \cosh \xi + t^\prime \sinh \xi + u_L^\prime \cosh \xi , \nonumber \\
    &=& R_L + u_L^\prime \cosh \xi , \label{irf3} \\
{\vec r}_\perp &=& {\vec R}_\perp^{\,\prime} + {\vec u}_\perp^{\,\prime} . \nonumber 
\end{eqnarray}
{}From this relation, the $\sigma$ and $\omega$ fields are given by  
\begin{eqnarray}
\sigma_I(t^\prime, {\vec u}^{\,\prime}) &=& \sigma(R_L(T) + u_L^\prime \cosh \xi , 
{\vec R}_\perp^{\,\prime}(T) + {\vec u}_L^{\,\prime}) ,  \label{sigma1} \\
\omega_I^\mu(t^\prime, {\vec u}^{\,\prime}) &=& \eta^\mu \omega(R_L(T) + u_L^\prime \cosh \xi , 
{\vec R}_\perp^{\,\prime}(T) + {\vec u}_L^{\,\prime}) ,  \label{omega1} 
\end{eqnarray}
where $\eta^\mu = (\cosh \xi, -{\hat v}\sinh \xi)$. 

For a while, suppose that the effect of finite size of the nucleon is negligible and that the mean 
fields in the IRF can be approximated by their values at ${\vec R}(T)$ 
in the NRF, for example, $\sigma_I \simeq \sigma({\vec R}(T))$.  As the typical time scale for a change in 
the motion of the quark is $\tau \sim 0.5$ fm, the relative change of the field during this time, 
$\Delta \sigma$, is 
\begin{equation}
\Delta \sigma = |\sigma({\vec R}(T+\tau)) - \sigma({\vec R}(T))|  
= ({\vec v}\cdot{\hat R})\tau \left( \frac{d\sigma}{dR}\right) . 
\label{sigmachange}
\end{equation}
It is then expected that the $\sigma$ field roughly follows the nuclear density, and as long as it is constant 
in the interior of the nucleus, $\Delta \sigma$ almost vanishes.  The variation of the density occurs 
mainly at the surface where it drops to zero from the normal nuclear matter density, $\rho_0$, over a distance 
$d$ of about 2 fm.  Therefore, one can 
estimate $|(d\sigma/dR)/\sigma |$ as approximately $1/d$ at the 
surface.  Furthermore, the factor $({\vec v}\cdot{\hat R})$ 
introduces a factor of $1/3$,   
because $\vec v$ is isotropic.  If we 
take $|{\vec v}| \simeq k_F/M_N \sim 0.36$ ($k_F \simeq 1.7$ fm$^{-1}$), 
the relative change of the $\sigma$ field is then estimated to be 
\begin{equation}
\frac{\Delta \sigma}{\sigma} \sim  \frac{0.5 \times 0.36}{3\times 2} \sim 3\% . 
\label{sigmachange2}
\end{equation}
This value is quite small, implying that the Born-Oppenheimer approximation 
certainly works well in the 
nucleus and that we can treat the position ${\vec R}(T)$ 
as a fixed parameter in 
solving the equation of motion for the quarks in matter. 
Clearly this amounts to neglecting terms of order $v$ 
in the argument of $\sigma$ and $\omega$.  
In order to be consistent we also neglect terms of order 
$v^2$ -- i.e., we replace $u_L^\prime \cosh \xi$ by $u_L^\prime$ 
in the argument. 

\subsubsection{\it Quantization at the quark level  
\label{subsubsec:quantum}}

In the IRF, the interaction Lagrangian density thus becomes 
\begin{equation}
{\cal L}_I = g_\sigma^q {\overline \psi_q}^\prime \psi_q^\prime(u^\prime) \sigma({\vec R} + {\vec u}^{\,\prime}) - 
g_\omega^q {\overline \psi_q}^{\,\prime}(t^\prime, {\vec u}^{\,\prime}) [\gamma_0 \cosh \xi + 
{\vec \gamma} \cdot {\hat v} \sinh \xi] 
\psi_q^\prime(t^\prime, {\vec u}^{\,\prime}) \omega({\vec R} + {\vec u}^{\,\prime}) ,  
\label{LI2}
\end{equation}
and the corresponding Hamiltonian is 
\begin{equation}
H = \int_{V_B} d{\vec u}^{\,\prime} 
{\overline \psi_q}^{\,\prime} [-i {\vec \gamma}\cdot{\vec \nabla} + m_q - 
g_\sigma^q \sigma({\vec R} + {\vec u}^{\,\prime}) 
+ g_\omega^q (\gamma_0 \cosh \xi + {\vec \gamma} \cdot {\hat v} \sinh \xi) \omega({\vec R} + {\vec u}^{\,\prime}) ] 
\psi_q^\prime(t^\prime, {\vec u}^{\,\prime}) + BV_B ,  
\label{H2}
\end{equation}
where the integral is performed within the bag volume $V_B$. 
The momentum operator is then given by 
\begin{equation}
{\vec P} = \int_{V_B} d{\vec u}^{\,\prime} 
\psi_q^{\,\prime \dagger} [-i {\vec \nabla} ] \psi_q^\prime .   
\label{P1}
\end{equation}
Since the mean fields appreciably vary only near the nuclear surface, it makes sense to separate the 
Hamiltonian into two pieces, $H = H_0 + H_1$: 
\begin{equation}
H_0 = \int_{V_B} d{\vec u}^{\,\prime} 
{\overline \psi_q}^{\,\prime} [-i {\vec \gamma}\cdot{\vec \nabla} + m_q - 
g_\sigma^q \sigma({\vec R}) 
+ g_\omega^q (\gamma_0 \cosh \xi + {\vec \gamma} \cdot {\hat v} \sinh \xi) \omega({\vec R}) ] 
\psi_q^\prime(t^\prime, {\vec u}^{\,\prime}) + BV_B ,  
\label{H0}
\end{equation}
\begin{equation}
H_1 = \int_{V_B} d{\vec u}^{\,\prime} 
{\overline \psi_q}^{\,\prime} [- g_\sigma^q (\sigma({\vec R} + {\vec u}^{\,\prime}) - \sigma({\vec R})) 
+ g_\omega^q (\gamma_0 \cosh \xi + {\vec \gamma} \cdot {\hat v} \sinh \xi) 
(\omega({\vec R} + {\vec u}^{\,\prime}) - \omega({\vec R})) ] 
\psi_q^\prime(t^\prime, {\vec u}^{\,\prime})  . 
\label{H11}
\end{equation}
Then, it is possible to consider the latter Hamiltonian $H_1$ as a perturbation. 

We prepare a complete and orthogonal set of eigenfunctions for 
the quark field.  They are denoted by $\phi^\alpha$, 
where $\alpha$ is a collective symbol labeling the quantum numbers: 
\begin{eqnarray}
h\phi^\alpha({\vec u}^{\,\prime}) &\equiv& (-i\gamma^0{\vec \gamma}\cdot{\vec \nabla} + 
m_q^\ast \gamma^0) \phi^\alpha({\vec u}^{\,\prime}) = \frac{\Omega_\alpha}{R_B} \phi^\alpha({\vec u}^{\,\prime}) , 
\label{comp1} \\
(1+i{\vec \gamma}\cdot{\hat u}^{\,\prime}) \phi^\alpha({\vec u}^{\,\prime}) &=& 0 ,  
\ \ \mbox{at} \ |{\vec u}^{\,\prime}| = R_B ,   \label{comp2} \\
\int_{V_B} d{\vec u}^{\,\prime} \phi^{\alpha \dagger} \phi^\beta &=& \delta^{\alpha \beta} ,  \label{comp3} 
\end{eqnarray}
with $m_q^\ast$ a parameter. The lowest, positive eigenfunction is then given by  
\begin{equation}
\phi^{0m}(t^\prime, {\vec u}^{\,\prime}) = \frac{\cal N}{4\pi} 
\left( \begin{array}{c}
j_0(xu^\prime/R_B) \\
i\beta_q {\vec \sigma}\cdot{\hat u}^{\,\prime}j_1(xu^\prime/R_B)
\end{array} \right) \chi_m , 
\label{lowest}
\end{equation}
with $u^\prime=|\vec{u}^{\,\prime}|$ and $\chi_m$ the spin function and 
\begin{eqnarray}
\Omega_0 &=& \sqrt{x^2+(m_q^\ast R_B)^2}, \ \ \beta_q = \sqrt{\frac{\Omega_0 - m_q^\ast R_B}{\Omega_0 + m_q^\ast R_B}} ,  
\label{omega0} \\
{\cal N}^{-2} &=& 2R_B^3 j_0^2(x)[\Omega_0(\Omega_0-1) + m_q^\ast R_B/2]/x^2  , 
\label{norm}
\end{eqnarray}
where $x$ is the eigenvalue for the lowest mode, 
which satisfies the boundary condition at the bag surface, 
$j_0(x) = \beta_q j_1(x)$. 

Using this set, the quark field can be expanded as 
\begin{equation}
\psi_q^\prime(t^\prime, {\vec u}^{\,\prime}) = \sum_\alpha e^{-i{\vec k}\cdot{\vec u}^{\,\prime}} 
\phi^\alpha({\vec u}^{\,\prime}) b_\alpha(t^\prime) , 
\label{qcond}
\end{equation}
with $b_\alpha$ an annihilation operator for the quark and 
${\vec k} = g^q_\omega \omega({\vec R}) {\hat v} \sinh \xi$, which ensures the correct momentum for 
a particle in the vector field. 
Since the quark field satisfies 
\begin{equation}
-i\gamma^0{\vec \gamma}\cdot{\vec \nabla} \psi_q^\prime = -i\gamma^0{\vec \gamma}\cdot{\vec k} \psi_q^\prime
+ \left( \frac{\Omega_\alpha}{R_B} - m_\alpha^\ast \gamma^0 \right) \psi_q^\prime , 
\label{qprime}
\end{equation}
the free Hamiltonian and momentum operators are, respectively 
\begin{eqnarray}
H_0 &=& \sum_\alpha \left( \frac{\Omega_\alpha }{R_B} \right) b_\alpha^\dagger b_\alpha 
- \sum_{\alpha \beta} \langle \alpha | ( g_\sigma^q \sigma({\vec R}) - m_q + m_q^\ast )\gamma _0 | \beta \rangle 
b_\alpha^\dagger b_\alpha + {\hat N}_q g_\omega^q \omega({\vec R}) \cosh \xi + BV_B , \label{H01} \\ 
{\vec P} &=& \sum_{\alpha \beta} \langle \alpha | -i{\vec \nabla} | \beta \rangle 
b_\alpha^\dagger b_\alpha - {\hat N}_q {\vec k} , \label{P2}
\end{eqnarray}
where ${\hat N}_q (= \sum_\alpha b_\alpha^\dagger b_\alpha)$  
is the quark number operator and 
\begin{equation}
\langle \alpha | A | \beta \rangle = \int_{V_B} d{\vec u}^{\,\prime} \phi ^{\alpha \dagger}({\vec u}^{\,\prime}) 
A \phi^\beta({\vec u}^{\,\prime}) . 
\label{expec}
\end{equation}

Choosing an effective quark mass as $m_q^\ast = m_q - g_\sigma^q \sigma({\vec R})$, we find the leading part of 
the energy and momentum operators in the IRF 
\begin{eqnarray}
H_0^I &=& \sum_\alpha \frac{\Omega_\alpha({\vec R})}{R_B} b_\alpha^\dagger b_\alpha 
+ {\hat N}_q g_\omega^q \omega({\vec R}) \cosh \xi + BV_B , \label{H02} \\ 
{\vec P}^I &=& \sum_{\alpha \beta} \langle \alpha | -i{\vec \nabla} | \beta \rangle 
b_\alpha^\dagger b_\alpha - {\hat N}_q  g_\omega^q \omega({\vec R}) {\hat v} \sinh \xi , \label{P3}
\end{eqnarray}
where the frequency $\Omega$ depends 
on ${\vec R}$ because the effective quark mass 
varies, depending on position through the $\sigma$ field. 
If we quantize the $b_\alpha$ in the usual way, 
the unperturbed Hamiltonian is diagonalized by number 
states $| N_\alpha , N_\beta , \cdots \rangle$ with 
$N_\alpha$ the eigenvalue of the number 
operator $b_\alpha^\dagger b_\alpha$ for the mode $\{\alpha\}$.  
Since we suppose that the nucleon should be described 
in terms of the three quarks in the lowest mode 
($\alpha =0$) and should remain in that configuration 
as ${\vec R}$ changes in matter, the gradient term 
in the momentum operator becomes zero because of 
parity conservation. Then, we obtain the energy and 
momentum in the IRF as
\begin{eqnarray}
E_0^I &=& M_N^\ast({\vec R}) + 3 g_\omega^q \omega({\vec R}) \cosh \xi , \label{H03} \\ 
{\vec P}^I &=& - 3 g_\omega^q \omega({\vec R}) {\hat v} \sinh \xi , \label{P41}
\end{eqnarray}
with the effective nucleon mass 
\begin{equation}
M_N^\ast({\vec R}) = \frac{3\Omega_0({\vec R})}{R_B} + BV_B . 
\label{nmass}
\end{equation}

Since we are treating the corrections to leading order in the velocity, 
the Lorentz transformation simply 
provides the expressions for the leading terms in the energy and momentum in the NRF as 
\begin{eqnarray}
E_0 &=& M_N^\ast({\vec R}) \cosh \xi + 3 g_\omega^q \omega({\vec R}) , \label{E01} \\ 
{\vec P} &=& M_N^\ast({\vec R}) {\vec v} \sinh \xi . \label{P4}
\end{eqnarray}
This implies 
\begin{equation}
E_0 = \sqrt{M_N^{\ast 2}({\vec R}) + {\vec P}^2} +  3 g_\omega^q \omega({\vec R}) . 
\label{E02}
\end{equation}
As usual, we here parameterize the sum of the center of mass (c.m.) correction and gluon fluctuation 
corrections to the bag energy by the familiar form, $-z_0/R_B$, where $z_0$ is assumed to be {\em independent} of 
the nuclear density (see the next subsection). Then, the effective nucleon mass in matter takes the form 
\begin{equation}
M_N^\ast({\vec R}) = \frac{3\Omega_0({\vec R})-z_0}{R_B} + BV_B , 
\label{EMass}
\end{equation}
and the equilibrium condition is required as 
\begin{equation}
\frac{dM_N^\ast({\vec R})}{dR_B} = 0 .  
\label{EMass2}
\end{equation}
This is again justified by the Born-Oppenheimer approximation, according 
to which the internal structure 
of the nucleon has enough time to adjust the varying 
external field so as to stay in its ground 
state.  We emphasize that the effective nucleon mass 
depends on position only through the scalar 
field. This result does not depend on the specific model of the nucleon, but 
is correct in {\em any} quark model in which the nucleon 
contains relativistic quarks which are linearly coupled to  
Lorentz-scalar and vector fields. 

\subsubsection{\it Correction due to the center of mass motion 
\label{subsubsec:cmcorrection}}

Let us now consider the dependence of the c.m. correction  
on the external fields~\cite{GUI-3}. 
It turns out that this correction is only weakly dependent 
on the external field strength for the 
densities of interest.  We estimate this c.m. correction 
for an external scalar field (note that 
a linearly coupled vector field does not 
alter the quark structure of the nucleon).  Our aim here is 
not to obtain an exact expression for the c.m. correction to 
the MIT bag, but only to look for its dependence 
on the external field.  Therefore, it is reasonable 
that we consider a model where the quark mass grows quadratically 
with the distance from the center of the bag, instead 
of the sharp boundary condition for the MIT bag. 
In this simple estimate, we further assume that the 
quark number is conserved, which allows us to 
formulate the problem in the first quantized form. 

We now consider a model defined by the following first-quantized Hamiltonian: 
\begin{equation}
H_{h.o.} = \sum_{i}^N \gamma_0(i) ({\vec \gamma}(i)\cdot{\vec p}_i + m({\vec r}_i)) , 
\ \ \ {\vec p}_i = -i {\vec \nabla}_i , 
\label{HOH}
\end{equation}
with the quark mass $m({\vec r}) = m_q^\ast + Kr^2$ ($K$ the strength of the harmonic oscillator potential), 
where $m_q^\ast (= m_q - g_\sigma^q \sigma)$ is the 
effective quark mass in the presence of the external scalar field.  

Using the quark number $N$, we can define the intrinsic coordinates $({\vec \rho}, {\vec \pi})$ as 
\begin{eqnarray}
{\vec \pi}_i &=& {\vec p}_i - \frac{{\vec P}}{N}, \ \ \ {\vec P} = \sum_i {\vec p}_i, \ \ \ 
\sum_i {\vec \pi}_i = 0 , \label{co1} \\ 
{\vec \rho}_i &=& {\vec r}_i - {\vec R}, \ \ \ {\vec R} = \frac{1}{N}\sum_i {\vec r}_i, \ \ \ 
\sum_i {\vec \rho}_i = 0 . \label{co2} 
\end{eqnarray}
Then we can separate the Hamiltonian into two pieces; 
\begin{eqnarray}
H_{h.o.} &=& H_{intr.} + H_{c.m.} , \label{Hho1} \\ 
H_{intr.} &=& \sum_i  \gamma_0(i) ({\vec \gamma}(i)\cdot{\vec \pi}_i + m({\vec \rho}_i)) ,  \label{Hintr} \\
H_{c.m.} &=& \frac{\vec P}{N}\cdot\sum_i \gamma_0(i) {\vec \gamma}(i) + \sum_i  \gamma_0(i) (m({\vec r}_i)
- m({\vec \rho}_i)) . \label{Hcm} 
\end{eqnarray}
Here we note that the intrinsic Hamiltonian, $H_{intr.}$, 
commutes with the total momentum, ${\vec P}$, and the 
coordinate of the center of mass, ${\vec R}$, and 
that $[H_{c.m.}, {\vec R}] = -i \sum_i \gamma_0(i) {\vec \gamma}(i)$, 
where the r.h.s. can be identified with the time derivative of ${\vec R}$.  

Now all we know are the eigenstates of $H_{h.o.}$, but we can consider $H_{c.m.}$ as a correction of order 
$1/N$ with respect to the leading term in the total energy. Therefore, we calculate its effect in first 
order perturbation method, that is 
\begin{equation}
E_{intr.} = E_{h.o.} - E_{c.m.} = E_{h.o.} - \langle A | H_{c.m.} | A \rangle , 
\label{Eintr}
\end{equation}
where $| A \rangle$ is the eigenstate with its energy $E_A$. Thus, the energy due to the c.m. 
correction is 
\begin{equation}
E_{c.m.} = \langle A | \frac{\vec P}{N}\cdot \sum_i \gamma_0(i){\vec \gamma}(i) + 
2K{\vec R}\cdot \sum_i \gamma_0(i){\vec r}_i - K {\vec R}^2 \sum_i \gamma_0(i) | A \rangle .  
\label{Ecm}
\end{equation}
Let $| \alpha \rangle$ be the one body solutions, that is 
\begin{equation}
\gamma_0 ( {\vec \gamma}\cdot{\vec p} 
+ m({\vec r}) ) \phi_\alpha = \Omega_\alpha \phi_\alpha . \label{equa} 
\end{equation}
%
Supposing that all the quarks are in the lowest mode, we then find 
\begin{equation}
E_{c.m.} = \Omega_0 - m_q^\ast \langle 0 | \gamma_0 | 0 \rangle + K \langle 0 | \gamma_0r^2 | 0 \rangle 
- \langle 0 | \gamma_0 | 0 \rangle \langle 0 | r^2 | 0 \rangle + {\cal O}(1/N) . \label{Ecmf}
\end{equation}

To determine the wave function we solve Eq.~(\ref{equa}) numerically and adjust the potential strength $K$ 
so as to give $\Omega_0 = 2.04$, i.e., the lowest energy of the free bag (in units of the bag radius $R_B = 1$ fm).  
We then find $K= 1.74$. 
We can now compute $E_{c.m.}$ numerically (see Eq.~(\ref{Ecmf})) as a function of $m_q^\ast$. We find that 
in the region of $-1.5 < m_q^\ast <0$ the value of $E_{c.m.}$ is almost constant and its dependence on the 
external scalar field is very weak.  Thus, for practical purposes, it is a very reasonable approximation to 
ignore the dependence of $E_{c.m.}$ on the external field~\cite{GUI-3}. 

\subsubsection{\it Correction due to $H_1$ and Thomas precession  
\label{subsubsec:h1}}

Here we return to the Hamiltonian Eq.~(\ref{H2}) 
and estimate the perturbation term, $H_1$ (Eq.~(\ref{H11})). 
Expanding the scalar and vector 
fields, $\sigma({\vec R}+{\vec u}^{\,\prime})$ and $\omega({\vec R}+{\vec u}^{\,\prime})$, in powers of 
${\vec u}^{\,\prime}$ 
\begin{eqnarray}
\sigma({\vec R}+{\vec u}^{\,\prime}) &=& \sigma({\vec R}) + {\vec u}^{\,\prime}\cdot{\vec \nabla}_R
\sigma({\vec R}) + \cdots , \label{sigma1f} \\ 
\omega({\vec R}+{\vec u}^{\,\prime}) &=& \omega({\vec R}) + {\vec u}^{\,\prime}\cdot{\vec \nabla}_R
\omega({\vec R}) + \cdots , \label{omega1f} 
\end{eqnarray}
and computing the effect up to first order, $H_1$ is given by 
\begin{equation}
H_1 = \int_{V_B} d{\vec u}^{\,\prime} 
{\overline \psi_q}^{\,\prime} {\vec \gamma} \psi_q^\prime(t^\prime, {\vec u}^{\,\prime}) \cdot {\hat v} 
g_\omega^q \sinh \xi {\vec u}^{\,\prime}\cdot{\vec \nabla}_R \omega({\vec R}) ,   \label{H12}
\end{equation}
where the scalar term and the vector term including $\gamma_0$ vanish because of parity conservation. Then, the 
perturbation can be rewritten as
\begin{equation}
\langle (0)^3 | H_1 | (0)^3 \rangle = g_\omega^q \sum_{\alpha \beta} 
\langle (0)^3 | b_\alpha^\dagger b_\beta | (0)^3 \rangle 
\langle \alpha | \gamma_0 {\vec \gamma}\cdot{\hat v} {\vec u}^{\,\prime} \sinh \xi |\beta \rangle 
\cdot{\vec \nabla}_R \omega({\vec R}) ,   \label{H13}
\end{equation}
where we set $\{\alpha \} = \{0, m_\alpha \}$ with $m_\alpha$ the spin projection of the quark in the lowest 
mode $\{0\}$. Then, we find 
\begin{eqnarray}
\langle \alpha | \gamma_0 {\vec \gamma}\cdot{\hat v} {\vec u}^{\,\prime} |\beta \rangle 
&=& \int d{\vec u}^{\,\prime} \phi^{0m_\alpha \ast} \gamma_0 ({\vec \gamma}\cdot{\hat v}) {\vec u}^{\,\prime} 
\phi^{0m_\beta} ,   \nonumber \\
&=& -I({\vec R}) \langle m_\alpha | \frac{{\vec \sigma}}{2} | m_\beta \rangle \times {\hat v} , \label{I1} 
\end{eqnarray}
with 
\begin{equation}
I({\vec R}) = \frac{4}{3} \int du^\prime u^{\prime 3} f(u^\prime)g(u^\prime) , \label{I2} 
\end{equation}
where $u^\prime=|\vec{u}^{\,\prime}|$, and 
$f$ and $g$ are respectively the upper and 
lower components of the quark wave function and the 
factor $4/3$ comes from the angular part of the integral 
with respect to ${\vec u}^{\,\prime}$. 
If we use the wave function of the MIT bag, we find 
\begin{equation}
I({\vec R}) = \frac{R_B}{3} \left[ \frac{4\Omega_0+2m_q^\ast R_B -3}{2\Omega_0(\Omega_0-1) +m_q^\ast R_B} 
\right] . \label{I3} 
\end{equation}

This integral, $I({\vec R})$, depends on ${\vec R}$ through the implicit dependences of the bag radius and 
eigenvalue on the local scalar field.  Its value can be related to 
the nucleon magnetic moment ${\vec \mu}$:
\begin{equation}
{\vec \mu} = \frac{1}{2} \int d{\vec r} \, {\vec r} \times \psi_q^\dagger(r) {\vec \alpha} \psi_q(r) .  \label{mag} 
\end{equation}
The values of $I$ in the free case, $I_0$, may then 
be expressed in terms of the nucleon isoscalar magnetic moment: 
$I_0 = 3\mu_s/M_N$ with $\mu_s=\mu_p+\mu_n$ 
and $\mu_p=2.79$ and $\mu_n=-1.91$ the experimental 
values (all in units of $\mu_N$). 
Using this value, $H_1$ is rewritten by 
\begin{equation}
\langle (0)^3 | H_1 | (0)^3 \rangle = \mu_s \left[ \frac{I({\vec R})M_N^\ast({\vec R})}{I_0M_N} \right] 
\frac{1}{M_N^{\ast 2}({\vec R}) R} \left[ \frac{d}{dR} 3g_\omega^q \omega({\vec R}) \right] {\vec S}\cdot{\vec L} ,   
\label{H14}
\end{equation}
with ${\vec S}$ the nucleon spin operator and ${\vec L}$ its angular momentum. 

This spin-orbit interaction is nothing but the interaction of 
the magnetic moment of the nucleon with the 
"magnetic" field of the $\omega$ meson seen from the rest frame of the nucleon. This induces a rotation of the spin 
as a function of time.  However, even if $\mu_s$ were zero, the spin would nevertheless rotate because of Thomas 
precession~\cite{JAC}, which is a relativistic effect independent of the structure
\begin{equation}
H_{prec} = -\frac{1}{2} {\vec v}\times \frac{d{\vec v}}{dt} \cdot {\vec S} ,  \label{Thomas}
\end{equation}
where the acceleration up to lowest order in the velocity is given by 
\begin{equation}
\frac{d{\vec v}}{dt} = - \frac{1}{M_N^\ast({\vec R})} {\vec \nabla} [M_N^\ast({\vec R}) + 
3g_\omega^q \omega({\vec R}) ] .  \label{accele}
\end{equation}
Combining this precession and the effect of $H_1$, we finally get the total spin orbit interaction to first 
order in the velocity 
\begin{equation}
H_1 + H_{prec} = V_{s.o.}({\vec R}) {\vec S} \cdot {\vec L} ,  \label{s.o.}
\end{equation}
where 
\begin{equation}
V_{s.o.}({\vec R}) = - \frac{1}{2M_N^{\ast 2}({\vec R})R} \left[ 
\left( \frac{d}{dR} M_N^\ast({\vec R}) \right) + (1-2\mu_s \eta_s({\vec R})) 
\left( \frac{d}{dR} 3g_\omega^q \omega({\vec R}) \right) \right] ,   
\label{s.o.2}
\end{equation}
and 
\begin{equation}
\eta_s({\vec R}) = \frac{I({\vec R})M_N^\ast({\vec R})}{I_0M_N} .    
\label{s.o.3}
\end{equation}
\subsubsection{\it Total Hamiltonian in the mean fields 
\label{subsubsec:totalH}}

It is not difficult to introduce the neutral $\rho$ meson field.  In the mean field approximation, only the 
neutral component is active. The interaction Lagrangian density may 
then be given by 
\begin{equation}
{\cal L}_I^\rho = - g_\rho^q {\overline \psi_q}^{\prime} 
\gamma_\mu \frac{\tau_3}{2} \psi_q^\prime(u^\prime) 
\rho_{I}^\mu(u^\prime) ,  \label{rho}
\end{equation}
where $\rho_{I}^\mu$ is the neutral $\rho$-meson field in 
the IRF and $\tau_3$ is the the third component of Pauli matrix 
acting on the quark field.  If the mean field value of the time component of the $\rho$ field is denoted by 
$b({\vec R})$ in the NRF, one can transpose the results for the $\omega$ field.  The only difference between 
the two fields comes from trivial isospin factors 
\begin{equation}
3g_\omega^q \rightarrow g_\rho^q \frac{\tau_3^N}{2} , \ \ \ \mu_s \rightarrow \mu_v = \mu_p - \mu_n ,     
\label{trans}
\end{equation}
where $\tau_3^N/2$ is the nucleon isospin operator. 

Because, in the NRF, 
$E = \cosh \xi E^{I} + \sinh \xi P_L^{I}$ and $P_L = \cosh \xi P_L^{I} + \sinh \xi E^{I}$, 
one finally finds that the NRF energy-momentum of the nucleon moving in the meson fields is 
\begin{eqnarray}
E &=& M_N^\ast({\vec R}) \cosh \xi + V({\vec R}) ,   \label{em1} \\
{\vec P} &=& M_N^\ast({\vec R}) {\hat v} \sinh \xi , \label{em2} 
\end{eqnarray}
with 
\begin{eqnarray}
V({\vec R}) &=& V_c({\vec R}) + V_{s.o.}({\vec R}) {\vec S}\cdot{\vec L},   \label{v1} \\
V_c({\vec R}) &=& 3g_\omega^q \omega({\vec R}) + g_\rho^q \frac{\tau_3^N}{2} b({\vec R})  , \label{v2} \\
V_{s.o.} &=& -\frac{1}{2M_N^{\ast 2}({\vec R}) R} \left( \Delta_\sigma +(1-2\mu_s\eta_s({\vec R})) \Delta_\omega 
+(1-2\mu_v\eta_v({\vec R})) \frac{\tau_3^N}{2} \Delta_\rho \right) , \label{v3} 
\end{eqnarray}
and 
\begin{equation}
\Delta_\sigma = \frac{d}{dR} M_N^\ast({\vec R}) , \ \ \Delta_\omega = \frac{d}{dR} 3g_\omega^q({\vec R}) , \ \ 
\Delta_\rho = \frac{d}{dR} g_\rho^q({\vec R})  .     
\label{deltas}
\end{equation}
For a point-like nucleon, one has $\mu_s = 1$ while the physical value is $\mu_s = 0.88$.  Thus, in so far as the 
$\omega$ contribution to the spin-orbit force is concerned, the point-like result has no problem. But, this is not 
the case for the $\rho$ contribution, because we still have $\mu_v = 1$ for the point-like particle 
but the observed value is $\mu_v = 4.7$. This structure is included 
in a very natural way in the QMC model. 

Now let us consider how the model is quantized in the 
non-relativistic framework. Until now the motion of the 
nucleon has been treated as classical, but it is necessary to quantize it. The simple way to proceed may be to 
find a Lagrangian which can realize the energy-momentum expressions 
Eqs.~(\ref{em1}) and~(\ref{em2}).  If we keep 
only terms up to those quadratic in the velocity, we find that the Lagrangian 
\begin{equation}
L({\vec R}, {\vec v}) = -M_N^\ast({\vec R}) \sqrt{1-v^2} - V_c({\vec R}) ,      
\label{Lcl}
\end{equation}
can produce the energy and momentum. Here we drop the spin dependent correction since it already involves 
the velocity. Then, the non-relativistic expression of the Lagrangian may be 
\begin{equation}
L_{nr}({\vec R}, {\vec v}) = \frac{1}{2}M_N^\ast({\vec R})v^2 - M_N^\ast({\vec R}) - V_c({\vec R}) ,      
\label{Lclnr}
\end{equation}
which leads to the Hamiltonian 
\begin{equation}
H_{nr}({\vec R}, {\vec P}) = {\vec P}\cdot\frac{1}{2M_N^\ast({\vec R})}{\vec P} + M_N^\ast({\vec R}) + V({\vec R}) ,      
\label{Hnr}
\end{equation}
where the spin-orbit interaction is re-inserted in $V({\vec R})$. Thus, the nuclear, quantum Hamiltonian for 
the nucleus with atomic number $A$ is given by 
\begin{equation}
H_{nr} = \sum_{i=1}^A H_{nr}({\vec R}_i, {\vec P}_i) , \ \ \ 
{\vec P_i} = -i {\vec \nabla}_i .  
\label{NHnr}
\end{equation}
\subsubsection{\it Equations for the meson fields
\label{subsubsec:mesons}}

The equations of motion for the meson-field operators ($\hat \sigma$, ${\hat \omega}^\nu$, ${\hat \rho}^\nu$) are 
given by  
\begin{eqnarray}
(\partial_\mu \partial^\mu + m_\sigma^2) {\hat \sigma} &=& g_\sigma^q {\overline \psi_q}\psi_q ,   \label{sig1} \\
(\partial_\mu \partial^\mu + m_\omega^2) {\hat \omega}^\nu &=& g_\omega^q {\overline \psi_q}\gamma^\nu \psi_q ,   
\label{omg1} \\
(\partial_\mu \partial^\mu + m_\rho^2) {\hat \rho}^\nu &=& g_\rho^q {\overline \psi_q}\gamma^\nu \frac{\tau_3}{2}\psi_q ,  
\label{rho1} 
\end{eqnarray}
where $m_\sigma$, $m_\omega$ and $m_\rho$ are respectively 
the masses of $\sigma$, $\omega$ and (neutral) $\rho$ mesons.   
We shall apply the mean field approximation to these meson fields.  
The mean fields are calculated as the expectation 
values with respect to the nuclear ground state $|A\rangle$: 
\begin{eqnarray}
\langle A | {\hat \sigma}(t, {\vec r}) | A \rangle &=& \sigma({\vec r}) ,   \label{sig2} \\
\langle A | {\hat \omega}^\nu(t, {\vec r}) | A \rangle &=& \delta_{\nu, 0}\omega({\vec r})  ,   \label{omg2} \\
\langle A | {\hat \rho}^\nu(t, {\vec r}) | A \rangle &=& \delta_{\nu, 0}b({\vec r})  .  \label{rho2} 
\end{eqnarray}

Next, we need the expressions for the sources in Eqs.~(\ref{sig1}), 
(\ref{omg1}) and~(\ref{rho1}) 
\begin{equation}
\langle A | {\overline \psi_q}\psi_q(t, {\vec r}) | A \rangle , \ \ \ 
\langle A | {\overline \psi_q}\gamma^\nu \psi_q(t, {\vec r}) | A \rangle , \ \ \ 
\langle A | {\overline \psi_q}\gamma^\nu \frac{\tau_3}{2}\psi_q(t, {\vec r}) | A \rangle .  \label{source} 
\end{equation}
Since the sources are given by the sums of the source created by each nucleon in MFA, 
we can write 
\begin{equation}
{\overline \psi_q}\psi_q(t, {\vec r}) = \sum_{i} \langle {\overline \psi_q}\psi_q(t, {\vec r}) \rangle_i , \ \ \ 
{\overline \psi_q}\gamma^\nu \psi_q(t, {\vec r}) = \sum_{i} \langle 
{\overline \psi_q}\gamma^\nu \psi_q(t, {\vec r}) \rangle_i , 
\label{source2} 
\end{equation}
where $\langle \cdots \rangle_i$ is the matrix element in the nucleon $i$ located at ${\vec R}$ at time $t$. 
(For a while, we shall not present the expression of the source for the $\rho$ meson because it is the same as that for 
the $\omega$ except for trivial isospin factor.) 
Because we apply the Born-Oppenheimer approximation within the model, 
the sources are given by $3$ quarks in the lowest 
state. In the IRF, one finds 
\begin{eqnarray}
\langle {\overline \psi_q}^{\prime}\psi_q^\prime(t^\prime, {\vec r}^{\,\prime}) \rangle_i &=& 
3\sum_m {\overline \phi}_i^{0,m}({\vec u}^{\,\prime}) \phi_i^{0,m}({\vec u}^{\,\prime}) 
\equiv 3 s_i({\vec u}^{\,\prime}) ,   
\label{sour1} \\
\langle {\overline \psi_q}^{\prime}\gamma^\nu \psi_q^\prime(t^\prime, {\vec r}^{\,\prime}) \rangle_i &=& 
3 \delta_{\nu, 0}\sum_m \phi_i^{\dagger 0,m}({\vec u}^{\,\prime}) \phi_i^{0,m}({\vec u}^{\,\prime}) 
\equiv  3 \delta_{\nu, 0} w_i({\vec u}^{\,\prime}) .   \label{sour22f} 
\end{eqnarray}
Note that the space component of the vector field vanishes because of parity conservation. 

The Lorentz transformation between the IRF and NRF leads to 
\begin{equation}
u_{i,L}^\prime = (r_L - R_{i,L}) \cosh \xi_i , \ \ \ {\vec u}_{i, \perp}^{\,\prime} 
= {\vec r}_{\perp} - {\vec R}_{i, \perp} ,  \label{ltu} 
\end{equation}
at the common time $t$ in the NRF.  Because of the Lorentz-scalar and vector characters, the sources are 
rewritten in the NRF as 
\begin{eqnarray}
\langle {\overline \psi_q}\psi_q(t, {\vec r}) \rangle_i &=& 
3 s_i((r_L - R_{i,L}) \cosh \xi_i, {\vec r}_{\perp} - {\vec R}_{i, \perp}) ,   
\label{sour21} \\
\langle {\overline \psi_q}\gamma^0 \psi_q(t, {\vec r}) \rangle_i &=& 
3 w_i((r_L - R_{i,L}) \cosh \xi_i, {\vec r}_{\perp} - {\vec R}_{i, \perp}) \cosh \xi_i ,   \label{sour22} \\
\langle {\overline \psi_q}{\vec \gamma} \psi_q(t, {\vec r}) \rangle_i &=& 
3 w_i((r_L - R_{i,L}) \cosh \xi_i, {\vec r}_{\perp} - {\vec R}_{i, \perp}) {\hat v} \sinh \xi_i .   \label{sour23} \\
\end{eqnarray}
These sources can be transformed 
\begin{eqnarray}
\langle {\overline \psi_q}\psi_q(t, {\vec r}) \rangle_i &=& \frac{3}{(2\pi)^3\cosh\xi_i} \int d{\vec k} \, e^{i{\vec k}\cdot
({\vec r} - {\vec R}_{i})} S({\vec k}, {\vec R}_i) ,   \label{sour31} \\
\langle {\overline \psi_q}\gamma^0 \psi_q(t, {\vec r}) \rangle_i &=& \frac{3}{(2\pi)^3} \int d{\vec k} \, e^{i{\vec k}\cdot
({\vec r} - {\vec R}_{i})} W({\vec k}, {\vec R}_i) ,   \label{sour32} \\
\langle {\overline \psi_q}{\vec \gamma} \psi_q(t, {\vec r}) \rangle_i 
&=& \frac{3}{(2\pi)^3}{\hat v} \int d{\vec k} \, e^{i{\vec k}\cdot
({\vec r} - {\vec R}_{i})} W({\vec k}, {\vec R}_i) ,   \label{sour33} 
\end{eqnarray}
with the sources in momentum space 
\begin{eqnarray}
S({\vec k}, {\vec R}_i)  = \int d{\vec u}\, e^{-i({\vec k}_\perp\cdot{\vec u}_\perp + k_Lu_L/\cosh\xi_i)} s_i({\vec u}) , 
\label{sourm1} \\
W({\vec k}, {\vec R}_i)  = \int d{\vec u}\, e^{-i({\vec k}_\perp\cdot{\vec u}_\perp + k_Lu_L/\cosh\xi_i)} w_i({\vec u}) . 
\label{sourm2} 
\end{eqnarray}
Thus, the mean field expressions for the meson sources are given by 
\begin{eqnarray}
\langle A | {\overline \psi_q}\psi_q(t, {\vec r}) | A \rangle 
&=& \frac{3}{(2\pi)^3} \int d{\vec k} \, e^{i{\vec k}\cdot{\vec r}} 
\langle A | \sum_i (\cosh\xi_i)^{-1} e^{-i{\vec k}\cdot{\vec R}_{i}} S({\vec k}, {\vec R}_i) | A \rangle , 
\label{sour41} \\
\langle A | {\overline \psi_q}\gamma^0\psi_q(t, {\vec r}) | A \rangle 
&=& \frac{3}{(2\pi)^3} \int d{\vec k}\, e^{i{\vec k}\cdot{\vec r}} 
\langle A | \sum_i e^{-i{\vec k}\cdot{\vec R}_{i}} W({\vec k}, {\vec R}_i) | A \rangle , 
\label{sour42} \\
\langle A | {\overline \psi_q} {\vec \gamma} \psi_q(t, {\vec r}) | A \rangle &=& 0 .  
\label{sour43} 
\end{eqnarray}
Note that the velocity vector averages to zero. 

The matrix element, $\langle A | \sum_i \exp(-i{\vec k}\cdot{\vec R}_{i}) \cdots | A \rangle$, in Eqs.~(\ref{sour41}) 
and~(\ref{sour42}) only remains when $k$ is less than, or of order of the inverse of the nuclear radius. 
Furthermore, since $u$ is bounded by the nucleon size, 
the argument of the exponential in Eqs.~(\ref{sourm1})  
and~(\ref{sourm2}) can be ignored when the model is 
applied to large enough nuclei. In this approximation, we can 
simplify the sources further 
\begin{eqnarray}
\langle A | {\overline \psi_q}\psi_q(t, {\vec r}) | A \rangle &=& 3 S({\vec r}) \rho_s({\vec r}) ,  \label{sour51} \\
\langle A | {\overline \psi_q}\gamma^\nu \psi_q(t, {\vec r}) | A \rangle 
&=& 3 \delta_{\nu, 0} \rho_B({\vec r}) , \label{sour52} \\
\langle A | {\overline \psi_q} \gamma^\nu \frac{\tau_3}{2} \psi_q(t, {\vec r}) | A \rangle 
&=& \delta_{\nu, 0} \rho_3({\vec r}) ,  
\label{sour53} 
\end{eqnarray}
with the scalar, baryon and isospin densities of the nucleon in the nucleus 
\begin{eqnarray}
\rho_s({\vec r}) &=& \langle A | \sum_i \frac{M_N^\ast({\vec R}_i)}{E_i-V({\vec R}_i)} 
\delta({\vec r}-{\vec R}_i) | A \rangle  ,  \label{crhos} \\
\rho_B({\vec r}) &=& \langle A | \sum_i \delta({\vec r}-{\vec R}_i) | A \rangle  ,  \label{crhoB} \\
\rho_3({\vec r}) &=& \langle A | \sum_i \frac{\tau_3^N}{2} \delta({\vec r}-{\vec R}_i) | A \rangle  .  \label{crho3} 
\end{eqnarray}
Here we deduced the source for the $\rho$ meson (the isospin density, $\rho_3$) from that for the $\omega$.  
In Eq.~(\ref{sour51}), $S({\vec r})$ for the 
$i$-th nucleon (at $\r$) is given by 
\begin{equation}
S({\vec r}) = S({\vec 0}, {\vec r}) = \int d{\vec u}\, s_i({\vec u}) 
= \frac{\Omega_0/2 + m_q^\ast R_B(\Omega_0-1)}{\Omega_0(\Omega_0-1) + m_q^\ast R_B/2} ,   \label{scalar} 
\end{equation}
in the MIT bag model. This has an implicit coordinate dependence through the local scalar field at ${\vec r}$. 

Since these sources are time independent, the equations for the meson fields are finally given by 
\begin{eqnarray}
(-\nabla_r^2 + m_\sigma^2) \sigma({\vec r}) &=& g_\sigma C({\vec r}) \rho_s({\vec r}) ,   \label{csigf} \\
(-\nabla_r^2 + m_\omega^2) \omega({\vec r}) &=& g_\omega \rho_B({\vec r}) ,   \label{comgf} \\
(-\nabla_r^2 + m_\rho^2) \rho({\vec r}) &=& g_\rho \rho_3({\vec r}) ,   \label{crhof} 
\end{eqnarray}
where the meson-nucleon coupling constants and $C$ are respectively defined by 
\begin{equation}
g_\sigma =3g_\sigma^q S(\sigma =0) , \ \ \ g_\omega =3g_\omega^q , \ \ \ g_\rho = g_\rho^q , 
\ \ \ C({\vec r}) = S({\vec r})/S(\sigma =0) .    \label{msdef} 
\end{equation}
Thus, the mean fields carry the energy 
\begin{equation}
E_{meson} = \frac{1}{2} \int d{\vec r} \, [ ({\vec \nabla} \sigma)^2 + m_\sigma^2 \sigma^2 
- ({\vec \nabla} \omega)^2 - m_\omega^2 \omega^2 
- ({\vec \nabla} b)^2 - m_\rho^2 b^2 ] .    \label{msenergy} 
\end{equation}
\subsubsection{\it Self-consistent procedure
\label{subsubsec:consistency}}

Let us summarize the model and review the procedure for the 
self-consistent calculation for finite nuclei: 
\begin{enumerate} 
\item Choose the bare quark mass, $m_q$, and construct a quark model of the nucleon so as to 
produce the free nucleon properties.  For example, if we take the MIT bag model, the bag parameters, $B$ 
and $z_0$, are fixed to fit the free nucleon mass and the bag radius. 
\item Assume that the coupling constants and the masses of the mesons are known. 
\item Calculate the nucleon properties, $I(\sigma)$ 
(Eq.~(\ref{I2})) and $S(\sigma)$ (Eq.~(\ref{scalar})), at 
a given value of $\sigma$. 
\item Guess initial forms of the densities, $\rho_s({\vec r})$, $\rho_B({\vec r})$, $\rho_3({\vec r})$, in 
Eqs.~(\ref{crhos})-(\ref{crho3}). 
\item For $\rho_s({\vec r})$, $\rho_B({\vec r})$ and 
$\rho_3({\vec r})$ fixed, solve Eqs.~(\ref{csigf})-(\ref{crhof}) 
for the meson fields. 
\item Evaluate the effective nucleon mass, 
$M_N^\ast$, in the nucleus (for example, Eq.~(\ref{EMass})) 
and the potential, $V({\vec r})$ (Eq.~(\ref{v1})). 
If we take the MIT bag model, the bag radius at each position in 
the nucleus is determined by the equilibrium condition, 
Eq.~(\ref{EMass2}).  For practical purposes, it is very 
convenient to make simple parameterizations for $M_N^\ast$ and $C({\vec r})$ as functions of the value of the 
$\sigma$ field at given position ${\vec r}$. The bag model enables us to 
make parameterizations which work 
very well at moderate densities (see section~\ref{sec:properties}). 
\item Solve the eigenvalue problem given by the nuclear Hamiltonian 
Eq.~(\ref{NHnr}) and compute the shell 
states from which the densities, $\rho_s({\vec r})$, $\rho_B({\vec r})$, $\rho_3({\vec r})$, can be evaluated.  
\item Go to 5 and iterate until self-consistency is achieved. 
\end{enumerate} 

\subsection{\it Relativistic model 
\label{subsec:relamodel}}
\subsubsection{\it Relativistic treatment 
\label{subsubsec:relativistic}}

To facilitate a comparison between QMC and 
the widely used Quantum Hadrodynamics 
(QHD)~\cite{QHD}, we re-formulate the model as 
a relativisitic field theory~\cite{SAI-3,GUI-3}. 
We make no attempt to justify the formulation of a local, relativisitic 
field theory at a fundamental level because it is not possible for 
composite hadrons. Our point is to try to 
express the present idea in the framework of relativisitic field 
theory at the mean field level.  To do so, 
we first write a relativisitic Lagrangian and check 
that it is equivalent to the non-relativisitic formulation 
in an appropriate approximation. 
Since it is vital to consider the $\sigma$ and $\omega$ fields but 
less important to 
include the $\rho$ field in a (isospin) symmetric 
nuclear medium, once again we omit the $\rho$ 
contribution for a while.  

Our basic result is that essentially the nucleon in a nuclear medium can be treated as a point-like particle 
with the effective mass $M_N^\ast(\sigma({\vec r}))$, which depends on position through the local scalar 
field, moving in the vector potential $g_\omega \omega({\vec r})$.  Since, as already pointed out, the spin-orbit 
force due to the $\omega$ is almost equivalent to that of a point-like Dirac nucleon, a possible Lagrangian 
density for the symmetric nuclear system may be written as 
\begin{equation}
{\cal L} = {\overline \psi} [i\gamma \cdot \partial -M_N^\ast({\hat \sigma}) -g_\omega {\hat \omega}^\mu \gamma_\mu ] \psi 
+ {\cal L}_{meson} ,    \label{lag1} 
\end{equation}
where $\psi$, ${\hat \sigma}$ and ${\hat \omega}$ are respectively the nucleon, $\sigma$ and $\omega$ field operators. 
The free meson Lagrangian density is 
\begin{equation}
{\cal L}_{meson} = \frac{1}{2} (\partial_\mu {\hat \sigma} \partial^\mu {\hat \sigma} - m_\sigma^2 {\hat \sigma}^2) 
- \frac{1}{2} \partial_\mu {\hat \omega}_\nu (\partial^\mu {\hat \omega}^\nu - \partial^\nu {\hat \omega}^\mu)  
+ \frac{1}{2} m_\omega^2 {\hat \omega}^\mu {\hat \omega}_\mu .     \label{mlag1} 
\end{equation}

If we separate the effective nucleon mass as 
\begin{equation}
M_N^\ast({\hat \sigma}) = M_N - g_\sigma({\hat \sigma}) {\hat \sigma} ,    \label{efnmas} 
\end{equation}
and define the $\sigma$-field dependent (or position dependent) coupling constant, $g_\sigma({\hat \sigma})$, 
one can check that $g_\sigma(\sigma =0)$ is the same as $g_\sigma$ defined in 
Eq.~(\ref{msdef}). Furthermore, 
if we rewrite the Lagrangian density 
\begin{equation}
{\cal L} = {\overline \psi} [i\gamma \cdot \partial -M_N + g_\sigma({\hat \sigma}) {\hat \sigma} - 
g_\omega {\hat \omega}^\mu \gamma_\mu ] \psi + {\cal L}_{meson} ,    \label{lag2} 
\end{equation}
it is very clear that the difference between the present model and QHD lies only in the fact that 
the substructure effect of the nucleon provides a {\em known} dependence of the scalar meson-nucleon 
coupling constant on the scalar field itself in a nuclear medium.  It should be stressed that 
this dependence is not the same as that 
adopted in recent density-dependent hadron field theories~\cite{DDFT}, 
in which the meson-nucleon vertices are {\em assumed} 
to {\em phenomenologically} depend on the baryonic densities. 

In MFA, we can replace the meson-field operators by their expectation values in 
the Lagrangian: ${\hat \sigma} \to \sigma({\r})$ and ${\hat \omega}^\mu({\r}) \to \delta_{\mu, 0} 
\omega({\vec r})$. Then, the nucleon and meson fields satisfy the following equations: 
\begin{eqnarray}
(i\gamma \cdot \partial - M_N^\ast(\sigma) - g_\omega \gamma_0 \omega ) \psi &=& 0 ,   \label{dirac} \\
(-\nabla_r^2 + m_\sigma^2) \sigma({\vec r}) &=& -\left( \frac{\partial}{\partial \sigma} M_N^\ast(\sigma) 
\right) \langle A| {\overline \psi}\psi({\vec r})|A\rangle ,   \label{sigeq} \\
(-\nabla_r^2 + m_\omega^2) \omega({\vec r}) &=& g_\omega \langle A| \psi^\dagger \psi({\vec r})|A\rangle ,  
\label{omgeq} 
\end{eqnarray}
where the derivative of the effective nucleon mass with respect to the scalar field is the response of the nucleon 
to the external scalar field - the so-called {\em scalar polarizability}.  
That is given in terms of the scalar density of a quark in the nucleon 
$S(\sigma({\vec r}))$ in Eq.~(\ref{scalar}):
\begin{equation}
\frac{d}{d\sigma} M_N^\ast(\sigma) = -C(\sigma) g_\sigma(\sigma =0) = - \frac{d}{d\sigma} (g_\sigma(\sigma)\sigma) . 
\label{nmderiv} 
\end{equation}
Note also that the sources in the r.h.s. of Eqs.~(\ref{sigeq}) 
and~(\ref{omgeq}) 
are, respectively, equal to the previously defined scalar ($\rho_s$) and baryon ($\rho_B$) sources. 

It is possible to see that the present 
relativistic formulation can go back to the previous non-relativisitic 
one (without the $\rho$ coupling) under 
the following conditions: (1) only terms of second order in the velocity 
are kept, (2) second derivatives of the meson fields 
are ignored, (3) each field is small in comparison with 
the nucleon mass, (4) the isoscalar magnetic 
moment, $\mu_s$, is supposed to be unity. 

To describe a nucleus with different numbers of protons and neutrons ($Z \neq N$), it is necessary to include the 
$\rho$-meson contribution. Any realistic treatment of nuclear structure also requires the Coulomb interaction. Thus, 
an interaction Lagrangian density to be added to Eq.~(\ref{lag2}) 
at the quark level may be 
\begin{equation}
{\cal L}_{\rho +\gamma} = - {\overline \psi_q}\left[g_\rho^q\frac{\tau_3}{2}\gamma_\mu \rho^\mu 
-e\left( \frac{1}{6} + \frac{\tau_3}{2} \right) \gamma_\mu A^\mu \right] \psi_q  ,  \label{lagRA} 
\end{equation}
(see also Eq.~(\ref{rho})). 
We saw in the previous section that this leads to both a central and spin-orbit potential for the nucleon in the 
NRF. Furthermore, for the isoscalar $\omega$ contribution the potentials are well represented by a (point) 
$\omega$-nucleon coupling.  However, for the $\rho$, the relativisitic formulation {\em at the nucleon level} 
requires a strong tensor coupling ($\sim \sigma^{\mu\nu} q_\nu$) if it is to reproduce the large isovector 
magnetic moment ($\mu_v = 4.7$).  In the present stage, we follow the usual 
procedure adopted in the Hartree treatment of QHD, namely, we only use a vector coupling for the $\rho$, with its 
coupling adjusted to give the bulk symmetry energy in MFA. (More precise discussion for the inclusion of the 
$\rho$ meson remains as a future problem.) In this case, it is easy to get the expressions for the $\rho$ and 
Coulomb fields by including trivial isospin factors: 
$3g_\omega^q \omega({\vec r}) \to g_\rho (\tau_3^N/2) b({\vec r})$ 
or $\to (e/2)(1+\tau_3^N) A({\vec r})$. 
This is the first, relativistic version of the quark-meson coupling model - we shall call this 
QMC-I.  

In summary, our effective Lagrangian density for the QMC-I model in MFA takes the form 
\begin{eqnarray}
{\cal L}_{QMC-I} &=& {\overline \psi} [i\gamma \cdot \partial -M_N^\ast(\sigma({\vec r}))  
- g_\omega \omega({\vec r}) \gamma_0 - g_\rho \frac{\tau_3^N}{2} b({\vec r}) \gamma_0 
- \frac{e}{2}(1+\tau_3^N) A({\vec r}) \gamma_0 ] \psi  \nonumber \\
&-& \frac{1}{2} [({\vec \nabla} \sigma({\vec r}))^2 + m_\sigma^2 \sigma({\vec r})^2 ]  
+ \frac{1}{2} [({\vec \nabla} \omega({\vec r}))^2 + m_\omega^2 \omega({\vec r})^2 ]  \nonumber \\
&+& \frac{1}{2} [({\vec \nabla} b({\vec r}))^2 + m_\rho^2 b({\vec r})^2 ] 
+ \frac{1}{2} ({\vec \nabla} A({\vec r}))^2  .    \label{LQMC-I} 
\end{eqnarray}

The mean field description can be improved by the inclusion of 
exchange contribution 
(Fock terms)~\cite{FOCK}. 
The inclusion of Fock terms allows us to explore the momentum 
dependence of meson-nucleon vertices and 
the role of pionic degrees of freedom in matter. 
It has been found that the Fock terms maintain the 
mean field predictions of the QMC model but that 
the Fock terms significantly increase the absolute values of 
the single-particle energy, 4-component scalar and vector potentials.
This is particularly relevant for the spin-orbit 
splitting in finite nuclei (see also section~\ref{subsec:variant}). 

\subsubsection{\it Properties of the in-medium nucleon mass
\label{subsubsec:nmass}}

We now consider the properties of the nucleon in a nuclear medium, 
based on the relativisitic formulation 
developed in the previous section. 
In general, the Dirac equation for the quark field in matter 
may be written as (see also Eq.~(\ref{comp1}))
\begin{equation}
[ i\gamma\cdot\partial - (m_q - V_s^q) - V_c({\vec r}) 
 - \gamma_0 V_v^q ] \psi_q({\vec r}) = 0 , 
\label{dirac2}
\end{equation}
where the quark feels the scalar $V_s^q$ (like the $\sigma$) and vector $V_v^q$ (like the $\omega$) 
potentials in nuclear matter, which are self-consistently generated 
by the medium, while $V_c$ represents the confinement potential.  
As discussed in section~\ref{subsubsec:classical}, 
the scalar and vector potentials depend on the position of 
the nucleon, but they may be supposed to be 
constant inside the nucleon because the potentials 
do not vary much across it -- this is the 
local density approximation (LDA) (see Ref.~\cite{BLU} for details).  

As the nucleon is static, the time-derivative operator in the Dirac equation can be 
replaced by the quark energy, $-i \epsilon_q$.  
By analogy with the procedure applied to the nucleon
in QHD~\cite{QHD}, the Dirac equation, Eq.~(\ref{dirac2}), 
can be rewritten in the same form as that in free space, 
with the effective quark mass $m_q^\ast (=m_q - V_s^q)$ and 
the energy $\epsilon^\ast (= \epsilon_q - V_v^q)$, instead of 
$m_q$ and $\epsilon_q$ in the free case. In other words, the vector interaction has again {\em no effect} 
on the nucleon structure except for an overall phase in the quark wave 
function, which gives a shift in the nucleon energy.  This fact 
{\em does not} depend on how to choose the confinement potential $V_c$.  
The nucleon energy (at rest), $E^*_N$, in the medium is 
then expressed by~\cite{SAI-3,SAI-8} 
\begin{equation}
E^*_N = M_N^\ast(V_s^q) + 3V_v^q , 
\label{efmasf}
\end{equation}
where the effective nucleon mass $M_N^\ast$ depends on {\em only} the 
scalar potential in the medium, as discussed in the non-relativisitic treatment. 
We here again stress that the basic result in the QMC model 
is that, in the scalar ($\sigma$) and vector ($\omega$) 
meson fields, the nucleon behaves essentially as a point-like 
particle with an effective mass 
$M_N^\ast$, which depends on the position through only the $\sigma$ 
field, moving in a vector potential generated by the $\omega$ meson.  
This feature should hold in {\em any model} in which the nucleon contains 
{\em relativistic} quarks and the (middle- and long-range) 
{\em attractive} 
and (short-range) {\em repulsive} nucleon-nucleon (N-N) forces have {\em Lorentz-scalar} and 
{\em vector} characters, respectively.  

We consider the nucleon mass in matter further.  The nucleon mass is a 
function of the scalar field.  Because the scalar field is small 
at low density the nucleon mass can be expanded in terms of $\sigma$ as 
\begin{equation}
M_N^\ast = M_N + \left( \frac{\partial M_N^\ast}{\partial \sigma} 
\right)_{\sigma=0} \sigma + \frac{1}{2} \left( \frac{\partial^2 M_N^\ast}
{\partial \sigma^2} \right)_{\sigma=0} \sigma^2 + \cdots . 
\label{nuclm}
\end{equation}
In the QMC model, the interaction Hamiltonian between the nucleon and the 
$\sigma$ field at the quark level is given by $H_{int} = - 3 g_{\sigma}^q 
\int d{\vec r} \ {\overline \psi}_q \sigma \psi_q$, and the derivative of 
$M_N^\ast$ with respect to $\sigma$ is thus given by the quark-scalar density in the nucleon, 
$-3g_\sigma^qS_N(\sigma)$ (see Eq.~(\ref{scalar})).  (Hereafter, we add the suffix $N$ to the scalar density 
$S$ and its ratio $C$ to specify the nucleon.) Because of the negative value of 
$\left( \frac{\partial M_N^\ast}{\partial \sigma} \right)$, 
the nucleon mass decreases in matter at low density.  

Furthermore, if we use Eq.~(\ref{nmderiv}), we find that 
\begin{equation}
M_N^\ast = M_N - g_{\sigma} \sigma - \frac{1}{2} g_{\sigma} 
C_N^\prime(0) \sigma^2 + \cdots . 
\label{nuclm2}
\end{equation}
In general, the scalar-density ratio, $C_N$, (or the scalar polarizability) is a decreasing function 
because the quark in matter is 
more relativistic than in free space.  Thus, $C_N^\prime(0)$ takes a 
negative value, and hence the third term in the right hand side 
of Eq.~(\ref{nuclm2}) makes the mass larger. 
If the nucleon were structureless $C_N$ would not depend on 
the scalar field, that is, $C_N$ would be constant ($C_N=1$).  Therefore, 
only the first two terms in the right hand side of Eq.~(\ref{nuclm2}) remain, 
which is exactly the same as the equation for the effective nucleon 
mass in QHD.  By taking the heavy-quark-mass limit in QMC we can reproduce 
the QHD results~\cite{SAI-1}. We recall that this decrease in $C_N$
constitutes a new saturation mechanism~\cite{GUI-1} -- different from pure QHD -- and
is the main reason why the scalar coupling constant is somewhat smaller
in QMC than in QHD. Finally, we note that it is possible 
to relate the mass reduction in the QMC model to 
the change of quark condensates in matter~\cite{SAI-1,SAI-q,SAI-qq} 
(see section~\ref{subsubsec:condensates}). 

\subsubsection{\it Effect of meson structure 
\label{subsubsec:mesonstr}}

Until now, we have studied only the effect of the nucleon substructure. It is, however, true that the mesons are also 
built of quarks and antiquarks, and that they may also change their properties in a nuclear medium. It may not be 
so difficult to include such effects in the present model~\cite{SAI-3}.  
To incorporate it, we suppose that the vector mesons 
are again described by a relativistic quark model with {\em common} scalar and vector mean fields, like the nucleon. 
Thus, the effective vector meson mass in matter, $m_v^\ast (v=\omega, \rho)$, again depends only on the value of 
scalar mean field in matter.  However, for the $\sigma$ meson it may not be easy to describe it 
by a simple quark model (like a bag) because it couples strongly 
to the pseudoscalar ($2 \pi$) channel, which requires a direct 
treatment of chiral symmetry in a nuclear medium.  Since, according to the 
Nambu--Jona-Lasinio model~\cite{Bernard,Hatsuda} or the Walecka 
model~\cite{SAI-9,SAI-omega}, one might expect the 
$\sigma$-meson mass in medium $m_{\sigma}^\ast$ to be less than in free 
space, we shall parameterize it using a quadratic 
function of the scalar field: 
\begin{equation}
\left( \frac{m_{\sigma}^\ast}{m_{\sigma}} \right) = 1 - a_{\sigma} 
(g_{\sigma} \sigma) + b_{\sigma} (g_{\sigma} \sigma)^2 , 
\label{sigmas}
\end{equation}
with two, additional new parameters $a_{\sigma}$ and $b_{\sigma}$.  

Using these effective meson masses, one can construct a new Lagrangian 
density for finite nuclei, 
which involves the structure effects of not only the nucleons but also the 
mesons, in the MFA: 
\begin{eqnarray}
{\cal L}_{QMC-II}&=& {\overline \psi} [i \gamma \cdot \partial 
- M_N^\ast - g_\omega \omega({\vec r}) \gamma_0 
- g_\rho \frac{\tau^N_3}{2} b({\vec r}) \gamma_0 
- \frac{e}{2} (1+\tau^N_3) A({\vec r}) \gamma_0 ] \psi \nn \\
&-& \frac{1}{2}[ ({\vec \nabla} \sigma({\vec r}))^2 + 
m_{\sigma}^{\ast 2}({\vec r})  \sigma({\vec r})^2 ] 
+ \frac{1}{2}[ ({\vec \nabla} \omega({\vec r}))^2 + m_{\omega}^{\ast 2}({\vec r})  
\omega({\vec r})^2 ] \nn \\
&+& \frac{1}{2}[ ({\vec \nabla} b({\vec r}))^2 + m_{\rho}^{\ast 2}({\vec r}) b({\vec r})^2 ] 
+ \frac{1}{2} ({\vec \nabla} A({\vec r}))^2 , 
\label{qmc-2}
\end{eqnarray}
where the masses of the mesons and the nucleon have a dependence 
on position through the scalar 
mean-field.  We call this model QMC-II~\cite{SAI-3}.  

At low density, the vector-meson mass can be expanded in the similar manner to the nucleon case (Eq.~(\ref{nuclm})): 
\begin{eqnarray}
m_v^\ast &=& m_v + \left( \frac{\partial m_v^\ast}{\partial \sigma} 
\right)_{\sigma=0} \sigma + \frac{1}{2} \left( \frac{\partial^2 m_v^\ast}
{\partial \sigma^2} \right)_{\sigma=0} \sigma^2 + \cdots , \nn \\
  &\simeq&  m_v - 2 g_{\sigma}^q S_v(0) \sigma - g_{\sigma}^q S_v^\prime(0) 
\sigma^2 , \nn \\
  &\equiv& m_v - \frac{2}{3} g_\sigma \Gamma_{v/N} \sigma 
  - \frac{1}{3} g_\sigma \Gamma_{v/N} C_v^\prime(0) \sigma^2 , 
\label{vmm}
\end{eqnarray}
where $m_v$ is the free mass and $S_v(\sigma)$ is the quark-scalar density in the vector meson, 
\begin{equation}
\left( \frac{\partial m_v^*}{\partial \sigma} \right) 
= - \frac{2}{3} g_\sigma \Gamma_{v/N} C_v(\sigma) , 
\label{deriv3}
\end{equation}
and $C_v(\sigma) = S_v(\sigma)/S_v(0)$.  In Eqs.~(\ref{vmm}) 
and~(\ref{deriv3}), we introduce a correction factor, $\Gamma_{v/N}$, 
which is given by $S_v(0)/S_N(0)$, because the coupling 
constant, $g_\sigma$, is defined specifically for the nucleon
by Eq.~(\ref{msdef}).  

Although the fields for the vector mesons satisfy 
the usual equations Eqs.~(\ref{comgf}) and~(\ref{crhof}) with 
their effective masses, instead of the free ones, 
the equation of motion for the $\sigma$ is also modified by the 
effect of meson substructure. Since the quark and 
antiquark inside the mesons interact with the $\sigma$ field, 
the source for the $\sigma$ is given by not only 
the derivative of the nucleon mass with respect to $\sigma$ 
but also those for the mesons in QMC-II. (Actual expression 
will be given in section~\ref{sec:properties}.)  

\subsubsection{\it Naturalness
\label{subsubsec:natural}}

Here we check whether the concept of
"{\em naturalness}" applies in the QMC model~\cite{SAI-5}. 
In general, an effective field theory at low energy will contain an infinite 
number of interaction terms, which incorporate the {\em compositeness} 
of hadrons. This  
is expected to involve numerous couplings which may not be renormalizable. 
Such an EFT requires an organizing principle to make sensible calculations.  

Manohar and Georgi~\cite{MAN} have proposed a systematic way to 
manage such complicated, effective field theories called ``naive
dimensional analysis'' (NDA).
NDA gives rules for assigning a coefficient of the 
appropriate size to any interaction term in an effective Lagrangian.  
After extracting the dimensional factors and 
some appropriate counting factors using NDA, the remaining 
{\em dimensionless} 
coefficients are all assumed to be of order {\em unity}.  This is the 
so-called {\em naturalness} assumption.  If naturalness is valid, the 
effective Lagrangian can be truncated at a given order with a reasonable 
bound on the truncation error for physical observables.  Then we can control 
the effective Lagrangian, at least at the tree level.  

NDA has been already applied to QHD~\cite{FUR-1}, 
where it was concluded that the 
relativistic Hartree approximation (RHA) in 
QHD leads to {\em unnaturally} large 
coefficients due to the treatment of the vacuum in terms of 
the excitation of $N{\overline N}$-pairs. 
This means that the loop expansion in QHD 
does not work as well as one would desire~\cite{PRA}. 

Here we use NDA to see whether the QMC model gives natural coefficients.
In brief, NDA tells us the following: for the strong interaction 
there are two relevant 
scales, namely, the pion-decay constant $f_\pi$ ($=93$ MeV) and a larger 
scale, $\Lambda \sim$ 1 GeV, which characterizes the mass scale of physics 
beyond the Goldstone bosons.  The NDA rules indicate how those scales should 
appear in a given term in the effective Lagrangian.  The rules are: 
\begin{enumerate}
\item include a factor of $ 1/f_\pi $ for each strongly interacting field, 
\item assign an overall normalization factor of $(f_\pi \Lambda)^2$, 
\item multiply by factors of $1/\Lambda$ to achieve dimension (mass)$^4$, 
\item include appropriate counting factors, e.g. 1/$n!$ for $\phi^n$ 
(where $\phi$ is a meson field). 
\end{enumerate}

Since the QMC Lagrangian in MFA is given in 
terms of the nucleon ($\psi$), scalar 
($\sigma$) and vector ($\omega$ and $\rho$) meson fields, we can scale a 
generic Lagrangian component as 
\begin{equation}
{\cal L} \sim c_{\ell m n p} \frac{1}{m! n! p!} 
\left( \frac{{\overline \psi} \Gamma (\tau/2) \psi}{f_\pi^2 \Lambda} \right)^{\ell} 
\left( \frac{\sigma}{f_\pi} \right)^m 
\left( \frac{\omega}{f_\pi} \right)^n 
\left( \frac{b}{f_\pi} \right)^p (f_\pi \Lambda)^2 , 
\label{general}
\end{equation}
where $\Gamma$ and $\tau$ stand for a combination of Dirac matrices and 
isospin operators.   The overall coupling constant $c_{\ell m n p}$ is 
dimensionless and of ${\cal O}(1)$ if naturalness holds.  

The QMC Lagrangian is given by
\begin{equation}
{\cal L}_{QMC} = {\cal L}_{free} + {\cal L}_{em} + {\cal L}_{QMC}^{int.} , 
\label{QMCgen}
\end{equation}
where ${\cal L}_{free}$ and ${\cal L}_{em}$ 
stand for the free Lagrangian for the nucleon and 
mesons and the electromagnetic interaction, respectively, while 
${\cal L}_{QMC}^{int.}$ involves (strong) interaction terms.  
For the QMC-I and QMC-II, ${\cal L}_{QMC}^{int.}$ is respectively given 
by Eqs.~(\ref{LQMC-I}) and~(\ref{qmc-2})
\begin{equation}
{\cal L}_{QMC-I}^{int.} = {\overline \psi} \left[ 
\left(M_N - M_N^{\ast}(\sigma) \right) 
- g_\omega \gamma_0 \omega - g_\rho \left( \frac{\tau_3^N}{2} \right) 
\gamma_0 b \right] \psi, 
\label{QMC-I}
\end{equation}
and 
\begin{eqnarray}
{\cal L}_{QMC-II}^{int.} &=& {\overline \psi} \left[ 
\left(M_N - M_N^{\ast}(\sigma) \right) 
- g_\omega \gamma_0 \omega - g_\rho \left( \frac{\tau_3^N}{2} \right) 
\gamma_0 b \right] \psi \nn \\
&-& \frac{1}{2} \left[ m_\sigma^{* 2}(\sigma)-m_\sigma^2 \right] \sigma^2 
+ \frac{1}{2} \left[ m_\omega^{* 2}(\sigma)-m_\omega^2 \right] \omega^2 \\
&+& \frac{1}{2} \left[ m_\rho^{* 2}(\sigma)-m_\rho^2 \right] b^2 , \nn 
\label{QMC-II}
\end{eqnarray}
where the effective masses in the medium depend only on the scalar field.  

If the MIT bag model is used to describe the hadrons, one finds that 
the mass at nuclear density $\rho_B$ is given by quite a simple 
form for the region of $\rho_B \leq 3 \rho_0$~\cite{SAI-3}.  
The reduction of the mass from the free value, 
$\delta M^*_j=M_j-M^*_j$, is then given by 
\begin{equation}
\delta M^*_j \simeq  \frac{n_q}{3} g_\sigma\left[ 1 - 
  \frac{a_j}{2} (g_\sigma \sigma) \right]\sigma , 
\label{efmas}
\end{equation}
where $j=N, \sigma, \omega, \rho$, and $n_q$ is the number of light 
quarks in the hadron $j$, and 
$a_j$ is a slope parameter for the hadron $j$, which is given by the 
second derivative of the mass with respect to the $\sigma$ field 
(for details, see section~\ref{subsubsec:scaling}). 

There are several coupling constants to be determined in the QMC-I and QMC-II models. 
In section~\ref{sec:properties}, we will explain in detail 
how these constants are fixed. Here, 
we merely present the values of the dimensionless coefficients, 
$c_{\ell m n p}$, in Eq.~(\ref{general}) for each 
interaction term in Table~\ref{tab:coef}.  
(In the QMC-II model, we choose three parameter sets for 
the effective $\sigma$-meson mass in matter (see Eq.~(\ref{sigmas})).  
See for details section~\ref{subsec:matter}.) 
Using Eqs.~(\ref{sigmas}) and~(\ref{efmas}), 
the QMC-I Lagrangian has four 
interaction terms, while the QMC-II Lagrangian offers 16 terms 
due to the internal structure of the nucleon {\em and} the mesons.  
\begin{table}
\begin{center}
\begin{minipage}[t]{16.5 cm}
\caption{Interaction terms and corresponding (dimensionless) coupling 
constants. We take $\Lambda = M_N$ in Eq.~(\ref{general}). A, B and C denote three types of 
parameterization of the $\sigma$ mass in a medium (see section~\protect\ref{subsec:matter}).}
\label{tab:coef}
\end{minipage}
\begin{tabular}{c|c|c|c|c|c}
\hline
term & $c_{\ell m n p}$ & QMC-I & A & B & C \\
\hline
${\overline  \psi} \sigma \psi$ & $c_{1100}$ & 0.82 & 0.69 & 0.70 & 0.69 \\
${\overline \psi} \sigma^2 \psi$ & $c_{1200}$ & -0.55 & -0.40 & -0.41 & -0.40 \\
${\overline \psi} \gamma_0 \omega \psi$ & $c_{1010}$ & -0.81 & 
                                                 -0.58 & -0.63 & -0.64 \\
${\overline  \psi} \left( \frac{\tau}{2} \right) \gamma_0 b \psi$ & $c_{1001}$ & 
                                         -0.92 & -0.83 & -0.81 & -0.80 \\
$ \sigma^3 $ & $c_{0300}$ & --- & 0.40 & 0.67 & 1.0 \\
$ \sigma^4 $ & $c_{0400}$ & --- & -3.6 & -2.2 & -4.4 \\
$ \sigma^5 $ & $c_{0500}$ & --- & 3.3 & 2.9 & 8.3 \\
$ \sigma^6 $ & $c_{0600}$ & --- & -22 & -5.7 & -22 \\
$ \sigma \omega^2 $ & $c_{0120}$ & --- & -0.77 & -0.78 & -0.76 \\
$ \sigma^2 \omega^2 $ & $c_{0220}$ & --- & 0.85 & 0.87 & 0.85 \\
$ \sigma^3 \omega^2 $ & $c_{0320}$ & --- & -0.71 & -0.73 & -0.70 \\
$ \sigma^4 \omega^2 $ & $c_{0420}$ & --- & 0.39 & 0.41 & 0.39 \\
$ \sigma b^2 $ & $c_{0102}$ & --- & -0.75 & -0.76 & -0.75 \\
$ \sigma^2 b^2 $ & $c_{0202}$ & --- & 0.84 & 0.86 & 0.84 \\
$ \sigma^3 b^2 $ & $c_{0302}$ & --- & -0.70 & -0.73 & -0.70 \\
$ \sigma^4 b^2 $ & $c_{0402}$ & --- & 0.39 & 0.41 & 0.39 \\
\hline
\end{tabular}
\end{center}
\end{table}

As seen in the table, the QMC-I model provides remarkably {\em natural} 
coupling constants, whose magnitudes lie in the range 0.5 -- 1.0.  
In QMC-II, 14 or 15 of the 16 coupling constants can be regarded as 
{\em natural}.  Only the large absolute values of $c_{0500}$ for set C and 
$c_{0600}$ for sets A - C are unnatural. 
Since the coefficients, $c_{0500}$ and $c_{0600}$, are respectively 
proportional to $a_\sigma b_\sigma$ and $b_\sigma^2$, we see that those 
unnaturally large numbers are associated with the 
parameterization of the $\sigma$ mass 
in matter.  In particular, a large value for the coefficient  
$b_\sigma$ leads to unnatural values for 
$c_{0500}$ and $c_{0600}$.  In the present estimate, 
the reduction of the $\sigma$-meson mass in matter was
the sole feature of the model which could not be calculated but was put
in by hand. There is nothing within the QMC model itself which requires
$b_\sigma$ to be so large.
Therefore, the QMC model itself can be regarded as a 
{\em natural} effective field theory for nuclei. 

\subsection{\it Relationship between the QMC model and conventional nuclear models
\label{subsec:conventional}}

We shall consider the relationship between the QMC model and conventional nuclear approach. 
To this end we re-formulate the QMC model of a nucleus as a many body problem in the non-relativistic framework. 
This allows us to take the limit corresponding
to a zero range force which can be compared to the familiar Skyrme force in conventional nuclear 
physics~\cite{GUI-2}. 

In the previous section~\ref{subsubsec:totalH}, we have found the following expression for the
classical energy of a nucleon with position (${\vec R}$) and momentum (${\vec P}$) 
(see Eq.~(\ref{Hnr})):
\begin{equation}
E_{N}(\vec{R}) =  \frac{\vec{P}^{2}}{2M_N^\ast(\vec{R})}+M_N^\ast(\vec{R})+
g_{\omega }\omega (\vec{R})+V_{so}  ,  
\label{Eq-QMC1} 
\end{equation}
To get the dynamical mass $M_N^\ast(\vec{R})$
one has to solve a quark model of the nucleon (e.g., the bag model) 
in the field $\sigma (\vec{R})$.
For the present purpose, it is sufficient to know that it is well approximated
by the expression (see Eq.~(\ref{nuclm2}))
\begin{equation}
\label{Eq-QMC4}
M_N^\ast(\vec{R})=M_N-g_{\sigma }\sigma (\vec{R})+
\frac{d}{2}\left( g_{\sigma }\sigma (\vec{R})\right)^{2} ,  
\end{equation}
where $d$ is a parameter -- the bag model gives $d=0.22R_{B}$. 
The last term, which represents the response of the
nucleon to the applied scalar field -- the scalar polarizability --    
is an essential element of the QMC model. From
the numerical studies we know that the approximation 
Eq.~(\ref{Eq-QMC4}) 
is quite accurate at moderate nuclear densities 
(see section~\ref{sec:properties}).  

The energy~(\ref{Eq-QMC1}) is for one particular nucleon
moving classically in the nuclear meson fields. 
The total energy of the system is then given by the sum
of the energy of each nucleon and the energy carried by
the fields (see Eq.~(\ref{msenergy})):
\begin{eqnarray}
E_{tot}&=&\sum _{i}E_{N}(\vec{R}_{i})+E_{meson},  \\
E_{meson}&=&\frac{1}{2}\int d\vec{r} \left[
\left( {\vec \nabla}\sigma \right)^{2} + 
m_{\sigma }^{2}\sigma^{2}-\left( {\vec \nabla}\omega 
\right)^{2}-m_{\omega }^{2}\omega^{2} \right]. \label{mesonenergy} 
\end{eqnarray}

To simplify the expression for $E_{N}(\vec{R})$, we estimate the
quantity $g_{\sigma }\sigma$ using the field equations 
$\delta E_{tot}/\delta \sigma (\vec{r})=0$. 
Neglecting the velocity dependent terms, setting 
$M_N^\ast \approx M_N-g_{\sigma }\sigma$ 
and neglecting $( {\vec \nabla}\sigma )^{2}$ 
with respect to $m_{\sigma }^{2}\sigma^{2}$, we find the total energy 
\begin{equation}
E_{tot}=E_{meson}+\sum _{i}\left( M_N +
\frac{\vec{P}^{2}_{i}}{2M_N}+V_{so}(i)\right) -
 \int d\vec{r}\, \rho ^{cl}_{s}\, \left( 
g_{\sigma }\sigma -\frac{d}{2}(g_{\sigma }\sigma )^{2}\right) + 
\int d\vec{r}\, \rho ^{cl}\, g_{\omega }\omega , 
\label{Eq-QMC10} 
\end{equation}
where we define the classical densities as $\rho^{cl}(\vec{r})=\sum _{i}\delta (\vec{r}-\vec{R}_{i})$ and 
$\rho^{cl}_{s}(\vec{r})=\sum _{i}(1-{\vec{P}^{2}_{i}}/2M_N^{2}) 
\delta (\vec{r}-\vec{R}_{i})$. 
This will be the starting point for the many body formulation of
the QMC model. 

To eliminate the meson fields from the energy, we use the equations for the mesons, 
$\delta E_{tot}/\delta \sigma (\vec{r})=\delta E_{tot}/\delta \omega (\vec{r})=0$, 
and leave a system whose dynamics depends only on the nucleon coordinates. 
We first remark that, roughly speaking, the meson fields should follow
the matter density. Therefore the typical scale for the ${\vec \nabla}$ 
operator acting on $\sigma$ or $\omega$ is the thickness 
of the nuclear surface, that is about $1$ fm. Therefore, 
it looks reasonable that we can consider the second derivative terms acting on the meson fields as perturbations. 
Then, starting from the lowest order approximation, we solve the equations for the 
meson fields iteratively, and neglect a small difference between $\rho^{cl}_{s}$ and $\rho^{cl}$ 
except in the leading term. When inserted into 
Eq.~(\ref{Eq-QMC10}), the series for the meson fields generates 
$N$-body forces in the Hamiltonian. 
To complete the effective Hamiltonian, we now include the effect of
the isovector $\rho$ meson, which can be done by analogy with
the $\omega$ meson.  

The quantum effective Hamiltonian finally takes the form 
\begin{eqnarray}
 H_{QMC}&=&\sum _{i}\frac{{\overleftarrow  \nabla}_{i}\cdot 
 {\overrightarrow \nabla}_{i}}{2M_N}+
 \frac{G_{\sigma }}{2M_N^{2}}\sum _{i\neq j}\overleftarrow{\nabla_{i}}
 \delta (\vec{R}_{ij})\cdot \overrightarrow{\nabla}_{i} 
 +\frac{1}{2}\sum _{i\neq j}\left[ {\vec \nabla}^{2}_{i}\delta (\vec{R}_{ij})\right] 
 \left( \frac{G_{\omega }}{m_{\omega }^{2}}-\frac{G_{\sigma }}{m_{\sigma }^{2}}
+\frac{G_{\rho }}{m_{\rho }^{2}}\frac{\vec{\tau }_{i}.\vec{\tau }_{j}}{4}\right) \nonumber \\
&+&\frac{1}{2}\sum _{i\neq j}\delta (\vec{R}_{ij})\left[ G_{\omega }-G_{\sigma }
+G_{\rho }\frac{\vec{\tau }_{i}.\vec{\tau }_{j}}{4}\right] 
+\frac{dG_{\sigma }^{2}}{2}\sum _{i\neq j\neq k}\delta^{2}(ijk)
-\frac{d^{2}G_{\sigma }^{3}}{2}\sum _{i\neq j\neq k\neq l}\delta ^{3}(ijkl) \nonumber \\
&+&\frac{i}{4M_N^{2}}\sum _{i\neq j}A_{ij}\overleftarrow{\nabla}_{i}
\delta (\vec{R}_{ij})\times \overrightarrow{\nabla}_{i}\cdot \vec{\sigma }_{i} ,
\label{Eq-QMC28} 
\end{eqnarray}
where $G_i=g_i^{2}/m_i^{2}$ ($i= \sigma, \omega, \rho$) and $A_{ij}=G_{\sigma }+(2\mu _{s}-1)G_{\omega }+
(2\mu _{v}-1)G_{\rho }\vec{\tau }_{i}\cdot \vec{\tau }_{j}/4$.  
Here $\vec{R}_{ij}=\vec{R}_{i}-\vec{R}_{j}$ and ${\vec \nabla}_{i}$ 
is the gradient with respect to $\vec{R}_{i}$ acting on the delta
function. To shorten the equations, we used the notation
$\delta^{2}(ijk)$ for $\delta (\vec{R}_{ij})\delta (\vec{R}_{jk})$ 
and analogously for $\delta ^{3}(ijkl)$. Furthermore, we have dropped the contact interactions
involving more than 4-bodies because their matrix elements vanish
for antisymmetrized states. 

To fix the free parameters, $G_i$, 
the volume and symmetry coefficients of 
the binding energy per nucleon of infinite
nuclear matter, $E_{B}/A=a_{1}+a_{4}(N-Z)^{2}/A^{2}$, are calculated and fitted so as to produce the experimental 
values. Using the bag model with the radius 
$R_{B}=0.8$ fm and the physical masses for the mesons and 
$m_{\sigma}= 600$ MeV, one gets, in fm$^{2}$,  
$G_{\sigma }=11.97$, $G_{\omega }=8.1$ and $G_{\rho }=6.46$. 
\begin{table}
\begin{center}
\begin{minipage}[t]{16.5 cm}
\caption{QMC predictions (with $m_\sigma =600$ MeV) compared with the Skyrme force. See text for details.}
\label{tab:Tab-QMB4}
\end{minipage}
\begin{tabular}{c|c|c|c}
\hline
& QMC & QMC(N=3) & SkIII \\
\hline
$t_0$(MeV\,fm$^3$) & -1082 & -1047 & -1129 \\
$x_0$ & 0.59 & 0.61 & 0.45 \\
$t_3$(MeV\,fm$^6$) & 14926 & 12513 & 14000 \\
$3t_1+5t_2$(MeV\,fm$^5$) & 475 & 451 & 710 \\
$5t_2-9t_1$(MeV\,fm$^5$) & -4330 & -4036 & -4030 \\
$W_0$(MeV\,fm$^5$) & 97 & 91 & 120 \\
$K$(MeV) & 327 & 364 & 355 \\
\hline
\end{tabular}
\end{center}
\end{table}

It is now possible to compare the present Hamiltonian with the effective Skyrme interaction. Since, in
our formulation, the medium effects are summarized in the 3- and 4-body forces, 
we consider Skyrme forces of the same type, that is, without density dependent interactions. They are
defined by a potential energy of the form 
\begin{eqnarray}
V&=&t_{3}\sum _{i<j<k}\delta (\vec{R}_{ij})\delta (\vec{R}_{jk})
+ \sum _{i<j} \left\{ t_{0}(1+x_{0}P_{\sigma })\delta (\vec{R}_{ij}) 
+\frac{1}{4}t_{2}\, \overleftarrow{\nabla }_{ij}\cdot \delta (\vec{R}_{ij})\overrightarrow{\nabla }_{ij} \right. 
\nonumber \\
&-& \left. \frac{1}{8}t_{1}\left[ \delta (\vec{R}_{ij}){\overrightarrow  \nabla}_{ij}^{2}+
\overleftarrow{\nabla}_{ij}^{2}\delta (\vec{R}_{ij})\right] 
+\frac{i}{4}W_{0}(\vec{\sigma }_{i}+\vec{\sigma }_{j})\cdot 
\overleftarrow{\nabla}_{ij}
\times \delta (\vec{R}_{ij})\overrightarrow{\nabla}_{ij}^2  \right\},
\label{Eq-QMC53} 
\end{eqnarray}
with $\nabla_{ij}=\nabla_{i}-\nabla_{j}$. There is no 4-body
force in Eq.~(\ref{Eq-QMC53}).  Comparison of Eq.~(\ref{Eq-QMC53})
with the QMC Hamiltonian, Eq.~(\ref{Eq-QMC28}), allows one to identify
\begin{equation}
t_{0}=-G_{\sigma }+G_{\omega }-\frac{G_{\rho }}{4}, \ \ \ 
t_{3}=3dG_{\sigma }^{2}, \ \ \ x_{0}=-\frac{G_{\rho }}{2t_{0}}  .
\label{Eq-QMC100}
\end{equation} 

To simplify the situation further, we restrict our considerations to doubly
closed shell nuclei, and assume that one can
neglect the difference between the radial wave functions of the single
particle states with $j=l+1/2$ and $j=l-1/2$.  Then, by comparing the Hartree-Fock Hamiltonian obtained
from $H_{QMC}$ and that of Ref.~\cite{VAU} corresponding to the Skyrme force, we obtain the relations 
\begin{eqnarray}
3t_{1}+5t_{2}&=&\frac{8G_{\sigma }}{M_N^{2}}+4\left( \frac{G_{\omega }}{m_{\omega }^{2}}
-\frac{G_{\sigma }}{m_{\sigma }^{2}}\right)
  +3\frac{G_{\rho }}{m_{\rho }^{2}} ,\label{EqQMC110}\\
5t_{2}-9t_{1}&=&\frac{2G_{\sigma }}{M_N^{2}}+28\left( \frac{G_{\omega }}{m_{\omega }^{2}}
-\frac{G_{\sigma }}{m_{\sigma
 }^{2}}\right) -3\frac{G_{\rho }}{m_{\rho }^{2}} , \label{5t2}\\
W_{0}&=&\frac{1}{12M_N^{2}}\left( 5G_{\sigma }+5(2\mu _{s}-1)G_{\omega }
+\frac{3}{4}(2\mu _{v}-1)G_{\rho }\right) . \label{w0}
\end{eqnarray}

In Table~\ref{tab:Tab-QMB4}, we compare our results with the parameters
of the force SkIII~\cite{FRI}, which is considered a good representative 
of density independent effective interactions. 
Instead of $t_1$ and $t_2$, we show the combinations $3t_{1}+5t_{2}$,
which controls the effective mass, and $5t_{2}-9t_{1}$, which controls
the shape of the nuclear surface~\cite{VAU}.  From the table, 
one sees that the
level of agreement with SkIII is very impressive. An important point is that the spin orbit strength 
$W_{0}$ comes out with approximately the correct value. 
The middle column (N=3) shows the results when
we switch off the 4-body force. The main change is a decrease of the
predicted 3-body force. Clearly this mocks up the effect of the attractive
4-body force which may then appear less important. However, this
is misleading if we look at the incompressibility of nuclear matter, $K$, 
which decreases by as much as $37$ MeV when we restore this 4-body force. 

Now one can recognize a remarkable agreement between the
phenomenologically successful Skyrme force (SkIII) 
and the effective interaction
corresponding to the QMC model - a result which  
suggests that the response of nucleon internal structure to the nuclear
medium does indeed play a vital role in nuclear structure.

\subsection{\it Relationship between the QMC model and Quantum Hadrodynamics 
\label{subsec:QMCQHD}}

Next, let us examine the relationship between the QMC-I model 
and QHD in detail~\cite{SAI-2}. 
The main difference between QMC and QHD at the {\em hadronic}
level lies in the dependence of the nucleon mass on the
scalar field in matter (see Eq.~(\ref{nuclm2})).  By re-defining the scalar
field~\cite{MUL,SAI-2}, the QMC Lagrangian density can be cast into a form
similar to that of a QHD-type mean-field model, in which the nucleon mass
depends on the scalar field linearly (as in QHD), 
with self-interactions of the
scalar field.  
It should be emphasized here that the QMC model 
explicitly provides how the quark structure changes 
inside the in-medium nucleon.  Such information is 
necessarily not part of any QHD-type 
mean-field model generated through the re-definition of the scalar field. 
 
In the QMC model, the nucleon mass in matter is given by a function
of $\sigma$, $M_{N,QMC}^\ast (\sigma)$, obtained from a quark model of
the nucleon, while in QHD the mass depends on a scalar
field linearly, $M_{N,QHD}^\ast = M_{N} - g_0 \phi$, where $\phi$ is a 
scalar field in a QHD-type model.  Hence, to transform the QMC into a
QHD-type model, we may re-define the scalar field of QMC as 
\begin{equation}
g_0 \phi(\sigma) = M_N - M_{N,QMC}^\ast(\sigma) ,
\label{redef}
\end{equation}
where $g_0$ is a constant chosen so as to normalize the scalar
field $\phi$ in the limit $\sigma \to 0$: $\phi(\sigma) = \sigma +
{\cal O}(\sigma^2)$. Thus, $g_0$ is given by 
$g_0 = - ( \partial M_{N,QMC}^\ast/\partial \sigma )_{\sigma = 0}$, and we find $g_0 = g_\sigma$. 

The contribution of the scalar field to the total energy, $E_{scl}$, can now 
be rewritten in terms of the new field $\phi$ as
\begin{equation}
E_{scl} = \frac{1}{2} \int d\vec{r} \, [({\vec \nabla} \sigma)^2 + m_{\sigma}^2
\sigma^2 ] 
= \int d\vec{r} \, \left[ \frac{1}{2}h(\phi)^2({\vec \nabla} \phi)^2 + U_s(\phi) \right] ,
\label{es}
\end{equation}
where $U_s$ describes self-interactions of the scalar field: 
\begin{equation}
U_s(\phi) = \frac{1}{2} m_{\sigma}^2 \sigma(\phi)^2 \ \ \ \mbox{and} \ \ \ 
h(\phi) = \left( \frac{\partial \sigma}{\partial \phi} \right)
= \frac{1}{m_\sigma \sqrt{2U_s(\phi)}}
\left( \frac{\partial U_s(\phi)}{\partial \phi} \right) .
\label{Us}
\end{equation}
Note that in uniformly distributed nuclear matter the derivative
term in $E_{scl}$ does not contribute.  
Thus this term only affects the properties of finite nuclei~\cite{MUL}.
Now, QMC {\em at the hadronic level} can be 
re-formulated in terms of the new scalar 
field $\phi$, and what we obtain has the same form as QHD, with the
non-linear scalar potential $U_s(\phi)$ and the coupling $h(\phi)$
to the gradient of the scalar field.  (Note that because this re-definition of
the scalar field does not involve the vector interaction, the energy due to 
the $\omega$ field is not modified.) 

In general, the in-medium nucleon mass may be given by a complicated
function of the scalar field.  However, in the QMC with the bag model, the mass can be
parameterized by a simple expression up to 
${\cal O}(g_\sigma^2)$ (see also Eq.~(\ref{nuclm2})): 
\begin{equation}
M_N^\ast / M_N \simeq 1 - a y + b y^2 ,
\label{mparam}
\end{equation}
where $y~(= g_\sigma \sigma / M_N)$ is a dimensionless scale and $a$
and $b$ are (dimensionless) parameters. This parameterization is 
accurate at moderate nuclear densities 
(see for details section~\ref{subsubsec:scaling}).   

Once the parameters $a$ and $b$ are fixed, we can easily re-define the
scalar field using Eq.~(\ref{redef}): 
\begin{equation}
g_0 = a g_\sigma, \ \ 
\phi(\sigma) = \sigma - d \sigma^2 \ \ \mbox{and} \ \ 
\sigma(\phi) = \frac{1-\sqrt{1-4d\phi}}{2d} ,
\label{qmcphi}
\end{equation}
with $d = bg_\sigma /aM_N$.  This satisfies the condition that $\sigma \to
0$ in the limit $\phi \to 0$.  The non-linear potential is thus calculated
as
\begin{eqnarray}
U_s(\phi) &=& \frac{m_\sigma^2}{2} \left( \frac{\sigma(\phi) - \phi}{d}
\right) , \nonumber \\
&=& \frac{m_\sigma^2}{2} \phi^2
+ g_\sigma r \left( \frac{m_\sigma^2}{M_N} \right) \phi^3
+ \frac{5}{2} g_\sigma^2 r^2 \left( \frac{m_\sigma}{M_N} \right)^2
\phi^4 + {\cal O}(g_\sigma^3) ,
\label{qmcnonlex}
\end{eqnarray}
where $r = b/a$.
By contrast, the standard form of the non-linear scalar potential is usually given
by
\begin{equation}
U_s(\phi) = \frac{1}{2} m_{\sigma}^2 \phi^2 + \frac{\kappa}{6} \phi^3
+ \frac{\lambda}{24} \phi^4 .
\label{stUs}
\end{equation}
It is well known that the non-linear scalar potential given in 
Eq.~(\ref{stUs}) is vital to reproduce the bulk properties of finite
nuclei and nuclear matter in relativisitic mean field models. 

We can estimate the parameters $\kappa$ and $\lambda$ in the QMC by
comparing Eqs.~(\ref{qmcnonlex}) and~(\ref{stUs}). Because 
the parameters $a$ and $b$ are respectively about unity and $0.2-0.5$ 
(see Table II in Ref.~\cite{SAI-2}), one finds $\kappa \sim 20 - 40$ (fm$^{-1}$) and
$\lambda \sim 80 - 400$.  Since the QMC predicts that both $a$ and $b$ are
{\em always} positive, we can expect that the quark
substructure of the in-medium nucleon provides a non-linear potential
with {\em positive} $\kappa$ and {\em positive} $\lambda$ in the
QHD-type mean-field model.  By contrast, the parameters $\kappa$
and $\lambda$ phenomenologically determined in RMF models 
take wide range of values (for example, Ref.~\cite{FUR-1,ZIM,RMF}), 
and the potentials in those models are then quite different from one another
for large $|\phi|$. However, from the point of 
view of field theory, the coefficient of 
quartic term should be positive.  The QMC model always 
yields a positive $\lambda$.

In general, the phenomenological potential may
consist of a part that is due to the quark substructure of the
nucleon, as well a piece involving inherent self-couplings 
of the scalar field in matter.
It would therefore be very intriguing if the potential due to 
the internal structure
of the nucleon could be inferred from analyses of experimental
data in the future.   

\subsection{\it Modified quark-meson coupling model
\label{subsec:modfqmc}}

As we have seen in the previous sections, 
the QMC model provides a simple and attractive 
framework to incorporate the quark structure of the nucleon in a nucleus. 
In the original QMC model, we chose to hold the bag constant 
at its free space value,   
even for nuclear matter.  When a
nucleon bag is put into the nuclear medium, the bag as a whole reacts to the
environment. As a result, the bag constant might be modified. 
There is little doubt that at sufficiently high densities, the bag  
eventually melts away and quarks and gluons become 
the appropriate degrees of freedom.  
Thus, it seems reasonable that the bag constant be modified and
decrease as the density increases. 
Moreover, the MIT bag constant may be related to the energy 
associated with chiral symmetry restoration, and 
the in-medium bag constant may drop relative to its free space value.  
This way of thinking leads to a modified version of the 
QMC model --  called the modified quark-meson 
coupling (MQMC) model~\cite{MUL,JIN-1,MUL-2,LU-1}. 
In principle, the parameter $z$ may also be modified in the nuclear medium.
However, unlike the bag constant, it is unclear how $z$ changes with
the density as $z$ is not directly related to chiral symmetry. Here 
we assume that the medium modification of $z$ is small at low and
moderate densities.

To reflect this physics, we introduce 
a direct coupling between the bag constant, $B$, in matter and the scalar mean field
\begin{equation}
{B\over B_0} = \left[ 1
- g_\sigma^B\, {4 \over \delta} {\sigma\over M_N} \right]^\delta\ ,
\label{an-dir}
\end{equation}
where $g_\sigma^B$ and $\delta$ are new, real positive parameters and 
$B_0$ is the free value of the bag constant.  Note 
that $g_\sigma^B$ differs from $g_\sigma^q$
(or $g_\sigma$) and that when $g_\sigma^B = 0$ the original QMC model is
recovered.  
This direct coupling is partially motivated from
considering a non-topological soliton model for the nucleon where a
scalar soliton field provides the confinement of the quarks. Roughly
speaking, the bag constant in the MIT bag model mimics the effect of
the scalar soliton field in the soliton model. 
In the limit $\delta \rightarrow
\infty$, Eq.~(\ref{an-dir}) reduces to an exponential form with a
single parameter $g_\sigma^B$, ${B / B_0} = \exp[- 4 g^B_\sigma \sigma/M_N]$.
In the limit of zero current quark mass and $g_\sigma=0$,
the nucleon mass scales like
$B^{1/4}$ from dimensional arguments. Then, from Eq.~(\ref{an-dir}) we get
$ M_N^*/M_N  = \left( B/B_0  \right)^{1/4} = \left[ 1
- 4g_\sigma^B \sigma / \delta M_N \right]^{\delta/4}$.
One observes that the linear $\sigma$-nucleon coupling is just
$g_\sigma^B$ while $\delta$ controls the non-linearities. For
$\delta=4$, the non-linearities vanish and the QHD-I is recovered 
but with a density dependent bag radius. 

Alternative is a scaling
model, which relates the in-medium bag constant to the in-medium
nucleon mass directly through 
\begin{equation}
{B\over B_0} = \left[ M_N^*\over M_N \right]^\kappa\ ,
\label{an-br}
\end{equation}
where $\kappa$ is a real positive parameter ($\kappa=0$ corresponds 
to the usual QMC model). 
Note that in this model, the effective nucleon mass $M_N^*$ and the bag
constant $B$ are determined self-consistently .

One notices that both Eqs.~(\ref{an-dir}) and~(\ref{an-br}) give rise
to a reduction of the bag constant in nuclear medium relative to its
free-space value. While the scaling model is characterized by a single
free parameter $\kappa$, it leads to a complicated and implicit 
relation between the bag constant and the scalar mean field. On the other
hand, the direct coupling model features a straightforward coupling
between the bag constant and the scalar mean field, which, however, 
introduces two free parameters, $g_\sigma^B$ and $\delta$.

The most important feature is that the reduction of $B$ relative to $B_0$ 
leads to a decrease of $M_N^*/M_N$ 
and an increase of the repulsive vector potential  
relative to their values in the original QMC model.  In the MQMC model, 
the reduction of the bag constant in nuclear medium provides a
new source of attraction as it effectively reduces $M_N^*$. Consequently, 
additional vector field strength is required to obtain the correct 
saturation properties of nuclear matter.

It can be seen that when the bag constant is
reduced significantly in nuclear matter, the resulting magnitudes for 
$M_N^* - M_N$ and $U_{\rm v} \equiv 
g_\omega {\omega}$ are qualitatively different from those 
obtained in the usual QMC model and could be $\sim 700$ MeV and $\sim 200$ MeV, 
respectively, for some parameter sets~\cite{JIN-1}. 
These values imply that large and canceling scalar 
and vector potentials for the nucleon can be produced in nuclear matter and they are
comparable to those suggested by Dirac phenomenology~\cite{HAM}.  
These potentials also imply a strong 
nucleon spin-orbit potential and the essential features of
relativistic nuclear phenomenology are recovered. The corresponding 
results for the nuclear matter incompressibility $K$ are, however,  
larger than the value in the usual QMC model but 
smaller than that in QHD-I. The relationship among the QMC, MQMC and QHD is clarified and discussed in 
Ref.~\cite{MUL,MUL-2,LU-1}. 

In the usual QMC model, the bag radius decreases slightly and the
quark root-mean-square (rms) radius increases slightly in nuclear
matter with respect to their free-space values.  When the bag constant 
drops relative to its free-space value, the bag pressure decreases and 
hence the bag radius increases in the medium.  When the reduction of the
bag constant is significant, the bag radius in saturated nuclear matter 
is $25 - 30\%$ larger than its free-space value. The quark rms radius 
also increases with density, with essentially the same rate as for the 
bag radius. This implies a ``swollen'' nucleon in nuclear medium. 
However, a $25 - 30\%$ increase in
the nucleon size is generally considered too large 
in comparison with the conclusion drawn from  
electron scattering data~\cite{SIC,JOU}.  
The MQMC model has also been used to study many nuclear phenomena. 

\subsection{\it Variants of the QMC model  
\label{subsec:variant}}

In the original QMC model, the MIT bag model was used to describe the nucleon 
structure.  Although the basic idea of the QMC model is not altered, it is possible to 
replace the bag model with the constituent quark model. 
Toki {\it et al.}~\cite{TOK} have proposed a quark mean field (QMF) model for nucleons in nuclei, 
where the constituent quarks interact with the meson fields created by other nucleons 
and hence change the nucleon properties in nuclei (see also Ref.~\cite{FOCK}).  They have found very good nuclear matter 
properties with the use of the nonlinear self-energy terms for the meson fields. 
In particular, the spin-orbit splitting in finite nuclei becomes large due to 
the large reduction of the in-medium nucleon mass.  In this picture, the nucleon size increases 
by about $7\%$ at the normal nuclear matter density. 
This QMF model has been applied to various nuclear phenomena, for example, the effect of density-dependent 
coupling constants for finite nuclei~\cite{TAN}, $\Lambda$-hyper 
nuclei~\cite{SHE-1}, 
pentaquark $\Theta^+$ in nuclear matter and 
$\Theta^+$ hypernuclei~\cite{SHE-2}. 

Krein {\it et al.}~\cite{FOCK} have also proposed a similar model, in which the constituent quark model is used to 
describe a nucleon. In addition to that, the exchange contribution and pionic effect have been calculated. 
As in QHD, the exchange effect is repulsive and the coupling of the mesons directly to the quarks in the 
nucleons introduces a new effect on the exchange energies that provides an extra repulsive contribution to the 
binding energy. In this approach, the pionic effect is not small. 

Recently, the QMC model has been extended to a model based on $SU(3)_L\times SU(3)_R$ symmetry and scale 
invariance~\cite{WAN}.  In this model it is possible to study strange hadronic matter, multi-strange hadronic system 
or strangelets, including $\Lambda$, $\Sigma$ and $\Xi$ hyperons. The properties of hyperons in matter have been 
discussed with several types of confining potentials. The phase transition corresponding to chiral symmetry 
restoration in high density nuclear matter is also investigated. 

Using the Nambu-Jona-Lasinio model to describe the nucleon as 
a quark-diquark state, it is also possible to discuss the stability of 
nuclear matter, based on the QMC idea~\cite{bentz}.
This model has been applied to interesting nuclear phenomena including 
the high energy lepton-nucleus scattering~\cite{bentz2,Cloet:2005rt}. 

\section{Properties of nuclei 
\label{sec:properties}}

In the previous section, we presented the basic idea of the QMC model, and saw various models relating to 
it. In this section, we are now in a position to report 
numerical results for various nuclear phenomena. 

\subsection{\it Nuclear matter properties  - saturation properties and incompressibility  
\label{subsec:matter}}

Let us first consider the simplest version of the QMC model (QMC-I), which includes only the nucleon, $\sigma$ 
and $\omega$ mesons, and calculate the properties for symmetric 
nuclear matter~\cite{GUI-1,SAI-1,SAI-3,SAI-8}. 
In this case, because of the uniform matter distribution, the sources of the meson fields 
are constant and can be related to the nucleon Fermi momentum 
$k_F$ as 
\begin{eqnarray}
\rho_B &=& \frac{4}{(2\pi)^3}\int d\vec{k}\ \theta (k_F - k) = 
\frac{2 k_F^3}{3\pi^2} , 
\label{rhoB}\\
\rho_s &=& \frac{4}{(2\pi)^3}\int d\vec{k} \ \theta (k_F - k)
\frac{M_N^\ast }{\sqrt{M_N^{\ast 2}+\vec{k}^2}},
\label{rhos}
\end{eqnarray}
where $M_N^{\ast}$ is the constant value of the effective nucleon 
mass at a given density.  We here adopt the MIT bag model, and the bag constant $B$ and the 
parameter $z_0$ (which accounts for the sum of the c.m. and 
gluon fluctuation corrections) are determined so as to produce the free nucleon mass under the stationary 
condition with respect to the free bag radius, $R_N$ 
(hereafter $R_N$ denotes the nucleon bag radius in free space, see Eqs.~(\ref{EMass}) and~(\ref{EMass2})).  
In the following we 
treat the free bag radius as a variable parameter to test 
the sensitivity of our 
results to the free nucleon size.  The results for $B$ and $z_0$ are 
shown in Table~\ref{tab:b,z}.  The free quark mass may be chosen to be $m_q$ = 0, 5, 
10 MeV. 
\begin{table}
\begin{center}
\begin{minipage}[t]{16.5 cm}
\caption{Bag constant and parameter $z_0$ for several bag 
radii and quark masses.}
\label{tab:b,z}
\end{minipage}
\begin{tabular}{c|ccc|ccc|ccc}
\hline
$m_q$(MeV) & & 0 & & & 5 & & & 10 & \\
\hline
$R_N$(fm) & 0.6 & 0.8 & 1.0 & 0.6 & 0.8 & 1.0 & 0.6 & 0.8 & 1.0 \\
\hline
$B^{1/4}$(MeV) &211.3&170.3&144.1&210.9&170.0&143.8&210.5&169.6&143.5 \\
$z_0$     &3.987&3.273&2.559&4.003&3.295&2.587&4.020&3.317&2.614 \\
\hline
\end{tabular}
\end{center}
\end{table}
Let (${\sigma}, {\omega}$) be the constant 
mean-values of the meson fields which are given by 
\begin{eqnarray}
{\omega}&=&\frac{g_\omega \rho_B}{m_\omega^2} , \label{omgf}\\
{\sigma}&=&\frac{g_\sigma }{m_\sigma^2}C_N({\sigma})
\frac{4}{(2\pi)^3}\int d\vec{k} \ \theta (k_F - k) 
\frac{M_N^{\ast}}{\sqrt{M_N^{\ast 2}+\vec{k}^2}} , \label{sigf}
\end{eqnarray}
where $C_N(\sigma)$ is now the constant value of the scalar density 
ratio defined by Eq.~(\ref{msdef}).  

Once the self-consistency equation for the ${\sigma}$ has been solved, 
one can evaluate the total energy per nucleon 
\begin{equation}
E^{tot}/A=\frac{4}{(2\pi)^3 \rho_B}\int d\vec{k} \ 
\theta (k_F - k) \sqrt{M_N^{\ast 2}+
\vec{k}^2}+\frac{m_\sigma^2 {\sigma}^2}{2 \rho_B}+
\frac{g_\omega^2\rho_B}{2m_\omega^2} .
\label{toten}
\end{equation}
We then determine the coupling constants, $g_{\sigma}$ and $g_{\omega}$, so as 
to fit the binding energy ($-15.7$ MeV) per nucleon at the saturation 
density, $\rho_0$ = 0.15 fm$^{-3}$ ($k_F^0$ = 1.305 fm$^{-1}$), for 
symmetric nuclear matter.  
The coupling constants and some calculated properties 
of nuclear matter (for $m_{\sigma}=550$ MeV and 
$m_{\omega}=783$ MeV) at the saturation density are listed in 
Table~\ref{tab:c.c.-I}. 
The last three columns show the relative changes (from their values at 
zero density) of the bag radius 
($\delta R_N^{\ast}/R_N$), the lowest eigenvalue ($\delta x_0^{\ast} /x_0$) 
and the rms radius of the nucleon calculated by the quark wave function 
($\delta r_q^{\ast}/r_q$) at saturation density. 
\begin{table}
\begin{center}
\begin{minipage}[t]{16.5 cm}
\caption{Coupling constants and calculated properties for 
symmetric nuclear matter at normal nuclear matter density (QMC-I).  
The effective nucleon mass, $M_N^{\ast}$, and the nuclear 
incompressibility, $K$, are quoted in MeV. 
The bottom row is for QHD~\cite{QHD}.}
\label{tab:c.c.-I}
\end{minipage}
\begin{tabular}{c|c|cccccccc}
\hline
$m_q$(MeV)&$R_N$(fm)&$g_{\sigma}^2/4\pi$&$g_{\omega}^2/4\pi$&
$M_N^{\ast}$&$K$&$\delta R_N^{\ast}/R_N$&$\delta x_0^{\ast} /x_0$&$\delta 
r_q^{\ast}/r_q$ \\
\hline
   &  0.6 & 5.84 & 6.29 & 730 & 293 & -0.02 & -0.13 & 0.01 \\
 0 &  0.8 & 5.38 & 5.26 & 756 & 278 & -0.02 & -0.16 & 0.02 \\
   &  1.0 & 5.04 & 4.50 & 774 & 266 & -0.02 & -0.19 & 0.02 \\
\hline
   &  0.6 & 5.86 & 6.34 & 729 & 295 & -0.02 & -0.13 & 0.01 \\
 5 &  0.8 & 5.40 & 5.31 & 755 & 280 & -0.02 & -0.16 & 0.02 \\
   &  1.0 & 5.07 & 4.56 & 773 & 267 & -0.02 & -0.19 & 0.02 \\
\hline
   &  0.6 & 5.87 & 6.37 & 728 & 295 & -0.02 & -0.13 & 0.02 \\
10 &  0.8 & 5.42 & 5.36 & 753 & 281 & -0.02 & -0.16 & 0.02 \\
   &  1.0 & 5.09 & 4.62 & 772 & 269 & -0.02 & -0.18 & 0.03 \\
\hline
 QHD &  --  & 7.29 & 10.8 & 522 & 540 & & &  \\
\hline
\end{tabular}
\end{center}
\end{table}

The most notable fact is that the calculated incompressibility, $K$, 
is well within the experimental range: $K = 200 \sim 
300$ MeV. Although the bag radius shrinks a little at finite density, the rms radius 
of the quark wave function actually {\em increases} by a few percent 
at saturation density. This seems consistent with 
the observed results~\cite{SIC,JOU} 
(see also section~\ref{subsec:formf}). 
The saturation mechanism of the binding energy of nuclear matter is novel. 
In QHD, the $\sigma$ and $\omega$ couplings to the nucleon are really constant, while in the 
QMC the $\sigma$-N coupling is reduced by the effect of the internal structure of the nucleon, that is 
responsible for the saturation~\cite{GUI-1,SAI-1}. 

Next let us consider the QMC-II model, including the effect of the meson structure 
(see section~\ref{subsubsec:mesonstr}). For asymmetric nuclear matter ($Z\neq N$), we take the Fermi momenta 
for protons and neutrons to be $k_{F_i}$ ($i=p$ or $n$). This is 
defined by $\rho_i = k_{F_i}^3 / (3\pi^2)$, where $\rho_i$ is the 
density of protons or neutrons, and the total baryon density, 
$\rho_B$, is then given by $\rho_p + \rho_n$.  From the Lagrangian density 
Eq.~(\ref{qmc-2}), 
the total energy per nucleon is then written 
\be
E^{tot}/A = \frac{2}{\rho_B (2\pi)^3}\sum_{i=p,n}\int^{k_{F_i}} 
d\vec{k} \ \sqrt{M_i^{\ast 2} + \vec{k}^2} + \frac{m_{\sigma}^{\ast 2}}
{2\rho_B}{\sigma}^2 
+ \frac{g_{\omega}^2}
{2m_{\omega}^{\ast 2}}\rho_B + \frac{g_{\rho}^2}{8m_{\rho}^{\ast 2}
\rho_B} \rho_3^2 , \label{tote}
\ee
where the value of the $\omega$ field is determined by baryon number 
conservation, Eq.~(\ref{omgf}), with the effective mass 
$m_\omega^\ast$ instead of the free mass, 
and the $\rho$ field value by the difference in proton and neutron 
densities, $\rho_3 = \rho_p - \rho_n$, as ${b} = g_{\rho} \rho_3 / 
(2m_{\rho}^{\ast 2})$. 

In contrast, the $\sigma$ field is given by a self-consistency 
condition (instead of Eq.~(\ref{sigf})): 
\bea
{\sigma} &=& \frac{2g_\sigma}{(2\pi)^3 m_{\sigma}^{\ast 2}} 
\left[ \sum_{i=p,n} C_i({\sigma}) 
\int^{k_{F_i}} d\vec{k} \ \frac{M_i^{\ast}}
{\sqrt{M_i^{\ast 2} + {\vec k}^2}} \right] 
 + g_\sigma \left( \frac{m_\sigma}{m_\sigma^\ast} \right) 
 \left[ a_\sigma - 2 b_\sigma (g_\sigma {\sigma} ) \right] {\sigma}^2  \nn \\
&-& \frac{2}{3} \left( \frac{g_\sigma}{m_\sigma^{\ast 2}} \right) 
\left[ \frac{g_\omega^2 \rho_B^2}
{m_\omega^{\ast 3}} \Gamma_{\omega/N} C_\omega({\sigma}) + \frac{g_\rho^2 \rho_3^2}{4 m_\rho^{\ast 3}} 
\Gamma_{\rho/N} C_\rho({\sigma}) \right]. 
\label{scc2}
\eea
Here we used Eq.~(\ref{sigmas}) to describe the $\sigma$ mass in matter. Note that $\Gamma_{v/N}$ ($v=\omega$ 
or $\rho$) is the correction factor to the scalar density ratio 
(see around Eq.~(\ref{deriv3})). 
In actual calculations using the bag model, 
we find that $\Gamma_{\omega, \rho/N}$ 
= 0.9996, which may be replaced by unity for practical 
purposes~\cite{SAI-3}.  

Next we must fix the two parameters in the parameterization for the 
$\sigma$ mass in matter (see Eq.~(\ref{sigmas})).  For example, we here 
consider three parameter sets: 
\begin{description}
\item[](A) $a_\sigma = 3.0 \times 10^{-4}$ (MeV$^{-1}$),  
$b_\sigma = 100 \times 10^{-8}$ (MeV$^{-2}$), 
\item[](B) $a_\sigma = 5.0 \times 10^{-4}$ (MeV$^{-1}$), 
$b_\sigma = 50 \times 10^{-8}$ (MeV$^{-2}$), 
\item[](C) $a_\sigma = 7.5 \times 10^{-4}$ (MeV$^{-1}$), 
$b_\sigma = 100 \times 10^{-8}$ (MeV$^{-2}$). 
\end{description}
The parameter sets A, B and 
C give about 2\%, 7\% and 10\% decreases of the $\sigma$ mass 
at saturation density, respectively.

We then determine the coupling constants.  As in the QMC-I model, 
$g_{\sigma}^2$ and $g_{\omega}^2$ are fixed to fit the binding energy 
at the saturation density for 
symmetric nuclear matter.  Furthermore, the $\rho$-meson coupling constant is 
used to reproduce the bulk symmetry energy, 35 MeV~\cite{HOR-1}.  
We take $m_\rho$ = 770 MeV. The coupling constants 
and some calculated properties for symmetric nuclear matter are 
listed in Table~\ref{tab:c.c.-II}.  
In Fig.~\ref{fig:EPN12}, we show the total energy per nucleon for 
iso-symmetric nuclear matter calculated in QMC-I and three types of QMC-II.  
\begin{figure}
\epsfysize=9.0cm
\begin{center}
\begin{minipage}[t]{8 cm}
\hspace*{-2.5cm}
\epsfig{file=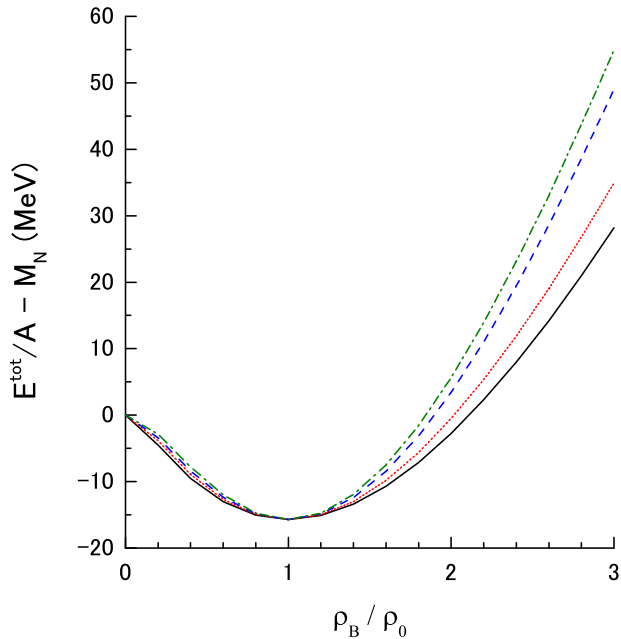,height=10cm}
\end{minipage}
\begin{minipage}[t]{16.5 cm}
\caption{Energy per nucleon for iso-symmetric nuclear matter (for $m_q = 5$ MeV and $R_N = 0.8$ fm). 
The solid curve is for QMC-I, while the dotted (dashed) [dot-dashed] curve is for type A (B) [C] 
of QMC-II.   
}
\label{fig:EPN12}
\end{minipage}
\end{center}
\end{figure}
\begin{table}[htbp]
\begin{center}
\begin{minipage}[t]{16.5 cm}
\caption{Coupling constants and calculated properties for 
symmetric nuclear matter at normal nuclear matter density (QMC-II).  We take $m_q$ = 5 MeV, 
$R_N$ = 0.8 fm and $m_\sigma =550$ MeV.  
The effective nucleon mass, $M_N^{\ast}$, and the nuclear 
compressibility, $K$, are quoted in MeV.}
\label{tab:c.c.-II}
\end{minipage}
\begin{tabular}[t]{c|cccccccc}
\hline
type & $g_{\sigma}^2/4\pi$&$g_{\omega}^2/4\pi$&$g_\rho^2/4\pi$&
$M_N^{\ast}$&$K$&$\delta R_N^{\ast}/R_N$&$\delta x_0^{\ast}/x_0$&$\delta 
r_q^{\ast}/r_q$ \\
\hline
 A &  3.84 & 2.70 & 5.54 & 801 & 325 & -0.01 & -0.11 & 0.02 \\
 B &  3.94 & 3.17 & 5.27 & 781 & 382 & -0.01 & -0.13 & 0.02 \\
 C &  3.84 & 3.31 & 5.18 & 775 & 433 & -0.02 & -0.14 & 0.02 \\
\hline
\end{tabular}
\end{center}
\end{table}
We note that the nuclear incompressibility is higher than that in QMC-I.  
However, it is still much lower than in QHD.  As in QMC-I, the bag radius of the 
nucleon shrinks a little, while its rms radius swells a little.  

We here comment on other non-relativisitic or relativistic calculations for nuclear matter. 
In non-relativistic approach, 
the Brueckner-Bethe-Goldstone formalism~\cite{BBG} or variation chain summation methods~\cite{VCS} with 
an effective two-body interaction like the Argonne $v_{18}$ are often used to study the nuclear bulk 
properties.  However, it is well known that a phenomenological three-body force like Urbana IX is vital to 
produce the empirical values of the saturation energy at $\rho_0$ and the incompressibility.  As  
discussed in section~\ref{subsec:conventional}, the QMC model in non-relativistic limit has already involved 
many-body interactions which are generated by the response of the nucleon to the applied scalar field -- 
the scalar polarizability.  Thus, it would be very interesting to compare the role of three-body forces in the 
conventional calculations with the contact interactions, 
Eq.~(\ref{Eq-QMC53}), in the non-relativistic QMC.  

The relativisitic Dirac-Brueckner approach is also often used to study the bulk nuclear matter properties.  
Using it, 
Dalen {\it et al.}~\cite{dalen} have calculated various properties of iso-asymmetric nuclear matter.  They have 
obtained a reasonable saturation property for symmetric nuclear matter and calculated the 
mean-field effective coupling constants, which decrease with increasing nuclear density. 
We note that the origin of such changes is different from that of the coupling-constant change in QMC. 
In contrast, 
using the relativistic Brueckner-Hartree-Fock approach with explicit intermediate negative energy states 
(in the Gross approximation), Jong and Lenske~\cite{Jong} have pointed out that the intermediate anti-nucleons 
provide a large vector repulsion and the effective mean-field coupling constants increase with increasing nuclear 
density.  Furthermore, to obtain a reasonable saturation result, it is vital to 
modify the nucleonic off-shell form factor, which may relate to the internal structure of the nucleon, compared 
with the free one. 
Although it may be uncertain whether the anti-nucleon degrees of freedom is relevant in the intermediate state, 
it would be very interesting that the in-medium form factor plays the important role even in 
the relativistic Brueckner-Hartree-Fock calculation. 

\subsubsection{\it A new scaling formula for hadron masses in nuclear matter 
\label{subsubsec:scaling}}

We have discussed the effective nucleon and $\omega$ meson 
masses in detail in sections~\ref{subsubsec:nmass} 
and~\ref{subsubsec:mesonstr}.  Here we describe a new scaling 
formula for hadron masses in matter that is predicted by the QMC model. 

In the QMC-II model, the $\sigma$ and $\omega$ masses are respectively given by Eqs.~(\ref{sigmas}) and~(\ref{vmm}). 
The nucleon mass is also expressed by Eq.~(\ref{nuclm2}). 
At small density, the scalar field is approximated 
\begin{equation}
g_\sigma {\sigma} \simeq 200 \ \mbox{(MeV)} \ \left( \frac{\rho_B}{\rho_0}\right) , \label{appv}
\end{equation}
and we find 
\begin{equation}
\left( \frac{M_N^\ast}{M_N} \right) \simeq 1 - 0.21 
\left( \frac{\rho_B}{\rho_0} \right) \ \ \ \mbox{and} \ \ \ 
\left( \frac{m_v^\ast}{m_v} \right) \simeq 1 - 
0.17 \left( \frac{\rho_B}{\rho_0} \right). 
\label{nstr2}
\end{equation}

Since the ratio of the quark-scalar densities for 
the nucleon, $C_N$ (see Eq.~(\ref{nuclm2})), 
is well approximated by a linear function of $g_\sigma \sigma$
\begin{equation}
C_N({\sigma}) = 1 - a_N \times (g_\sigma {\sigma}) , \label{cnpar}
\end{equation}
Eq.~(\ref{nmderiv}) can be solved easily and the nucleon 
mass is then expressed as   
\begin{equation}
M_N^\ast = M_N - g_\sigma 
\left[ 1 - \frac{a_N}{2}(g_\sigma {\sigma}) \right]\sigma . \label{nmapp}
\end{equation}
It is also true that the ratio for the vector meson can be  
well described by a similar, linear function of $g_\sigma \sigma$: 
\begin{equation}
C_v(\sigma) = 1 - a_v \times (g_{\sigma} \sigma) . 
\label{paramCV}
\end{equation}

In the QMC model, it is possible to calculate masses of other hadrons in matter.  
In particular, there is a considerable interest in studying the masses 
of hyperons in medium -- e.g. $\Lambda$, $\Sigma$ and $\Xi$.  For the hyperons 
themselves we again use the MIT bag model and  
assume that the strange quark in the hyperon does not directly couple 
to the scalar field in MFA, as one would expect if the $\sigma$ meson
represented a two-pion-exchange potential. It is also assumed
that the addition of a single hyperon to nuclear matter of density 
$\rho_B$ does not alter the values of the scalar and vector mean-fields -- 
that is, we take the local-density approximation to the hyperons.  For example, 
the mass of the strange quark $m_s$ is taken to be $m_s$ = 250 MeV, and 
new $z$-parameters in the mass formula are again introduced to 
reproduce the free hyperon masses: $z_\Lambda$ = 3.131, $z_\Sigma$ = 2.810, 
and $z_\Xi$ = 2.860.  Using those parameters, we can calculate the 
masses of $\Lambda$, $\Sigma$ and $\Xi$ in nuclear matter 
and finite nuclei~\cite{SAI-3}.

In general, we find that the effective hadron mass in medium is given by
\begin{eqnarray}
M_j^* &=& M_j + \left( \frac{\partial M_j^*}{\partial \sigma} 
\right)_{\sigma=0} \sigma + \frac{1}{2} \left( \frac{\partial^2 M_j^*}
{\partial \sigma^2} \right)_{\sigma=0} \sigma^2 + \cdots , \nn \\
  &\simeq& M_j - \frac{n_q}{3} g_\sigma \Gamma_{j/N} \sigma 
  - \frac{n_q}{6} g_\sigma \Gamma_{j/N} C_j^\prime(0) \sigma^2 , 
\label{hm}
\end{eqnarray}
where $j$ stands for $N$, $\omega$, $\rho$, $\Lambda$, $\Sigma$, $\Xi$, etc., 
$n_q$ is the number of light quarks in the hadron $j$, 
$\Gamma_{j/N} = S_j(0)/S_N(0)$ with the quark-scalar density, $S_j$, in 
$j$, and the scalar density ratio, $C_j(\sigma) = S_j(\sigma)/S_j(0)$.  

Using Eqs.~(\ref{appv}) and~(\ref{hm}), the hyperon masses at 
low density are given by 
\begin{equation}
\left( \frac{M_\Lambda^*}{M_\Lambda} \right) \simeq 1 - 
0.12 \left( \frac{\rho_B}{\rho_0} \right) , \ \ 
\left( \frac{M_\Sigma^*}{M_\Sigma} \right) \simeq 1 - 
0.11 \left( \frac{\rho_B}{\rho_0} \right)  
\ \ \mbox{and} \ \ 
\left( \frac{M_\Xi^{\ast}}{M_\Xi} \right) \simeq 1 - 
0.05 \left( \frac{\rho_B}{\rho_0} \right) , 
\label{lambm2}
\end{equation}
where we take $\Gamma_{\Lambda,\Sigma,\Xi/N} =1$, because we find that 
the $\Gamma$ factor for the hyperons is again quite close to unity. 

The linear approximation to the quark-scalar density 
ratio, $C_j$, is again relevant to not only the nucleon and vector mesons but also the 
hyperons: 
\begin{equation}
C_j(\sigma) = 1 - a_j \times (g_{\sigma} \sigma) , 
\label{paramH}
\end{equation}
where $a_j$ is a slope parameter for the hadron $j$.  We list them in 
Table~\ref{tab:slope}. 
\begin{table}
\begin{center}
\begin{minipage}[t]{16.5 cm}
\caption{Slope parameters for the hadrons ($\times 10^{-4}$ 
MeV$^{-1}$).  Types A, B, C correspond to those in the effective $\sigma$ mass. }
\label{tab:slope}
\end{minipage}
\begin{tabular}[t]{c|cccccc}
\hline
type & $a_N$ & $a_\omega$ & $a_\rho$ & $a_\Lambda$ & $a_\Sigma$ & 
$a_\Xi$ \\
\hline
 A &  9.01 & 8.63 & 8.59 & 9.27 & 9.52 & 9.41 \\
 B &  8.98 & 8.63 & 8.58 & 9.29 & 9.53 & 9.43 \\
 C &  8.97 & 8.63 & 8.58 & 9.29 & 9.53 & 9.43 \\
\hline
\end{tabular}
\end{center}
\end{table}
We should note that the dependence of $a_j$ on the hadron is quite weak, 
and it ranges around $8.6 \sim 9.5 \times 10^{-4}$ (MeV$^{-1}$).  

If we ignore the weak dependence of $a_j$ on the hadron and take 
$\Gamma_{j/N}=1$ in Eq.~(\ref{hm}), the effective hadron mass can be 
rewritten in a quite simple form: 
\begin{equation}
M_j^{\ast} \simeq M_j - \frac{n_q}{3} g_\sigma\left[ 1 - 
  \frac{a_j}{2} (g_\sigma \sigma) \right]\sigma , 
\label{hm3}
\end{equation}
where $a \simeq 9.0 \times 10^{-4}$ (MeV$^{-1}$).  This mass formula can 
reproduce the hadron masses in 
matter quite well over a wide range of $\rho_B$, up to $\sim 3 \rho_0$.
(See also Eq.~(\ref{efmas}).)

Since the scalar field is common to all hadrons, Eq.~(\ref{hm3}) leads to 
a new, simple scaling relationship among the hadron masses: 
\begin{equation}
\left( \frac{\delta m_v^{\ast}}{\delta M_N^{\ast}} \right) \simeq 
\left( \frac{\delta M_\Lambda^{\ast}}{\delta M_N^{\ast}} \right) \simeq 
\left( \frac{\delta M_\Sigma^{\ast}}{\delta M_N^{\ast}} \right) \simeq 
\frac{2}{3} \ \ \ \mbox{ and } \ \ \ 
\left( \frac{\delta M_\Xi^{\ast}}{\delta M_N^{\ast}} \right) \simeq 
\frac{1}{3} , 
\label{scale}
\end{equation}
where $\delta M_j^{\ast} \equiv M_j - M_j^{\ast}$.  The factors, 
$\frac{2}{3}$ 
and $\frac{1}{3}$, in Eq.~(\ref{scale}) come from the ratio of the number of 
light quarks in the hadron  
$j$ to that in the nucleon.  This means that the hadron 
mass modification in nuclear medium (the scalar potential) 
is practically determined by only the number of light quarks, 
which feel the common scalar field generated by the 
surrounding nucleons in medium, 
and the corresponding strength of the scalar field 
(see also section~\ref{subsec:scbA}).  

Finally, we note recent experiments concerning the change of 
the in-medium meson masses. 
In an helicity analysis of subthreshold $\rho^0$ 
production on $^2$H, $^3$He and 
$^{12}$C at low photo-production energies~\cite{HUB}, the data support an 
in-medium modification of the $\rho^0$ invariant mass, 
with the result being  
consistent with the QMC model~\cite{HUB,SAI-7}.  
Recently, the ELSA tagged photon facility~\cite{elsa} and 
KEK~\cite{kek} have independently measured the in-medium $\omega$ meson mass. 
Using the Crystal Barrel/TAPS experiment at ELSA, 
the in-medium modification of $\omega$ 
meson was studied via the reaction 
$\gamma + A \to \omega + X \to \pi^0 \gamma + X^\prime$, and 
a significant change toward lower masses for $\omega$ 
mesons produced on the Nb target was 
observed.  For momenta less than $500$ MeV/c, they 
concluded that the in-medium $\omega$ mass was 
about $722$ MeV at $0.6 \rho_0$. This is just 
the value predicted by QMC-II~\cite{SAI-3}. From the $12$ GeV $p + A$ 
reactions at KEK, 
it was found that the in-medium mass is reduced by 
about $m_\omega^\ast / m_\omega \sim 1 - 
0.1 (\rho_B / \rho_0)$, which is also consistent with the QMC model.  

\subsubsection{\it Variations of quark and gluon condensates in nuclear matter 
\label{subsubsec:condensates}}

The QCD ground state is highly non-trivial, and the strong condensates of 
scalar quark-antiquark pairs $\langle {\bar q} q \rangle$ and gluon fields 
$\langle G_{\mu \nu}^a G^{a \mu \nu} \rangle$ may play important roles in 
a wide range of low-energy hadronic phenomena~\cite{shif,yaz,druk}.  
Therefore, it is quite interesting to study the density dependence of the 
condensates in nuclear matter~\cite{baldo}.  
The vacuum values of the lowest-dimensional 
quark and gluon condensates are typically given by~\cite{yaz} 
\bge
Q_0 \equiv \langle {\bar q} q \rangle_0 \simeq - (225 \pm 25 \mbox{MeV})^3, 
\label{qcon0}
\ene
\bge
G_0 \equiv \langle G_{\mu \nu}^a G^{a \mu \nu} \rangle_0 \simeq (360 \pm 20 
\mbox{MeV})^4. 
\label{gcon0}
\ene

Drukarev {\it et al.}~\cite{druk}, Cohen {\it et al.}~\cite{cohen2} and Lutz {\it et 
al.}~\cite{lutz2} have shown that the leading 
dependence on the nuclear density, $\rho_B$, of 
the quark condensate in nuclear 
matter, $Q(\rho_B)$, is given by the model-independent form: 
\bge
\frac{Q(\rho_B)}{Q_0} \simeq 1 - \frac{\sigma_N}{f_\pi^2 m_\pi^2} \rho_B , 
\label{indep}
\ene
where $\sigma_N$ is the pion-nucleon sigma term (empirically $\sigma_N \simeq 
45$ MeV~\cite{gass}), $m_\pi$ is the pion mass (138 MeV) and $f_\pi \simeq 
93$ MeV, the pion decay constant.  
Further, the strange quark content in the nucleon at finite density (and 
temperature) was studied in Ref.~\cite{tsushimcond} using 
the Nambu--Jona-Lasinio (NJL) model, supplemented by an instanton 
induced interaction involving the in-medium quark condensates. 
The gluon condensate at finite density,  
$G(\rho_B)$, has also been discussed in Ref.~\cite{cohen2}. 

We here consider the variations of quark and gluon condensates 
in nuclear matter within 
QMC-II (see also Refs.~\cite{SAI-1,SAI-4,SAI-3,SAI-q,SAI-qq}). The total 
energy per nucleon is given by Eq.~(\ref{tote}). 
The density-dependent quark condensate $Q(\rho_B)$ is formally derived by 
applying the Hellmann--Feynman theorem to the chiral-symmetry-breaking quark 
mass term of the total Hamiltonian.  One finds the relation for the 
quark condensate in nuclear matter at the baryon density $\rho_B$: 
\bge
m_q (Q(\rho_B) - Q_0) =  m_q \frac{d}{dm_q}  {\cal E}(\rho_B)  , 
\label{rel}
\ene
where ${\cal E}(\rho_B) = \rho_B (E^{tot}/A)$ (see Eq.~(\ref{tote})) 
and $m_q$ is 
the average, current quark mass of the $u$ and $d$ quarks.
The resulting value for $Q/Q_0$ as a function of density is 
shown in Fig.~\ref{fig:qcon} -- dashed line. (We have chosen the quark mass to 
be $m_q = 5$ MeV and the bag radius of the free nucleon 
$R_N = 0.8$ fm, but the result is quite insensitive to these choices.)
\begin{figure}
\epsfysize=9.0cm
\begin{center}
\begin{minipage}[t]{8 cm}
\hspace*{-2.5cm}
\epsfig{file=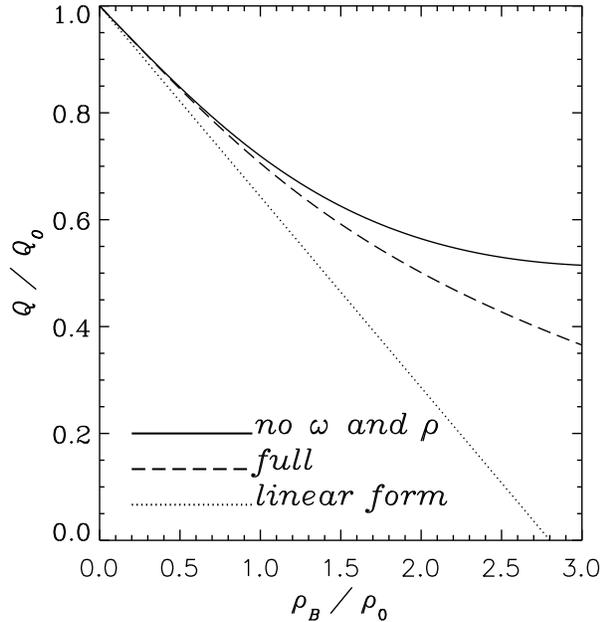,height=9cm}
\end{minipage}
\begin{minipage}[t]{16.5 cm}
\caption{Quark condensate at finite density using parameter type B of QMC-II.  
The dashed and dotted curves 
are respectively for the full calculation in symmetric nuclear matter 
and the linear approximation, Eq.~(\protect\ref{qcon3}).  
The solid curve is the result corrected by removing the spurious 
$\omega$ and $\rho$ contributions.
}
\label{fig:qcon}
\end{minipage}
\end{center}
\end{figure}

Using Eq.~(\ref{rel}), the Gell-Mann--Oakes--Renner relation 
and the explicit expression for the self-consistency condition
of the $\sigma$ field in nuclear matter 
Eq.~(\ref{scc2}) lead to the following explicit relation for the 
ratio of $Q(\rho_B)$ to $Q_0$:  
\bg
\frac{Q(\rho_B)}{Q_0} = 1 &-& \left( \frac{\sigma_N}{m_\pi^2 f_\pi^2} 
\right)  \left( \frac{m_\sigma^*}{g_\sigma} \right)^2 (g_\sigma 
{\sigma}) \left[ 1 - \frac{1}{g_\sigma} \left( \frac{d g_\sigma^q}{d m_q} 
\right) (g_\sigma {\sigma}) \right] \nn \\
&-& \left( \frac{\sigma_N \rho_0}{6S_N(0) m_\pi^2 f_\pi^2} \right) 
\rho_r^2 \left[ \frac{\rho_0}{m_\omega^{* 2}} 
\left( \frac{d g_\omega^2}{d m_q} \right) + 
\frac{\rho_0}{4 m_\rho^{* 2}} (2 f_p -1)^2 
\left( \frac{d g_\rho^2}{d m_q} \right) \right] , 
\label{qcon1}
\en
where $\rho_r = \rho_B/\rho_0$ and $f_p = \rho_p/\rho_B$.  
Because $m_q$ enters only in the combination $m_q - g^q_\sigma
\sigma$, which is generally regarded as the
chiral-symmetry-breaking term in nuclear medium, we were able to
evaluate $({d m_\sigma^*}/{d m_q})$ in terms of 
the derivative of $m_\sigma^*$ with respect to the applied
scalar field $\sigma$.

In Eq.~(\ref{qcon1}) we have followed the usual convention of identifying 
$3 m_q S_N(0)$, which is the sigma commutator in the free 
MIT bag, as the experimental pion-nucleon sigma term, $\sigma_N$. 
It is well known that the meson cloud of the nucleon (mainly the pions),
as well as its strange quark content contribute 
significantly to $\sigma_N$~\cite{jaffe}.  
However, because we are concerned primarily with the variation of $Q$ in matter 
from its free value, $Q_0$, it should be reasonable to replace $\sigma_N$ in 
Eq.~(\ref{qcon1}) by its 
empirical value. (We note that the main
variation of $Q$ in medium is generated by the $\sigma$ mean-field.)

Clearly, from Eq.~(\ref{qcon1}), the leading dependence of the quark 
condensate on the density is given by the scalar field:
\bge
\frac{Q(\rho_B)}{Q_0} \simeq 1 - \frac{\sigma_N}{m_\pi^2 f_\pi^2} 
\left( \frac{m_\sigma}{g_\sigma} \right)^2 (g_\sigma {\sigma}) . 
\label{qcon2}
\ene
One can easily show that Eq.~(\ref{qcon2}) reduces to the model-independent 
result, Eq.~(\ref{indep}), to leading order in the density, so that 
for small $\rho_r$ one has (for type B of QMC-II):  
\bge
Q(\rho_B) / Q_0 \simeq 1 - 0.357 \rho_r . 
\label{qcon3}
\ene
This is shown as the dotted line in Fig.~\ref{fig:qcon}.

Equation (\ref{qcon1}) also involves deviations of the 
quark-meson coupling constants
with respect to $m_q$. In principle, if one could derive these coupling 
constants from QCD, their dependence on $m_q$ would be given.
Within the present model there is no reason to believe that the couplings 
should vary with $m_q$. This is especially so for the vector couplings 
since they involve conserved vector currents. On the other hand, 
we require that our model reproduces the correct saturation energy 
and density of nuclear matter whatever parameters are chosen for the 
free nucleon. As a consequence, the coupling constants depend on 
$m_q$ in a way that has nothing to do with chiral symmetry breaking.
(For example, for type B of QMC-II, we find  
$g_\sigma^q = 4.891 - 0.005880  m_q + 1.200 \times 10^{-5} m_q^2, \ \ 
g_\omega^2 = 39.59 + 0.03828 m_q + 1.144 \times 10^{-3} m_q^2$ and 
$g_\rho^2 = 66.3 - 0.02 m_q$, with $m_q$ in MeV.)  

In order to extract a physically meaningful result for $Q/Q_0$ we 
should therefore remove the spurious contributions associated 
with $\frac{dg^q_M}{dm_q}$ 
$(M=\sigma,\rho,\omega)$ in Eq.~(\ref{qcon1}).
In fact, the variation of $g^q_\sigma$ with $m_q$ is extremely small 
so we need only correct the $\omega$ and $\rho$ contributions. 
The final, corrected result is shown as the solid line in 
Fig.\ref{fig:qcon}.
Even in MFA, our calculations 
show that the higher-order contributions in the nuclear density become 
very important and that they weaken the chiral 
symmetry restoration at high density.
In QMC-II, the $\sigma$ field in nuclear matter 
is suppressed at high density (for example, $g_\sigma {\sigma} \simeq$ 
200 (300) MeV at $\rho_0$ ($3 \rho_0$)) because the quark scalar density, 
$S_N$, decreases significantly as the density rises, as a result of the 
change in the quark structure 
of the bound nucleon (see section~\ref{subsec:matter}).  Since the reduction of the quark 
condensate is mainly controlled by the scalar field, it 
is much smaller than in the simple, linear approximation, 
Eq.~(\ref{qcon3}). From the difference between the solid and dashed curves we see that the 
correction for the dependence of the coupling constants on $m_q$ 
is significant and this should be born in mind in any phenomenological 
treatment. 

We should note here that, from extensive studies of chiral perturbation 
theory for nuclear matter, especially the work of 
Birse~\cite{birse2}, a reduction of the quark condensate from its 
vacuum value may {\em not\/} be enough to conclude 
that the chiral symmetry has been partially restored -- in particular, 
if part of the change in $\langle {\bar q}q 
\rangle$ arises from low-momentum pions.  We note also that higher-order condensates 
may play an increasingly important role as the quark condensate tends to zero.  

Next let us consider the in-medium gluon condensate.  
Cohen et al.~\cite{cohen2} 
also developed a model-independent prediction of the gluon condensate that is 
valid to first order in the nuclear density through an application of the trace anomaly 
and the Hellmann--Feynman theorem.  Following their approach, the 
ratio of the gluon condensate in nuclear matter $G(\rho_B)$ to 
that in vacuum $G_0$ is given by
\bge
G(\rho_B) / G_0 \simeq 1 - \left( \frac{8}{9 G_0} \right) 
\left[ {\cal E}(\rho_B) 
- 2 m_q (Q(\rho_B) - Q_0) - m_s (Q_s(\rho_B) - Q_{s0}) \right] , 
\label{gcon1}
\ene
where $m_s$ is the strange-quark mass and $Q_s(\rho_B)$ ($Q_{s0}$) is the 
strange-quark condensate in nuclear matter (in vacuum).  Up to first 
order in the density, the change of the strange-quark condensate may be 
written in terms of the strange quark content 
of the nucleon in free space, ${\cal S}$: 
\bge
m_s (Q_s(\rho_B) - Q_{s0}) = {\cal S} \rho_B + {\cal O}(\rho_B^2) . 
\label{stran}
\ene
The strange quark content is commonly specified by the dimensionless 
quantity, $y$, defined by 
\bge
y \equiv \frac{ 2 \langle {\bar s}s \rangle_N}
  {\langle {\bar u}u + {\bar d}d \rangle_N} , 
\label{yyy}
\ene
which leads to ${\cal S} = (m_s/2m_q) \sigma_N y$.  
Roughly speaking, $y$ represents the probability to find $s$ or ${\bar s}$ in 
the nucleon and is a measure of the OZI-rule violation.  If 
$m_s/m_q \simeq 25$ and $y \simeq 0.45$~\cite{cohen2}, we get 
${\cal S} \simeq 250$ MeV.  We note, however, that $y \simeq 0.45$ is an 
extreme value. 
In the analysis below, we shall take care to examine the sensitivity 
to the full range of variation of $y$.

At very low $\rho_B$, ${\cal E}(\rho_B)$ can be expanded as~\cite{QHD} 
\bge
{\cal E}(\rho_B) = M_N \rho_B \left[ 1 + 
     \frac{3}{10 M_N^2} \left( \frac{3\pi^2}{2} \right)^{2/3} \rho_B^{2/3} 
     \right] + {\cal O}(\rho_B^2) ,
\label{eee}
\ene
where the second term in the bracket is the non-relativistic Fermi-gas energy.  
Using the approximate form, 
$g_\sigma {\sigma} \simeq 214$ (MeV) $\times \rho_r$  
(for type B),  
we find 
\bge
2 m_q (Q(\rho_B) - Q_0) = 214 (\mbox{MeV}) \times 
\sigma_N \left( \frac{m_\sigma}{g_\sigma} 
  \right)^2 \rho_r + {\cal O}(\rho_r^2). 
\label{qqq}
\ene
Choosing the central value of $G_0$ in Eq.~(\ref{gcon0}), we then get the 
in-medium gluon condensate at low $\rho_B$ (for type B): 
\bge
G(\rho_B)/G_0 = 1 - ( 0.03892 \rho_r + 0.001292 \rho_r^{5/3} ) + 
{\cal O}(\rho_r^2) . 
\label{gappr}
\ene
\begin{figure}
\epsfysize=9.0cm
\begin{center}
\begin{minipage}[t]{8 cm}
\hspace*{-2.5cm}
\epsfig{file=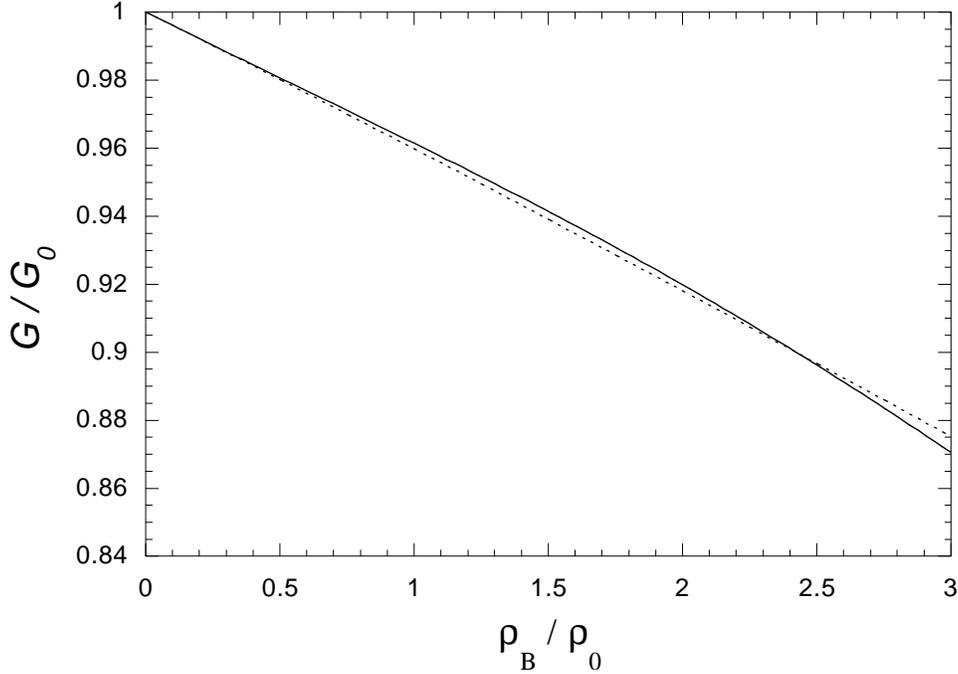,height=9cm}
\end{minipage}
\begin{minipage}[t]{16.5 cm}
\caption{
Gluon condensate at finite density using parameter type B.  
The solid and dotted curves 
are respectively for the full calculation in symmetric nuclear matter 
and the approximation, Eq.~(\protect\ref{gappr}).
}
\label{fig:gcon}
\end{minipage}
\end{center}
\end{figure}
Our numerical results for the full calculation as well as the approximate 
calculation with Eq.~(\ref{gappr}) are shown in Fig.~\ref{fig:gcon} 
(for $m_q=5$ MeV and $R_N=$ 0.8 fm).
The reduction of the 
gluon condensate at finite density is not large; for example, it 
is reduced by only 4\% at $\rho_0$.  
The approximation of Eq.~(\ref{gappr}) works very well for a wide range of the 
nuclear density, which may imply that the effect of 
higher-order contributions (in powers of the density)
is small for the gluon condensate.  However, one should keep in mind 
that the  
in-medium gluon condensate evaluated here 
contains a large uncertainty,
originating from the uncertainty in the value for the 
strange quark content of the nucleon in free space.  
We note that if we assume a 
vanishing strange quark content of the nucleon 
in free space (${\cal S}=0$ or $y=0$), the gluon condensate would be 
reduced by about 6\% at $\rho_0$.  

Finally, we relate the quark condensate to the variation of the hadron mass 
in nuclear matter.  In QMC-II, the hadron mass at low $\rho_B$ is simply given 
in terms of the scalar field (see Eq.~(\ref{hm3})): 
\bge
M_j^* \simeq M_j - \frac{n_q}{3} (g_\sigma \sigma) , 
\label{efmas_qcon}
\ene
where $j$(= N, $\omega, \rho, \Lambda$, etc.).  Since the quark condensate at low $\rho_B$ is 
also determined by the scalar field, we find a 
simple relation between the variations of the hadron mass and the quark 
condensate: 
\bge
\delta M_j^{\ast} \simeq \left( \frac{m_\pi^2 f_\pi^2}{3 \sigma_N} \right) 
\left( \frac{g_\sigma}{m_\sigma} \right)^2 n_q 
\left( 1 - \frac{Q(\rho_B)}{Q_0} \right) \approx 200 (\mbox{MeV}) \times 
n_q \left( 1 - \frac{Q(\rho_B)}{Q_0} \right) ,  
\label{relat}
\ene
where $\delta M_j^* = M_j - M_j^{\ast}$.  

However, as shown in Ref.~\cite{birse2},  
we know that the nucleon mass in matter cannot depend in any simple way 
on the quark condensate alone because the leading non-analytic 
contribution (LNAC) to the pion-nucleon sigma term -- the term of 
order $m_\pi^3$ -- should {\em not\/} appear in the nucleon-nucleon 
interaction~\cite{ch}.  To discuss this problem further, we have to include 
pions self-consistently in the QMC model, which is a future study. 

We note that in Ref.~\cite{SAI-4} the effect of $\delta$ meson on the quark condensates 
and a measure of isospin symmetry breaking in the in-medium quark condensates are 
also discussed in details. 

\subsubsection{\it Nuclear matter at finite temperature and neutron stars 
\label{subsubsec:appl}}

Song and Su extended the QMC model to include the 
effect of finite temperature~\cite{SON}, 
and applied the resulting EOS for nuclear matter to the liquid-gas phase 
transition. In their calculations, they simply added the thermal 
distribution functions for the nucleon and anti-nucleon to the energy-momentum expressions and the 
self-consistency condition. They found that 
the EOS at finite temperature is much softer 
than that in QHD and is comparable to that in the 
Zimanyi-Moszkowski model~\cite{ZIM}. Furthermore, they 
have found that the critical temperature of the liquid-gas phase transition becomes lower in the 
QMC than in QHD. 

Panda {\it et al.} have also studied hot nuclear matter within the framework of the QMC model~\cite{PAN-1}. 
In their calculations, the possible single-particle quark and antiquark energies at finite 
temperature $T$ are summed up in the bag energy. The quark (antiquark) energy in the orbit $(n, \kappa)$ 
($n$ specifies the number of nodes) is estimated as 
\begin{equation}
\epsilon_{\pm}^{n\kappa}=\Omega^{n\kappa}\pm g_\omega^q\omega R_N,
\end{equation}
where 
\begin{equation}
\Omega^{n\kappa}=(x^2_{n\kappa}+R_N^2{m^*_q}^2)^{1/2}
\end{equation}
with $m^*_q=m_q-g_\sigma^q\sigma$.
The eigenvalue $x_{n\kappa}$ is given by the boundary condition at the bag surface. 
The total energy from the quarks and antiquarks then reads 
\begin{equation}
E_{tot}=3\sum_{n,\kappa}\frac{\Omega^{n\kappa}}{R_N}\Big[
\frac{1}{e^{(\epsilon_+^{n\kappa}/R_N-\mu_q)/T}+1}+
\frac{1}{e^{(\epsilon_-^{n\kappa}/R_N+\mu_q)/T}+1}\Big],
\end{equation}
with $\mu_q$ the chemical potential for the quark. The bag energy now becomes
\begin{equation}
E_{bag}=E_{tot}-\frac{z_0}{R_N}+\frac{4\pi}{3}R_N^3B,
\end{equation}
which reduces to the usual form at zero temperature.
They also adopted the MQMC model to include the medium-dependent bag parameter: 
\begin{equation}
B=B_0 \exp \left( -\frac{4 g_\sigma^B \sigma}{M_N}\right),
\label{bagc}
\end{equation}
with $g_\sigma^B$ as an additional parameter (see section~\ref{subsec:modfqmc}).

They have found that the scalar mean field at zero density
attains a nonzero value at a temperature of 200 MeV similar to the
Walecka model calculations for nuclear matter~\cite{FUR-2}, 
which is indicative of a phase transition to a system with 
baryon-antibaryon pairs. There is a softening in the
phase transition in this case as compared to the earlier calculations. 
This is because the thermal contributions from the quarks are dominant and
lead to a rise of the effective nucleon mass with temperature.
The nucleon bag shrinks in size with 
increasing temperature. The nucleon mass at finite temperature
and zero baryon density is then appreciably different from that in vacuum.
This approach has been applied to various 
nuclear and stellar problems: for example, rotating 
neutron stars in the Komatsu-Eriguchi-Hachisu method~\cite{PAN-2}   
and hybrid stars which consist of both hadron and quark matter~\cite{PAN-3}. 
Recently they have also studied kaon condensation in compact 
stars~\cite{PAN-4}.

Zakout {\it et al.} have developed a QMC model at finite temperature 
in a similar manner.  They have, however, included the effect of 
dilatons in their model~\cite{ZAK-1}.  
Applying the effective potential with dilatons to nuclear matter, 
the nucleon properties at finite temperature are found to be 
appreciably different from those in cold nuclear 
matter. The dilaton potential improves the shape of 
the saturation curve at $T=0$ and significantly affects the properties 
of hot nuclear matter. Hot hypernuclear matter~\cite{ZAK-2}, neutron 
stars~\cite{ZAK-3} and relativistic heavy iron 
reactions~\cite{ZAK-4} have also been investigated in their approach. 

\subsection{Finite nuclei - $^{16}$O, $^{40}$Ca, $^{48}$Ca, $^{90}$Zr, $^{208}$Pb 
\label{subsec:fnt}}
To describe a finite nucleus with different numbers of protons and neutrons 
($Z \neq N$), it is necessary to include the contributions of the $\rho$
meson explicitly. Any realistic treatment also needs the inclusion of 
Coulomb force. The variation of the Lagrangian for QMC-I 
(see Eq.~(\ref{LQMC-I})) results in the   
following equations for static, spherically symmetric nuclei: 
\bea
\frac{d^2}{dr^2} \sigma(r) + \frac{2}{r} \frac{d}{dr} \sigma(r) 
    - m_\sigma^2 \sigma(r) &=& - g_\sigma C_N(\sigma(r)) \rho_s(r) \nn \\
    &\equiv& - g_\sigma C_N(\sigma(r)) \sum_\alpha^{occ} d_\alpha(r) 
    (|G_\alpha(r)|^2 - |F_\alpha(r)|^2), \label{sig3} \\
\frac{d^2}{dr^2} \omega(r) + \frac{2}{r} \frac{d}{dr} \omega(r) 
    - m_\omega^2 \omega(r) &=& - g_\omega \rho_B(r) \nn \\
    &\equiv& - g_\omega \sum_\alpha^{occ} d_\alpha(r)
    (|G_\alpha(r)|^2 + |F_\alpha(r)|^2), \label{omg3} \\
\frac{d^2}{dr^2} b(r) + \frac{2}{r} \frac{d}{dr} b(r) 
    - m_\rho^2 b(r) &=& - \frac{g_\rho}{2} \rho_3(r) \nn \\
    &\equiv& - \frac{g_\rho}{2} \sum_\alpha^{occ} 
    d_\alpha(r) (-)^{t_\alpha -1/2} 
    (|G_\alpha(r)|^2 + |F_\alpha(r)|^2), \label{rho3} \\
\frac{d^2}{dr^2} A(r) + \frac{2}{r} \frac{d}{dr} A(r) 
    &=& - e \rho_p(r) \nn \\
    &\equiv& - e \sum_\alpha^{occ} d_\alpha(r) 
    (t_\alpha + \frac{1}{2}) 
    (|G_\alpha(r)|^2 + |F_\alpha(r)|^2), \label{phtn3} 
\eea
where $d_\alpha(r)= (2j_\alpha+1)/4\pi r^2$ and 
\bea
\frac{d}{dr} G_\alpha(r) + \frac{\kappa}{r} G_\alpha(r) - 
\left[ \epsilon_\alpha - g_\omega \omega(r) - t_\alpha g_\rho b(r)
\right. 
&-& \left. (t_\alpha + \frac{1}{2}) e A(r) + M_N \right. \nn \\
&-& \left. g_\sigma(\sigma(r)) \sigma(r) \right] F_\alpha(r) = 0 , 
\label{qwave1} \\
\frac{d}{dr} F_\alpha(r) - \frac{\kappa}{r} F_\alpha(r) + 
\left[ \epsilon_\alpha - g_\omega \omega(r) - t_\alpha g_\rho b(r)
\right.
&-& \left. (t_\alpha + \frac{1}{2}) e A(r) - M_N \right. \nn \\
&+& \left. g_\sigma (\sigma(r)) \sigma (r) \right] G_\alpha(r) = 0 . 
\label{qwave2} 
\eea
Here $iG_\alpha(r)/r$ and $-F_\alpha(r)/r$ are respectively the 
radial part of the upper and lower 
components of the solution to the Dirac equation for the nucleon 
($\epsilon_\alpha$ being 
the energy) under the normalization condition: 
\be
\int dr (|G_\alpha(r)|^2 + |F_\alpha(r)|^2) =1 . \label{normf}
\ee
As usual, $\kappa$ specifies the angular quantum numbers and $t_\alpha$ 
the eigenvalue of the isospin operator, $\tau^N_3/2$.  $C_N(\sigma)$ and 
$g_\sigma (\sigma)$ are practically given by 
Eqs.~(\ref{efnmas}), (\ref{cnpar}) and (\ref{nmapp}), i.e., 
\be
g_\sigma (\sigma({\vec r})) =  g_\sigma \left[ 1 - \frac{a_N}{2} 
            g_\sigma \sigma({\vec r}) \right] . \label{gsgm}
\ee
The total energy of the system is then given by 
\bea
E_{tot} &=& \sum_\alpha^{occ} (2j_\alpha + 1) \epsilon_\alpha 
 - \frac{1}{2} \int d{\vec r} \ [ -g_\sigma C_N(\sigma (r)) \sigma(r) 
\rho_s(r) \nn \\
 &+& g_\omega \omega(r) \rho_B(r) + \frac{1}{2} g_\rho b(r) \rho_3(r) 
 + eA(r) \rho_p(r) ].  \label{ftoten}
\eea

For the QMC-II model, the Lagrangian density, Eq.~(\ref{qmc-2}), leads 
\bea
\frac{d^2}{dr^2} \sigma(r) + \frac{2}{r} \frac{d}{dr} \sigma(r) 
    - m_\sigma^{\ast 2} \sigma(r) &=& - g_\sigma C_N \rho_s(r) 
    - m_\sigma m_\sigma^{\ast} g_\sigma [ a_\sigma - 2 b_\sigma g_\sigma 
      \sigma(r) ] \sigma(r)^2 \nn \\ 
    &+& \frac{2}{3} g_\sigma [ m_\omega^{\ast} \Gamma_{\omega/N} 
      C_\omega \omega(r)^2 
    + m_\rho^{\ast} \Gamma_{\rho/N} C_\rho b(r)^2 ] , \label{scmot} \\
\frac{d^2}{dr^2} \omega(r) + \frac{2}{r} \frac{d}{dr} \omega(r) 
    - m_\omega^{\ast 2} \omega(r) &=& - g_\omega \rho_B(r) , \\
\label{vcmot}
\frac{d^2}{dr^2} b(r) + \frac{2}{r} \frac{d}{dr} b(r) 
    - m_\rho^{\ast 2} b(r) &=& - \frac{g_\rho}{2} \rho_3(r) .  
\label{rhmot}
\eea
The photon field and nucleon wave functions are again given by 
Eqs.~(\ref{phtn3})-(\ref{qwave2}). 
Then, the total energy reads 
\bea
E_{tot} &=& \sum_\alpha^{occ} (2j_\alpha + 1) \epsilon_\alpha 
 - \frac{1}{2} \int d{\vec r} \ [ -g_\sigma D(\sigma (r)) \sigma(r) \nn \\
 &+& g_\omega \omega(r) \rho_B(r) + \frac{1}{2} g_\rho b(r) \rho_3(r) 
 + eA(r) \rho_p(r) ] ,  \label{ftotenf}
\eea
where
\bea
D(\sigma(r)) &=& C_N \rho_s(r) 
    + m_\sigma m_\sigma^{\ast} [ a_\sigma - 2 b_\sigma g_\sigma 
      \sigma(r) ] \sigma(r)^2 \nn \\
    &-& \frac{2}{3} \left[ m_\omega^{\ast} \Gamma_{\omega/N} 
      C_\omega \omega(r)^2 + m_\rho^{\ast} \Gamma_{\rho/N} C_\rho 
    b(r)^2 \right] . 
\label{dd}
\eea

We are now in a position to show results for finite 
nuclei. We calculate the properties of $^{16}$O, 
$^{40}$Ca, $^{48}$Ca, $^{90}$Zr and $^{208}$Pb.  Firstly, we consider the QMC-I model~\cite{SAI-8}. 
Equations~(\ref{sig3}) to~(\ref{normf}) give a set of coupled non-linear 
differential equations, which may be solved by a standard iteration 
procedure.  The numerical calculation was 
carried out by modifying the technique described by Horowitz 
{\it et al.}~\cite{HOR-1,HOR-2}.  The calculation is achieved in at most 
20 iterations when it is performed with a maximum radius of 12 (15) fm on 
a mesh of 0.04 fm for medium mass (Pb) nuclei.  

There are seven parameters to be determined: $g_\sigma$, $g_\omega$, 
$g_\rho$, $e$, $m_\sigma$, $m_\omega$ and $m_\rho$.  We first fix $m_\omega$, $m_\rho$ and 
$e^2/4\pi (= 1/137.036)$ to be the experimental values.  The coupling constants $g_\sigma$ and 
$g_\omega$ were fixed so as to produce the nuclear matter properties 
with $m_\sigma$ = 550 MeV in section~\ref{subsec:matter}.  
The $\sigma$ meson mass, however, determines the range of the attractive 
interaction and changes in $m_\sigma$ affect the nuclear-surface slope and 
its thickness.  Therefore, we adjust $m_\sigma$ to produce the root-mean-square (rms) charge radius of 
$^{40}$Ca: $r_{ch}$($^{40}$Ca) = 3.48 fm, the experimental 
value~\cite{SIN}.  
We here notice that variations of $m_\sigma$ at fixed  
($g_\sigma / m_\sigma$) 
have no effect on the infinite nuclear matter properties.  Therefore, 
keeping the ratio ($g_\sigma / m_\sigma$) constant we vary $m_\sigma$ to fit 
the rms charge radius of $^{40}$Ca. We 
expect that $m_\sigma$ ranges around 400 $\sim$ 550 MeV~\cite{BLU,MAC}.  
The last parameter $g_\rho$ is adjusted to yield the bulk symmetry 
energy per baryon of 35 MeV~\cite{HOR-1}.  We summarize the parameters in 
Table~\ref{tab:c.c.2}.
\begin{table}
\begin{center}
\begin{minipage}[t]{16.5 cm}
\caption{Model parameters for finite nuclei (QMC-I).}
\label{tab:c.c.2}
\end{minipage}
\begin{tabular}[t]{cccccc}
\hline
$m_q$(MeV)&$R_N$(fm)&$g_{\sigma}^2/4\pi$&$g_{\omega}^2/4\pi$&
$g_{\rho}^2/4\pi$&$m_\sigma$(MeV) \\
\hline
   &  0.6 & 3.55 & 6.29 & 6.79 & 429 \\
 0 &  0.8 & 2.94 & 5.26 & 6.93 & 407 \\
   &  1.0 & 2.51 & 4.50 & 7.03 & 388 \\
\hline
   &  0.6 & 3.68 & 6.34 & 6.78 & 436 \\
 5 &  0.8 & 3.12 & 5.31 & 6.93 & 418 \\
   &  1.0 & 2.69 & 4.56 & 7.02 & 401 \\
\hline
   &  0.6 & 3.81 & 6.37 & 6.78 & 443 \\
10 &  0.8 & 3.28 & 5.36 & 6.92 & 428 \\
   &  1.0 & 2.91 & 4.62 & 7.02 & 416 \\
\hline
\end{tabular}
\end{center}
\end{table}
\begin{figure}
\epsfysize=9.0cm
\begin{center}
\begin{minipage}[t]{8 cm}
\hspace*{-3cm}
\epsfig{file=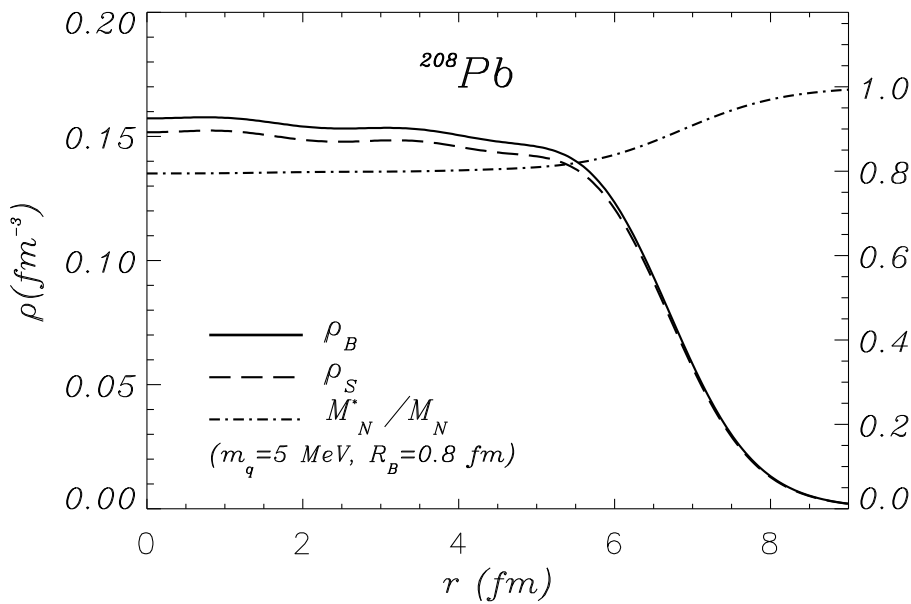,height=9cm}
\end{minipage}
\begin{minipage}[t]{16.5 cm}
\caption{Model predictions for the effective nucleon mass, the baryon and 
scalar densities in $^{208}$Pb (for $m_q$ = 5 MeV and $R_N$ = 0.8 fm).  
The scale on the right vertical axis is for $M_N^{\ast}/M_N$
(from Ref.~\cite{SAI-8}.).}
\label{fig:pbd58}
\end{minipage}
\end{center}
\end{figure}

In Fig.~\ref{fig:pbd58}, we show the baryon and scalar densities as well as 
the effective nucleon mass in $^{208}$Pb.  We expect that the baryon density 
in the interior of lead would be close to the saturation density of infinite 
nuclear matter, $\rho_0$.  As seen in the figure, the calculated baryon density  
at the center is quite close to 0.15 fm$^{-1}$, which supports our choice of 
the parameters.  

Next we show the charge density distributions calculated, $\rho_{ch}$, 
in comparison with those of QHD~\cite{QHD,HOR-1} and the experimental 
data in Figs.~\ref{fig:o58}$-$\ref{fig:ch48a05}. 
Having solved Eqs.~(\ref{sig3})-(\ref{normf}), we obtain the 
{\em point-}proton and neutron densities in a nucleus.  
Thus, we should estimate 
the effect of proton form factor on the densities. 
(Note that our calculation shows the effect of neutron form factor 
is eventually small and ignored in the results.) 
One can calculate the charge
density by a convolution of the point-proton density, $\rho_p({\vec r})$, with 
the proton charge distribution, $\rho^p_{ch}({\vec r})$: 
\be
\rho_{ch}({\vec r}) = \int d{\vec r}\, ' \ \rho^p_{ch}({\vec r} - 
{\vec r}\, ') \rho_p({\vec r}\, ') , \label{charge}
\ee
where we have used a Gaussian form for $\rho^p_{ch}$ 
\be
\rho^p_{ch}({\vec r}) = (\beta/\pi)^{3/2} \exp (-\beta r^2) . 
\label{gauss}
\ee
The parameter $\beta$, which determines the proton size, is chosen so 
as to reproduce the experimental rms charge radius of the proton, 0.82 fm 
(i.e., $\beta$ = 2.231 fm$^{-2}$).  
In the QMC model, the rms radius of the nucleon 
in nuclear matter increases a little.  However, since this 
amount is quite small, it can be ignored in the numerical calculations 
of nuclear parameters.  
\begin{figure}
\epsfysize=9.0cm
\begin{center}
\begin{minipage}[t]{8 cm}
\hspace*{-3cm}
\epsfig{file=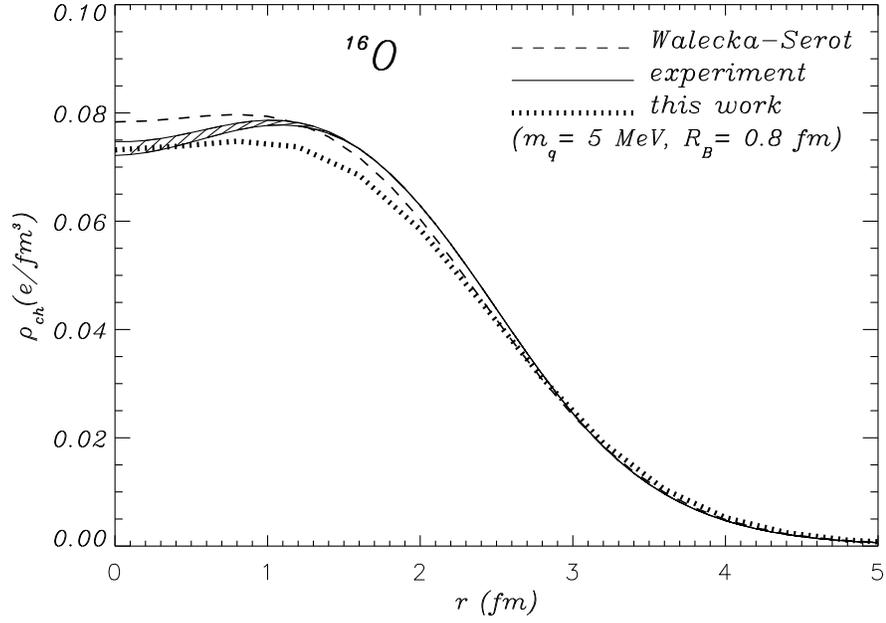,height=9cm}
\end{minipage}
\begin{minipage}[t]{16.5 cm}
\caption{Same as Fig.~\ref{fig:ca58} but for $^{16}$O
(from Ref.~\cite{SAI-8}).}
\label{fig:o58}
\end{minipage}
\end{center}
\end{figure}
\begin{figure}
\epsfysize=9.0cm
\begin{center}
\begin{minipage}[t]{8 cm}
\hspace*{-3cm}
\epsfig{file=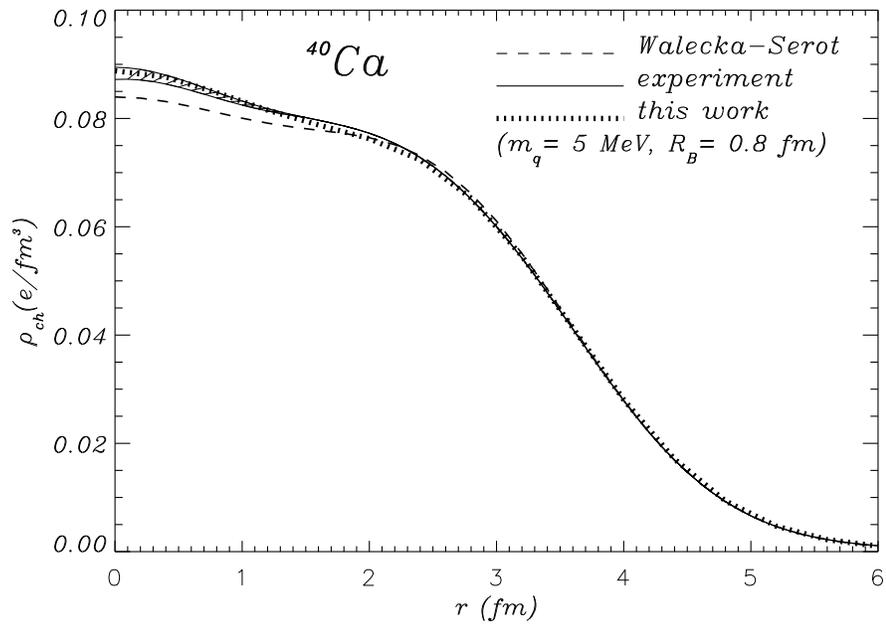,height=9cm}
\end{minipage}
\begin{minipage}[t]{16.5 cm}
\caption{Charge density distribution for $^{40}$Ca (for $m_q$ = 5 MeV and 
$R_N$ = 0.8 fm) compared with the experimental data (hatched area) 
and that of QHD (from Ref.~\cite{SAI-8}).}
\label{fig:ca58}
\end{minipage}
\end{center}
\end{figure}
\begin{figure}
\epsfysize=9.0cm
\begin{center}
\begin{minipage}[t]{8 cm}
\hspace*{-3cm}
\epsfig{file=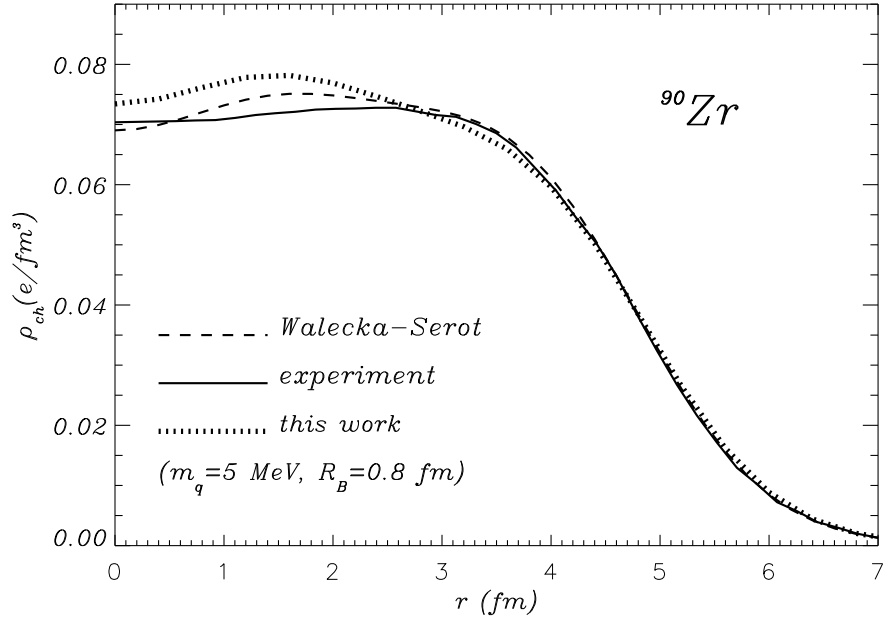,height=9cm}
\end{minipage}
\begin{minipage}[t]{16.5 cm}
\caption{Same as Fig.~\ref{fig:ca58} but for $^{90}$Zr
(from Ref.~\cite{SAI-8}).}
\label{fig:zr58}
\end{minipage}
\end{center}
\end{figure}
\begin{figure}
\epsfysize=9.0cm
\begin{center}
\begin{minipage}[t]{8 cm}
\hspace*{-3cm}
\epsfig{file=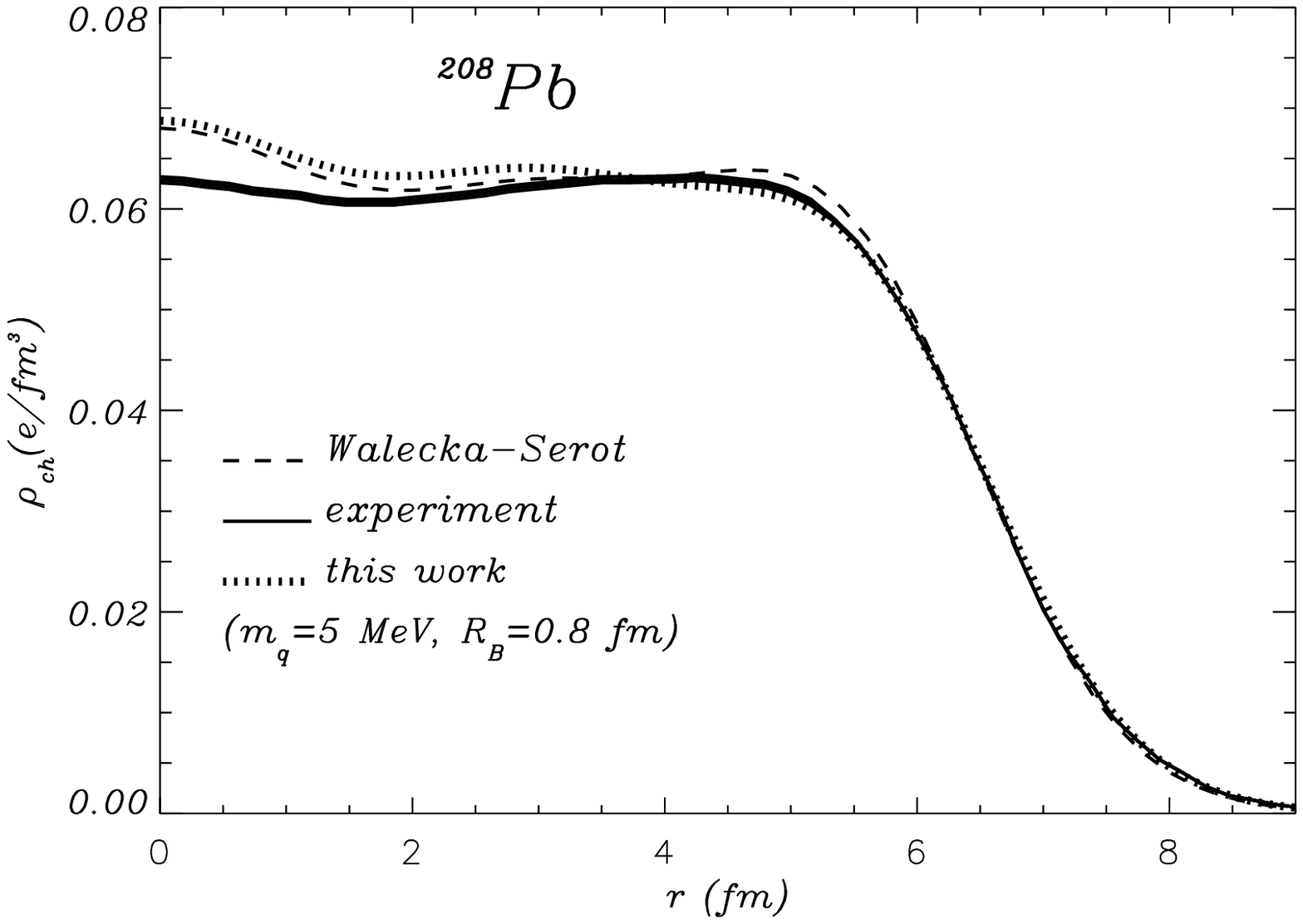,height=9cm}
\end{minipage}
\begin{minipage}[t]{16.5 cm}
\caption{Same as Fig.~\ref{fig:ca58} but for $^{208}$Pb
(from Ref.~\cite{SAI-8}).}
\label{fig:pb58}
\end{minipage}
\end{center}
\end{figure}
\begin{figure}
\epsfysize=9.0cm
\begin{center}
\begin{minipage}[t]{8 cm}
\hspace*{-3cm}
\epsfig{file=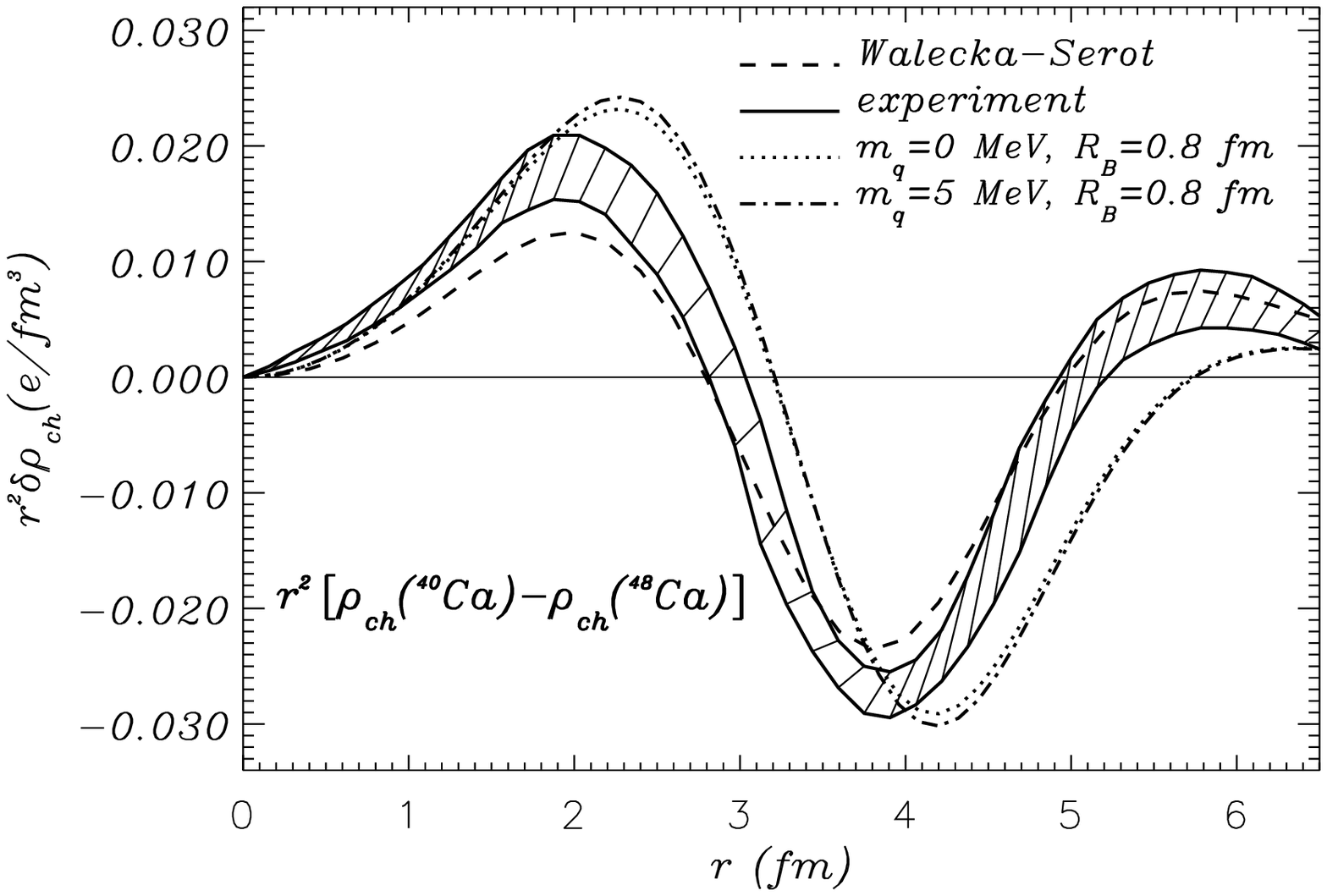,height=9cm}
\end{minipage}
\begin{minipage}[t]{16.5 cm}
\caption{Isotope shift between $\rho_{ch}$($^{40}$Ca) and 
$\rho_{ch}$($^{48}$Ca) compared with the experimental data and that of QHD 
(for $m_q$ = 0 and 5 MeV with $R_N$ = 0.8 fm)
(from Ref.~\cite{SAI-8}).}
\label{fig:ch48a05}
\end{minipage}
\end{center}
\end{figure}

In Fig.~\ref{fig:ca58}, the charge density distribution for $^{40}$Ca is 
presented. The experimental data is taken from Ref.~\cite{CA40}.  
Once the rms charge radius of $^{40}$Ca is fitted by adjusting $m_\sigma$, 
the QMC model can reproduce $\rho_{ch}$($^{40}$Ca) quite well.  
As seen in the figures, the calculated charge densities lie almost within 
the experimental area.  We note that the dependence of 
$\rho_{ch}$($^{40}$Ca) on the bag radius is quite weak~\cite{SAI-8}.  

It is interesting to see the quantum oscillations of the interior 
density in lead (see Fig.~\ref{fig:pb58}). This is probably a consequence of
using a pure shell model, with no configuration mixing. 
The dependence of $\rho_{ch}$($^{208}$Pb) on the 
quark mass is again not strong. 
We note that the dependence of $\rho_{ch}$($^{208}$Pb)
on the free nucleon size is also quite 
weak~\cite{SAI-8}.  As seen in the figures, our model gives charge densities 
very close to those of QHD and still somewhat larger in the central 
region than those observed experimentally~\cite{PB208}.  

In Figs.~\ref{fig:o58} and \ref{fig:zr58}, 
we show respectively the charge density 
distributions for $^{16}$O and $^{90}$Zr.  For zirconium the calculated 
$\rho_{ch}$ lies in between that of the non-relativistic density-dependent 
Hartree-Fock calculations~\cite{FRIA} and that of QHD~\cite{HOR-1}.  
The experimental data for oxygen and zirconium are taken 
from Refs.~\cite{O16} and~\cite{ZR}.  (For both cases the dependence of 
$\rho_{ch}$ on $m_q$ and $R_N$ is again weak.)  

To see the isotope shift in charge density we have plotted $r^2$ times the 
difference between $\rho_{ch}$($^{40}$Ca) and $\rho_{ch}$($^{48}$Ca) in 
Fig.~\ref{fig:ch48a05}.  Its dependence on the bag radius is 
weak for a small quark mass while it becomes a little stronger 
for $m_q$ = 10 MeV~\cite{SAI-8}.  
The experimental data is taken from Ref.~\cite{CA48}. (Note that in
this case we also checked that including the charge distribution of the
neutron had a small effect.)

In Figs.~\ref{fig:can58}$-$\ref{fig:pbn58}, we present the point-neutron density 
distributions, $\rho_n$, in calcium and lead.  
For $^{40}$Ca, since the dependence of $\rho_n$ on $m_q$ and $R_N$ is again 
fairly weak, only the result for $m_q$ = 5 MeV and $R_N$ = 0.8 fm is 
shown, together with the empirical fit~\cite{CA40N} to proton 
scattering data.  
\begin{figure}
\epsfysize=9.0cm
\begin{center}
\begin{minipage}[t]{8 cm}
\hspace*{-3cm}
\epsfig{file=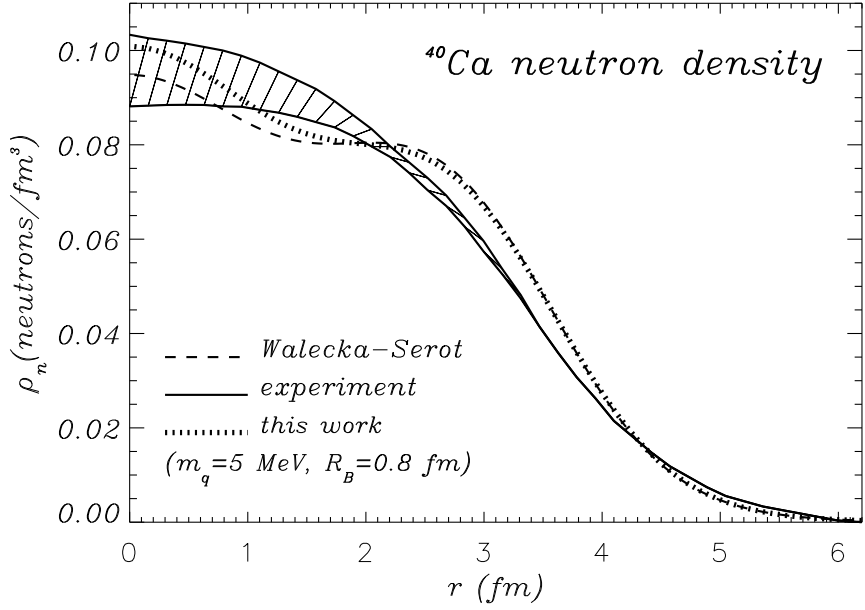,height=9cm}
\end{minipage}
\begin{minipage}[t]{16.5 cm}
\caption{Point-neutron density distribution in $^{40}$Ca 
(for $m_q$ = 5 MeV and 
$R_N$ = 0.8 fm) compared with that of QHD and the empirical fit
(from Ref.~\cite{SAI-8}).}
\label{fig:can58}
\end{minipage}
\end{center}
\end{figure}
\begin{figure}
\epsfysize=9.0cm
\begin{center}
\begin{minipage}[t]{8 cm}
\hspace*{-3cm}
\epsfig{file=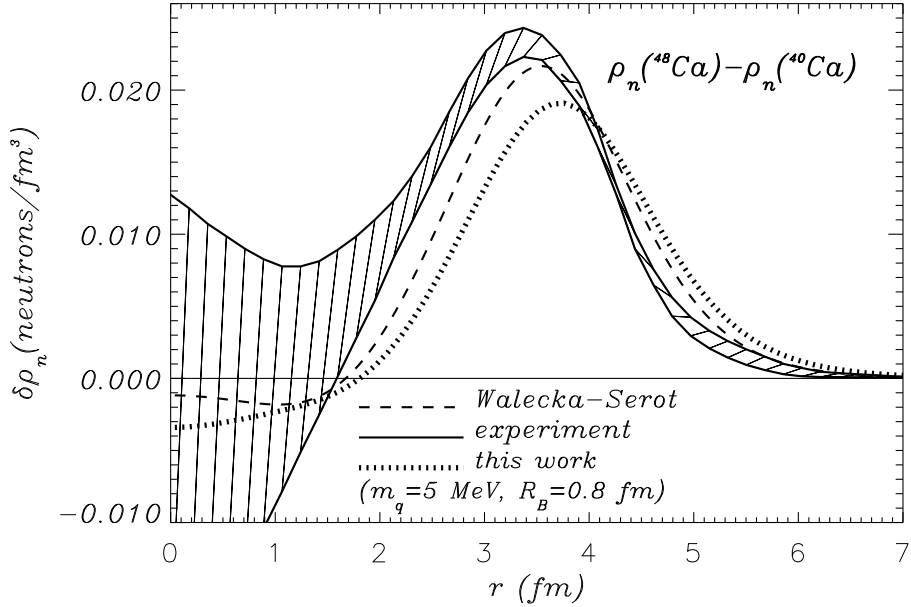,height=9cm}
\end{minipage}
\begin{minipage}[t]{16.5 cm}
\caption{Difference between $\rho_{n}$($^{48}$Ca) and $\rho_{n}$($^{40}$Ca) 
compared with that of QHD and the empirical fit 
(for $m_q$ = 5 and $R_N$ = 0.8 fm) (from Ref.~\cite{SAI-8}).}
\label{fig:nd58}
\end{minipage}
\end{center}
\end{figure}
\begin{figure}
\epsfysize=9.0cm
\begin{center}
\begin{minipage}[t]{8 cm}
\hspace*{-3cm}
\epsfig{file=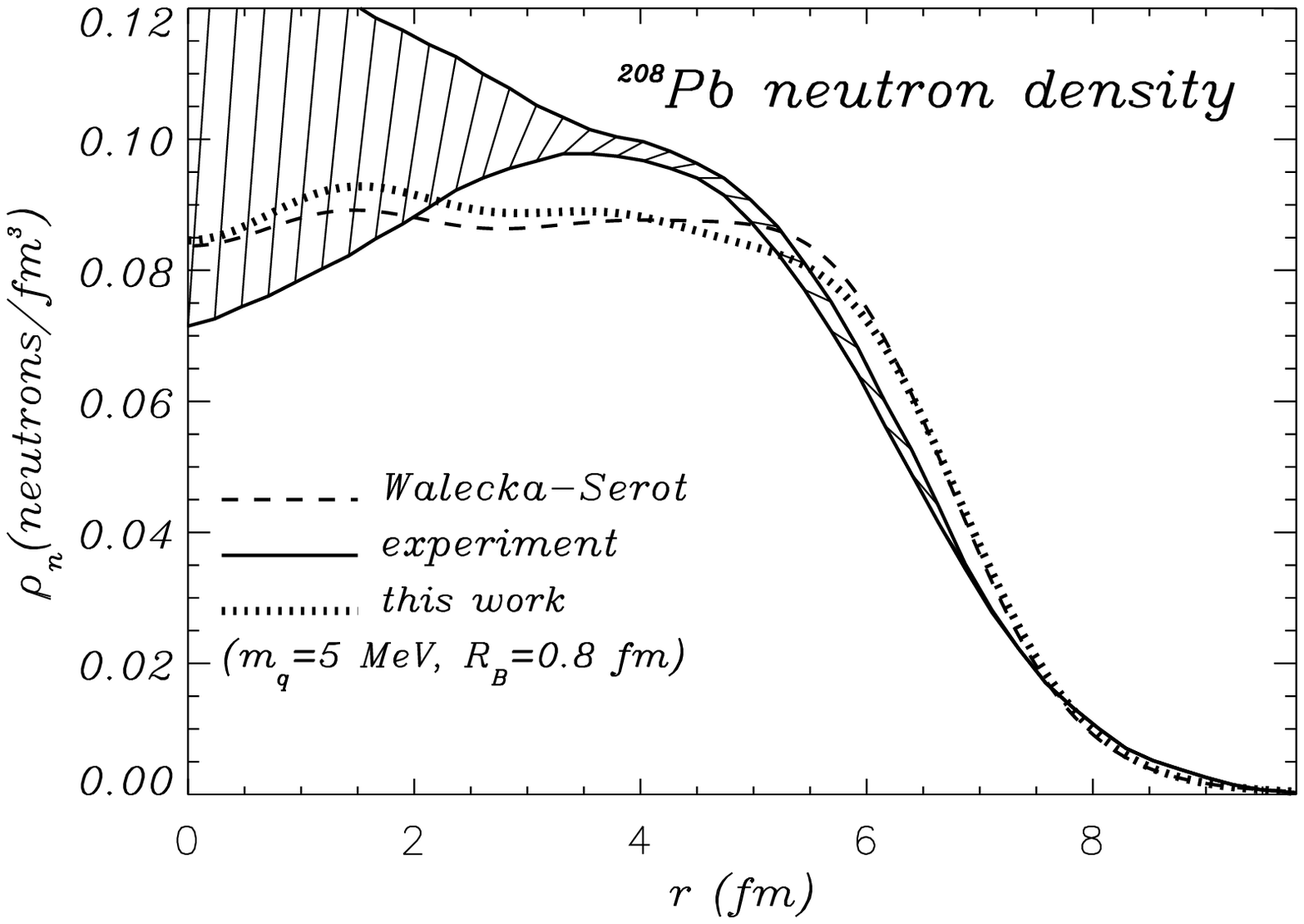,height=9cm}
\end{minipage}
\begin{minipage}[t]{16.5 cm}
\caption{Same as Fig.~\ref{fig:can58} but for $^{208}$Pb
(from Ref.~\cite{SAI-8}).}
\label{fig:pbn58}
\end{minipage}
\end{center}
\end{figure}
We again find reasonable agreement with the data.  
For the isotope shift of $\rho_n$($^{48}$Ca)$-\rho_n$($^{40}$Ca), 
the calculated difference is closer to those of non-relativistic results 
than to those of QHD.  
The neutron density distribution in lead is shown in Fig.~\ref{fig:pbn58}.  
Its behavior is again similar to that of QHD.  

Next, let us show several numerical results for finite nuclei 
in the QMC-II model~\cite{SAI-3}.
A set of coupled, nonlinear equations for the nucleon and meson fields are already obtained. 
The coupling constants $g_\sigma$, $g_\omega$ and $g_\rho$ were fixed to 
describe the nuclear matter properties and 
the bulk symmetry energy per baryon of 35 MeV (see Table~\ref{tab:c.c.-II}).  
\begin{table}
\begin{center}
\begin{minipage}[t]{16.5 cm}
\caption{Model parameters for finite nuclei for $m_q$ = 5 MeV and 
$R_N$ = 0.8 fm (QMC-II).}
\label{tab:c.c.-IIf}
\end{minipage}
\begin{tabular}[t]{c|cccc}
\hline
 type & $g_{\sigma}^2/4\pi$ & $g_{\omega}^2/4\pi$ & 
$g_{\rho}^2/4\pi$ & $m_\sigma$(MeV) \\
\hline
 A & 1.67 & 2.70 & 5.54 & 363 \\
\hline
 B & 2.01 & 3.17 & 5.27 & 393 \\
\hline
 C & 2.19 & 3.31 & 5.18 & 416 \\
\hline
\end{tabular}
\end{center}
\end{table}
Since the $\sigma$ mass however determines the range of the attractive 
interaction, the nuclear-surface slope and its thickness, as in the QMC-I model, 
we again adjust $m_\sigma$ to fit the measured rms charge radius 
of $^{40}$Ca, $r_{ch}$($^{40}$Ca) = 3.48 fm~\cite{SIN}. 
(Notice that variations of $m_\sigma$ at fixed ($g_\sigma / m_\sigma$) again have no effect on the infinite nuclear matter 
properties.)  We summarize the parameters in Table~\ref{tab:c.c.-IIf}.  

\begin{figure}
\epsfysize=9.0cm
\begin{center}
\begin{minipage}[t]{8 cm}
\hspace*{-3cm}
\epsfig{file=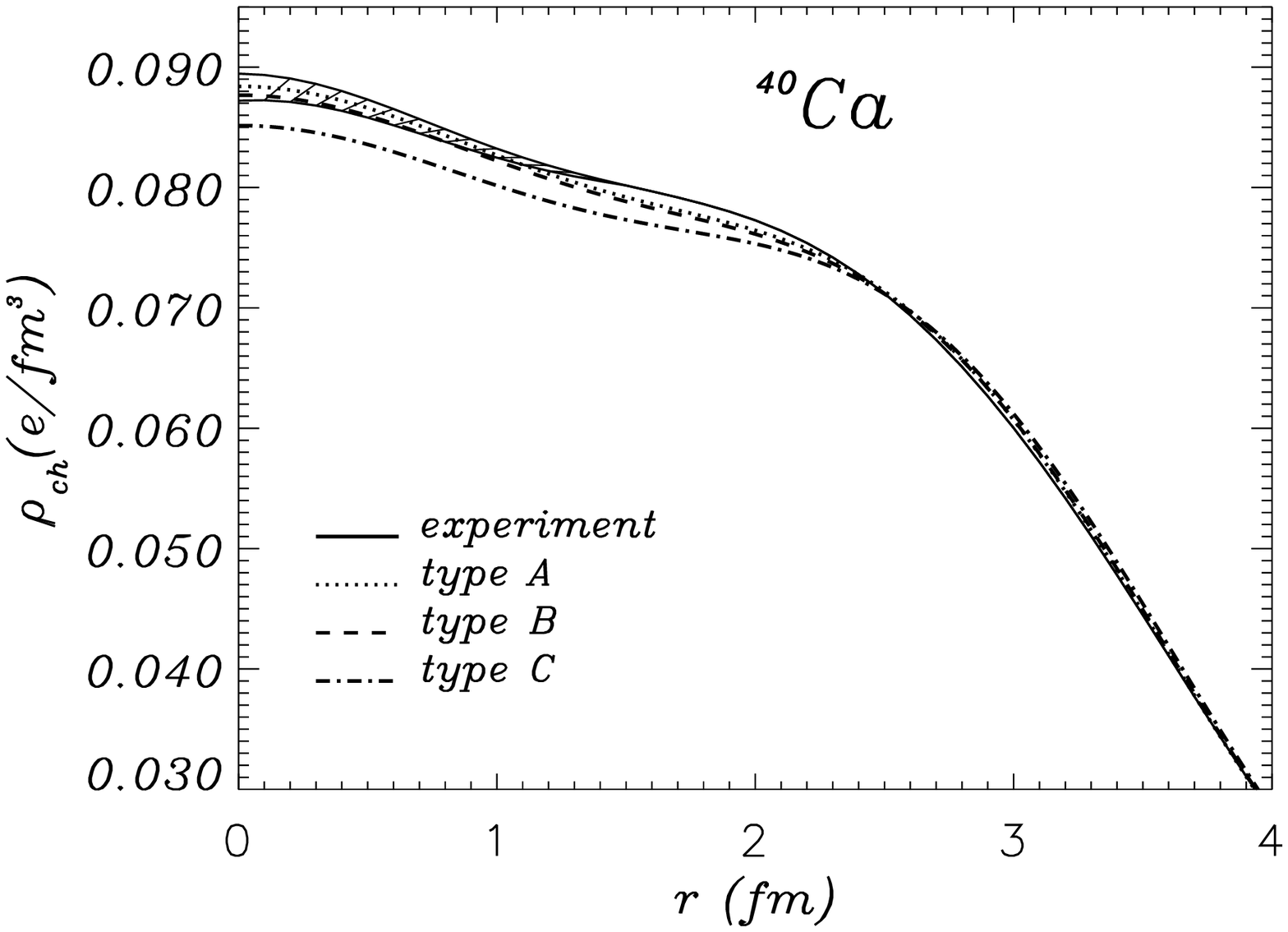,height=9cm}
\end{minipage}
\begin{minipage}[t]{16.5 cm}
\caption{Charge density distribution for $^{40}$Ca (QMC-II) 
compared with the experimental data~\cite{CA40} (from Ref.~\cite{SAI-3}).}
\label{fig:chca40II}
\end{minipage}
\end{center}
\end{figure}
\begin{figure}
\epsfysize=9.0cm
\begin{center}
\begin{minipage}[t]{8 cm}
\hspace*{-3cm}
\epsfig{file=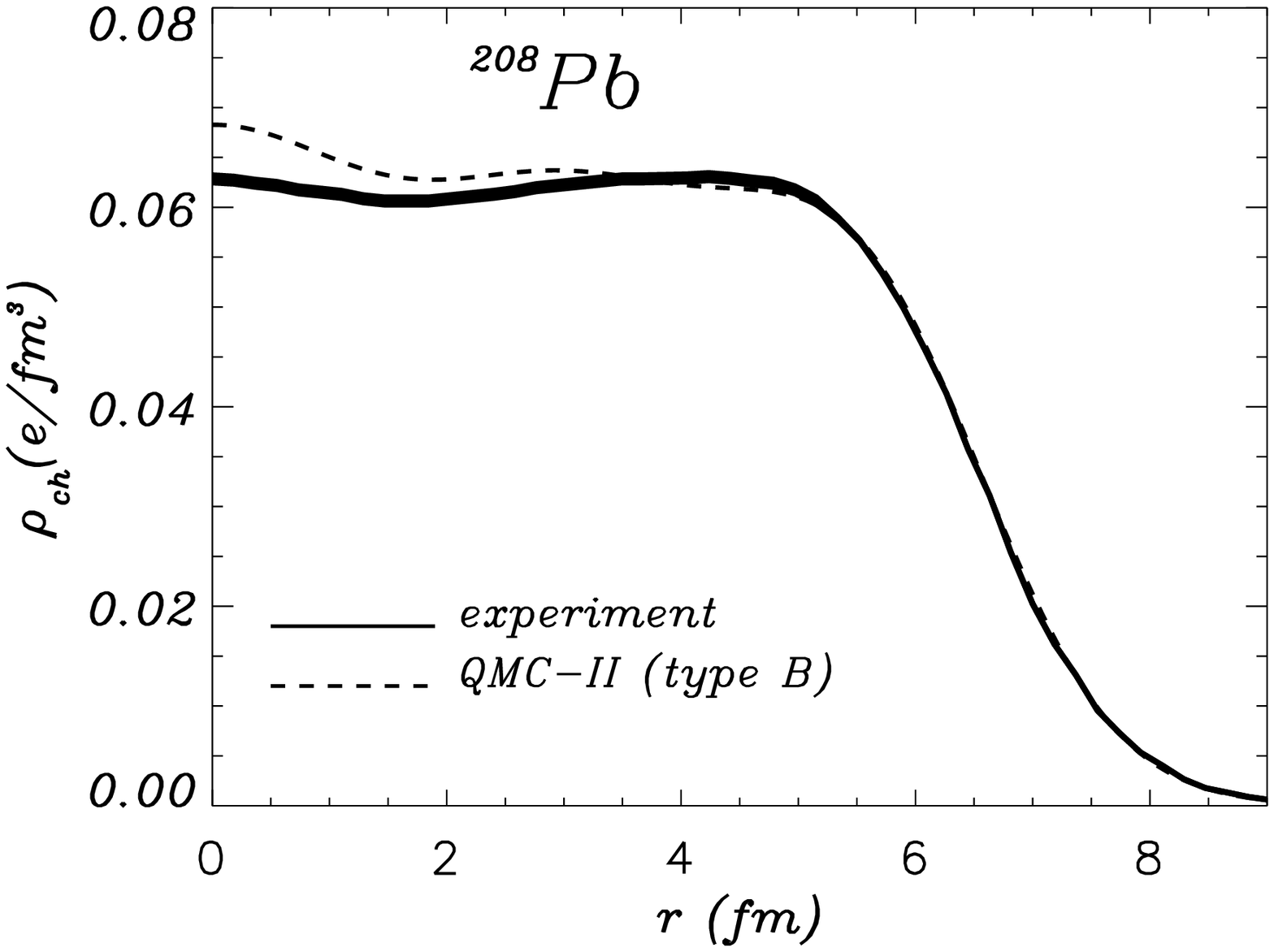,height=9cm}
\end{minipage}
\begin{minipage}[t]{16.5 cm}
\caption{Same as Fig.~\ref{fig:chca40II} but for $^{208}$Pb.  
The parameter set B is used. 
The experimental data are taken from Ref.~\cite{PB208}
(from Ref.~\cite{SAI-3}).}
\label{fig:chpbII}
\end{minipage}
\end{center}
\end{figure}

We show the charge density distributions calculated, $\rho_{ch}$, 
of $^{40}$Ca and $^{208}$Pb in comparison with those of the experimental 
data in Figs.~\ref{fig:chca40II} and~\ref{fig:chpbII}.    
As in the QMC-I model, we have used the convolution of the point-proton 
density with the proton charge distribution to calculate the charge distribution.  For $^{40}$Ca 
the QMC-II model with parameter sets A and B give similar charge 
distributions to those in QMC-I, while the result of QMC-II with  
parameter set C is closer to that in QHD.  From 
Fig.~\ref{fig:chpbII}, we see that the present model also yields a  
charge distribution for $^{208}$Pb which is similar to those calculated using QMC-I or QHD.  

\begin{figure}
\epsfysize=9.0cm
\begin{center}
\begin{minipage}[t]{8 cm}
\hspace*{-2.5cm}
\epsfig{file=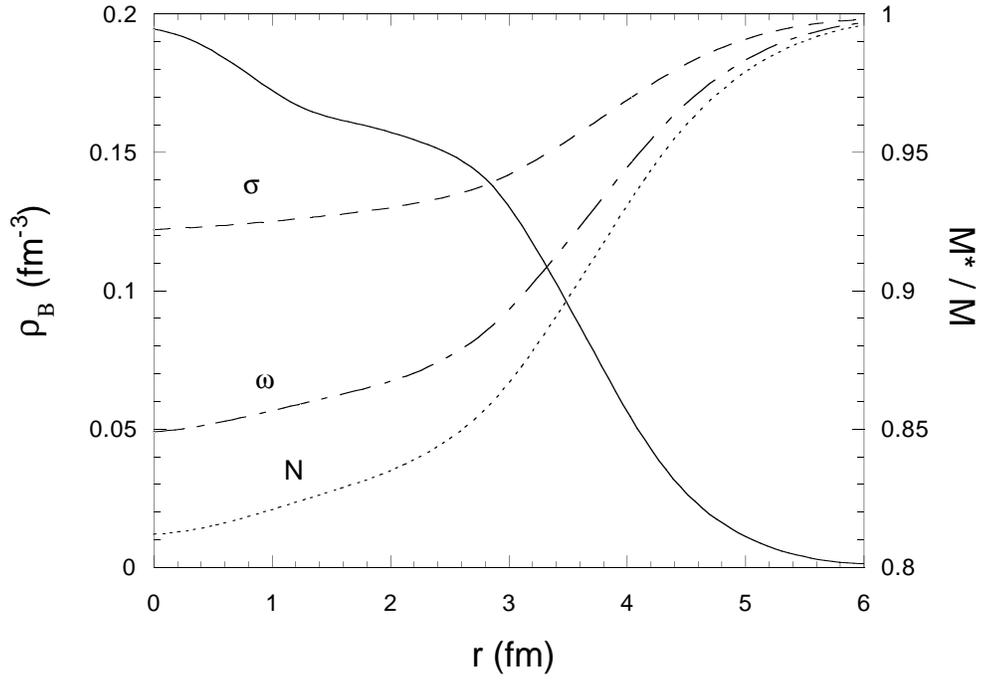,height=9cm}
\end{minipage}
\begin{minipage}[t]{16.5 cm}
\caption{Changes of the nucleon, $\sigma$ and $\omega$ meson masses 
in $^{40}$Ca.  The nuclear baryon density is also illustrated (solid curve).  
The right (left) scale is for the effective mass (the baryon density). 
The parameter set B is used (from Ref.~\cite{SAI-3}).}
\label{fig:hmca}
\end{minipage}
\end{center}
\end{figure}
\begin{figure}
\epsfysize=9.0cm
\begin{center}
\begin{minipage}[t]{8 cm}
\hspace*{-2.5cm}
\epsfig{file=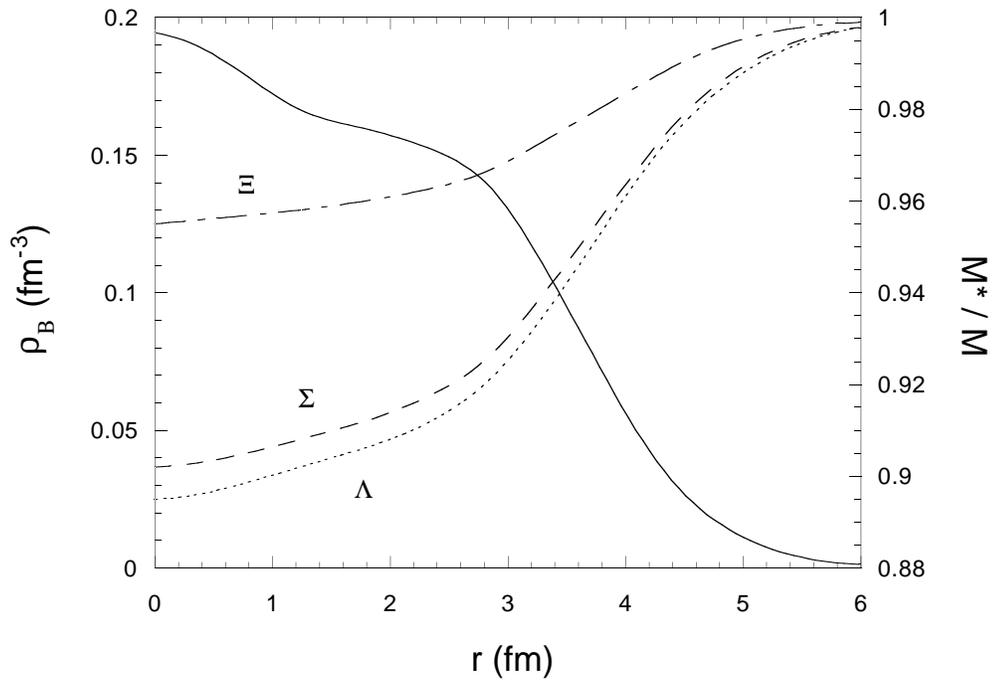,height=9cm}
\end{minipage}
\begin{minipage}[t]{16.5 cm}
\caption{Same as Fig.~\ref{fig:hmca} but for the hyperon 
($\Lambda$, $\Sigma$ and $\Xi$) masses  
in $^{40}$Ca (from Ref.~\cite{SAI-3}).}
\label{fig:hypca}
\end{minipage}
\end{center}
\end{figure}

In Figs.~\ref{fig:hmca} and~\ref{fig:hypca}, we present the changes of the 
nucleon, $\sigma$, $\omega$ and hyperon ($\Lambda$, $\Sigma$ and $\Xi$) masses in $^{40}$Ca. 
Because the interior density of $^{40}$Ca is higher than $\rho_0$, 
the effective hadron masses 
at the center become quite small.  
Using the local-density approximation and Eq.~(\ref{hm3}), 
the QMC-II model allows us to 
calculate the changes of the hyperon ($\Lambda$, $\Sigma$ and $\Xi$) masses in a nucleus. 
Our quantitative calculations for 
the changes of the hyperon masses in finite nuclei may be quite important 
in forthcoming experiments concerning hypernuclei.  

Table~\ref{tab:sum} gives a summary of the calculated binding 
energy per nucleon ($E_B/A$), rms charge radii and the difference 
between nuclear rms radii for neutrons and protons ($r_n - r_p$) 
for several closed-shell nuclei.  
\begin{table}
\begin{center}
\begin{minipage}[t]{16.5 cm}
\caption{Binding energy per nucleon $E_B/A$ (in MeV), rms charge radius 
$r_{ch}$ (in fm) and difference between $r_n$ and $r_p$ (in fm).  
$m_q$ = 5 MeV and $R_N$ = 0.8 fm. ($^*$ fit)}
\label{tab:sum}
\end{minipage}
\begin{tabular}[t]{c|ccc|ccc|ccc}
\hline
 & & $-E_B/A$ & & & $r_{ch}$ & & & $r_n-r_p$ & \\
\hline
Model & QMC-I & QMC-II & Exp. & QMC-I & QMC-II & Exp. & QMC-I & QMC-II & Exp. \\
\hline
$^{16}$O &5.84&5.11&7.98&2.79&2.77&2.73&-0.03&-0.03&0.0 \\
$^{40}$Ca&7.36&6.54&8.45&3.48$^*$&3.48$^*$&3.48&-0.05&-0.05&0.05$\pm$0.05\\
$^{48}$Ca&7.26&6.27&8.57&3.52&3.47&3.53&0.23&0.24&0.2$\pm$0.05 \\
$^{90}$Zr&7.79&6.99&8.66&4.27&4.26&4.28&0.11&0.12&0.05$\pm$0.1 \\
$^{208}$Pb&7.25&6.52&7.86&5.49&5.46&5.49&0.26&0.27&0.16$\pm$0.05 \\
\hline
\end{tabular}
\end{center}
\end{table}
Since the calculated properties do not depend strongly on $m_q$ and $R_N$, 
we only list the values for the QMC-I and QMC-II models with $m_q$ = 5 MeV 
and $R_N$ = 0.8 fm.  References for the experimental values can be found in 
Ref.~\cite{HOR-1}.  While there are still some discrepancies between 
the results and data, the present models provide quite reasonable results.  
We note that the QMC model gives much larger binding energies 
(larger absolute values) per nucleon 
than those of QHD while still reproducing 
the rms charge radii for medium and heavy nuclei quite well.  

\begin{table}
\begin{center}
\begin{minipage}[t]{16.5 cm}
\caption{Calculated proton and neutron spectra of $^{40}$Ca for 
QMC-I and QMC-II (type B) compared with 
the experimental data.  We choose $m_q$ = 5 MeV and $R_N$ = 0.8 fm. All energies are in MeV. 
}
\label{tab:spca40}
\end{minipage}
\begin{tabular}[t]{c|ccc|ccc}
\hline
 & \multicolumn{3}{c}{neutron} & 
\multicolumn{3}{c}{proton} \\
\cline{2-7} 
Shell & QMC-I & QMC-II & Expt. & QMC-I & QMC-II & Expt. \\
\hline
$1s_{1/2}$ & 43.1 & 41.1 & 51.9 & 35.2 & 33.2 & 50$\pm$10 \\
$1p_{3/2}$ & 31.4 & 30.0 & 36.6 & 23.8 & 22.3 & 34$\pm$6 \\
$1p_{1/2}$ & 30.2 & 29.0 & 34.5 & 22.5 & 21.4 & 34$\pm$6 \\
$1d_{5/2}$ & 19.1 & 18.0 & 21.6 & 11.7 & 10.6 & 15.5 \\
$2s_{1/2}$ & 15.8 & 14.7 & 18.9 &  8.5 &  7.4 & 10.9 \\
$1d_{3/2}$ & 17.0 & 16.4 & 18.4 &  9.7 &  9.0 & 8.3  \\
\hline
\end{tabular}
\end{center}
\end{table}
In Table~\ref{tab:spca40}, the calculated spectrum of $^{40}$Ca is presented.
In Fig.~\ref{fig:pbsp2}, for example, the spectra of $^{208}$Pb calculated by the QMC-I is presented.
\begin{figure}
\epsfysize=9.0cm
\begin{center}
\begin{minipage}[t]{8 cm}
\hspace*{-1.5cm}
\epsfig{file=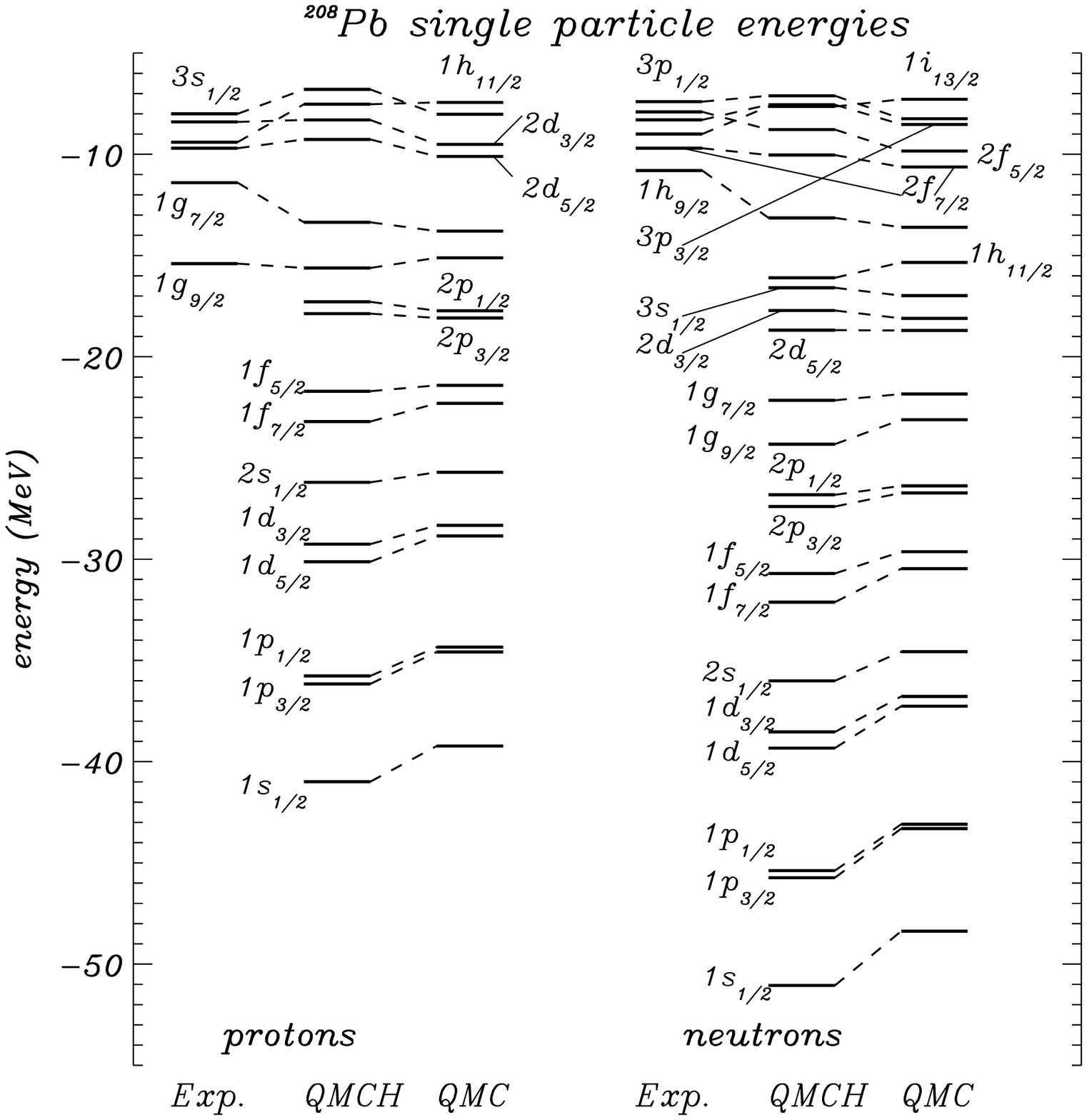,height=10cm}
\end{minipage}
\begin{minipage}[t]{16.5 cm}
\caption{Energy spectrum for $^{208}$Pb predicted by QMC-I. 
QMC(H) denotes the case for $m_q = 5 (300)$ MeV and $R_N = 0.8$ fm. }
\label{fig:pbsp2}
\end{minipage}
\end{center}
\end{figure}
Because of the relatively smaller scalar and vector fields in the present 
model than in QHD or RMF models, the spin-orbit splittings are smaller. 
The improvement over QHD in the binding energy per nucleon comes 
at the expense of a reduction in the spin-orbit force.  
We should note that there is a strong correlation between the 
effective nucleon mass and the spin-orbit force.  

As a test of the sensitivity of the spin-orbit splitting to features of the model, 
we consider the case of a larger quark mass.  For example, 
we have calculated the case $m_q$ = 300 MeV (and $R_N$ = 0.8 fm) which 
is a typical constituent quark mass. 
The calculated spectrum for $m_q$ = 300 MeV is also illustrated 
in Fig.~\ref{fig:pbsp2} (QMCH).  
In this case, the various parameters were 
$g_\sigma^2/4\pi$ = 5.58, $g_\omega^2/4\pi$ = 8.51 (to 
satisfy the saturation condition), $g_\rho^2/4\pi$ = 6.45, 
$m_\sigma$ = 497 MeV (to fit the rms charge radius of $^{40}$Ca), 
$K$ = 334 MeV and $M_N^*$ = 674 MeV at saturation density.  
We should record that the bag radius in this case 
increases by 7\% at saturation density (which is relatively large).  
The slope parameter in Eq.~(\ref{cnpar}) is $a=3.9 \times 10^{-4}$.  
One can expect that a heavy quark mass gives a spectrum closer to 
those of QHD, because the effective nucleon mass is smaller than the value in 
case of the light quark mass.  We can see from the figure 
that the calculated spectrum is somewhat closer to the experimental data. 
We note that the charge density distributions for $^{40}$Ca and $^{208}$Pb 
are also reproduced well in this case.  

The problem concerning the spin-orbit force in the QMC model has also been
studied in Refs.~\cite{BLU,JIN-1}.  This relatively small spin-orbit 
splittings may be somewhat recovered by including the exchange contributions
(Fock terms)~\cite{FOCK}. Alternative is the MQMC 
model~\cite{MUL,JIN-1,MUL-2,LU-1}, in which we can get a small 
effective nucleon mass varying the bag constant in matter 
(see section~\ref{subsec:modfqmc}) and produce large spin-orbit splittings
correspondingly. In next section, we will again study the 
spin-orbit force in exotic nuclei in detail.

Recent RMF models (for example, see Refs.~\cite{RMF,NIKS}) are very successful 
to reproduce single-particle energy spectra of various nuclei (including spin-orbit 
splittings), because of 
several parameters for the nonlinear self-interaction terms.  Furthermore, the 
density-dependent relativistic hadron field (DDRH) 
model~\cite{DDFT} is attractive (see also Ref.~\cite{Tjon}), 
because such an approach is also successful and thought to be related to the 
underlying microscopic description of nuclear interactions.  However, we should 
emphasize that the idea of the QMC model is completely different from those 
density-dependent nuclear models, and that the vertices in QMC depend on 
the scalar field in matter, and this is caused by the substructure of nucleon.  

\subsection{\it Strange, charm and bottom hadrons in nucleus  
\label{subsec:scbA}}

In this section, we start from the mean-field Lagrangian density for 
strange, charm and bottom hypernuclei, and study first the properties 
of these hadrons in nuclear medium. 
Then, we study the properties of various hypernuclei 
with strange, charm and bottom hyperons.

\subsubsection{\it Strange, charm, and bottom hadron properties 
in nuclear matter
\label{subsubsec:scbmatter}}

Until recently, theoretical studies of hadron properties 
in a nuclear medium have been restricted to just a few approaches, 
such as QCD sum rules and QHD-type relativistic mean-field models. 
For strange hadrons including the possible 
$\Theta^+$ pentaquark (in QMC~\cite{Ryu}), some 
studies~\cite{Jin,Klingl1,Navarra} were made using QCD sum rules. 
A similar story holds for heavy baryons with 
charm and bottom quarks in nuclear medium, 
where some studies for  
$J/\Psi$~\cite{Hayashigaki1,Klingl2,Kim} 
and $D (\Dbar)$~\cite{Hayashigaki2} were also made using QCD sum rules. 

Next we turn to the study of the properties of hadrons 
in symmetric nuclear matter 
using the QMC (QMC-I) model. 
One of the advantages of the QMC model is that it can treat
various hadrons in nuclear medium in a simple, systematic manner based on
SU(6) quark model as long as the hadrons contain light quarks.
One simply uses the same coupling constants between the
light quarks and the isoscalar-scalar ($\sigma$), 
isoscalar-vector ($\omega$) and
isovector-vector ($\rho$) fields.
One does not need to introduce new coupling constants 
which depend on the hadron species in the QMC model.

To study first the properties of strange, charm, and bottom hadrons 
in symmetric nuclear matter,
we start from a mean-field Lagrangian density for the corresponding  
hypernuclei~\cite{Tsushima_hyp,Tsushima_bc,Tsushima_hypbc}. 
We assume that the hypernuclei are static and spherically symmetric.  
Practically, we treat a hypernucleus as a closed shell nuclear core
plus one strange, charm or bottom baryon  
($\Lambda,\Sigma,\Xi,\Lambda_c,\Sigma_c,\Xi_c,\Lambda_b$),  
ignoring the effects of non-sphericity due to the embedded baryon.
(Hereafter we will refer each of these baryons as simply the "hyperon".)
The existence of such a hyperon inside or outside of the nuclear core,  
in particular, a very heavy hyperon $\Lambda_b$, will
break spherical symmetry and one should include this effect 
in a truly rigorous treatment. We neglected it in this introductory study 
because the effect is expected to be small 
for spectroscopic calculations
for medium to large baryon number 
hypernuclei~\cite{fur,coh}. 
However, we include the response of the nuclear core
arising from the self-consistent
calculation for the hypernucleus, which is a purely 
relativistic effect~\cite{coh,coh2}.
(For the system of infinitely large (symmetric) nuclear matter, 
one can safely ignore the effect of the presence of such a hyperon.)
Furthermore, we adopt the mean-field approximation in the same way that it   
was applied for the normal nuclei studied in section~\ref{subsec:fnt}.
Thus, we ignore also the $\rho NN$ tensor coupling below. 

A relativistic Lagrangian density for hypernuclei in the QMC model 
is given by~\cite{Tsushima_hyp,Tsushima_bc,Tsushima_hypbc}:
\begin{eqnarray}
{\cal L}^{Y}_{QMC} &=& {\cal L}_{QMC} + {\cal L}^Y_{QMC},
\label{eq:LagYQMC} \\
{\cal L}_{QMC} &\equiv&  \overline{\psi}_N(\vec{r})
\left[ i \gamma \cdot \partial
- M_N^*(\sigma) - (\, g_\omega \omega(\vec{r})
+ g_\rho \frac{\tau^N_3}{2} b(\vec{r})
+ \frac{e}{2} (1+\tau^N_3) A(\vec{r}) \,) \gamma_0
\right] \psi_N(\vec{r}) \quad \nn \\
  & & - \frac{1}{2}[ (\nabla \sigma(\vec{r}))^2 +
m_{\sigma}^2 \sigma(\vec{r})^2 ]
+ \frac{1}{2}[ (\nabla \omega(\vec{r}))^2 + m_{\omega}^2
\omega(\vec{r})^2 ] \nn \\
 & & + \frac{1}{2}[ (\nabla b(\vec{r}))^2 + m_{\rho}^2 b(\vec{r})^2 ]
+ \frac{1}{2} (\nabla A(\vec{r}))^2, \label{eq:LagN} \\
{\cal L}^Y_{QMC} &\equiv&
\overline{\psi}_Y(\vec{r})
\left[ i \gamma \cdot \partial
- M_Y^*(\sigma)
- (\, g^Y_\omega \omega(\vec{r})
+ g^Y_\rho I^Y_3 b(\vec{r})
+ e Q_Y A(\vec{r}) \,) \gamma_0
\right] \psi_Y(\vec{r}), 
\nn\\
& & (Y = \Lambda,\Sigma^{0,\pm},\Xi^{0,+},
\Lambda^+_c,\Sigma_c^{0,+,++},\Xi_c^{0,+},\Lambda_b),
\label{eq:LagY}
\end{eqnarray}
where $\psi_N(\vec{r})$ and $\psi_Y(\vec{r})$
are respectively the nucleon and the hyperon 
(strange, charm or bottom baryon) fields.
Other notations are given in section~\ref{subsubsec:relativistic}.

In an approximation where the $\sigma$, $\omega$ and $\rho$ fields couple
only to the $u$ and $d$ light quarks,
the coupling constants for the hyperon, 
are obtained as $g^Y_\omega = (n_q/3) g_\omega$, and
$g^Y_\rho \equiv g_\rho = g_\rho^q$, with $n_q$ being the total number of
valence light quarks in the hyperon $Y$. $I^Y_3$ and $Q_Y$
are the third component of the hyperon isospin operator and its electric
charge in units of the proton charge, $e$, respectively.
The field dependent $\sigma$-$N$ and $\sigma$-$Y$
coupling strengths,
$g_\sigma(\sigma) \equiv g^N_\sigma(\sigma)$ and  $g^Y_\sigma(\sigma)$,
appearing in Eqs.~(\ref{eq:LagN}) and~(\ref{eq:LagY}), are defined by
\bg
M_N^*(\sigma) &\equiv& M_N - g_\sigma(\sigma)
\sigma(\vec{r}) ,  \\
M_Y^*(\sigma) &\equiv& M_Y - g^Y_\sigma(\sigma)
\sigma(\vec{r}) , \label{effective_mass}
\en
where $M_N$ ($M_Y$) is the free nucleon (hyperon) mass 
(see also Eq.~(\ref{efnmas})).
Note that the dependence of these coupling strengths on the applied
scalar field must be calculated self-consistently within the quark
model~\cite{GUI-1,SAI-1,GUI-3,SAI-8,Tsushima_hyp}.
Hence, unlike QHD~\cite{QHD}, even though
$g^Y_\sigma(\sigma) / g_\sigma(\sigma)$ may be
2/3 or 1/3 depending on the number of light quarks in the hyperon 
in free space, $\sigma = 0$ (even this is true only when their bag 
radii in free space are exactly the same), this will not necessarily 
be the case in a nuclear medium. 

In the following, we consider the limit of
infinitely large, uniform (symmetric) nuclear matter,
where all scalar and vector fields become constant.
In this limit, we can treat any single hadron (denoted by $h$)  
embedded in the nuclear medium 
in the same way as for a hyperon. One simply may replace  
${\cal L}^Y_{QMC}$ in Eq.~(\ref{eq:LagY}) 
by the corresponding Lagrangian density for the hadron $h$.  

The Dirac equations for the quarks and antiquarks
($q=u$ or $d$, and $Q=s,c$ or $b$, hereafter)
in the bag of hadron $h$ in nuclear matter at the position  
$x=(t,\vec{r})$ are given by~\cite{Tsushima_k,Tsushima_d}:
\begin{eqnarray}
\left[ i \gamma \cdot \partial_x -
(m_q - V^q_\sigma)
\mp \gamma^0
\left( V^q_\omega +
\frac{1}{2} V^q_\rho
\right) \right]
\left( \begin{array}{c} \psi_u(x)  \\
\psi_{\bar{u}}(x) \\ \end{array} \right) &=& 0,
\label{diracu}\\
\left[ i \gamma \cdot \partial_x -
(m_q - V^q_\sigma)
\mp \gamma^0
\left( V^q_\omega -
\frac{1}{2} V^q_\rho
\right) \right]
\left( \begin{array}{c} \psi_d(x)  \\
\psi_{\bar{d}}(x) \\ \end{array} \right) &=& 0,
\label{diracd}\\
\left[ i \gamma \cdot \partial_x - m_{Q} \right]
\psi_{Q} (x)\,\, ({\rm or}\,\, \psi_{\Qbar}(x)) &=& 0,  
\qquad (|\vec{r}|\le {\rm bag~radius}), 
\label{diracQ}
\end{eqnarray}
where we neglect the Coulomb force, and assume SU(2) symmetry for 
the light quarks ($q=u=d$).
The constant mean-field potentials in nuclear matter are defined by, 
$V^q_\omega \equiv g^q_\omega \omega$ and
$V^q_\rho \equiv g^q_\rho b$,
with $g^q_\sigma$, $g^q_\omega$ and
$g^q_\rho$ the corresponding quark-meson coupling constants.
                                                                             
The normalized, static solution for the ground state quarks or antiquarks
with flavor $f$ in the hadron $h$, may be written,
$\psi_f (x) = N_f e^{- i \epsilon_f t / R_h^*}
\psi_f (\vec{r})$,
where $N_f$ and $\psi_f(\vec{r})$
are the normalization factor and
corresponding spin and spatial part of the wave function.
The bag radius in medium for a hadron $h$, $R_h^*$,
is determined through the
stability condition for the mass of the hadron against the
variation of the bag radius~\cite{GUI-1,SAI-1,GUI-3}.
The eigenenergies in units of $1/R_h^*$ are given by,
\bge
\left( \begin{array}{c}
\epsilon_u \\
\epsilon_{\bar{u}}
\end{array} \right)
= \Omega_q^* \pm R_h^* \left(
V^q_\omega
+ \frac{1}{2} V^q_\rho \right),\,\,
\left( \begin{array}{c} \epsilon_d \\
\epsilon_{\bar{d}}
\end{array} \right)
= \Omega_q^* \pm R_h^* \left(
V^q_\omega
- \frac{1}{2} V^q_\rho \right),\,\,
\epsilon_{Q}
= \epsilon_{\Qbar} =
\Omega_{Q}.
\label{energy}
\ene
                                                                                
The hadron masses
in a nuclear medium $m^*_h$ (free mass $m_h$),
are calculated by
\begin{eqnarray}
m_h^* &=& \sum_{j=q,\bar{q},Q,\Qbar}
\frac{ n_j\Omega_j^* - z_h}{R_h^*}
+ {4\over 3}\pi R_h^{* 3} B,\quad
\left. \frac{\partial m_h^*}
{\partial R_h}\right|_{R_h = R_h^*} = 0,
\label{hmass}
\end{eqnarray}
where $\Omega_q^*=\Omega_{\bar{q}}^*
=[x_q^2 + (R_h^* m_q^*)^2]^{1/2}$, with
$m_q^*=m_q{-}g^q_\sigma \sigma$,
$\Omega_Q^*=\Omega_{\Qbar}^*=[x_Q^2 + (R_h^* m_Q)^2]^{1/2}$,
and $x_{q,Q}$ being the lowest bag eigenfrequencies.
$n_q (n_{\qbar})$ and $n_Q (n_{\Qbar})$ 
are the quark (antiquark)
numbers for the quark flavors $q$ and $Q$, respectively.
The MIT bag quantities, $z_h$, $B$, $x_{q,Q}$,
and $m_{q,Q}$ are the parameters for the sum of the c.m. and gluon
fluctuation effects, bag constant, lowest eigenvalues for the quarks
$q$ or $Q$, respectively, and the corresponding current quark masses.
$z_N$ and $B$ ($z_h$) are fixed by fitting the nucleon
(the hadron) mass in free space. For the current quark masses
we use $(m_{u,d},m_s,m_c,m_b) = (5,250,1300,4200)$ MeV,
where the values for $m_c$ and $m_b$ are
the averaged values from Refs.~\cite{PDG96} and~\cite{PDG00}, respectively,
and these values were used in
Refs.~\cite{Tsushima_bc,Tsushima_hypbc}.
Then, we obtain the bag constant $B = (170$ MeV$)^4$.
The quark-meson coupling constants, which are determined so as
to reproduce the saturation properties of symmetric nuclear matter are,
($g^q_\sigma, g^q_\omega, g^q_\rho$) = ($5.69, 2.72, 9.33$),
where $g_\sigma \equiv g^N_\sigma \equiv
3 g^q_\sigma S_N(0) = 3 \times 5.69 \times 0.483
= 8.23$~\cite{SAI-8}. These are summarized in Table~\ref{coupcc}. 
(See also Eq.~(\ref{msdef}) and section~\ref{subsec:matter}.)
The parameters $z_h$, and the bag radii $R_h$ for various hadrons
in free space, and some quantities calculated at normal
nuclear mater density $\rho_0 = 0.15$ fm$^{-3}$ are listed
in Table~\ref{bagparambc},
together with the free space masses~\cite{PDG96,PDG00,PDG98,PDG02}.
%
\begin{table}
\begin{center}
\begin{minipage}[t]{16.5cm}
\caption{Current quark masses (input), coupling constants 
and the bag constant.}
\label{coupcc}
\end{minipage}
\begin{tabular}[t]{r|r||l|l}
\hline
$m_{u,d}$ &5    MeV &$g^q_\sigma$ &5.69\\
$m_s$     &250  MeV &$g^q_\omega$ &2.72\\
$m_c$     &1300 MeV &$g^q_\rho$   &9.33\\
$m_b$     &4200 MeV &$B^{1/4}$    &170 MeV\\
\end{tabular}
\end{center}
\end{table}
\begin{table}
\begin{center}
\begin{minipage}[t]{16.5cm}
\caption{The bag parameters,
various hadron masses and the bag radii in free space
[at normal nuclear matter density, $\rho_0=0.15$ fm$^{-3}$]
$z_h, R_h$ and $M_h$ [$M_h^*$ and $R_h^*$].
$M_h$ and $R_N = 0.8$ fm in free space are inputs.
Note that the quantities for the physical 
$\omega$, $\phi$, $\eta$ and $\eta'$ 
are calculated including the octet-singlet mixing effect, 
and that $\omega$ and $\rho$ 
below are standing for the physical particles and are different from those 
appearing in the Lagrangian density of QMC. 
(See section~\ref{subsec:mesonA} for details.)
}
\label{bagparambc}
\end{minipage}
\begin{tabular}{c|ccc|cc}
\hline
h &$z_h$ &$M_h$ (MeV) &$R_h$ (fm) &$M_h^*$ (MeV) &$R_h^*$ (fm)\\
\hline
$N$           &3.295 &939.0  &0.800 &754.5  &0.786\\
$\Lambda$     &3.131 &1115.7 &0.806 &992.7  &0.803\\
$\Sigma$      &2.810 &1193.1 &0.827 &1070.4 &0.824\\
$\Xi$         &2.860 &1318.1 &0.820 &1256.7 &0.818\\
$\Lambda_c$   &1.766 &2284.9 &0.846 &2162.5 &0.843\\
$\Sigma_c$    &1.033 &2452.0 &0.885 &2330.2 &0.882\\
$\Xi_c$       &1.564 &2469.1 &0.853 &2408.0 &0.851\\
$\Lambda_b$   &-0.643&5624.0 &0.930 &5502.9 &0.928\\
\hline
$\omega$      &1.866 &781.9  &0.753 &658.7  &0.749\\
$\rho$        &1.907 &770.0  &0.749 &646.2  &0.746\\
$K$           &3.295 &493.7  &0.574 &430.4  &0.572\\
$K^*$         &1.949 &893.9  &0.740 &831.9  &0.738\\
$\eta$        &3.131 &547.5  &0.603 &483.9  &0.600\\
$\eta'$       &1.711 &957.8  &0.760 &896.5  &0.758\\
$\phi$        &1.979 &1019.4 &0.732 &1018.9 &0.732\\
$D$           &1.389 &1866.9 &0.731 &1804.9 &0.730\\
$D^*$         &0.849 &2000.8 &0.774 &1946.7 &0.772\\
$B$           &-1.136&5279.2 &0.854 &5218.1 &0.852\\
\end{tabular}
\end{center}
\end{table}
%

                                                                                
However, in studies of the kaon system, we found that it was
necessary to increase the strength of the vector
coupling to the light quarks in the $K^+$ (by a factor of
$1.4^2$, i.e., $g_{K\omega}^q \equiv 1.4^2 g^q_\omega$)
in order to reproduce the empirically extracted $K^+$-nucleus
interaction~\cite{Tsushima_k}. This may be related to the fact that
kaon is a pseudo-Goldstone boson, where treatment of the Goldstone
bosons in a naive quark model is usually unsatisfactory.
We also assume this, $g^q_\omega \to 1.4^2 g^q_\omega$,
{}for the $D$, $\Dbar$~\cite{Tsushima_d} and  
$B$ and $\Bbar$ mesons to obtain an upper limit on the corresponding binding.
The scalar ($V^{h}_s$) and vector ($V^{h}_v$) potentials
felt by the hadrons $h$,
in nuclear matter are given by,
\bg
V^h_s &=& m^*_h - m_h,
\label{spot}\\
V^h_v &=&
  (n_q - n_{\bar{q}}) {V}^q_\omega + I^h_3 V^q_\rho,
\qquad (V^q_\omega \to \tilde{V}^q_\omega \equiv 1.4^2 {V}^q_\omega\,\,
{\rm for}\, K,\Kbar,D,\Dbar,B,\Bbar),
\label{vpot}\\
&\simeq& 41.8\times (n_q - n_{\bar{q}}) \left(\frac{\rho_B}{\rho_0}\right) 
+ 42.4\times I^h_3 \left(\frac{\rho_3}{\rho_0}\right) 
\quad{\rm (MeV)\quad with\quad} \rho_0=0.15\quad ({\rm fm}^{-3}), 
\label{vpotemp}
\en
where $I^h_3$ is the third component of isospin projection
of the hadron $h$, and $\rho_B=\rho_p+\rho_n$ ($\rho_3=\rho_p-\rho_n$) 
the baryon (isovector baryon) density with $\rho_p$ and 
$\rho_n$ being the proton and neutron densities, respectively. 
Thus, the vector potential felt by a heavy baryon
with charm and bottom quarks, is equal to that of the strange hyperon with
the same light quark configuration in QMC.
(See also section~\ref{subsubsec:scaling} concerning the scalar potential.)

In Figs.~\ref{mesonmass} and~\ref{baryonmass} we show ratios of
effective masses (free masses + scalar potentials)
versus those of the free particles, for
mesons and baryons, respectively.
\begin{figure}
\epsfysize=9.0cm
\begin{center}
\begin{minipage}[t]{8cm}
\hspace*{-2cm}
\epsfig{file=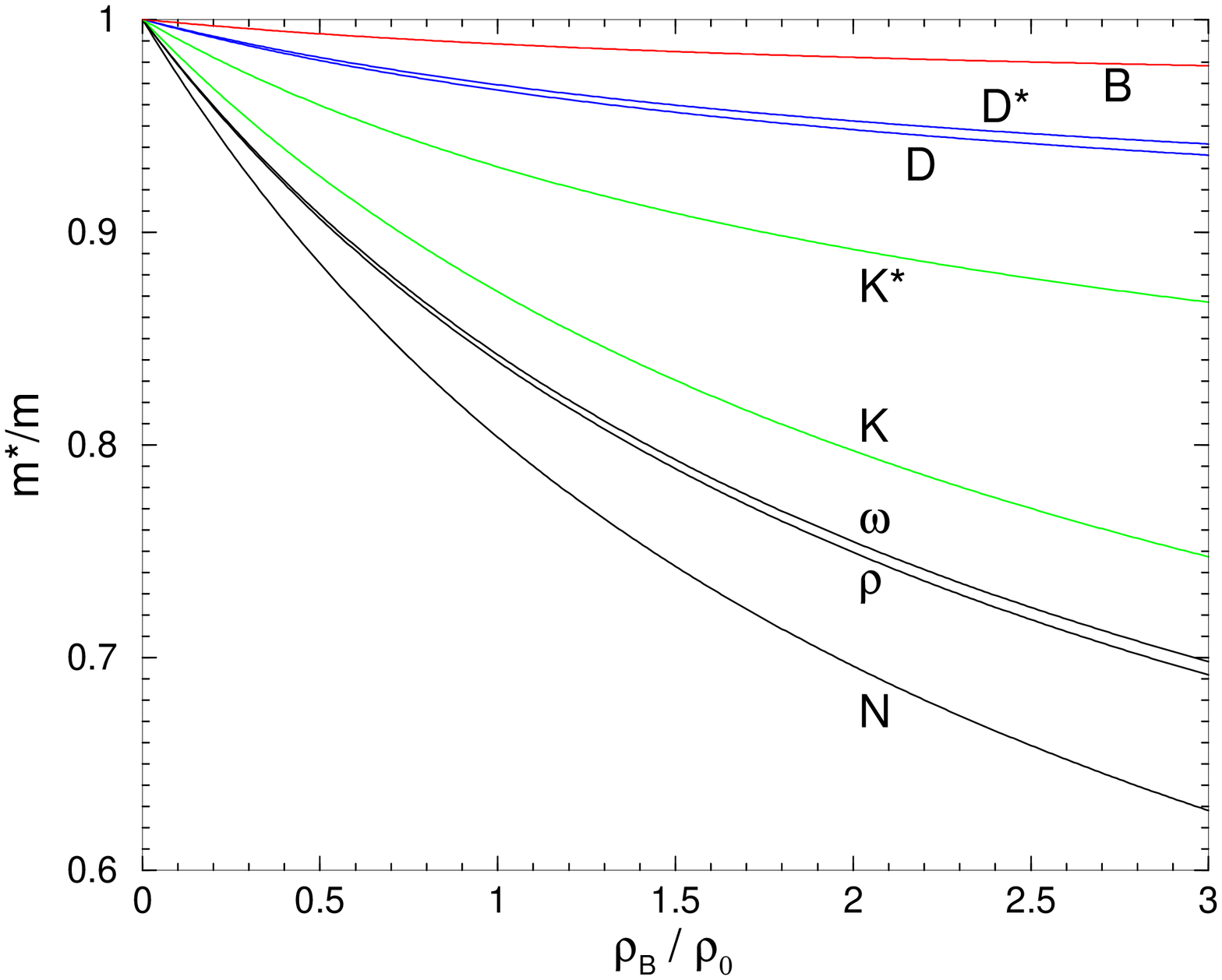,width=9cm,angle=-90}
\end{minipage}
\begin{minipage}[t]{16.5cm}
\caption{Effective mass ratios for mesons in symmetric nuclear matter
($\rho_0 = 0.15$ fm$^{-3}$).
$\omega$ and $\rho$ stand for physical mesons which are treated in
the quark model, and should not be confused with the fields appearing
in the QMC model (from Ref.~\cite{Tsushima_bc}).}
\label{mesonmass}
\end{minipage}
\end{center}
\end{figure}
\begin{figure}
\epsfysize=9.0cm
\begin{center}
\begin{minipage}[t]{8cm}
\hspace*{-2cm}
\epsfig{file=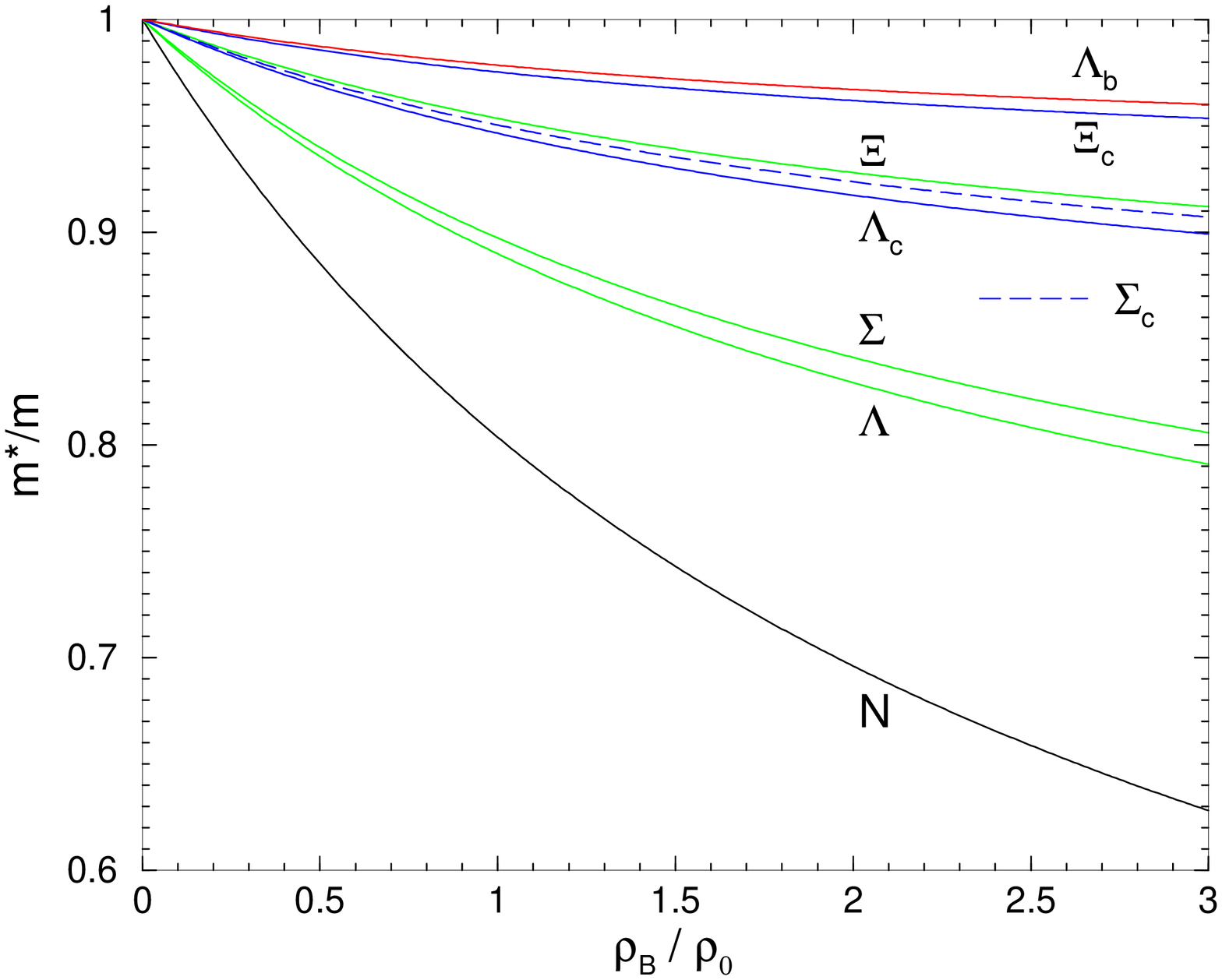,width=9cm,angle=-90}
\end{minipage}
\begin{minipage}[t]{16.5cm}
\caption{Effective mass ratios for baryons in symmetric nuclear matter. 
(from Ref.~\cite{Tsushima_bc}).}
\label{baryonmass}
\end{minipage}
\end{center}
\end{figure}
With increasing density the ratios decrease as expected,
but this decrease in magnitude is, from larger to smaller:
hadrons with only light
quarks, with one strange quark, with one charm quark, and with one
bottom quark. This is because their masses in free space
are in the order from light to heavy. Thus, the net ratios for the
decrease in masses (developing of scalar masses) compared to that of
the free masses becomes smaller.
This may be regarded as a measure of the role of light
quarks in each hadron system in nuclear matter, in the sense of how much 
they contribute a partial restoration of chiral
symmetry in the hadron. In Fig.~\ref{mesonmass} one can notice a 
somewhat anomalous behavior of the ratio for the kaon ($K$) mass, which is 
related to what we meant by its pseudo-Goldstone boson nature -- 
i.e., its mass in free space is relatively light, $m_K \simeq 495$ MeV,
and the relative reduction in its mass in-medium is large.

Perhaps it is much more quantitative and direct to compare
the scalar potentials (see Eq.~(\ref{spot})) 
{}felt by each hadron in nuclear matter, with the 
calculated results shown in Fig.~\ref{spotential}.
\begin{figure}
\epsfysize=9.0cm
\begin{center}
\begin{minipage}[t]{8cm}
\hspace*{-2cm}
\epsfig{file=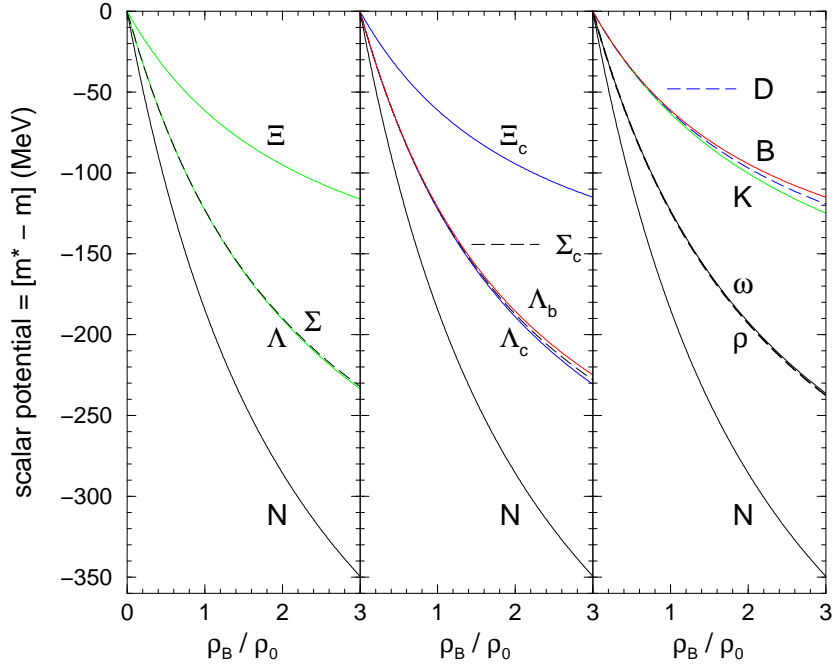,width=9cm,angle=-90}
\end{minipage}
\begin{minipage}[t]{16.5cm}
\caption{Scalar potentials for various hadrons in symmetric nuclear matter. 
(from Ref.~\cite{Tsushima_bc}).}
\label{spotential}
\end{minipage}
\end{center}
\end{figure}
These results confirm that
the scalar potential felt by the hadron $h$, $V_s^h$,
follows a simple light quark number scaling rule:
\bge
V_s^h \simeq \left(\frac{n_q + n_{\qbar}}{3}\right) V_s^N 
= - \left(\frac{n_q + n_{\qbar}}{3}\right) \delta M^*_N 
= - \left(\frac{n_q + n_{\qbar}}{3}\right) (M_N - M^*_N),
\ene
where $n_q$ ($n_{\qbar}$) is the number of light quarks (antiquarks) in
hadron $h$ and $V_s^N$ is the scalar potential felt by the nucleon 
(see Eq.~(\ref{spot})).
This {\it scaling formula} for hadrons in nuclear matter 
is already discussed in section~\ref{subsubsec:scaling}.
It is interesting to notice that, the baryons with charm or bottom quarks
($\Xi_c$ has the quark configuration, $qsc$), show very similar
features to those of the hyperons with one or two strange quarks.
Then, we can expect that the hyperons with charm or bottom
quarks, will also form charmed (bottom) hypernuclei as the strange 
hyperons do. (See Eq.~(\ref{vpot}), and recall that the repulsive vector 
potentials are the same for the corresponding strange hyperons 
with the same light quark configurations.)

In addition, as we discuss later 
(in section~\ref{subsec:mesonA}) in detail, 
the $B^-$ meson will also certainly form meson-nucleus bound states,
because it has the valence structure $\ubar b$ and feels a strong attractive
vector potential in addition to the attractive
Coulomb force. This makes it much easier to bind in a
nucleus compared to the $D^0$~\cite{Tsushima_d}, which is $c \ubar$
and is blind to the Coulomb force.
This reminds us of a situation
of the kaonic ($K^- (\ubar s)$) atom~\cite{katom1,katom2}.
A study of $B^- (\ubar b)$ atoms would be a fruitful experimental
program. Such atoms will have the
meson much closer to the nucleus and will thus probe
even smaller changes in the
nuclear density. This will provide complementary information to
that given by $D^- (\bar{c} d)$-nucleus bound states, which give 
information on the vector potential in a nucleus~\cite{Tsushima_d} 
-- again, see section~\ref{subsec:mesonA}. 

\subsubsection{\it Mean-field equations of motion for 
strange, charm, and bottom hypernuclei
\label{subsubsec:hypMFE}}

Next we consider the predictions for strange, charm and bottom hypernuclei.
The Lagrangian density Eq.~(\ref{eq:LagYQMC}) leads to a set of
equations of motion for the hypernuclear system:
\begin{eqnarray}
& &[i\gamma \cdot \partial -M^*_N(\sigma)-
(\, g_\omega \omega(\vec{r}) + g_\rho \frac{\tau^N_3}{2} b(\vec{r})
 + \frac{e}{2} (1+\tau^N_3) A(\vec{r}) \,)
\gamma_0 ] \psi_N(\vec{r}) = 0, \label{eqdiracn}\\
& &[i\gamma \cdot \partial - M^*_Y(\sigma)-
(\, g^Y_\omega \omega(\vec{r}) + g_\rho I^Y_3 b(\vec{r})
+ e Q_Y A(\vec{r}) \,)
\gamma_0 ] \psi_Y(\vec{r}) = 0, \label{eqdiracy}\\
& &(-\nabla^2_r+m^2_\sigma)\sigma(\vec{r}) =
- \left[\frac{\partial M_N^*(\sigma)}{\partial \sigma}\right]\rho_s(\vec{r})
- \left[\frac{\partial M_Y^*(\sigma)}{\partial \sigma}\right]\rho^Y_s(\vec{r}),
\nn \\
& & \hspace{7.5em} \equiv g_\sigma Y_N(\sigma) \rho_s(\vec{r})
    + g^Y_\sigma C_Y(\sigma) \rho^Y_s(\vec{r}) , \label{eqsigma}\\
& &(-\nabla^2_r+m^2_\omega) \omega(\vec{r}) =
g_\omega \rho_B(\vec{r}) + g^Y_\omega
\rho^Y_B(\vec{r}) ,\label{eqomega}\\
& &(-\nabla^2_r+m^2_\rho) b(\vec{r}) =
\frac{g_\rho}{2}\rho_3(\vec{r}) + g^Y_\rho I^Y_3 \rho^Y_B(\vec{r}),
 \label{eqrho}\\
& &(-\nabla^2_r) A(\vec{r}) =
e \rho_p(\vec{r})
+ e Q_Y \rho^Y_B(\vec{r}) ,\label{eqcoulomb}
\end{eqnarray}
where, $\rho_s(\vec{r})$ ($\rho^Y_s(\vec{r})$), $\rho_B(\vec{r})$
($\rho^Y_B(\vec{r})$), $\rho_3(\vec{r})$ and
$\rho_p(\vec{r})$ are the scalar, baryon, third component of isovector,
and proton densities at position $\vec{r}$ in
the hypernucleus~\cite{GUI-3,SAI-8,Tsushima_hyp}.
On the right hand side of Eq.~(\ref{eqsigma}),
$- [{\partial M_N^*(\sigma)}/{\partial \sigma}] \equiv
g_\sigma C_N(\sigma)$ and
$- [{\partial M_Y^*(\sigma)}/{\partial \sigma}] \equiv
g^Y_\sigma C_Y(\sigma)$, where $g_\sigma \equiv g_\sigma (\sigma=0)$ and
$g^Y_\sigma \equiv g^Y_\sigma (\sigma=0)$,
are new, characteristic features of QMC. 
At the hadronic level, the entire information
on the quark dynamics is condensed into the effective couplings
$C_{N,Y}(\sigma)$ of Eq.~(\ref{eqsigma}) 
(see section~\ref{subsec:relamodel}). 
Furthermore, when $C_{N,Y}(\sigma) = 1$, which corresponds to
a structureless nucleon or hyperon, the equations of motion
given by Eqs.~(\ref{eqdiracn})-(\ref{eqcoulomb})
can be identified with those derived
from QHD~\cite{mar,ruf,Jennings},
except for the terms arising from the tensor coupling and the non-linear
scalar and/or vector field interactions, introduced beyond the naive QHD.

The explicit expressions for coupled differential
equations to obtain various fields and hyperon and nucleon wave
functions can be obtained~\cite{Tsushima_hyp} in a similar manner 
to those obtained for finite nuclei 
(see section~\ref{subsec:fnt}), except for the additional modifications 
due to the embedded hyperon. 
It has been found that the function $C_j({\sigma})
(j=N,\Lambda,\Sigma,\Xi,\Lambda_c,\Sigma_c,\Xi_c,\Lambda_b)$
can be parameterized as a linear
form in the $\sigma$ field, $g_{\sigma}{\sigma}$, for practical
calculations~\cite{GUI-3,SAI-8,Tsushima_hyp,Tsushima_hypbc}:
\begin{equation}
C_j ({\sigma}) = 1 - a_j
\times (g_{\sigma} {\sigma}),\hspace{1em}
(j=N,\Lambda,\Sigma,\Xi,\Lambda_c,\Sigma_c,\Xi_c,\Lambda_b).
\label{cynsigma}
\end{equation}
The values obtained for $a_j$ are listed in Table~\ref{slope}.
(See also section~\ref{subsubsec:scaling} for 
some of the values calculated in QMC-II.)
This parameterization works very well up to
about three times normal nuclear matter density $\rho_B \simeq 3 \rho_0$.
Then, the effective masses for the baryon, $j$, in nuclear matter
are well approximated by~\cite{GUI-3,SAI-8,Tsushima_hyp}:
\bge
M^*_j \simeq M_j - \frac{n_q}{3} g_\sigma 
\left[1-\frac{a_j}{2}(g_\sigma {\sigma})\right]\sigma,
\hspace{2ex}(j=N,\Lambda,\Sigma,\Xi,\Lambda_c,\Sigma_c,\Xi_c,\Lambda_b),
\label{Mstar}
\ene
with $n_q$ being the number of light quarks in the baryon $j$ 
(see Eq.~(\ref{spot}) and section~\ref{subsubsec:scaling}).
The field strength, $g_\sigma {\sigma}$, 
versus baryon density can be found in Ref.~\cite{GUI-3}.
%
\begin{table}
\begin{center}
\begin{minipage}[t]{16.5cm}
\caption{The slope parameters, $a_j\,\,
(j=N,\Lambda,\Sigma,\Xi,\Lambda_c,\Sigma_c,\Xi_c,\Lambda_b)$. See also Table~\ref{tab:slope}.}
\label{slope}
\end{minipage}
\begin{tabular}[t]{c|c||c|c}
\hline
$a_j$ &$\times 10^{-4}$ MeV$^{-1}$ &$a_j$ &$\times 10^{-4}$ MeV$^{-1}$ \\
\hline
$a_N$           &8.8  &$a_{\Lambda_b}$   &10.9 \\
$a_\Lambda$     &9.3  &$a_{\Lambda_c}$ &9.8 \\
$a_{\Sigma}$    &9.5  &$a_{\Sigma_c}$    &10.3 \\
$a_{\Xi}$       &9.4  &$a_{\Xi_c}$       &9.9 \\
\end{tabular}
\end{center}
\end{table}
%
                                                                                
%
\subsubsection{\it Spin-orbit potential in the QMC model - strange hypernuclei
\label{subsubsec:hypSOpot}}

Here, we focus on the spin-orbit potential for a hyperon 
in the QMC model, where the internal structure of the hyperon 
gives one of the characteristic features not present in  
QHD-type models. As an illustrative example, we first discuss the spin-orbit
potential for the $\Lambda$.

The origin of the spin orbit force for a composite nucleon moving
through scalar and vector fields which vary with position was explained
in section~\ref{subsubsec:h1}.
The situation for the
$\Lambda$ is different in that, in an
SU(6) quark model, the $u$ and $d$ quarks are coupled
to spin zero, so that
the spin of the $\Lambda$ is carried by the $s$ quark.
As the $\sigma$-meson in QMC is viewed as a convenient parameterization of
two-pion-exchange and the $\omega$ and $\rho$ are non-strange, it seems
reasonable to assume that
the $\sigma$, $\omega$ and $\rho$ mesons couple only to
the $u$ and $d$ quarks. 
The direct contributions to the spin-orbit interaction
from these mesons then vanish due to the flavor-spin structure.
Thus, the spin-orbit interaction,
$V^\Lambda_{S.O.}(r) \vec{S}\cdot\vec{L}$,
at the position $\vec{r}$ for the $\Lambda$ in a hypernucleus
arises entirely from Thomas precession:
\begin{equation}
V^\Lambda_{s.o.}(r) \vec{S}\cdot\vec{L}
= - \frac{1}{2} {\vec{v}_\Lambda} \times
\frac{d \vec{v_\Lambda}}{dt} \cdot \vec{S}
= - \frac{1}{2 M^{* 2}_\Lambda (r) r}
\, \left( \frac{d}{dr} [ M^*_\Lambda (r)
+ g^\Lambda_\omega \omega(r) ] \right) \vec{S}\cdot\vec{L} ,
\label{so}
\end{equation}
where, $\vec{v}_\Lambda = \vec{p}_\Lambda/M^*_\Lambda$,
is the velocity of the $\Lambda$ in the rest frame of the $\Lambda$
hypernucleus, and the acceleration,
$d\vec{v}_\Lambda/dt$, is obtained from the Hamilton equations of motion
applied to the leading order Hamiltonian, as explained in 
section~\ref{subsubsec:h1}.
Because the contributions from the effective mass of the $\Lambda$,
$M^*_\Lambda (r)$, and the vector potential, $g^\Lambda_\omega \omega(r)$, 
are approximately equal and opposite in sign,
we quite naturally expect a very
small spin-orbit interaction for the $\Lambda$ in a hypernucleus.
(See section~\ref{subsec:fnt} for the spin-orbit splittings in normal 
nuclei.)
Although the spin-orbit splittings for the nucleon calculated
in QMC are already somewhat smaller  
than those calculated in QHD~\cite{QHD}, 
we can expect much smaller spin-orbit splittings for the $\Lambda$
in QMC. In order to include the spin-orbit potential of Eq.~(\ref{so}) 
approximately correctly, we added perturbatively the correction 
due to the vector potential,
$ -\frac{2}{2 M^{* 2}_\Lambda (r) r}\, 
\left( \frac{d}{dr} g^\Lambda_\omega \omega(r) \right) \vec{S}\cdot\vec{L}$,
to the single-particle energies for the specified shell state 
obtained with the Dirac equation Eq.~(\ref{eqdiracy}), 
by evaluating it with the corresponding shell-state wave function obtained 
for the $\Lambda$. (See also section~\ref{subsec:fnt}.)
This is necessary because the Dirac
equation corresponding to Eq.~(\ref{eqdiracy})
leads to a spin-orbit force which does not correspond to the underlying
quark model, namely:
\begin{equation}
V^\Lambda_{s.o.}(r) \vec{S}\cdot\vec{L}
= - \frac{1}{2 M^{* 2}_\Lambda (r) r}
\, \left( \frac{d}{dr} [ M^*_\Lambda (r)
- g^\Lambda_\omega \omega(r) ] \right) \vec{S}\cdot\vec{L}.
\label{sodirac}
\end{equation}
This correction to the spin-orbit force, which appears naturally in the
QMC model, may also be modeled at the hadronic level of the Dirac equation by
adding a tensor interaction, motivated by the quark
model~\cite{Jennings2,Cohen}.
In addition, one boson exchange model with underlying approximate
SU(3) symmetry in strong interactions, also leads to weaker
spin-orbit forces for the (strange) hyperon-nucleon ($YN$)
than that for the nucleon-nucleon ($NN$)~\cite{Gal}.
The very weak spin-orbit interaction for  
$\Lambda$ hypernuclei, which had been
phenomenologically suggested by Bouyssy and H\"{u}fner~\cite{bou},
was first explained by Brockman
and Weise~\cite{bro} in a relativistic Hartree model,
and directly confirmed later by experiment~\cite{bru}.

In the QMC model, the general expression for the spin-orbit
potential felt by the nucleon or hyperon, $j$
($j = N, \Lambda, \Sigma, \Xi$), may be expressed as~\cite{Tsushima_hyp} 
(see also section~\ref{subsubsec:totalH})
\bg
V^j_{s.o.}(r) \vec{S}\cdot\vec{L} &=& \frac{-1}{2M_j^{* 2}(r) r}
\left[ \Delta^j_\sigma
  +(G^s_j - 6 F^s_j \mu_s\eta_j(r))\Delta^j_\omega
+ (G^v_j - \frac{6}{5} F^v_j \mu_v\eta_j(r))\Delta^j_\rho \right]
\vec{S}\cdot\vec{L}, \quad
\label{spinorbit}
\en
with
\bg
\Delta^j_\sigma &=& \frac{d}{dr} M^*_j(r), \quad
\Delta^j_\omega = \frac{d}{dr} \left(\frac{1}{3}\right) g_\omega \omega(r)
= \Delta^q_\omega = \frac{d}{dr} g^q_\omega \omega(r), \quad
\Delta^j_\rho =\frac{d}{dr} g_\rho b(r), \label{delta} \\
\nn \\
G^s_j &=& \langle j| \sum_{i=u,d} 1(i) |j \rangle, \qquad
G^v_j = \langle j| \sum_{i=u,d} \frac{1}{2} \tau_3 (i) |j \rangle , \label{gsv} \\
F^s_j &=& \frac{\langle j| \sum_{i=u,d} \frac{1}{2} \vec{\sigma}(i) |j\rangle }
{\langle j| \frac{1}{2} \vec{\sigma}^j |j\rangle }, \qquad
F^v_j = \frac{\langle j| \sum_{i=u,d} \frac{1}{2} \vec{\sigma}(i)
\frac{1}{2} \tau_3(i) |j\rangle }
{\langle j| \frac{1}{2} \vec{\sigma}^j |j\rangle }, \label{fsv} \\
\nn \\
\mu_s &=& \frac{1}{3} M_N I_0 = \frac{1}{5} \mu_v, \qquad
\eta_j(r) = \frac{I_j^* M_j^* (r)}{I_0 M_N},  \label{mueta} \\
I_0 &=& \frac{R_N}{3} \frac{4 \Omega_N + 2 m_q R_N - 3}
{2 \Omega_N (\Omega_N - 1) + m_q R_N}, 
\qquad
I_j^* = \frac{R_j^*}{3} \frac{4 \Omega_j^* (\sigma) +
2 m^*_q(\sigma) R_j^* - 3}
{2 \Omega_j^* (\sigma) (\Omega_j^* (\sigma) - 1)
+ m^*_q(\sigma) R_j^*}. \label{int0j}
\en
Here $\vec{r}$ is the position of the baryon $j$ in the
hypernucleus (nucleus), and
the terms proportional to $\mu_s$ and $\mu_v$ are the anomalous
contributions from the light quarks
due to the finite size of the hyperon (nucleon).
They are related to the magnetic moments
of the proton, $\mu_p$, and the neutron, $\mu_n$, as
$\mu_s = \mu_p + \mu_n$ and $\mu_v = \mu_p - \mu_n$, with the experimental
values, $\mu_p = 2.79$ and $\mu_n = -1.91$ (in nuclear magnetons).
Note that $\mu_s$ and $\mu_v$, together with the quantities
$I_0$ and $I_j^*$ of Eq.~(\ref{int0j}),
are calculated self-consistently in the QMC model.
In addition, recall that the in-medium bag radius, $R_j^*$, and
the lowest bag eigenenergy for the light quarks, $\Omega_j^*/R_j^*$, depend  
on $j$ $(j = N,\Lambda,\Sigma,\Xi,\Lambda_c,\Sigma_c,\Xi_c,\Lambda_b)$
(see Table~\ref{bagparambc}).
The explicit expressions for the spin-orbit potentials
for the octet baryons in the QMC model are given by:
\bg
V^p_{s.o.}(r) &=& \frac{-1}{2M_N^{* 2}(r) r}
\left[ \Delta^N_\sigma
  + 3 (1 - 2\mu_s\eta_N(r))\Delta^N_\omega
+ \frac{1}{2}(1 - 2\mu_v\eta_N(r))\Delta_\rho \right],
\label{sop} \\
V^n_{s.o.}(r) &=& \frac{-1}{2M_N^{* 2}(r) r}
\left[ \Delta^N_\sigma
  + 3 (1 - 2\mu_s\eta_N(r))\Delta^N_\omega
- \frac{1}{2}(1 - 2\mu_v\eta_N(r))\Delta_\rho \right],
\label{son} \\
V^\Lambda_{s.o.}(r) &=& \frac{-1}{2M_\Lambda^{* 2}(r) r}
\left[ \Delta^\Lambda_\sigma
  + 2 \Delta^\Lambda_\omega \right],
\label{sola} \\
V^{\Sigma^+}_{s.o.}(r) &=& \frac{-1}{2M_\Sigma^{* 2}(r) r}
\left[ \Delta^\Sigma_\sigma
  + 2 (1 - 4\mu_s\eta_\Sigma(r))\Delta^\Sigma_\omega
+ (1 - \frac{4}{5}\mu_v\eta_\Sigma(r))\Delta_\rho \right],
\label{sos+} \\
V^{\Sigma^0}_{s.o.}(r) &=& \frac{-1}{2M_\Sigma^{* 2}(r) r}
\left[ \Delta^\Sigma_\sigma
  + 2 (1 - 4\mu_s\eta_\Sigma(r))\Delta^\Sigma_\omega \right],
\label{sos0} \\
V^{\Sigma^-}_{s.o.}(r) &=& \frac{-1}{2M_\Sigma^{* 2}(r) r}
\left[ \Delta^\Sigma_\sigma
  + 2 (1 - 4\mu_s\eta_\Sigma(r))\Delta^\Sigma_\omega
- (1 - \frac{4}{5}\mu_v\eta_\Sigma(r))\Delta_\rho \right],
\label{sos-} \\
V^{\Xi^0}_{s.o.}(r) &=& \frac{-1}{2M_\Xi^{* 2}(r) r}
\left[ \Delta^\Xi_\sigma
  +(1 + 2\mu_s\eta_\Xi(r))\Delta^\Xi_\omega
+ \frac{1}{2}(1 + \frac{2}{5}\mu_v\eta_\Xi(r))\Delta_\rho \right],
\label{sox0} \\
V^{\Xi^-}_{s.o.}(r) &=& \frac{-1}{2M_\Xi^{* 2}(r) r}
\left[ \Delta^\Xi_\sigma
  +(1 + 2\mu_s\eta_\Xi(r))\Delta^\Xi_\omega
- \frac{1}{2}(1 + \frac{2}{5}\mu_v\eta_\Xi(r))\Delta_\rho \right].
\label{sox-}
\en
The spin-orbit potentials for  
$\Lambda_c,\Sigma_c,\Xi_c$ and $\Lambda_b$ can be obtained in a similar way.
However, because of the heavier (effective) masses for them and   
because the contribution from the $\sigma$ and $\omega$ fields for 
the the spin-orbit potentials are very similar in magnitude   
to those for the strange hyperons, we expect that the 
contribution of the spin-orbit potential for charm and bottom 
hypernuclei should be negligible.

\subsubsection{\it Pauli blocking and channel coupling effects
\label{subsubsec:hypPauli}}

Next, we discuss the effects of Pauli blocking 
at the quark level, as well as the channel coupling.
For the numerical results, these effects are included in the 
self-consistent calculation.
When a hyperon sits in an 
orbital with the same quantum numbers as one already occupied
by nucleons we need to include
the effect of Pauli blocking at the quark level.
Furthermore, there is an additional correction
associated with the channel coupling, $\Sigma N - \Lambda N$.
These effects will be included in specific ways at the hadronic level. 
More consistent treatments at the quark level have not yet been studied 
within the QMC model.

First, we consider the Pauli blocking effect.
It seems natural to assume that this effect works
repulsively in a way that the strength is proportional
to the light quark baryonic (number) density of the core nucleons.
One might then expect that
the light quarks in the hyperon should  
feel a stronger repulsion at the position where the
baryon density is large. As a consequence,
the wave function of the hyperon (quark)
will be suppressed in this region.
Therefore, we assume that the Pauli blocking effect is simply 
proportional to the baryonic density.
Then, the Dirac equation for the hyperon Y, Eq.~(\ref{eqdiracy}),
is modified by
\bge
[i\gamma \cdot \partial - M^*_Y(\sigma)-
(\, \lambda_Y \rho_B(\vec{r}) + g^Y_\omega \omega(\vec{r})
+ g_\rho I^Y_3 b(\vec{r})
+ e Q_Y A(\vec{r}) \,)
\gamma_0 ] \psi_Y(\vec{r}) = 0, \label{eqmdiracy}
\ene
where, $\rho_B(\vec{r})$ is the baryonic density at the position $\vec{r}$
in the hypernucleus due to the core nucleons, and $\lambda_Y$ is a
constant to be determined empirically. 
In the present treatment, we chose this
constant $\lambda_Y$ for $Y=\Lambda$, 
in order to reproduce the empirical single-particle energy for the
$1s_{1/2}$ state in $^{209}_\Lambda$Pb, -27.0 MeV~\cite{aji}.
                                                                                
One might also imagine that this fitted value includes the attractive
$\Lambda N \rightarrow \Sigma N$
channel coupling effect for the $\Lambda$ single-particle energies,
because the value fitted is the experimentally observed one.
However, for $\Sigma$ hypernuclei, the repulsive
$\Sigma N \rightarrow \Lambda N$ channel coupling effect must be included
in addition to this effective Pauli blocking,
in a way to reproduce the relative repulsive energy shift in the
single-particle energies for the $\Sigma$.
The fitted value for the constant $\lambda_\Lambda$ is
$\lambda_\Lambda = 60.25$ MeV (fm)$^3$.
Then for the $\Sigma$ and $\Xi$ hypernuclei, we take  
the constants, $\lambda_{\Sigma,\Xi}$,
corresponding to the effective Pauli blocking effect as,
$\lambda_\Sigma = \lambda_\Lambda$,
and $\lambda_\Xi = \frac{1}{2} \lambda_\Lambda$,
by counting the total number of $u$ and $d$ quarks in these hyperons.
                                                                                
Next, we consider the channel coupling (strong conversion) effect
additional to the Pauli blocking at the quark level.
It is expected that the channel couplings,
$\Sigma N - \Lambda N$ and
$\Xi N - \Lambda \Lambda$, generally exist in hypernuclei,
with the former considered to be especially
important~\cite{joh,dov}.
We consider first the $\Sigma N - \Lambda N$ channel coupling.
We estimate this effect using the Nijmegen potential~\cite{nij}
as follows. Including solely the effective Pauli blocking potential,
$\lambda_\Sigma \rho_B(r)$ 
($\lambda_\Sigma = \lambda_\Lambda = 60.25$ MeV (fm)$^3$),
we obtain the $1s_{1/2}$ single-particle energy for 
in $^{\Sigma^0}_{209}$Pb, -26.9 MeV ($\simeq$ - 27.0 MeV of observed value).
This value does not contain entirely the effect of the correct 
channel coupling to the $\Lambda$.
On the other hand, in a conventional first-order
Brueckner calculation based on the standard choice of the single-particle
potentials (cf. Ref.~\cite{nmc} for details) the 
binding energy for the $\Sigma$ in nuclear matter with
the Nijmegen potential~\cite{nij} is, 12.6 MeV, for the case
where the correct channel
coupling effect to the $\Lambda$ is omitted, 
namely, without the Pauli-projector
in the $\Lambda N$ channel. When the channel coupling to the $\Lambda N$
is recovered and the Pauli-projector in the $\Lambda N$ channel is included,
the binding energy for the $\Sigma$ in nuclear matter decreases to 5.3 MeV.
Then, the decrease in the calculated binding energy for the $\Sigma$,
12.6 - 5.3 = + 7.3 MeV, may be taken  
as the net effect of the $\Sigma N - \Lambda N$ channel
coupling for the $\Sigma$.
We include the effect by assuming the same form as that
applied for the effective Pauli blocking via $\lambda_\Sigma \rho_B(r)$,
and readjust the parameter $\lambda_\Sigma = \lambda_\Lambda$ to
$\tilde{\lambda}_\Sigma \neq \lambda_\Sigma$ to reproduce
this difference in the
single-particle energy for the $1s_{1/2}$ in $^{\Sigma^0}_{209}$Pb,
namely, $-19.6 = -26.9 + 7.3$ MeV. 
Here we should point out that the baryon density calculated 
in section~\ref{subsec:fnt}, 
shows that the density around the center of the $^{208}$Pb nucleus is
consistently close to that of nuclear matter within the model.
The value obtained for $\tilde{\lambda}_\Sigma$ in this way is,
$\tilde{\lambda}_\Sigma$ = 110.6 MeV (fm)$^3$.
                                                                                
As for the $\Xi N - \Lambda \Lambda$ channel coupling, the studies of
Afnan and Gibson~\cite{afn2} show that the effect
is very small for the calculated 
binding energy for $^6_{\Lambda \Lambda}$He.
Although their estimate is not for large atomic number hypernuclei 
nor nuclear matter, we neglect the 
$\Xi N - \Lambda \Lambda$ channel coupling effect in the calculation.
For the effective Pauli blocking and the channel coupling effects 
for the corresponding charm and bottom hypernuclei, 
we apply exactly the same forms and the coupling constants 
as those obtained for the $\Lambda$ and $\Sigma$. 

\subsubsection{\it Results for strange hypernuclei
\label{subsubsec:hyps}}

First, we present the results for strange hypernuclei.
In Tables~\ref{spep1} and~\ref{spep2}, we list the 
single-particle energies calculated for $^{17}_Y$O, $^{41}_Y$Ca,
$^{49}_Y$Ca, $^{91}_Y$Zr and $^{209}_Y$Pb 
($Y=\Lambda,\Sigma^{\pm,0},\Xi^{-,0}$) hypernuclei,
together with the experimental data~\cite{aji,chr}
for the $\Lambda$ hypernuclei.
(A recent, extensive review on the progress of the $\Lambda$ 
hypernuclear spectroscopy is made in Ref.~\cite{Hashimoto}.)
We have searched for the single-particle states up to the highest
level of the core neutrons in each hypernucleus, since the
deeper levels are usually easier to observe in experiment.
Concerning the single-particle energy levels for the $\Lambda$ hypernuclei,
the QMC model supplemented by the effective Pauli blocking effect
employed at the hadronic level, reproduces the data reasonably well.
{}For the reasons explained earlier, small spin-orbit splittings 
are found for the $\Lambda$ hypernuclei. 

\begin{table}
\begin{center}
\begin{minipage}[t]{16.5cm}
\caption{Single-particle energies (in MeV)
for $^{17}_Y$O, $^{41}_Y$Ca and $^{49}_Y$Ca
($Y=\Lambda,\Sigma^{\pm,0},\Xi^{-,0}$)
strange hypernuclei, calculated
with the effective Pauli blocking and the $\Sigma N - \Lambda N$ channel
coupling. Experimental data are taken from Ref.~\cite{chr}.
Spin-orbit splittings are not well determined by the experiments.}
\label{spep1}
\end{minipage}
\begin{tabular}[t]{c|ccccccc}
\hline \hline
&$^{16}_\Lambda$O (Expt.)
&$^{17}_\Lambda$O    &$^{17}_{\Sigma^-}$O
&$^{17}_{\Sigma^0}$O &$^{17}_{\Sigma^+}$O
&$^{17}_{\Xi^-}$O    &$^{17}_{\Xi^0}$O\\
\hline \hline
$1s_{1/2}$&-12.5      &-14.1 &-17.2 &-9.6  &-3.3  &-9.9  &-4.5 \\
$1p_{3/2}$&-2.5 ($1p$)&-5.1  &-8.7  &-3.2  &---   &-3.4  &---  \\
$1p_{1/2}$&-2.5 ($1p$)&-5.0  &-8.0  &-2.6  &---   &-3.4  &---  \\ \\
\hline \hline
&$^{40}_\Lambda$Ca (Expt.)
&$^{41}_\Lambda$Ca    &$^{41}_{\Sigma^-}$Ca
&$^{41}_{\Sigma^0}$Ca &$^{41}_{\Sigma^+}$Ca
&$^{41}_{\Xi^-}$Ca    &$^{41}_{\Xi^0}$Ca\\
\hline \hline
$1s_{1/2}$&-20.0       &-19.5 &-23.5 &-13.4 &-4.1  &-17.0 &-8.1 \\
$1p_{3/2}$&-12.0 ($1p$)&-12.3 &-17.1 &-8.3  &---   &-11.2 &-3.3 \\
$1p_{1/2}$&-12.0 ($1p$)&-12.3 &-16.5 &-7.7  &---   &-11.3 &-3.4 \\
$1d_{5/2}$&            &-4.7  &-10.6 &-2.6  &---   &-5.5  &---  \\
$2s_{1/2}$&            &-3.5  &-9.3  &-1.2  &---   &-5.4  &---  \\
$1d_{3/2}$&            &-4.6  &-9.7  &-1.9  &---   &-5.6  &---  \\ \\
\hline \hline
&---
&$^{49}_\Lambda$Ca    &$^{49}_{\Sigma^-}$Ca
&$^{49}_{\Sigma^0}$Ca &$^{49}_{\Sigma^+}$Ca
&$^{49}_{\Xi^-}$Ca    &$^{49}_{\Xi^0}$Ca\\
\hline \hline
$1s_{1/2}$&           &-21.0 &-19.3 &-14.6 &-11.5 &-14.7 &-12.0\\
$1p_{3/2}$&           &-13.9 &-11.4 &-9.4  &-7.5  &-8.7  &-7.4 \\
$1p_{1/2}$&           &-13.8 &-10.9 &-8.9  &-7.0  &-8.8  &-7.4 \\
$1d_{5/2}$&           &-6.5  &-5.8  &-3.8  &-2.0  &-3.8  &-2.1 \\
$2s_{1/2}$&           &-5.4  &-6.7  &-2.6  &---   &-4.6  &-1.1 \\
$1d_{3/2}$&           &-6.4  &-5.2  &-3.1  &-1.2  &-3.8  &-2.2 \\
$1f_{7/2}$&           &---   &-1.2  &---   &---   &---   &---  \\
\end{tabular}
\end{center}
\end{table}
%
\begin{table}
\begin{center}
\begin{minipage}[t]{16.5cm}
\caption{Same as Table~\ref{spe1} but for $^{91}_Y$Zr and $^{208}_Y$Pb
($Y=\Lambda,\Sigma^{\pm,0},\Xi^{-,0}$).
Experimental data are taken from Ref.~\cite{aji}.
Spin-orbit splittings are not well determined by the experiments.
Double asterisks, $^{**}$, indicate the value used for fitting.}
\label{spep2}
\end{minipage}
\begin{tabular}[t]{c|ccccccc}
\hline \hline
&$^{89}_\Lambda$Yb (Expt.)
&$^{91}_\Lambda$Zr    &$^{91}_{\Sigma^-}$Zr
&$^{91}_{\Sigma^0}$Zr &$^{91}_{\Sigma^+}$Zr
&$^{91}_{\Xi^-}$Zr    &$^{91}_{\Xi^0}$Zr\\
\hline \hline
$1s_{1/2}$&-22.5       &-23.9 &-27.3 &-16.8 &-8.1  &-22.7 &-13.3\\
$1p_{3/2}$&-16.0 ($1p$)&-18.4 &-20.8 &-12.7 &-5.0  &-17.4 &-9.7 \\
$1p_{1/2}$&-16.0 ($1p$)&-18.4 &-20.5 &-12.4 &-4.7  &-17.4 &-9.7 \\
$1d_{5/2}$&-9.0  ($1d$)&-12.3 &-15.4 &-8.1  &-0.9  &-12.3 &-5.4 \\
$2s_{1/2}$&            &-10.8 &-15.6 &-6.5  &---   &-12.4 &-3.9 \\
$1d_{3/2}$&-9.0  ($1d$)&-12.3 &-14.8 &-7.5  &-0.3  &-12.4 &-5.5 \\
$1f_{7/2}$&-2.0  ($1f$)&-5.9  &-10.2 &-3.1  &---   &-7.5  &-0.7 \\
$2p_{3/2}$&            &-4.2  &-10.1 &---   &---   &-7.9  &---  \\
$1f_{5/2}$&-2.0  ($1f$)&-5.8  &-9.4  &-2.3  &---   &-7.6  &-0.8 \\
$2p_{1/2}$&            &-4.1  &-9.9  &---   &---   &-7.9  &---  \\
$1g_{9/2}$&            &---   &-5.2  &---   &---   &-3.3  &---  \\ \\
\hline \hline
&$^{208}_\Lambda$Pb (Expt.)
&$^{209}_\Lambda$Pb    &$^{209}_{\Sigma^-}$Pb
&$^{209}_{\Sigma^0}$Pb &$^{209}_{\Sigma^+}$Pb
&$^{209}_{\Xi^-}$Pb    &$^{209}_{\Xi^0}$Pb\\
\hline \hline
$1s_{1/2}$&-27.0      &-27.0$^{**}$ &-29.7 &-19.6$^{**}$ &-10.0 &-29.0 &-19.2\\
$1p_{3/2}$&-22.0 ($1p$)&-23.4 &-25.9 &-16.7 &-7.7  &-25.3 &-16.3\\
$1p_{1/2}$&-22.0 ($1p$)&-23.4 &-25.8 &-16.5 &-7.5  &-25.4 &-16.3\\
$1d_{5/2}$&-17.0 ($1d$)&-19.1 &-22.1 &-13.3 &-4.6  &-21.6 &-12.9\\
$2s_{1/2}$&            &-17.6 &-21.7 &-12.0 &-2.1  &-21.2 &-12.0\\
$1d_{3/2}$&-17.0 ($1d$)&-19.1 &-21.8 &-13.0 &-4.2  &-21.6 &-12.9\\
$1f_{7/2}$&-12.0 ($1f$)&-14.4 &-18.2 &-9.5  &-0.9  &-17.6 &-9.2 \\
$2p_{3/2}$&            &-12.4 &-17.4 &-7.8  &---   &-17.1 &-8.0 \\
$1f_{5/2}$&-12.0 ($1f$)&-14.3 &-17.8 &-9.0  &-0.4  &-17.7 &-9.2 \\
$2p_{1/2}$&            &-12.4 &-17.2 &-7.6  &---   &-17.1 &-8.0 \\
$1g_{9/2}$&-7.0  ($1g$)&-9.3  &-14.3 &-5.5  &---   &-13.6 &-5.2 \\
$1g_{7/2}$&-7.0  ($1g$)&-9.2  &-13.6 &-4.8  &---   &-13.7 &-5.2 \\
$1h_{11/2}$&           &-3.9  &-4.9  &-1.2  &---   &-9.7  &-1.0 \\
$2d_{5/2}$&            &-7.0  &-7.5  &---   &---   &-13.3 &-3.8 \\
$2d_{3/2}$&            &-7.0  &-7.5  &---   &---   &-13.3 &-3.8 \\
$1h_{9/2}$&            &-3.8  &-4.8  &---   &---   &-9.8  &-1.0 \\
$3s_{1/2}$&            &-6.1  &-7.8  &---   &---   &-8.3  &-3.1 \\
$2f_{7/2}$&            &-1.7  &-5.8  &---   &---   &-6.2  &---  \\
$3p_{3/2}$&            &-1.0  &-3.4  &---   &---   &-6.6  &---  \\
$2f_{5/2}$&            &-1.7  &-5.8  &---   &---   &-6.3  &---  \\
$3p_{1/2}$&            &-1.0  &-3.3  &---   &---   &-6.6  &---  \\
$1i_{13/2}$&           &---   &---   &---   &---   &-3.7  &---  \\
\end{tabular}
\end{center}
\end{table}

It should be mentioned here that in the case of larger mass number
$\Sigma$ hypernuclei, no narrow states have been observed experimentally,
although $^4_{\Sigma}$He was confirmed~\cite{sigmahyp}.
This experimental analysis~\cite{Sigmaexpt} supports the suggestion
made by Harada~\cite{Harada}, that a large isospin dependent
$\Sigma$-nucleus potential term may exist and the 1/A (A:baryon number)
dependence of that term reduces the likelihood of observing
bound states for $A > 5$. The absence of heavier $\Sigma$ hypernuclei 
may also arise from the width of $\Sigma$ states associated 
with strong $\Sigma - \Lambda$ conversion.
Nevertheless, it may be interesting to compare the single-particle
energies obtained for the charged hyperons, $\Sigma^\pm$ and $\Xi^-$, and
those of the neutral hyperons, $\Sigma^0$ and $\Xi^0$.
The present results imply that the Coulomb force is
important for forming (or unforming) a bound state of the
hyperon in hypernuclei.
This was also discussed by Yamazaki et al.~\cite{yam}, in the context of
the light $\Sigma^-$ hypernuclei.

In Table~\ref{prmst} we show
the calculated binding energy per baryon,
root-mean-square (rms) charge radius ($r_{ch}$) and 
rms radii of the strange hyperon ($r_Y$), neutron ($r_n$) 
and proton ($r_p$) distributions.
\begin{table}
\begin{center}
\begin{minipage}[t]{16.5cm}
\caption{Binding energy per baryon, $E_B/A$ (in MeV), rms charge radius 
($r_{ch}$), and rms radii of the strange hyperon ($r_Y$),
neutron ($r_n$), and proton ($r_p$) (in fm) for $^{17}_Y$O, $^{41}_Y$Ca,
$^{49}_Y$Ca, $^{91}_Y$Zr and $^{209}_Y$Pb strange hypernuclei.
The configuration of the strange hyperon,
$Y$, is $1s_{1/2}$ for all hypernuclei.
For comparison, we also list the corresponding results
for normal finite nuclei.
Double asterisks, $^{**}$, indicate
the value used for fitting.}
\label{prmst}
\end{minipage}
\begin{tabular}[t]{cccccc}
\\
\hline \hline
hypernuclei & $-E_B/A$ & $r_{ch}$ & $r_Y$ & $r_n$ & $r_p$ \\
\hline \hline
$^{17}_\Lambda$O     &6.37&2.84&2.49&2.59&2.72\\
$^{17}_{\Sigma^+}$O  &5.91&2.82&3.15&2.61&2.70\\
$^{17}_{\Sigma^0}$O  &6.10&2.83&2.79&2.60&2.71\\
$^{17}_{\Sigma^-}$O  &6.31&2.84&2.49&2.59&2.72\\
$^{17}_{\Xi^0}$O     &5.86&2.80&2.98&2.62&2.68\\
$^{17}_{\Xi^-}$O     &6.02&2.81&2.65&2.61&2.69\\
\hline
$^{16}$O             &5.84&2.79&--- &2.64&2.67\\
\hline \hline
$^{41}_\Lambda$Ca    &7.58&3.51&2.81&3.31&3.42\\
$^{41}_{\Sigma^+}$Ca &7.33&3.50&3.43&3.32&3.41\\
$^{41}_{\Sigma^0}$Ca &7.44&3.51&3.14&3.31&3.41\\
$^{41}_{\Sigma^-}$Ca &7.60&3.51&2.87&3.31&3.41\\
$^{41}_{\Xi^0}$Ca    &7.34&3.49&3.09&3.32&3.40\\
$^{41}_{\Xi^-}$Ca    &7.45&3.50&2.84&3.32&3.40\\
\hline
$^{40}$Ca            &7.36&3.48$^{**}$&--- &3.33&3.38\\
\hline \hline
$^{49}_\Lambda$Ca    &7.58&3.54&2.84&3.63&3.45\\
$^{49}_{\Sigma^+}$Ca &6.32&3.57&3.47&3.71&3.48\\
$^{49}_{\Sigma^0}$Ca &7.40&3.54&3.14&3.64&3.45\\
$^{49}_{\Sigma^-}$Ca &7.48&3.55&2.60&3.63&3.45\\
$^{49}_{\Xi^0}$Ca    &7.32&3.53&3.14&3.65&3.43\\
$^{49}_{\Xi^-}$Ca    &7.35&3.53&2.79&3.65&3.44\\
\hline
$^{48}$Ca            &7.27&3.52&--- &3.66&3.42\\
\hline \hline
$^{91}_\Lambda$Zr    &7.95&4.29&3.25&4.29&4.21\\
$^{91}_{\Sigma^+}$Zr &7.82&4.28&4.01&4.30&4.20\\
$^{91}_{\Sigma^0}$Zr &7.87&4.29&3.56&4.30&4.21\\
$^{91}_{\Sigma^-}$Zr &7.92&4.29&2.89&4.29&4.21\\
$^{91}_{\Xi^0}$Zr    &7.83&4.28&3.54&4.30&4.20\\
$^{91}_{\Xi^-}$Zr    &7.87&4.28&2.98&4.30&4.20\\
\hline
$^{90}$Zr            &7.79&4.27&--- &4.31&4.19\\
\hline \hline
$^{209}_\Lambda$Pb    &7.35&5.49&3.99&5.67&5.43\\
$^{209}_{\Sigma^+}$Pb &7.28&5.49&4.64&5.68&5.43\\
$^{209}_{\Sigma^0}$Pb &7.31&5.49&4.26&5.67&5.43\\
$^{209}_{\Sigma^-}$Pb &7.34&5.49&3.80&5.67&5.43\\
$^{209}_{\Xi^0}$Pb    &7.30&5.49&3.96&5.68&5.43\\
$^{209}_{\Xi^-}$Pb    &7.32&5.49&3.59&5.68&5.43\\
\hline
$^{208}$Pb            &7.25&5.49&--- &5.68&5.43\\
\end{tabular}
\end{center}
\end{table}
The results listed in Table~\ref{prmst} are calculated for 
the $1s_{1/2}$ hyperon configuration in all cases.
Here, one can again notice the important role of the Coulomb force.
For example, the binding energy per baryon,
$E_B/A$, for the $\Sigma^-$ ($\Sigma^+$)
hypernuclei is typically the largest (smallest) among the same atomic number
strange hypernuclei, while the rms radii for the hyperon, 
$r_{\Sigma^-}$ ($r_{\Sigma^+}$) is mostly the smallest (largest) among
the hypernuclei of the same atomic number.

To get an idea of the mass number dependence for the strange 
hypernuclei calculated, we show the absolute values of the scalar 
and vector potentials for the hyperons 
in $^{17}_Y$O, $^{41}_Y$Ca and $^{209}_Y$Pb
($Y = \Lambda,\Sigma^0,\Xi^0$) in Figs.~\ref{fig:potLa} 
and~\ref{fig:potX0}. 

\begin{figure}
\epsfysize=9.0cm
\begin{center}
\begin{minipage}[t]{8cm}
\hspace*{-2cm}
\epsfig{file=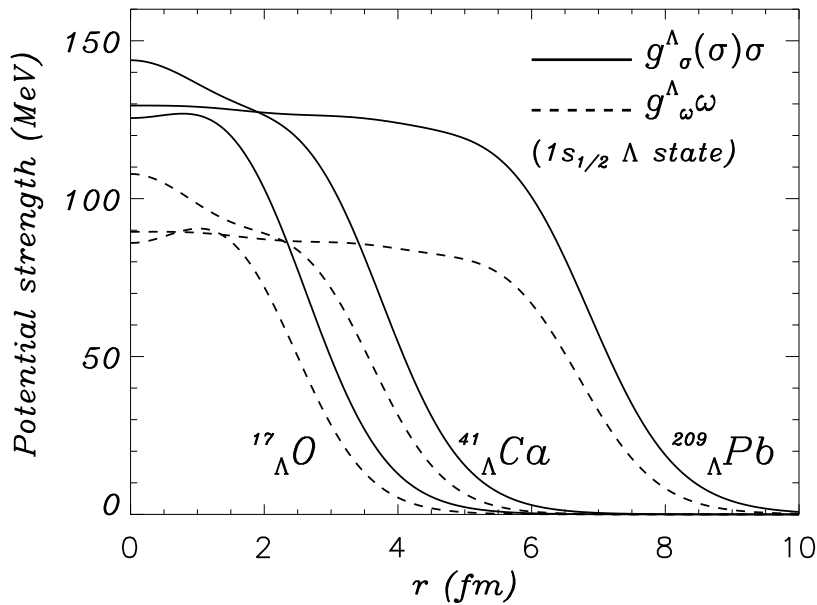,height=9cm}
\end{minipage}
\begin{minipage}[t]{16.5cm}
\caption{Potential strengths, $g^\Lambda_\sigma(\sigma)$
and $g^\Lambda_\omega$, for $\Lambda$ hypernuclei,
$^{17}_\Lambda$O, $^{41}_\Lambda$Ca and
$^{209}_\Lambda$Pb for $1s_{1/2} \Lambda$ state
(from Ref.~\cite{Tsushima_hyp}).}
\label{fig:potLa}
\end{minipage}
\end{center}
\end{figure}
\begin{figure}
\epsfysize=9.0cm
\begin{center}
\begin{minipage}[t]{8cm}
\hspace*{-2cm}
\epsfig{file=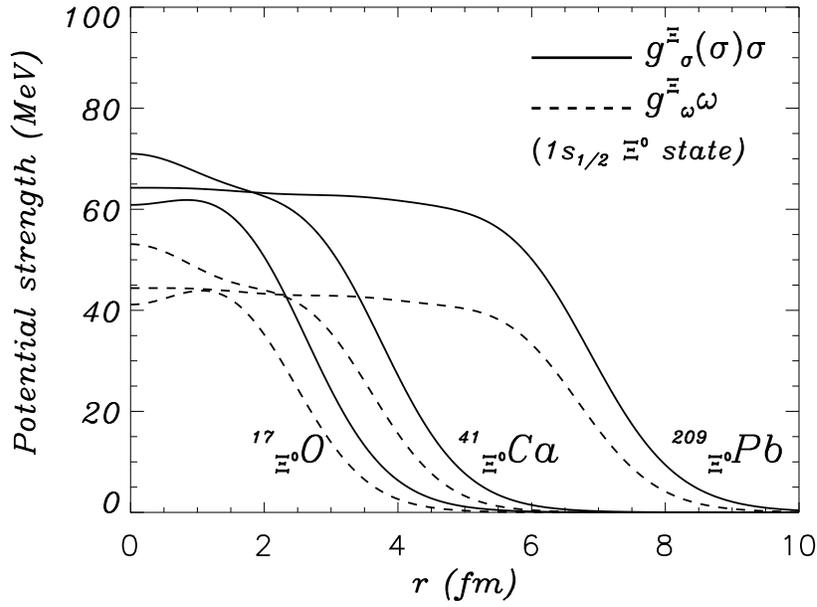,height=9cm}
\end{minipage}
\begin{minipage}[t]{16.5cm}
\caption{Same as Fig.~\ref{fig:potLa} but for the $\Xi^0$ hypernuclei 
(from Ref.~\cite{Tsushima_hyp}).}
\label{fig:potX0}
\end{minipage}
\end{center}
\end{figure}

We make one final remark concerning the $\Sigma$ hypernuclei.  
Recent experiments~\cite{Sigmaexpt}
suggest that a strongly repulsive $\Sigma$-nucleus potential with a
non-zero size of the imaginary part is best able to reproduce the
measured spectra for middle and large baryon number hypernuclei.
If we wish to be consistent with this result,
we need to introduce an even stronger repulsive potential,
either by the channel coupling effect, or based on the
quark model analysis to reproduce the $\Sigma$-atom data
without resorting to the channel coupling effect~\cite{Sigma_atom} 
for the $\Sigma$ ($\Sigma_c$) hypernuclei.
In this case, the repulsion should be strong enough
that it entails no bound state
for the $\Sigma$ hypernuclei with medium to large baryon 
number. This will need a more elaborate investigation
in the future, using further accumulated experimental data and analyses.
Thus, at present, our results for $\Sigma$ hypernuclei are incomplete 
and we need further investigation.

\subsubsection{\it Results for charm and bottom hypernuclei
\label{subsubsec:hypbc}}

Next, we present results for charm and bottom hypernuclei.
The formation of $\Lambda_c$ and/or $\Lambda_b$ hypernuclei
was first predicted in the mid-1970s by Tyapkin~\cite{Tyapkin},
and Dover and Kahana~\cite{Dover}.
Later on, theoretical studies~\cite{Gibson,Bando}
as well as suggestions concerning the 
possibility of experimental observation~\cite{bcexp} were made.
The study of charm and bottom hypernuclei can be 
made in the same way as those made for strange hypernuclei.
Namely, the treatment of spin-orbit force, and 
the Pauli blocking and the channel coupling effects
are included in the same way as those  
explained in sections~\ref{subsubsec:hypSOpot} 
and~\ref{subsubsec:hypPauli}.
However, since the charm and bottom baryons are heavy compared to 
the typical magnitude of the potentials needed to form a nucleus, the  
contribution from the spin-orbit piece of the single-particle energies
is negligible. As will be shown later, its typical contribution is 
of order $0.01$ MeV, and even the largest case is $\simeq 0.5$ MeV, 
{}for all the charm and bottom hypernuclear single-particle energies calculated.
This can be understood when one considers the limit,
$M^*_j \to \infty$ in Eq.~(\ref{spinorbit}) for the cases 
of charm and bottom hypernuclei, where the quantity inside 
the square brackets varies smoothly from an order of hundred 
MeV to zero near the surface of the hypernucleus, 
and the derivative with respect to $r$ is finite.
Thus, the spin-orbit splittings are expected to be tiny.

Concerning the channel coupling for $\Sigma_c N - \Lambda_c N$, 
we treat it in the same phenomenological 
way as that for $\Sigma N - \Lambda N$,
although the effect is expected to be smaller, 
since the mass difference for the former is larger.  
Thus, the channel coupling effect included for the 
$\Sigma_c$ hypernuclei should be regarded as the limiting case.
In addition, recall that the recent experimental results show 
no evidence for narrow width $\Sigma$ hypernuclei for the medium to  
large mass region. This suggests that one should improve the present 
treatment of the Pauli blocking and the channel 
coupling effects in the QMC model, for the $\Sigma_c$ (and 
$\Sigma$) hypernuclei.

In Tables~\ref{spe1} and~\ref{spe2} we give the 
single particle energies calculated  
for $^{17}_Y$O, $^{41}_Y$Ca, $^{49}_Y$Ca, $^{91}_Y$Zr and $^{209}_Y$Pb
($Y=\Lambda,\Lambda_c^+,\Sigma_c^{0,+,++},\Xi_c^{0,+},\Lambda_b$)
hypernuclei, together with the experimental data~\cite{chr,aji}
for the $\Lambda$ hypernuclei.
\begin{table}
\begin{center}
\begin{minipage}[t]{16.5cm}
\caption{Single-particle energies (in MeV)
for $^{17}_Y$O, $^{41}_Y$Ca and $^{49}_Y$Ca
($Y = \Lambda, \Lambda_c, \Sigma_c, \Xi_c, \Lambda_b$).
Experimental data are taken from Ref.~\cite{chr}.
Spin-orbit splittings for the $\Lambda$ hypernuclei
are not well determined by the experiments.}
\label{spe1}
\end{minipage}
\begin{tabular}[t]{c|ccccccccc}
\hline \hline
&$^{16}_\Lambda$O (Expt.)
&$^{17}_\Lambda$O    &$^{17}_{\Lambda^+_c}$O
&$^{17}_{\Sigma^0_c}$O &$^{17}_{\Sigma^+_c}$O &$^{17}_{\Sigma^{++}_c}$O
&$^{17}_{\Xi^0_c}$O    &$^{17}_{\Xi^+_c}$O &$^{17}_{\Lambda_b}$O\\
\hline \hline
$1s_{1/2}$&-12.5      &-14.1&-12.8&---  &-7.5&---&-7.9&-2.1&-19.6\\
$1p_{3/2}$&-2.5 ($1p$)&-5.1 &-7.3 &-10.7&-4.0&---&-3.5&--- &-16.5\\
$1p_{1/2}$&-2.5 ($1p$)&-5.0 &-7.3 &-10.2&-3.6&---&-3.5&--- &-16.5\\ \\
\hline \hline
&$^{40}_\Lambda$Ca (Expt.)
&$^{41}_\Lambda$Ca    &$^{41}_{\Lambda^+_c}$Ca
&$^{41}_{\Sigma^0_c}$Ca &$^{41}_{\Sigma^+_c}$Ca &$^{41}_{\Sigma^{++}_c}$Ca
&$^{41}_{\Xi^0_c}$Ca    &$^{41}_{\Xi^+_c}$Ca &$^{41}_{\Lambda_b}$Ca\\
\hline \hline
$1s_{1/2}$&-20.0       &-19.5&-12.8&-16.3&-6.3&---&-9.9&-1.0&-23.0\\
$1p_{3/2}$&-12.0 ($1p$)&-12.3&-9.2 &-13.2&-4.1&---&-6.7&--- &-20.9\\
$1p_{1/2}$&-12.0 ($1p$)&-12.3&-9.1 &-12.9&-3.8&---&-6.7&--- &-20.9\\
$1d_{5/2}$&            &-4.7 &-4.8 &-9.9 &--- &---&-3.3&--- &-18.4\\
$2s_{1/2}$&            &-3.5 &-3.4 &-9.3 &--- &---&-2.8&--- &-17.4\\
$1d_{3/2}$&            &-4.6 &-4.8 &-9.4 &--- &---&-3.3&--- &-18.4\\ \\
\hline \hline
&---
&$^{49}_\Lambda$Ca    &$^{49}_{\Lambda^+_c}$Ca
&$^{49}_{\Sigma^0_c}$Ca &$^{49}_{\Sigma^+_c}$Ca &$^{49}_{\Sigma^{++}_c}$Ca
&$^{49}_{\Xi_c^0}$Ca    &$^{49}_{\Xi_c^+}$Ca &$^{49}_{\Lambda_b}$Ca\\
\hline \hline
$1s_{1/2}$&           &-21.0&-14.3&--- &-7.4&-5.0&-7.8&-5.1&-24.4\\
$1p_{3/2}$&           &-13.9&-10.6&-7.2&-5.1&-3.4&-4.1&-2.8&-22.2\\
$1p_{1/2}$&           &-13.8&-10.6&-7.0&-4.9&-3.2&-4.1&-2.8&-22.2\\
$1d_{5/2}$&           &-6.5 &-6.5 &-4.0&-2.3&--- &-0.9&--- &-19.5\\
$2s_{1/2}$&           &-5.4 &-5.3 &-4.9&--- &--- &-1.5&--- &-18.8\\
$1d_{3/2}$&           &-6.4 &-6.4 &-3.6&-1.9&--- &-1.0&--- &-19.5\\
$1f_{7/2}$&           &---  &-2.0 &--- &--- &--- &--- &--- &-16.8\\
\end{tabular}
\end{center}
\end{table}
\begin{table}
\begin{center}
\begin{minipage}[t]{16.5cm}
\caption{Same as Table~\ref{spe1} but for $^{91}_Y$Zr and $^{208}_Y$Pb
($Y = \Lambda, \Lambda_c, \Sigma_c, \Xi_c, \Lambda_b$).
Experimental data are taken from Ref.~\cite{aji}.
The values given inside the brackets for $^{91}_{\Sigma^0_c}$Zr are those
obtained by switching off both the Pauli blocking and the channel
coupling effects simultaneously.
}
\label{spe2}
\end{minipage}
\begin{tabular}[t]{c|ccccccccc}
\hline \hline
&$^{89}_\Lambda$Yb (Expt.)
&$^{91}_\Lambda$Zr    &$^{91}_{\Lambda^+_c}$Zr
&$^{91}_{\Sigma^0_c}$Zr &$^{91}_{\Sigma^+_c}$Zr &$^{91}_{\Sigma^{++}_c}$Zr
&$^{91}_{\Xi^0_c}$Zr    &$^{91}_{\Xi^+_c}$Zr &$^{91}_{\Lambda_b}$Zr\\
\hline \hline
$1s_{1/2}$&-22.5       &-23.9&-10.8&--- (---)  &-3.7 &---&-9.3&---&-25.7\\
$1p_{3/2}$&-16.0 ($1p$)&-18.4&-8.7 &-10.2 (-26.5) &-2.3 &---&-6.6&---&-24.2\\
$1p_{1/2}$&-16.0 ($1p$)&-18.4&-8.7 &-10.1 (-26.4) &-2.1 &---&-6.7&---&-24.2\\
$1d_{5/2}$&-9.0  ($1d$)&-12.3&-5.8 &-7.6  (-21.8) &---  &---&-4.0&---&-22.4\\
$2s_{1/2}$&            &-10.8&-3.9 &-8.1  (-23.0) &---  &---&-3.9&---&-21.6\\
$1d_{3/2}$&-9.0  ($1d$)&-12.3&-5.8 &-7.3  (-21.6)&---  &---&-4.0&---&-22.4\\
$1f_{7/2}$&-2.0  ($1f$)&-5.9 &-2.4 &-5.1  (-17.1) &---  &---&-1.3&---&-20.4\\
$2p_{3/2}$&            &-4.2 &---  &-5.0  (-16.5) &---  &---&-1.3&---&-19.5\\
$1f_{5/2}$&-2.0  ($1f$)&-5.8 &-2.4 &-4.7  (-16.8) &---  &---&-1.4&---&-20.4\\
$2p_{1/2}$&            &-4.1 &---  &-4.9  (-16.3) &---  &---&-1.3&---&-19.5\\
$1g_{9/2}$&            &---  &---  &-2.4  (-12.4) &---  &---&--- &---&-18.1\\ \\%
\hline \hline
&$^{208}_\Lambda$Pb (Expt.)
&$^{209}_\Lambda$Pb    &$^{209}_{\Lambda^+_c}$Pb
&$^{209}_{\Sigma^0_c}$Pb &$^{209}_{\Sigma^+_c}$Pb &$^{209}_{\Sigma^{++}_c}$Pb
&$^{209}_{\Xi^0_c}$Pb    &$^{209}_{\Xi^+_c}$Pb &$^{209}_{\Lambda_b}$Pb\\
\hline \hline
$1s_{1/2}$&-27.0       &-27.0&-5.2&-7.5&---&---&-6.7&---&-27.4\\
$1p_{3/2}$&-22.0 ($1p$)&-23.4&-4.1&-6.6&---&---&-5.4&---&-26.6\\
$1p_{1/2}$&-22.0 ($1p$)&-23.4&-4.0&-6.5&---&---&-5.5&---&-26.6\\
$1d_{5/2}$&-17.0 ($1d$)&-19.1&-2.4&-5.3&---&---&-3.9&---&-25.4\\
$2s_{1/2}$&            &-17.6&--- &--- &---&---&-3.3&---&-24.7\\
$1d_{3/2}$&-17.0 ($1d$)&-19.1&-2.4&-5.1&---&---&-4.0&---&-25.4\\
$1f_{7/2}$&-12.0 ($1f$)&-14.4&--- &-3.8&---&---&-2.2&---&-24.1\\
$2p_{3/2}$&            &-12.4&--- &--- &---&---&-1.4&---&-23.2\\
$1f_{5/2}$&-12.0 ($1f$)&-14.3&--- &-3.5&---&---&-2.3&---&-24.1\\
$2p_{1/2}$&            &-12.4&--- &--- &---&---&-1.5&---&-23.2\\
$1g_{9/2}$&-7.0  ($1g$)&-9.3 &--- &-2.1&---&---&--- &---&-22.6\\
$1g_{7/2}$&-7.0  ($1g$)&-9.2 &--- &-1.8&---&---&--- &---&-22.6\\
$1h_{11/2}$&           &-3.9 &--- &--- &---&---&--- &---&-21.0\\
$2d_{5/2}$&            &-7.0 &--- &--- &---&---&--- &---&-21.7\\
$2d_{3/2}$&            &-7.0 &--- &--- &---&---&--- &---&-21.7\\
$1h_{9/2}$&            &-3.8 &--- &--- &---&---&--- &---&-21.0\\
$3s_{1/2}$&            &-6.1 &--- &--- &---&---&--- &---&-21.3\\
$2f_{7/2}$&            &-1.7 &--- &--- &---&---&--- &---&-20.1\\
$3p_{3/2}$&            &-1.0 &--- &--- &---&---&--- &---&-19.6\\
$2f_{5/2}$&            &-1.7 &--- &--- &---&---&--- &---&-20.1\\
$3p_{1/2}$&            &-1.0 &--- &--- &---&---&--- &---&-19.6\\
$1i_{13/2}$&           &---  &--- &--- &---&---&--- &---&-19.3\\
\end{tabular}
\end{center}
\end{table}
As for the strange hypernuclei, 
we searched for single-particle states up to the same level as the highest
state of the core neutrons in each hypernucleus, since the
deeper levels are usually easier to observe in experiment.
After a first look at the results shown in Tables~\ref{spe1} and~\ref{spe2}
we notice:
\begin{enumerate}
\item $\Sigma_c^{++}$ and $\Xi_c^+$ hypernuclei are very unlikely to be
formed under normal circumstances, while 
solid conclusions may not be drawn about $^{49}_{\Sigma_c^{++},\Xi_c^+}$Ca.
Although these results imply the formation of these hypernuclei,
those numbers may be regarded within the uncertainties of the model and
the approximations made in the calculation.
\item $\Sigma_c^0$ and $\Sigma_c^+$ hypernuclei may have some
possibility to be formed.
However, due to a peculiar feature that the correct
$1s_{1/2}$ state is not found in
$^{17}_{\Sigma^0_c}$O, $^{49}_{\Sigma^0_c}$Ca and
$^{91}_{\Sigma^0_c}$Zr, one needs to take this statement with some caution.   
(For detailed discussion, see Ref.~\cite{Tsushima_hypbc} and below.)
\item $\Lambda_c^+, \Xi_c^0$ and $\Lambda_b$ hypernuclei
are expected to be quite likely to be formed in a realistic situation.
However, for the $\Lambda_b$ hypernuclei, it will be very
difficult to achieve such a resolution to distinguish
the states experimentally.
The question of experimental observation is not tackled realistically.
\item The Coulomb force plays a crucial role in forming (unforming)
hypernuclei, by comparing the single-particle energies among/between
the members within each multiplet,
($\Lambda,\Lambda_c^+,\Lambda_b$), ($\Sigma_c^0,\Sigma_c^+,\Sigma_c^{++}$),
and ($\Xi_c^0,\Xi_c^+$) hypernuclei.
\end{enumerate}
We comment shortly on the $^{17}_{\Sigma_c^0}$O, 
$^{49}_{\Sigma_c^0}$Ca and $^{91}_{\Sigma_c^0}$Zr hypernuclei.
The $1s_{1/2}$ state for these hypernuclei were not found.
It was concluded in Ref.~\cite{Tsushima_hypbc} that this is caused by the 
strong isospin dependent $\rho$ mean field in the very central 
region of these hypernuclei. Some related discussion will be given later.

Next, in Figs.~\ref{Capot},~\ref{Pbpot},~\ref{CaSXpot} and~\ref{PbSXpot}
we show the potential strengths
for $1s_{1/2}$ state in 
$^{41}_Y$Ca and $^{209}_Y$Pb
($Y = \Lambda,\Lambda_c^+,\Lambda_b,\Sigma_c^{0,+},\Xi_c^0$).
\begin{figure}
\epsfysize=9.0cm
\begin{center}
\begin{minipage}[t]{8cm}
\epsfig{file=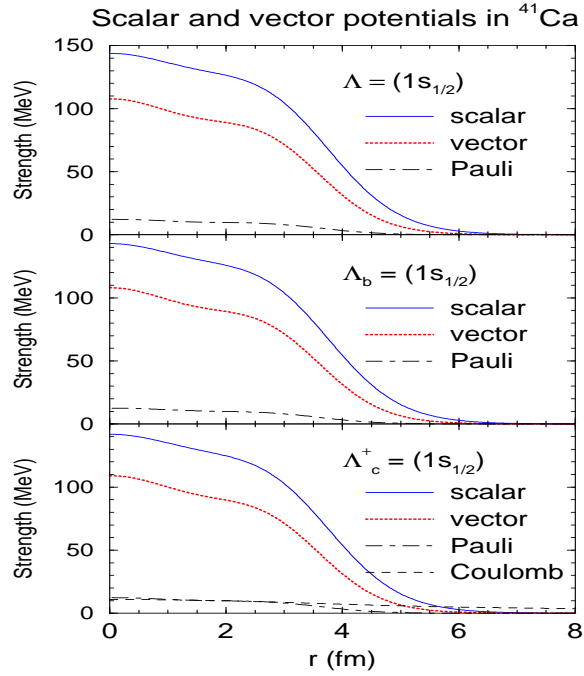,width=9cm,height=8cm,angle=-90}
\end{minipage}
\begin{minipage}[t]{16.5cm}
\caption{Potential strengths for $1s_{1/2}$ state felt by
the $\Lambda,\Lambda^+_c$ and $\Lambda_b$
in $^{41}_Y$Ca ($Y = \Lambda,\Lambda^+_c,\Lambda_b$).
"Pauli" stands for the effective, repulsive, potential representing
the Pauli blocking at the quark level plus
the $\Sigma_{c,b} N - \Lambda_{c,b} N$ channel coupling,
introduced phenomenologically at the baryon 
level~\cite{Tsushima_hyp} (see section~\ref{subsubsec:hypPauli}) 
(from Ref.~\cite{Tsushima_hypbc}).}
\label{Capot}
\end{minipage}
\end{center}
\end{figure}
\begin{figure}
\epsfysize=9.0cm
\begin{center}
\begin{minipage}[t]{8cm}
\epsfig{file=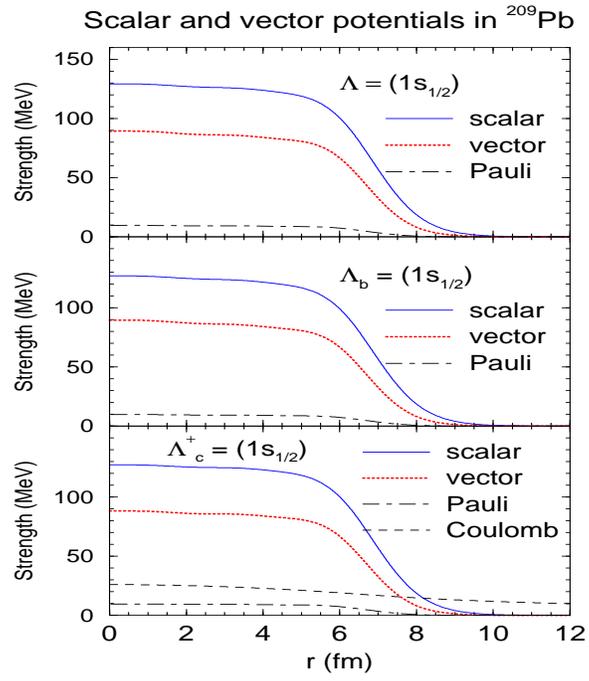,width=9cm,height=8cm,angle=-90}
\end{minipage}
\begin{minipage}[t]{16.5cm}
\caption{Same as Fig.~\ref{Capot}, but for $^{209}_Y$Pb
(from Ref.~\cite{Tsushima_hypbc}).}
\label{Pbpot}
\end{minipage}
\end{center}
\end{figure}
\begin{figure}
\epsfysize=9.0cm
\begin{center}
\begin{minipage}[t]{8cm}
\epsfig{file=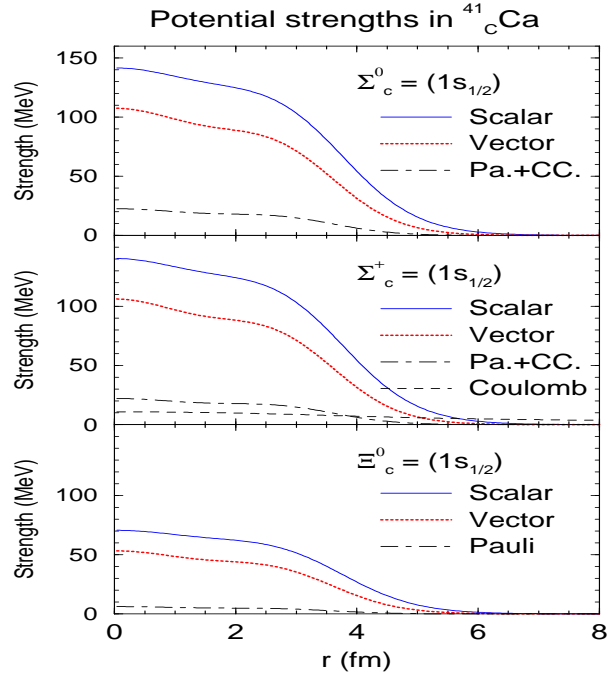,width=9cm,height=8cm,angle=-90}
\end{minipage}
\begin{minipage}[t]{16.5cm}
\caption{Potential strengths for the $\Sigma_c^{0,+}$ and $\Xi_c^0$
in $^{41}_Y$Ca ($Y = \Sigma_c^{0,+},\Xi_c^0$) for
$1s_{1/2}$ state (from Ref.~\cite{Tsushima_hypbc}).}
\label{CaSXpot}
\end{minipage}
\end{center}
\end{figure}
\begin{figure}
\epsfysize=9.0cm
\begin{center}
\begin{minipage}[t]{8cm}
\epsfig{file=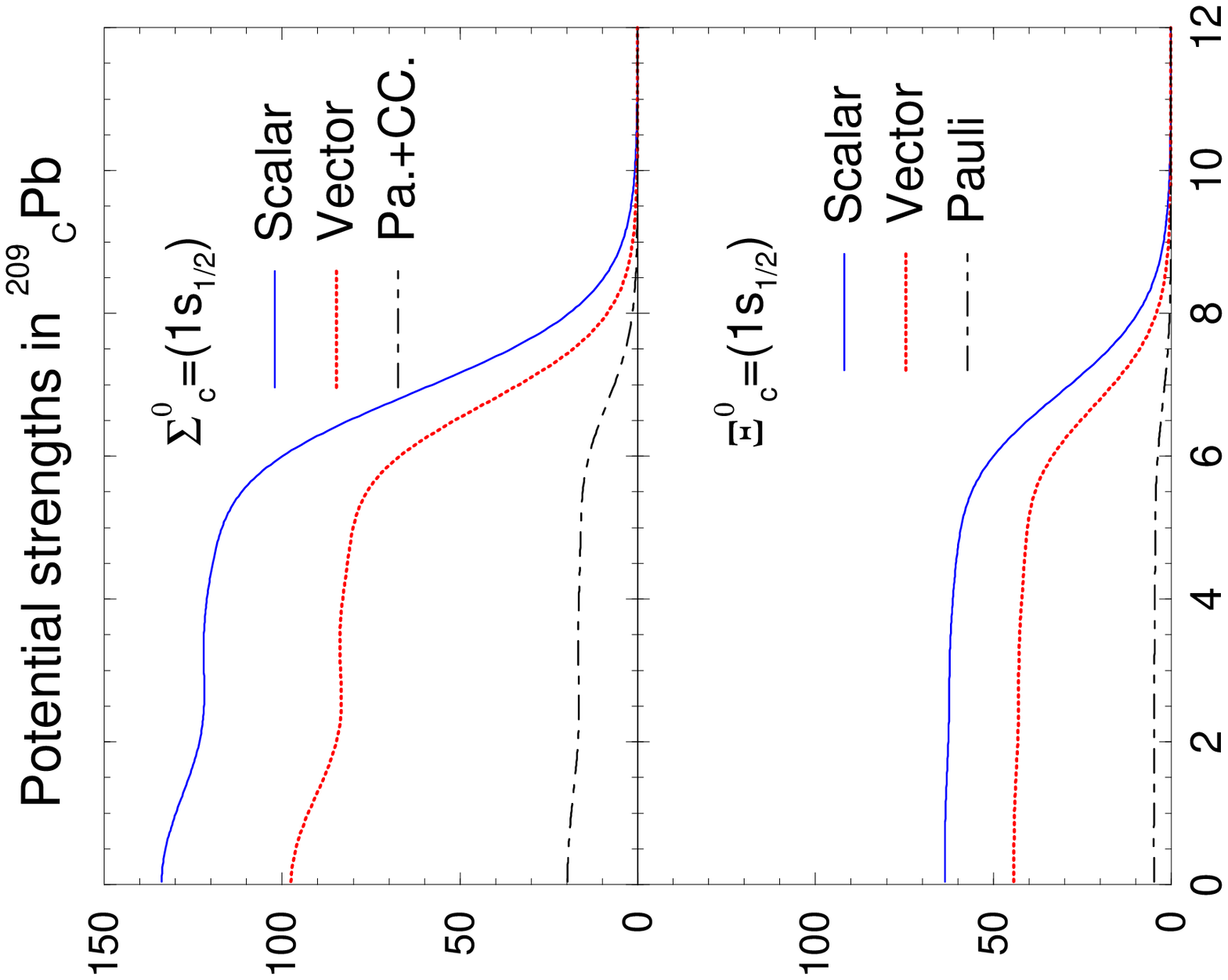,width=9cm,height=8cm,angle=-90}
\end{minipage}
\begin{minipage}[t]{16.5cm}
\caption{Same as Fig.~\ref{CaSXpot} but for $^{209}_C$Pb 
($C=\Sigma_c^0,\Xi_c^0$) 
(from Ref.~\cite{Tsushima_hypbc}).}
\label{PbSXpot}
\end{minipage}
\end{center}
\end{figure}
Recall that the effects are different for the $\Lambda_c$ and
$\Sigma_c$ as they are true for the $\Lambda$ and $\Sigma$.
As in the limit of nuclear matter in section 2.2, the scalar and vector
potentials for the heavy baryons are also quite similar to the
strange hyperons with the same light quark
numbers in the corresponding hypernuclei.
This feature also holds for the $\Sigma_c$ and
$\Sigma$ in the corresponding hypernuclei, except for a
contribution due to the difference in charges, where the Coulomb force
affects the baryon density distributions and then
the vector and scalar potentials are also slightly modified.
Thus, as far as the total baryon density distributions and
the scalar and vector potentials are concerned, among/between the members
within each multiplet,
($\Lambda,\Lambda_c,\Lambda_b$), ($\Sigma,\Sigma_c$)
and ($\Xi,\Xi_c$) hypernuclei, they show quite similar features.
However, in realistic nuclei, the Coulomb force plays a crucial
role as mentioned before, and the single-particle energies thus obtained
show very different features within each hypernuclear multiplet.
Of course, the mass differences within the multiplet are also the
dominant source for the differences in the 
single-particle energies calculated.
                                                                                
The baryon density distributions for hyperon $Y$ 
for $1s_{1/2}$ state, in $^{41}_Y$Ca and $^{209}_Y$Pb 
($Y =\Lambda, \Lambda_c^+, \Lambda_b, \Sigma_c^{0,+}, \Xi_c^0$), 
are shown in Figs.~\ref{HBdensity} and~\ref{HBSXdensity}.
\begin{figure}
\begin{center}
\begin{minipage}[t]{8cm}
\epsfig{file=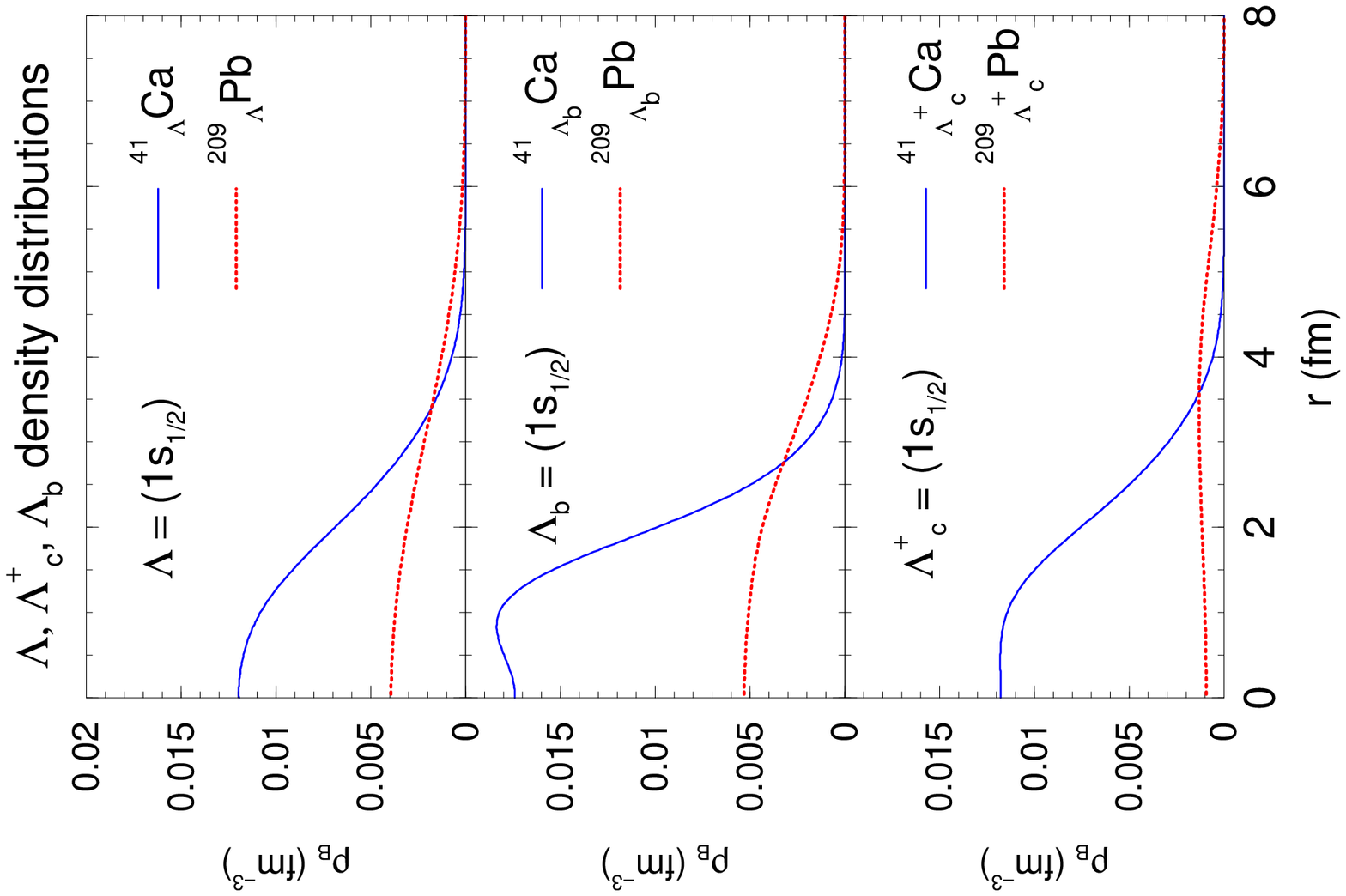,width=9cm,height=8cm,angle=-90}
\end{minipage}
\begin{minipage}[t]{16.5cm}
\caption{$\Lambda,\Lambda^+_c$ and $\Lambda_b$
baryon (probability) density distributions for $1s_{1/2}$ state
in $^{41}_Y$Ca and
$^{209}_Y$Pb ($Y = \Lambda,\Lambda^+_c,\Lambda_b$) 
(from Ref.~\cite{Tsushima_hypbc}).}
\label{HBdensity}
\end{minipage}
\end{center}
\end{figure}
\begin{figure}
\begin{center}
\begin{minipage}[t]{8cm}
\epsfig{file=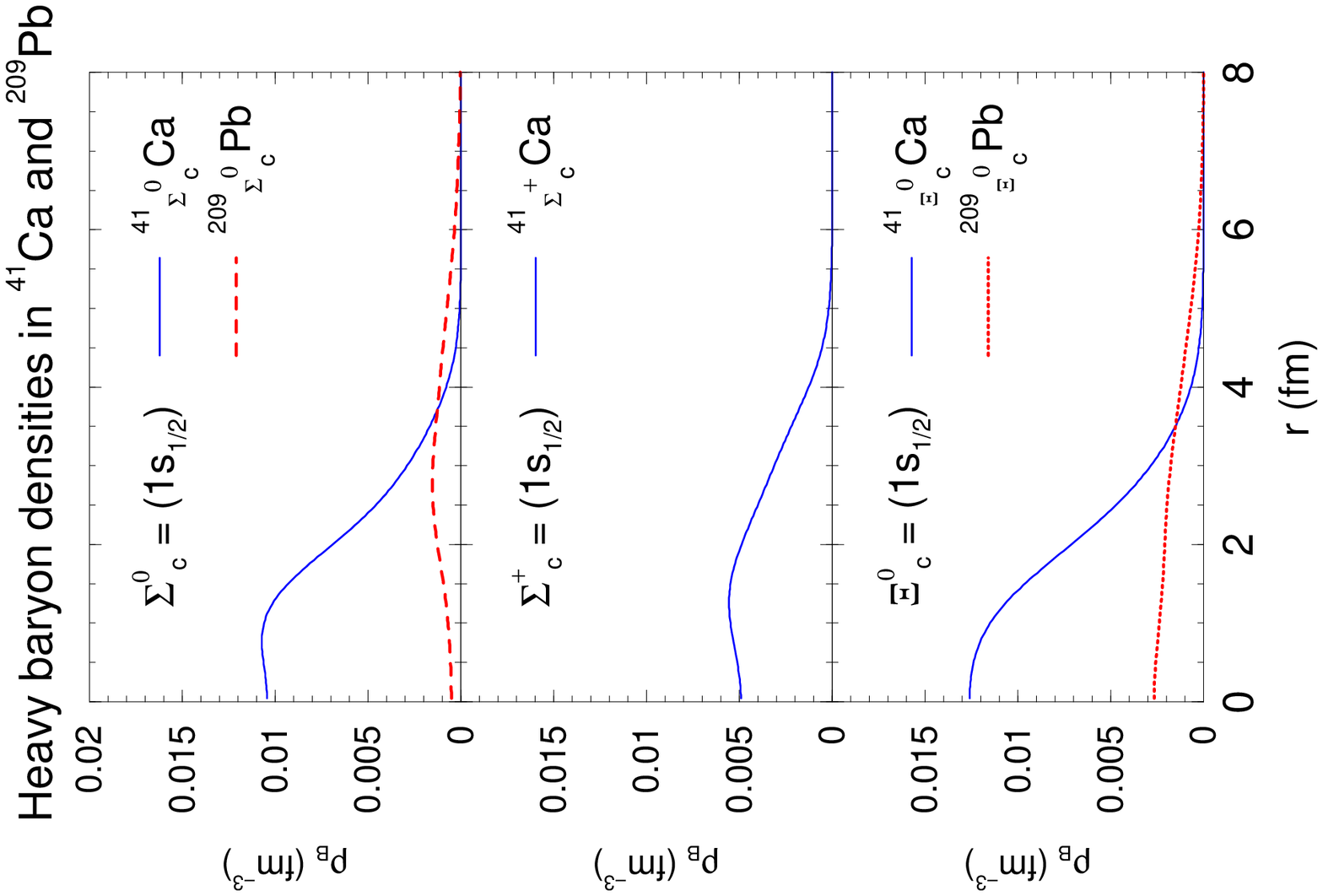,width=9cm,height=9cm,angle=-90}
\end{minipage}
\begin{minipage}[t]{16.5cm}
\caption{$\Sigma_c^{0,+}$ and $\Xi_c^0$
baryon (probability) density distributions for $1s_{1/2}$ state
in $^{41}_Y$Ca and
$^{209}_Y$Pb ($Y=\Sigma_c^{0,+},\Xi_c^0$) 
(from Ref.~\cite{Tsushima_hypbc}).}
\label{HBSXdensity}
\end{minipage}
\end{center}
\end{figure}
One notices in Fig.~\ref{HBdensity} that the heavier hyperon, 
$\Lambda_b$ is much more localized in the central region of the 
hypernucleus than $\Lambda_c^+$ and $\Lambda$.
In addition, the Coulomb force pushes the $\Lambda_c^+$ away 
from the central region in the $^{209}_{\Lambda_c^+}$Pb hypernucleus. 
Next, it is interesting to compare the
$\Sigma_c^0$ and $\Xi_c^0$ probability density
distributions in Fig.~\ref{HBSXdensity}.
Due to the Pauli blocking and the channel coupling effects,
the probability density distributions  
for the $\Sigma_c^0$ are pushed away from the origin
compared to those for
the $\Xi_c^0$ in both $^{41}_{\Sigma_c^0}$Ca and $^{209}_{\Sigma_c^0}$Pb.
In particular, the $\Sigma_c^0$ density distribution
in $^{209}_{\Sigma_c^0}$Pb is really pushed
away from the central region, and thus loses much of the character of
a typical $1s_{1/2}$ state wave function.
This fact shows that the isospin dependent $\rho$-meson mean field
for the $\Sigma^0_c$ $1s_{1/2}$ state in the central region of the neutron rich
nuclei is very strong. 
This may explain why the correct $1s_{1/2}$ state is not
found in $^{49}_{\Sigma^0_c}$Ca and $^{91}_{\Sigma^0_c}$Zr,
where the (isospin asymmetric) baryon density distributions
in central region of these nuclei are expected to be larger than
those for $^{209}_{\Sigma_c^0}$Pb -- as in the case of normal nuclei.
(In the case of $^{17}_{\Sigma_c^0}$O, we expect that the size of
the nucleus is
much smaller and the $\Sigma^0_c$ is more sensitive to the isospin
dependent $\rho$-meson mean field, due to the higher baryon density
and the limitation of the mean field approximation, than those for
the $^{209}_{\Sigma_c^0}$Pb.)
On the contrary, it is rather surprising that the $\Xi_c^0$
probability density distribution is higher and localized more  
in the central region than those for the $\Sigma_c^{0,+}$,
although one can naively expect that they have an opposite characteristic,
because of the smaller scalar attractive potential for the $\Xi_c^0$.
Thus, the effects of
the Pauli blocking, and particularly the channel coupling, play
an important role in giving rise to these different features.

Finally, we show in Table~\ref{prmstbc}
the binding energy calculated per baryon, $E_B/A$,
rms charge radius ($r_{ch}$), and rms radii of
the hyperons ($r_Y, Y=\Lambda,\Lambda_c,\Sigma_c,\Xi_c,\Lambda_b$), 
neutron ($r_n$) and proton ($r_p$) for the $1s_{1/2}$ hyperon configuration.
\begin{table}
\begin{center}
\begin{minipage}[t]{16.5cm}
\caption{Binding energy per baryon, $E_B/A$ (in MeV), rms charge radius 
($r_{ch}$), and rms radii of the $\Lambda$ and heavy baryons, $r_Y$,
neutron, $r_n$, and proton, $r_p$ (in fm) for $^{17}_Y$O, $^{41}_Y$Ca,
$^{49}_Y$Ca, $^{91}_Y$Zr and $^{209}_Y$Pb
($Y = \Lambda, \Lambda_c, \Sigma_c, \Xi_c, \Lambda_b$).
The configurations of the $\Lambda$ and heavy baryon
$Y$, are $1s_{1/2}$ for all hypernuclei.
}
\label{prmstbc}
\end{minipage}
\begin{tabular}[t]{cccccc}
\\
\hline \hline
hypernuclei & $-E_B/A$ & $r_{ch}$ & $r_Y$ & $r_n$ & $r_p$ \\
\hline \hline
$^{17}_\Lambda$O        &6.37&2.84&2.49&2.59&2.72\\
$^{17}_{\Lambda^+_c}$O  &6.42&2.85&2.19&2.58&2.73\\
$^{17}_{\Sigma^+_c}$O   &6.10&2.83&2.57&2.60&2.71\\
$^{17}_{\Xi^0_c}$O      &6.01&2.81&2.34&2.61&2.69\\
$^{17}_{\Xi^+_c}$O      &5.84&2.81&2.70&2.61&2.69\\
$^{17}_{\Lambda_b}$O    &6.69&2.87&1.81&2.57&2.75\\
\hline \hline
$^{41}_\Lambda$Ca       &7.58&3.51&2.81&3.31&3.42\\
$^{41}_{\Lambda^+_c}$Ca &7.58&3.51&2.66&3.31&3.42\\
$^{41}_{\Sigma^0_c}$Ca  &7.54&3.51&2.79&3.31&3.42\\
$^{41}_{\Sigma^+_c}$Ca  &7.42&3.50&3.21&3.31&3.41\\
$^{41}_{\Xi^0_c}$Ca     &7.43&3.49&2.76&3.32&3.40\\
$^{41}_{\Xi^+_c}$Ca     &7.32&3.49&3.11&3.32&3.39\\
$^{41}_{\Lambda_b}$Ca   &7.72&3.53&2.22&3.30&3.43\\
\hline \hline
$^{49}_\Lambda$Ca         &7.58&3.54&2.84&3.63&3.45\\
$^{49}_{\Lambda^+_c}$Ca   &7.54&3.54&2.67&3.63&3.45\\
$^{49}_{\Sigma^+_c}$Ca    &7.39&3.54&3.17&3.64&3.44\\
$^{49}_{\Sigma^{++}_c}$Ca &6.29&3.57&3.62&3.71&3.47\\
$^{49}_{\Xi^0_c}$Ca       &7.34&3.53&2.57&3.65&3.43\\
$^{49}_{\Xi^+_c}$Ca       &7.32&3.53&3.19&3.65&3.43\\
$^{49}_{\Lambda_b}$Ca     &7.64&3.56&2.16&3.63&3.46\\
\hline \hline
$^{91}_\Lambda$Zr       &7.95&4.29&3.25&4.29&4.21\\
$^{91}_{\Lambda^+_c}$Zr &7.85&4.29&3.34&4.30&4.21\\
$^{91}_{\Sigma^+_c}$Zr  &7.77&4.29&4.03&4.30&4.21\\
$^{91}_{\Xi^0_c}$Zr     &7.78&4.28&3.01&4.30&4.20\\
$^{91}_{\Lambda_b}$Zr   &7.93&4.30&2.63&4.29&4.22\\
$^{209}_\Lambda$Pb       &7.35&5.49&3.99&5.67&5.43\\
$^{209}_{\Lambda^+_c}$Pb &7.26&5.49&4.74&5.68&5.43\\
$^{209}_{\Sigma^0_c}$Pb  &6.95&5.51&4.88&5.70&5.45\\
$^{209}_{\Xi^0_c}$Pb     &7.24&5.49&4.44&5.68&5.43\\
$^{209}_{\Lambda_b}$Pb   &7.48&5.49&3.45&5.66&5.43\\
\hline \hline
\end{tabular}
\end{center}
\end{table}
At a first glance, it is very clear that the rms radius for the
$\Lambda_b$ is very small compared to those for other baryons
within hypernuclei with the same baryon number --  as one would expect,
because of its heavy mass. On the other hand, $r_{ch}$ and/or $r_p$
for the $\Lambda_b$ hypernuclei are usually the largest of all the
hypernuclei calculated. This implies that the protons in the
core nucleus are relatively more pushed away than in the other
hypernuclei, although such a feature is not seen for $r_n$.
The radii, $r_{ch}$, $r_n$ and $r_p$,
may be grouped in similar magnitudes
for all heavy baryon hypernuclei and $\Lambda$ hypernuclei
within the same baryon number multiplet
($^{41}_Y$Ca and $^{49}_Y$Ca in Table~\ref{prmstbc} may be grouped
together), reflecting the fact that
the effect of the embedded baryon on these
quantities are of order, $\simeq M_Y/[(A-1)M_N+M_Y]$.
As for the binding energy per baryon, the energy of
$\Lambda_b$ hypernuclei is usually the largest among hypernuclei with the
same baryon number.
One of the largest contributions for this is the single-particle
energy of the $1s_{1/2}$ state, even after divided by the total
baryon number.

Although there can be numerous speculations
on the implications of the present results we would like to
emphasize that our calculations indicate that
the $\Lambda^+_c,\Xi_c^0$, and $\Lambda_b$
hypernuclei would exist in realistic experimental conditions,
but there may be lesser possibilities for
the $\Sigma_c^0$ and $\Sigma_c^+$ hypernuclei.
Furthermore, it is very unlikely that the $\Sigma_c^{++}$ and $\Xi_c^+$
hypernuclei will be formed.
Experiments at facilities like Japan Hadron Facility (JHF) would provide
quantitative input to gain a better understanding of the
interactions of heavy baryons with nuclear
matter.
Experiments at colliders such as RHIC, LHC and
Fermilab could provide additional data to establish the
formation and decay of such heavy baryon hypernuclei. A combination of
these data inputs and a careful analysis, with the present
calculations being considered as a first step, would give
a valuable information about the physical implications for
the presence of heavy quarks in finite nuclei or dense nuclear matter.

\subsection{\it Meson-nucleus bound states 
\label{subsec:mesonA}}

In this section we discuss the meson-nucleus bound states for 
$\omega,\eta,\eta'$~\cite{Tsushima_etao,Tsushima_etap} first, and next 
$D(\Dbar)$ mesons~\cite{Tsushima_d}.
Studies of meson-nucleus bound states in different approaches 
have also been made   
in Refs.~\cite{Liu,Oset,Rakityansky,Hayano_mesic,Weise,Hirenzaki,
Nagahiro,Garcia}.
To avoid confusion with the isoscalar-vector ($\omega$) field appearing 
in the Lagrangian density of the QMC-I model, we emphasize that the 
$\omega$ meson appearing in this section is a physical one. 

As mentioned in section~\ref{subsubsec:scbmatter}, we can treat  
mesons in the same way as hyperons in hypernuclei,   
as long as the mesons contain light quarks (antiquarks).
(However, the pion cannot be treated in a satisfactory manner 
in a constituent quark model usually, because of its highly collective 
nature and low mass, originating in its Goldstone boson nature.)
To study the meson-nucleus bound states for 
$\omega,\eta,\eta'$ and $D (\Dbar)$ mesons, one may solve  
the Klein-Goldon equation for a given potential inside a nucleus, 
instead of solving the Dirac equation for a hyperon in the case for 
the hypernuclei. 
(For the $\omega$ meson, solving the Proca equation becomes 
equivalent to solving the Klein-Goldon equation with the Lorentz 
condition, as will be explained later.)

First, we consider the $\eta,\eta'$ and $\omega$ meson-nucleus bound states.
The physical states of the $\eta,\eta',\phi$ and $\omega$  
mesons are the superpositions of the octet and singlet states:
\bg
\xi  &=& \xi_8 \cos\theta_{P,V} - \xi_1 \sin\theta_{P,V},\quad
\xi' = \xi_8 \sin\theta_{P,V} + \xi_1 \cos\theta_{P,V},
\label{mixing1}\\
{\rm with}\hspace{2cm} \nn
\\
\xi_1 &=& \frac{1}{\sqrt{3}}\; (u\ubar + d\dbar + s\sbar),\quad
\xi_8 = \frac{1}{\sqrt{6}}\; (u\ubar + d\dbar - 2 s\sbar),
\label{mixing2}
\en
\noindent
where ($\xi$, $\xi'$) denotes ($\eta$, $\eta'$) or ($\phi$, $\omega$),
with the mixing angles
$\theta_P$ or $\theta_V$, respectively~\cite{PDG96,PDG00,PDG98,PDG02}.
Then, the masses for these mesons in
the nucleus at position $\vr$, are calculated 
using the local density approximation:
\bg
m_{\eta,\phi}^*(\vr) &=& 
\frac{2[a_{P,V}^2\Omega_q^*(\vr)+b_{P,V}^2\Omega^*_s(\vr)]
-z_{\eta,\phi}}{R_{\eta,\phi}^*}
+ {4\over 3}\pi R_{\eta,\phi}^{* 3} B,
\label{meta}\\
m_{\eta',\omega}^*(\vr) &=&
\frac{2[b_{P,V}^2\Omega_q^*(\vr)+a_{P,V}^2\Omega^*_s(\vr)]
-z_{\eta',\omega}}{R_{\eta',\omega}^*}
+ {4\over 3}\pi R_{\eta',\omega}^{* 3} B,
\label{momega}\\
\left.\frac{\partial m_j^*(\vr)}
{\partial R_j}\right|_{R_j = R_j^*} &=& 0, \quad\quad 
(j = \eta,\eta',\phi,\omega),
\label{equil}\\
{\rm with}\qquad \nn \\
a_{P,V} &=& \frac{1}{\sqrt{3}} \cos\theta_{P,V}
- \sqrt{\frac{2}{3}} \sin\theta_{P,V},\quad
b_{P,V} = \sqrt{\frac{2}{3}} \cos\theta_{P,V}
+ \frac{1}{\sqrt{3}} \sin\theta_{P,V}.
\label{abdeff}
\en
In practice, we use $\theta_P = - 10^\circ$ and
$\theta_V = 39^\circ$~\cite{PDG96,PDG00,PDG98,PDG02}, 
neglecting any possible mass
dependence and imaginary parts. We also assume that the values of 
the mixing angles do not change in medium.
Then, the in-medium modification of the $\phi$ meson mass is negligible, 
since the $\phi-\omega$ mixing is nearly ideal.
Thus, the $\phi$ meson is not expected to form nuclear bound states 
in the QMC model and we will not discuss on the $\phi$-nucleus 
bound states below.

In Eqs.~(\ref{meta}) and~(\ref{momega}), $z_{\eta,\eta',\phi,\omega}$ 
are the same as before, they parameterize the sum of 
the center-of-mass and gluon fluctuation effects, 
and are assumed to be independent of
density. The quark-meson coupling constants,
$g^q_\sigma$, $g^q_\omega$ and $g^q_\rho$, are given in 
Table~\ref{coupcc}, and other inputs, parameters, and some of 
the quantities calculated are listed in Table~\ref{bagparambc}.

Because the vector potentials for
the same flavor of quark and antiquark cancel each other,
the potentials for these mesons are given respectively by
$m^*_{\eta,\eta',\phi,\omega} - m_{\eta,\eta',\phi,\omega}$,
where they will depend only on the distance from the center of the
nucleus, $r = |\vr|$.
Before showing the potentials calculated for the 
$\eta,\eta'$ and $\omega$ mesons,
we first show in Figs.~\ref{etaomass} and~\ref{etapdmass} 
their effective masses divided by their free values, 
as well as those corresponding to 
$\omega = \frac{1}{\sqrt{2}} (u\ubar + d\dbar)$ 
(ideal mixing) and $\eta_{1,8} = \xi_{1,8}$ 
in Eq.~(\ref{mixing2}) for symmetric nuclear matter.
\begin{figure}
\epsfysize=9.0cm
\begin{center}
\begin{minipage}[t]{8cm}
\hspace*{-1cm}
\epsfig{file=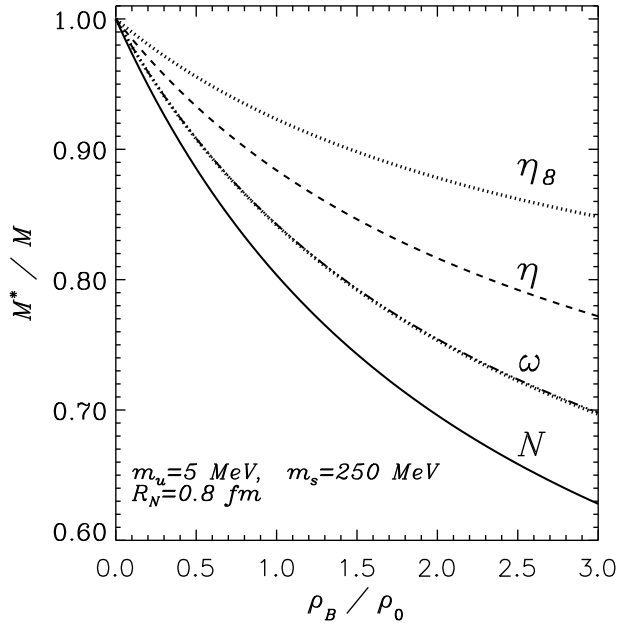,height=9cm}
\end{minipage}
\begin{minipage}[t]{16.5cm}
\caption{Effective masses of the nucleon, physical $\eta$ and $\omega$
mesons divided by their free values, and those correspond 
to the SU(3) quark model basis (the dotted lines), namely
$\omega = \frac{1}{\sqrt{2}} (u\ubar + d\dbar)$ (ideal mixing)
and $\eta_8 = \frac{1}{\sqrt{6}} (u\ubar + d\dbar - 2 s\sbar)$.
The two cases for the $\omega$ meson are almost degenerate
($\rho_0 =$ 0.15 fm$^{-3}$) -- from Ref.~\cite{Tsushima_etao}.}
\label{etaomass}
\end{minipage}
\end{center}
\end{figure}
\begin{figure}
\epsfysize=9.0cm
\begin{center}
\begin{minipage}[t]{8cm}
\hspace*{-1cm}
\epsfig{file=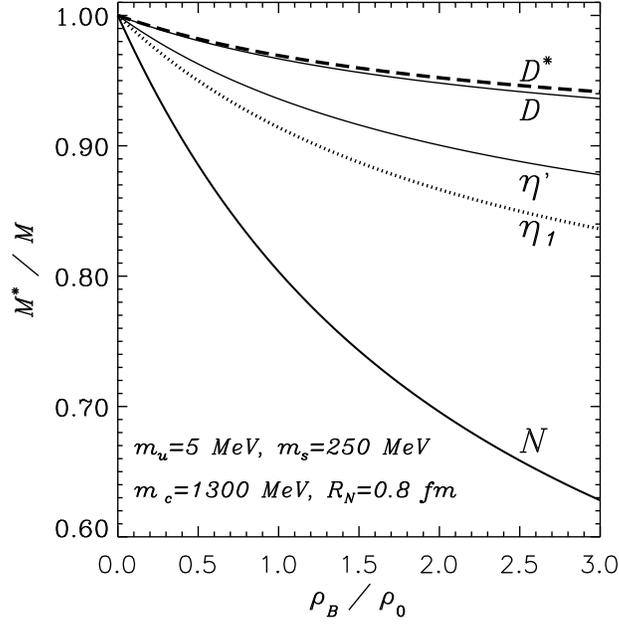,height=9cm}
\end{minipage}
\begin{minipage}[t]{16.5cm}
\caption{Same as caption of Fig.~\ref{etaomass}, but for 
physical $\eta',D$ and $D^*$ mesons, and that of the $\eta_1$, 
(the dotted lines), namely
$\eta_1 = \frac{1}{\sqrt{3}} (u\ubar + d\dbar + s\sbar)$
(from Ref.~\cite{Tsushima_JJHF}).}
\label{etapdmass}
\end{minipage}
\end{center}
\end{figure}
One can easily see that the effect of the singlet-octet mixing
is negligible for the $\omega$ mass in matter,
whereas it is important for the $\eta$ and $\eta'$ masses.

Next, as examples, we show the potentials for the mesons in
$^{26}$Mg and $^{208}$Pb nuclei, respectively, in Figs.~\ref{ompotmg} 
and~\ref{ompotpb}. 
Note that, although $^{6}$He, $^{11}$B and $^{26}$Mg are not spherical,
we have neglected the effect of deformation, which is expected to be
small and irrelevant for the present discussion. (We do not expect that
deformation should alter the calculated potentials by more than
a few MeV near the center of the deformed nucleus, because
the baryon (scalar) density there is also expected
to be more or less the same as that for a spherical nucleus --
close to normal nuclear matter density.)
\begin{figure}
\epsfysize=9.0cm
\begin{center}
\begin{minipage}[t]{8cm}
\hspace*{-1cm}
\epsfig{file=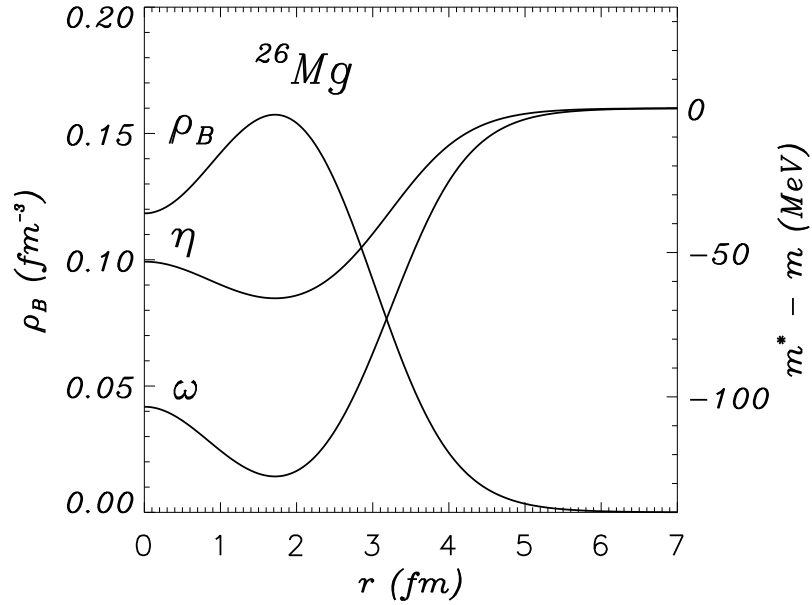,height=9cm}
\end{minipage}
\begin{minipage}[t]{16.5cm}
\caption{Potentials for the $\eta$ and $\omega$ mesons,
($m^*_\eta(r) - m_\eta$) and ($m^*_\omega(r) - m_\omega$), calculated
in QMC-I for $^{26}$Mg (from Ref.~\cite{Tsushima_etao}).}
\label{ompotmg}
\end{minipage}
\end{center}
\end{figure}
%
\begin{figure}
\epsfysize=9.0cm
\begin{center}
\begin{minipage}[t]{8cm}
\hspace*{-1cm}
\epsfig{file=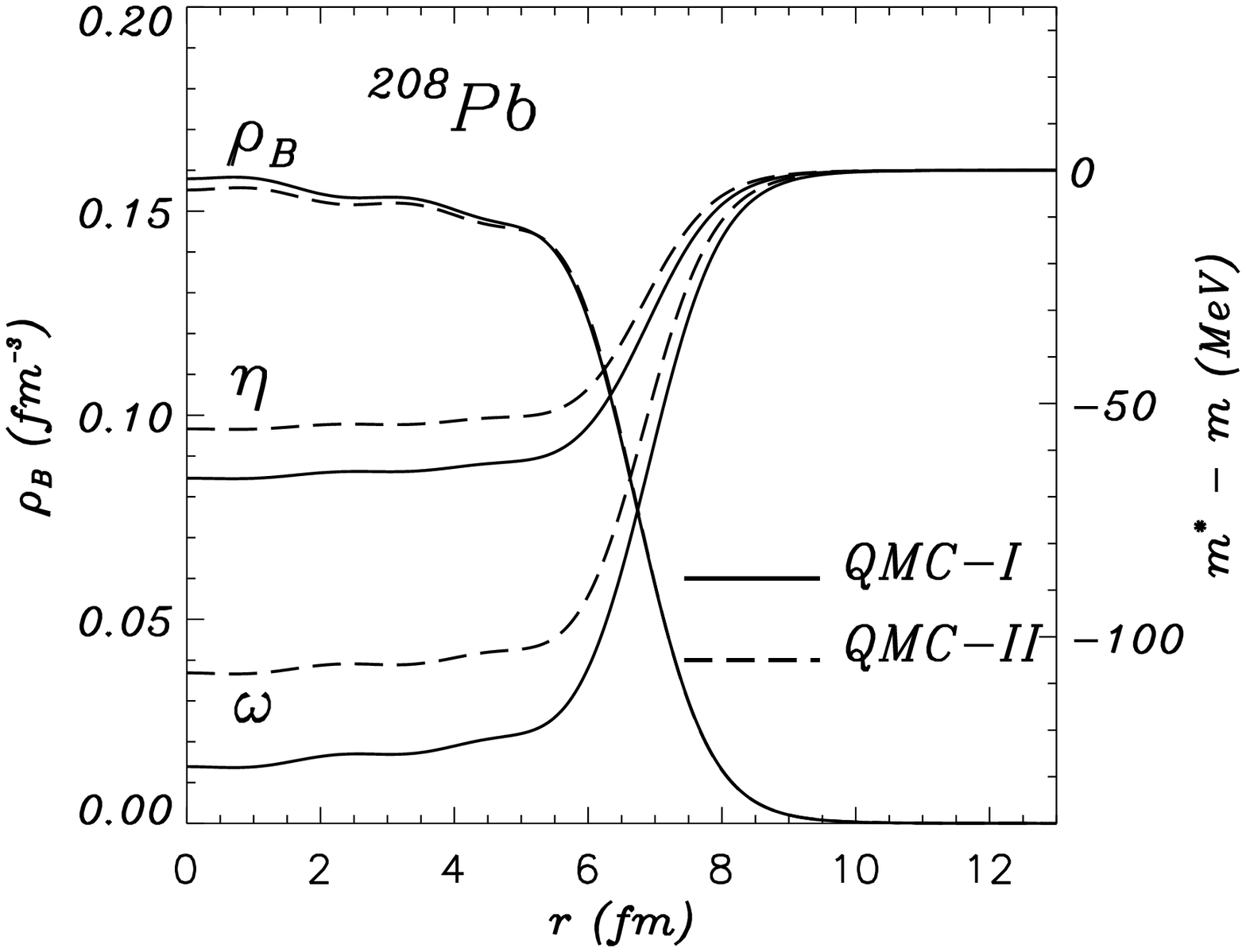,height=9cm}
\end{minipage}
\begin{minipage}[t]{16.5cm}
\caption{Same as Fig.~\ref{ompotmg} but for $^{208}$Pb.
Potentials calculated in QMC-II are also shown 
(from Ref.~\cite{Tsushima_etao}). 
(See section~\ref{subsec:matter} for QMC-II.)}
\label{ompotpb}
\end{minipage}
\end{center}
\end{figure}
Around the center of each nucleus, the depth of the potentials 
are typically 60 and 130 MeV for the $\eta$ and $\omega$ mesons, respectively.
In addition, we show the calculated potentials using QMC-II~\cite{SAI-3}
in Fig.~\ref{ompotpb}, for $^{208}$Pb, in order to estimate
the ambiguities due to different versions of the QMC model.
At the center of $^{208}$Pb, the potentials calculated using QMC-II are
about 20 MeV shallower than those calculated using QMC-I.
Note that in QMC-II the physical $\omega$ meson is identified 
with the isoscalar-vector $\omega$ field which 
mediates the interactions, while in QMC-I, it is the SU(6) quark model 
state and different from the isoscalar-vector $\omega$ field.
The potential for the $\eta$ in QMC-II is calculated using the scalar 
$\sigma$ field distribution in $^{208}$Pb obtained 
in the parameter set B.   

It is interesting to compare the potentials, or equivalently 
the effective masses for the $\eta$ and $\omega$ 
in nuclear matter obtained in different 
approaches. In QMC-I, we get:
\bea
\frac{m^*_\eta}{m_\eta} 
&\simeq& 1 - 0.12 \left(\frac{\rho_B}{\rho_0}\right),
\label{eta_effmass}
\\  
\frac{m^*_\omega}{m_\omega} 
&\simeq& 1 - 0.16 \left(\frac{\rho_B}{\rho_0}\right).  
\label{omega_effmass}
\eea
(See also Eq.~(\ref{nstr2}) for the $\omega$ obtained in QMC-II 
with using the approximated formula for the scalar field $\sigma$.)
For the $\eta$, Waas and Weise~\cite{Waas_eta} with 
a chiral SU(3) meson baryon Lagrangian, and 
Inoue and Oset~\cite{Inoue} in a chiral unitary approach, 
both obtained:
\be
\frac{m^*_\eta}{m_\eta} 
\simeq 1 - 0.05 \left(\frac{\rho_B}{\rho_0}\right),
\label{eta_Inoue}
\ee
which shows a slightly smaller downward shift for the effective 
$\eta$-meson mass, and thus entails 
a shallower potential than that obtained in QMC-I.
However, Zhong, Peng and Ning~\cite{Zhong} obtained a stronger 
downward shift compared to Eqs.~(\ref{eta_effmass}) 
and~(\ref{eta_Inoue}), based on chiral perturbation theory.
For the $\omega$, both based on QCD sum rules, Hatsuda and Lee~\cite{Hatsuda1}
obtained:
\be
\frac{m^*_\omega}{m_\omega}
\simeq 1 - (0.18 \pm 0.054) \left(\frac{\rho_B}{\rho_0}\right),
\label{omaga_Hatsuda}
\ee
while, Klingl, Kaiser and Weise~\cite{Klingl1} obtained:
\be
\frac{m^*_\omega}{m_\omega}
\simeq 1 - (0.16 \pm 0.6) \left(\frac{\rho_B}{\rho_0}\right).
\label{omaga_Klingl}
\ee
They are very similar to those obtained in QMC-I and QMC-II.
Thus, although approaches are different, they all predict  
the downward shift of the effective $\eta$ and $\omega$ masses 
in nuclear matter. 
For the $\omega$, this has been indeed confirmed independently 
by the ELSA tagged photon facility~\cite{elsa} and KEK~\cite{kek} 
as mentioned in section~\ref{subsubsec:scaling}. 
Then, one can expect a much realistic situation to find 
the $\omega$-nucleus bound states in near future.

Because the typical momentum of the bound $\omega$ is low, it should be
a very good approximation to neglect the possible energy difference
between the longitudinal and transverse components of the
$\omega$~\cite{SAI-omega}.
Then, after imposing the Lorentz condition, $\partial_\mu \phi^\mu = 0$,
solving the Proca equation becomes equivalent to solving the Klein-Gordon
equation the same as for the $\eta$ and $\eta'$ mesons:
\bge
\left[ \nabla^2 + E^2_j - m^{*2}_j(r) \right]\,
\phi_j(\vr) = 0, \qquad (j=\eta,\eta',\omega),
\label{kgequation1}
\ene
where $E_j$ is the total energy of the meson.
An additional complication, which has so far been ignored, is the
meson absorption in the nucleus, which requires a complex potential.
At the moment, we have not been able to calculate the imaginary part
of the potential (equivalently, the in-medium widths of the mesons)
self-consistently within the model.
In order to make a more realistic estimate for the meson-nucleus bound states,
we include the widths of the mesons in the nucleus
by assuming a specific form:
\bg
\tilde{m}^*_j(r) &=&
m^*_j(r) - \frac{i}{2}
\left[ (m_j - m^*_j(r))
\gamma_j + \Gamma^0_j \right], \qquad (j=\eta,\eta',\omega),
\label{imaginary}\\
&\equiv& m^*_j(r) - \frac{i}{2} \Gamma^*_j (r),
\label{width}
\en
where, $m_j$ and $\Gamma^0_j$ are the corresponding masses and widths
in free space, and
$\gamma_j$ are treated as phenomenological
parameters to describe the in-medium meson widths,
$\Gamma^*_j(r)$.
We have calculated the single-particle energies for several values
of the parameter, $\gamma_j$,
which cover the estimated ranges~\cite{Tsushima_etao}. 
However, we will present the results 
for the values, $(\gamma_\eta,\gamma_\omega)=(0.5,0.2)$, 
which are expected to be realistic.
For the $\eta'$, although it is possible that the width in a nucleus 
becomes broader, we have neglected the width for simplicity in the 
initial calculation~\cite{Tsushima_etap}, and present the results 
for $\gamma_{\eta'}=0$. (See Ref.~\cite{Nagahiro} for a recent   
estimate of the $\eta'$ width in a nucleus.)

In case the recoilless condition is achieved for meson production,   
as in the GSI experiment, we may expect that the energy
dependence of the potentials would not be strong~\cite{pion}.
Thus we actually solve the following, modified Klein-Gordon equations:
\bge
\left[
\nabla^2 + E^2_j - \tilde{m}^{*2}_j(r)
\right]\, \phi_j(\vr) = 0, \qquad (j=\eta,\omega).
\label{kgequation2}
\ene
This is carried out in momentum
space by the method developed in Ref.~\cite{landau}.
To confirm the results, we also calculated the
single-particle energies by solving
the Schr\"{o}dinger equation.
Calculated (complex) single-particle energies for 
the $\eta,\eta'$ and $\omega$ mesons, $E^*_j = E_j + m_j - i \Gamma_j/2, 
(j=\eta,\eta',\omega)$, obtained by solving the Klein-Gordon equation, 
are listed in Table~\ref{ometaenergy}. 
For a comparison, we also list the QHD results~\cite{QHDomega} 
for the $\omega$ single-particle nuclear state energies, 
calculated using the potential estimated by QHD.

\begin{table}
\begin{center}
\begin{minipage}[t]{16.5cm}
\caption{
Calculated meson-nucleus bound state energies,  
$E_j = Re (E^*_j - m_j)$, and widths, 
$\Gamma_j\,(j=\omega,\eta,\eta')$, (in MeV)
in QMC~\cite{Tsushima_etao} and those for the $\omega$ in
QHD, including the effect of $\sigma$-$\omega$ mixing~\cite{QHDomega}. 
The complex eigenenergies are given by,
$E^*_j = E_j + m_j - i \Gamma_j/2$. 
(* not calculated)
}
\label{ometaenergy}
\end{minipage}
\begin{tabular}[t]{lc|cc||c||cc|cc}
\hline \hline
& &$\Gamma^0_\eta=0$& &$\Gamma^0_{\eta'}=0$ 
&$\Gamma^0_\omega=8.43$& (MeV)& $\Gamma^0_\omega=8.43$ &(MeV)\\
& &$\gamma_\eta=0.5$ &(QMC) &(QMC) &$ \gamma_\omega$=0.2
&(QMC) &$\gamma_\omega=0.2$ &(QHD)\\
\hline \hline
& &$E_\eta$ &$\Gamma_\eta$ &$E_{\eta'}$ &$E_\omega$ &$\Gamma_\omega$
&$E_\omega$ &$\Gamma_\omega$\\
\hline
$^{6}_j$He &1s &-10.7&14.5 & * &-55.6&24.7 &-97.4&33.5 \\
\hline
$^{11}_j$B &1s &-24.5&22.8 & * &-80.8&28.8 &-129&38.5 \\
\hline
$^{26}_j$Mg &1s &-38.8&28.5 & * &-99.7&31.1 &-144&39.8 \\
            &1p &-17.8&23.1 & * &-78.5&29.4 &-121&37.8 \\
            &2s & --- & --- & * &-42.8&24.8 &-80.7&33.2  \\
\hline \hline
$^{16}_j$O &1s &-32.6&26.7 &-41.3 &-93.4&30.6 &-134&38.7 \\
           &1p &-7.72&18.3 &-22.8 &-64.7&27.8 &-103&35.5 \\
\hline
$^{40}_j$Ca &1s &-46.0&31.7 &-51.8 &-111&33.1  &-148&40.1 \\
            &1p &-26.8&26.8 &-38.5 &-90.8&31.0 &-129&38.3 \\
            &2s &-4.61&17.7 &-21.9 &-65.5&28.9 &-99.8&35.6  \\
\hline
$^{90}_j$Zr &1s &-52.9&33.2 &-56.0 &-117&33.4  &-154&40.6 \\
            &1p &-40.0&30.5 &-47.7 &-105&32.3  &-143&39.8 \\
            &2s &-21.7&26.1 &-35.4 &-86.4&30.7 &-123&38.0 \\
\hline
$^{208}_j$Pb &1s &-56.3&33.2 &-57.5 &-118&33.1 &-157&40.8 \\
             &1p &-48.3&31.8 &-52.6 &-111&32.5 &-151&40.5 \\
             &2s &-35.9&29.6 &-44.9 &-100&31.7 &-139&39.5 \\
\hline \hline
\end{tabular}
\end{center}
\end{table}

Although the specific form for the widths of the mesons in medium
could not be calculated in this model yet,
our results suggest that one should find
$\eta$- $\eta'$- and $\omega$-nucleus bound states. From 
the point of view of uncertainties arising from differences
between QMC-I and QMC-II,
the present results for both the single-particle energies
and calculated full widths should be no more than 20 \%
smaller in absolute value according to the estimate from the
potential for the $\omega$ in $^{208}$Pb in Fig.~\ref{ompotpb}.
Thus, for a heavy nucleus and relatively
wide range of the in-medium meson widths, it seems inevitable that one
should find such $\eta$- and $\omega$-nucleus bound  states.
Note that the correction to the real part of the single-particle
energies from the width, $\Gamma$, can be estimated non-relativistically,
to be of order of $\sim \Gamma^2/8m_j$
(repulsive), which is a few MeV if we use $\Gamma \simeq 100$ MeV.

Next, we discuss the $D$ and $\Dbar$ meson-nucleus bound states, 
where the treatment is different from that for the 
$\eta,\eta'$ and $\omega$ cases due to the non-zero 
vector potential for them. 
In particular, the long range Coulomb interaction in the case of the 
$D^-$-nucleus bound state, needs special treatment.
We again calculate the effective masses, 
$m^*_j(\vr)$ ($j=D,\Dbar$), and mean field potentials, 
$V^q_{\sigma,\omega,\rho}(\vr)$, at position $\vr$ in the nucleus 
using the local density approximation.
Note that the widths of the $D$ and $\overline{D}$ mesons in free space 
can be treated as zero in practice and we assume this for 
$D^-$ and $\overline{D^0}$ mesons in nuclei, because of their 
light quark content. We do not expect strong absorption for them. 
On the other hand, we expect that there should be strong absorption 
for $D^+$ and $D^0$ mesons and this is ignored in our main discussions below.
However, to get some idea, nevertheless, we calculate $D^0$ meson 
nuclear states, by forcing their widths to be zero.
Thus, the highlight in the following is the $D^-$-nucleus bound states.
The scalar and vector potentials, which depend only on 
the distance from the center of the nucleus, $r = |\vr|$, are given by:
\bg
V^j_s(r)
&=& m^*_j(r) - m_j,
\label{sdpot}\\
V^{D^-}_v(r) &=&
 V^q_\omega(r) - \frac{1}{2}V^q_\rho(r) - A(r),
\label{vdpot1}\\
V^{\overline{D^0}}_v(r) &=&
 V^q_\omega(r) + \frac{1}{2}V^q_\rho(r),
\label{vdpot2}\\
V^{D^0}_v(r) &=&
-\left( V^q_\omega(r) + \frac{1}{2}V^q_\rho(r) \right),
\label{vdpot3}
\en
where $A(r)$ is the Coulomb interaction between the meson and the nucleus.
Note that the $\rho$-meson mean field potential, $V^q_\rho(r)$, is negative
in a nucleus with a neutron excess, such as $^{208}$Pb.
For the larger $\omega$ meson coupling, suggested by $K^+$-nucleus scattering,
$V^q_\omega(r)$ is replaced by $\tilde{V}^q_\omega(r) = 1.4^2 V^q_\omega(r)$.
(See Eq.~(\ref{vpot}).)
In Fig.~\ref{dmespot} we show the sum of the potentials
for the $D^-$ in $^{208}$Pb for the two choices of
[$V^{D^-}_s(r)+V^{D^-}_v(r)$] (the dashed line corresponds to
$\tilde{V}^q_\omega(r)$
and the dotted line to ${V}^q_\omega(r)$).
%
\begin{figure}
\begin{center}
\begin{minipage}[t]{8cm}
\hspace*{-1cm}
\epsfig{file=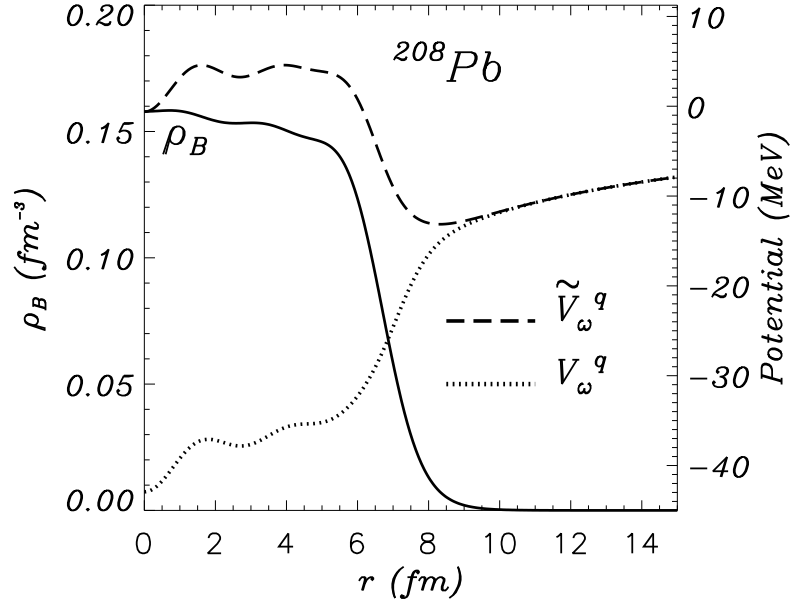,height=9cm}
\end{minipage}
\begin{minipage}[t]{16.5cm}
\caption{Sum of the scalar, vector and Coulomb potentials for the $D^-$
meson in $^{208}$Pb for two cases,
$(m^*_{D^-}(r) - m_{D^-}) + \tilde{V}^q_\omega(r)
+ \frac{1}{2} V^q_\rho(r) - A(r)$ (the dashed line) and
$(m^*_{D^-}(r) - m_{D^-}) + V^q_\omega(r)
+ \frac{1}{2} V^q_\rho(r) - A(r)$ (the dotted line),
where $\tilde{V}^q_\omega(r) = 1.4^2 V^q_\omega(r)$
(from Ref.~\cite{Tsushima_d}).}
\label{dmespot}
\end{minipage}
\end{center}
\end{figure}
Because the $D^-$ meson is heavy and may be described well
in the (non-relativistic) Schr\"{o}dinger equation, one
expects the existence of the $_{D^-}^{208}$Pb bound states
just from inspection of the naive sum of the potentials,
in a way which does not distinguish the vector or
scalar Lorentz character.

Now we calculate the bound state energies for the
$D$ and $\Dbar$ in $^{208}$Pb nucleus,
using the potentials calculated in QMC.
There are several variants of the dynamical equation for a bound meson-nucleus
system. Consistent with the mean field picture of QMC, we actually
solve the Klein-Gordon equation:
\bg
[ \nabla^2 &+& (E_j - V^j_v(r))^2- m^{*2}_j(r) ]\,
\phi_j(\vr) = 0,  
\label{kgequation}
\en
where $E_j$ is the total energy of the meson
(the binding energy is $E_j-m_j$).
To deal with the long range Coulomb potential, we first expand the quadratic
term (the zeroth component of Lorentz vector) as,
$(E_j - V^j_v(r))^2 = E_j^2 + A^2(r) + V^2_{\omega\rho}(r) +
2 A(r)V_{\omega\rho}(r) - 2 E_j (A(r)+V_{\omega\rho}(r))$,
where $V_{\omega\rho}(r)$ is the combined potential due to
$\omega$ and $\rho$ mesons
($V_{\omega\rho}(r) = V_\omega^q(r)-{1\over 2}V_\rho^q(r)$ for $D^-$).
Then, Eq.~(\ref{kgequation}) can be rewritten as an effective
Schr\"odinger-like equation,
\begin{equation}
\left[ - {\nabla^2\over 2m_j} + V_j(E_j,r)\right] \Phi_j(r)
= {E_j^2-m_j^2 \over 2m_j}\Phi_j(r)
\label{schrodinger}
\end{equation}
where $\Phi_j(r) = 2m_j\phi_j(r)$ and
$V_j(E_j,r)$ is an effective energy-dependent
potential which can be split
into three pieces (Coulomb, vector and scalar parts),
\begin{equation}
V_j(E_j,r) = {E_j\over m_j} A(r) + {2E_jV^j_{\omega\rho}(r)
- (A(r)+V^j_{\omega\rho}(r))^2\over 2m_j} + {{m^*_j}^2(r)-m^2_j\over 2m_j}.
\end{equation}
Note that only the first term in this equation is a long range
interaction and thus needs special treatment, while the second and third
terms are  short range interactions.
We would like to emphasize that no reduction has been made to derive
the Schr\"odinger-like equation, so that all relativistic corrections
are included in our calculation.
The calculated  meson-nucleus bound state energies for $^{208}$Pb,
are listed in Table~\ref{denergy}.

\begin{table}
\begin{center}
\begin{minipage}[t]{16.5cm}
\caption{
$D^-, \overline{D^0}$ and $D^0$ bound state energies (in MeV).
The widths are all set to zero.
Note that the $D^0$ bound state  
energies calculated with $\tilde{V}^q_\omega$ will be much larger 
(in absolute value) than
those calculated with $V^q_\omega$.
}
\label{denergy}
\end{minipage}
\begin{tabular}[t]{l|cccccc}
state  &$D^- (\tilde{V}^q_\omega)$ &$D^- (V^q_\omega)$
&$D^- (V^q_\omega$, no Coulomb) &$\overline{D^0} (\tilde{V}^q_\omega)$
&$\overline{D^0} (V^q_\omega)$ &$D^0 (V^q_\omega)$ \\
\hline
1s &-10.6 &-35.2 &-11.2 &unbound &-25.4 &-96.2\\
1p &-10.2 &-32.1 &-10.0 &unbound &-23.1 &-93.0\\
2s & -7.7 &-30.0 & -6.6 &unbound &-19.7 &-88.5\\
\end{tabular}
\end{center}
\end{table}

The results show that both the $D^-$ and $\overline{D^0}$ 
are bound in $^{208}$Pb
with the usual $\omega$ coupling constant.
For the $D^-$ the Coulomb force
provides roughly 24 MeV of  binding for the $1s$ state, and is strong enough
to bind the system even with the much more repulsive $\omega$ coupling
(viz., $1.4^2V_\omega^q$).
The $\overline{D^0}$ with the stronger $\omega$ coupling is not bound.
Note that the difference between $\overline{D^0}$ 
and $D^-$ without the Coulomb force
is due to the interaction with the
$\rho$ meson, which is attractive for the $\overline{D^0}$ but repulsive for
the $D^-$. For completeness, we also calculated the binding energies
for the $D^0$ setting a possible in-medium width to be zero. 
It is deeply bound because the $\omega$ interaction
with the light antiquarks is attractive. However, the expected
large width associated with strong absorption may render it
experimentally inaccessible. It is an extremely important experimental
challenge to see whether it can be detected.

We show the eigenfunctions obtained for the Schr\"odinger-like equation 
in Fig.~\ref{dmeswf}, together with the baryon density distribution
in $^{208}$Pb.
\begin{figure}
\begin{center}
\begin{minipage}[t]{8cm}
\hspace*{-1.5cm}
\epsfig{file=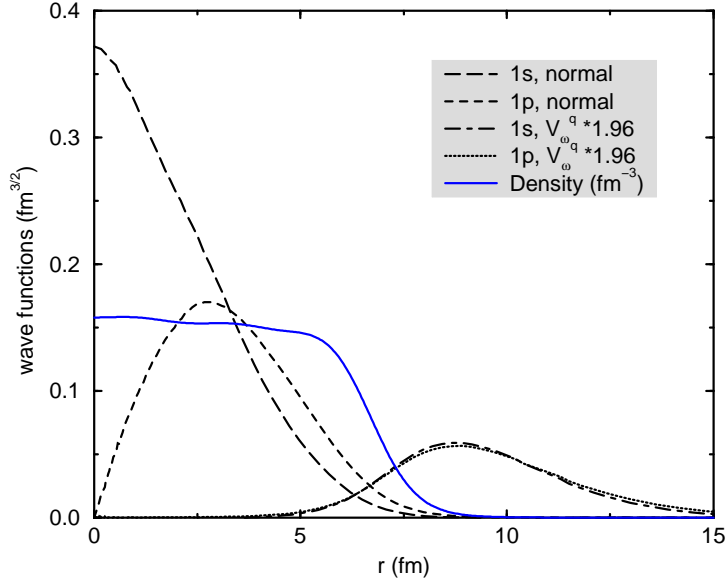,height=9cm}
\end{minipage}
\begin{minipage}[t]{16.5cm}
\caption{The Schr\"odinger-like bound state wave functions of the $D^-$ meson
in $^{208}$Pb, for two different $\omega$ meson coupling strengths.
See also the caption of Fig.~\ref{dmespot}.
The wave function is normalized as follows:
$\int_0^{\infty}\!dr\,4\pi r^2 |\Phi(r)|^2 = 1$
(from Ref.~\cite{Tsushima_d}).}
\label{dmeswf}
\end{minipage}
\end{center}
\end{figure}
For the usual $\omega$ coupling, the eigenstates ($1s$ and $1p$)
are well within the nucleus, and behave as expected at the origin.
For the stronger $\omega$ coupling, however, the $D^-$ meson
is considerably pushed out of the nucleus. In this case,
the bound state (an atomic state) is formed
solely due to the Coulomb force.
An experimental determination of whether this is a nuclear state or an atomic
state would give a strong constraint on the $\omega$ coupling.
We note, however, that
because it is very difficult to produce $D$-mesic nuclei with
small momentum transfer, and the $D$-meson production
cross section is small compared with the background from other channels,
it will be a challenging task to detect such bound states
experimentally~\cite{hayaex}.

In spite of possible model-dependent uncertainties, our results suggest that
the $D^-$ meson should be bound in $^{208}$Pb through two quite
different mechanisms, namely, the scalar and attractive
$\sigma$ mean field even without the assistance of the Coulomb force
in the case of the normal vector potential ($V^q_\omega$), and
solely due to the Coulomb force in the case of the stronger
vector potential ($\tilde{V}^q_\omega$).
(We recall that the kaon is a pseudo-Goldstone boson
and expected to be difficult to treat properly with the usual bag model.
Thus, the analysis of Ref.~\cite{Tsushima_k} 
on the vector potential for the light
quarks inside the kaon bag may not be applicable to the light
quarks inside the D-meson.)

Thus, whether or not the $\overline{D^0}$-$^{208}$Pb bound states exist
would give new information as to whether the interactions of
light quarks in a heavy meson are the same as those in a nucleon.
The enormous difference
between the binding energies of the $D^0$ ($\sim
100$ MeV) and the $\overline{D^0}$ 
($\sim 10$ MeV) is a simple consequence of the
presence of a strong Lorentz vector mean-field, while the existence of
any binding at all would give us important information concerning the
role of the Lorentz scalar $\sigma$ field (and hence dynamical symmetry
breaking) in heavy quark systems. In spite of the perceived experimental
difficulties,
we feel that the search for these bound systems should have a very high
priority.

\section{Effects of nucleon substructure on lepton-nucleus scatterings
\label{sec:leptonA}}

In this section, we summarize the applications of the QMC model 
to lepton-nucleus scattering, in which nucleon substructure effects
in a medium may emerge. The highlight of this section is the 
comparison between the QMC model prediction and the experimental 
results for the bound proton electromagnetic form factors 
measured at the Thomas Jefferson Laboratory (JLab)~\cite{POLE}.

\subsection{\it Nucleon form factors 
\label{subsec:formf}}
\subsubsection{\it Electromagnetic form factors  
\label{subsec:EMff}}

For an on-shell nucleon, the electric ($G_E$) and magnetic ($G_M$)
form factors can be conveniently defined in the Breit frame by
\begin{eqnarray}
\bra {N_{s'}({\vec{q}\over 2}) }  J^0(0) \ket {N_s(-{\vec{q}\over 2})}
&=& \chi_{s'}^\dagger\,\chi_s\, G_E(Q^2), 
\label{GE} \\
\bra {N_{s'}({\vec{q}\over 2}) } \vec{J}(0) \ket {N_s(-{\vec{q}\over 2})} &=&
\chi_{s'}^\dagger\, {i\vec{\sigma}\times\vec{q}\over 2 M_N}\,\chi_s\,
G_M(Q^2), 
\label{GM}
\end{eqnarray}
where $\chi_s$ and $\chi^{\dagger}_{s'}$ are Pauli spinors for
the initial and final nucleons respectively,
$\vec{q}$ is the three momentum transfer,
and $q^2 = -Q^2 = -\vec{q}^{\,2}$.
The major advantage of the Breit frame is that $G_E$ and $G_M$
are explicitly  decoupled and can be determined by the time and space
components of the electromagnetic current ($J^{\mu}$), respectively.
Note that, in the above definitions, 
both the initial and final states
are physical states which incorporate meson clouds.

The electromagnetic current of the quark is given by 
\begin{eqnarray}
j^\mu(x) &=& \sum_q Q_q e \overline{\psi}_q(x) \gamma^\mu \psi_q(x),
\label{current}
\end{eqnarray}
where $\psi_q(x)$ is the quark field operator
for the flavor $q$ and $Q_q$ is its charge in units of the proton charge, $e$.
This is represented by a diagram (a) in Fig.~\ref{medfig1}.
\begin{figure}
\begin{center}
\begin{minipage}[t]{8cm}
\hspace*{-3cm}
\epsfig{file=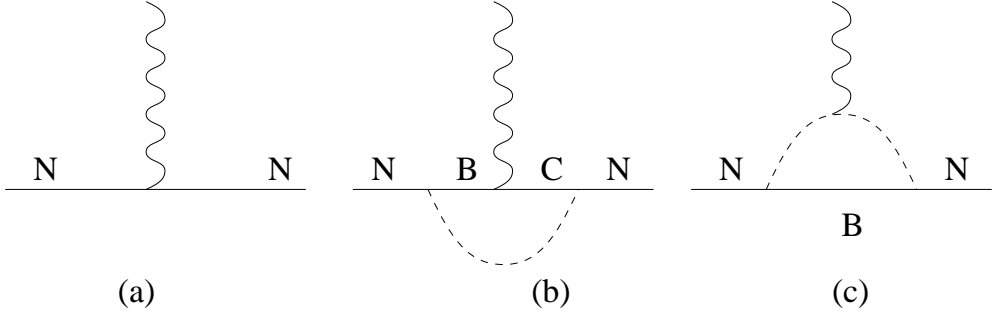,height=4cm}
\end{minipage}
\begin{minipage}[t]{16.5cm}
\caption{Diagrams illustrating various contributions included in
the calculation of the nucleon form factors (up to one pion loop).
The intermediate baryons $B$ and $C$ are restricted
to the $N$ and $\Delta$.}
\label{medfig1}
\end{minipage}
\end{center}
\end{figure}
The momentum eigenstate of a baryon is constructed
by the Peierls-Thouless projection method~\cite{Lu97,PT62},
\begin{equation}
\Psi_{\rm{PT}}(\vec{x}_1, \vec{x}_2, \vec{x}_3; \vec{p}) =
N_{\rm{PT}} e^{i\vec{p} \cdot \vec{x}_{\rm{c.m.}} }
\psi_q(\vec{x}_1 - \vec{x}_{\rm{c.m.}}) \psi_q(\vec{x}_2 - \vec{x}_{\rm{c.m.}})
\psi_q(\vec{x}_3 - \vec{x}_{\rm{c.m.}}), \label{PTWF}
\end{equation}
where $N_{\rm{PT}}$ is a normalization  constant,
$\vec{p}$  the total momentum of the baryon, and
$\vec{x}_{\rm{c.m.}} = (\vec{x}_1 + \vec{x}_2 + \vec{x}_3)/3$
is the center of mass of the baryon (we assume equal mass quarks here).
                                                                                
Using Eqs.~(\ref{current}) and~(\ref{PTWF}), the nucleon electromagnetic
form factors for the proton's 
quark core can be calculated by~\cite{Lu97}
\begin{eqnarray}
G_E(Q^2) &=& \int\! d^3r j_0(Qr)\rho_q(r)K(r)/D_{\rm PT}, \label{PTE}\\
G_M(Q^2) &=& (2 M_N/Q)\int\! d^3r j_1(Qr) 
\beta_q j_0(\Omega_q r/R_N)j_1(\Omega_q r/R_N)K(r)/D_{\rm PT}, \label{PTM}\\
D_{\rm PT} &=& \int\! d^3r \rho_q(r) K(r),
\end{eqnarray}
where $D_{\rm PT}$ is the normalization factor,
$\rho_q(r) \equiv j_0^2(\Omega_q r/R_N) 
+ \beta_q^2 j_1^2(\Omega_q r/R_N)$, and
$K(r) \equiv \int\! d^3x \, \rho_q(\vec{x}) \rho_q(-\vec{x} - \vec{r})$
is the recoil function to account for the correlation of the
two spectator quarks.

Apart from the center-of-mass correction, it is also vital to include
Lorentz contraction of the bag for the form factors at moderate
momentum transfer~\cite{Lu97,LP70}.
In the Breit frame, the photon-quark interaction can be
reasonably treated as instantaneous. The final form of the form
factors can be obtained through 
a simple rescaling~\cite{Lu97}, i.e.,
\begin{eqnarray}
G_E(Q^2) &=& \left(\frac{M_N}{E_N}\right)^2 G^{\rm sph}_E\left(\left(\frac{M_N}{E_N}\right)^2 Q^2\right), 
\label{GEcore}\\
G_M(Q^2) &=& \left(\frac{M_N}{E_N}\right)^2 G^{\rm sph}_M\left(\left(\frac{M_N}{E_N}\right)^2 Q^2\right),
\label{GMcore}
\end{eqnarray}
where $E_N=\sqrt{M_N^2 + Q^2/4}$ and
$G_{M,E}^{\rm sph}(Q^2)$ are the form factors calculated
with the static spherical bag wave function.
The scaling factor in the argument arises from the coordinate transformation
of the struck quark and
the factor in the front, $(M_N/E_N)^2$,  comes from  the reduction
of the integral measure of two spectator quarks in the Breit frame~\cite{LP70}.
For the case of a bound nucleon (a nucleon in nuclear matter)  
we use the substitution~\cite{Lu-eN}, $(M_N/E_N) \to (M^*_N/E^*_N)$ 
in Eqs.~(\ref{GEcore}) and~(\ref{GMcore}). 

As is well-known, a realistic picture of the nucleon should include
the surrounding meson cloud.
Following the cloudy bag model (CBM)~\cite{CBM,TT83},
we limit our consideration on the meson cloud correction to the
most important component, namely the pion cloud.
For the pion cloud contribution we include up to one pion loop, 
namely, diagrams (b) and (c) in Fig.~\ref{medfig1}.
As in free space, the pion field is a Goldstone boson field 
and acts to restore the chiral symmetry.
The Lagrangian related to the pion field and its interaction,
within the pseudoscalar quark-pion coupling scheme, is
\begin{equation}
{\cal L}_{\pi q} =
 {1\over 2} (\partial_\mu \vec{\pi})^2
        - {1\over 2} m^2_\pi \vec{\pi}^2
        - {i\over 2f_\pi} {\overline \psi_q} \gamma_5 \vec{\tau} \cdot
        \vec{\pi} \psi_q \delta_S,
\end{equation}
where $\delta_S$ is  a surface delta function of the bag,
$m_\pi$  the pion mass and $f_\pi$  the pion decay constant.
The electromagnetic current of the pion is
\begin{equation}
j^\mu_\pi(x) = -i e [ \pi^\dagger(x) \partial^\mu \pi(x)
               -\pi(x) \partial^\mu \pi^\dagger(x)],
\end{equation}
where
$ \pi(x) = {1\over \sqrt{2}}[\pi_1(x) + i\pi_2(x)]$
either destroys a negatively charged pion
or creates a positively charged one.
The detailed expressions for their contributions can be
found in Ref.~\cite{Lu97} with the following substitutions
$m_\pi \rightarrow m_\pi^*$,
$m_B   \rightarrow m_B^*$,
and $f_{\pi AB} \rightarrow f_{\pi AB}^*$.
                                                                                
In principle, the existence of the $\pi$ and $\Delta$ inside the nuclear
medium will also lead to some modification of their properties.
Since the pion is well approximated as a Goldstone boson,
the explicit chiral symmetry
breaking is small in free space,
and it should be somewhat smaller in nuclear medium~\cite{BR}.
While the pion mass would be slightly smaller in the medium,
because the pion field has little
effect on the form factors (other than $G_{\rm En}$), we use $m_\pi^*= m_\pi$.
As the $\Delta$ is treated on the same footing as the nucleon in the CBM,
its mass should vary in a similar manner as that of the nucleon. Thus
we assume that the in-medium and free space $N-\Delta$ mass splitting are
approximately equal, i.e.,
$m_\Delta^* - M_N^* \simeq m_\Delta - M_N$.
The physical $\pi AB$ coupling constant is obtained by
$ f^{AB} \simeq \left({f^{AB}_0\over f^{NN}_0}\right) f^{NN}$.
There are uncertain corrections to the bare coupling constant $f^{NN}_0$,
such as the nonzero quark mass and the correction for
spurious center of mass motion.
Therefore, we use the renormalized coupling constant in our
calculation, $f^{NN} \simeq 3.03$,
which corresponds to the usual $\pi NN$ coupling constant,
$f^2_{\pi NN}\simeq 0.081$. In the medium, the $\pi NN$ coupling constant
might be expected to decrease  slightly due to the enhancement of the
lower component of the quark wave function, but we shall
ignore this density dependence in the present treatment
and use $f_{\pi NN}^* \simeq f_{\pi NN}$.
                                                                                
The corrections for the center-of-mass motion and Lorentz contraction
lead to significantly better agreement with data than was obtained in
the original, static CBM calculations. For a detailed comparison with data we refer to Ref.~\cite{Lu97}. 
This approach is sometimes called the improved cloudy bag model (ICBM). 
Our main results, namely the
density dependence of the form factors in matter,
relative to those in free space,
are shown in Fig.~\ref{medfig3}.
\begin{figure}
\begin{center}
\begin{minipage}[t]{8cm}
\hspace*{-2cm}
\epsfig{file=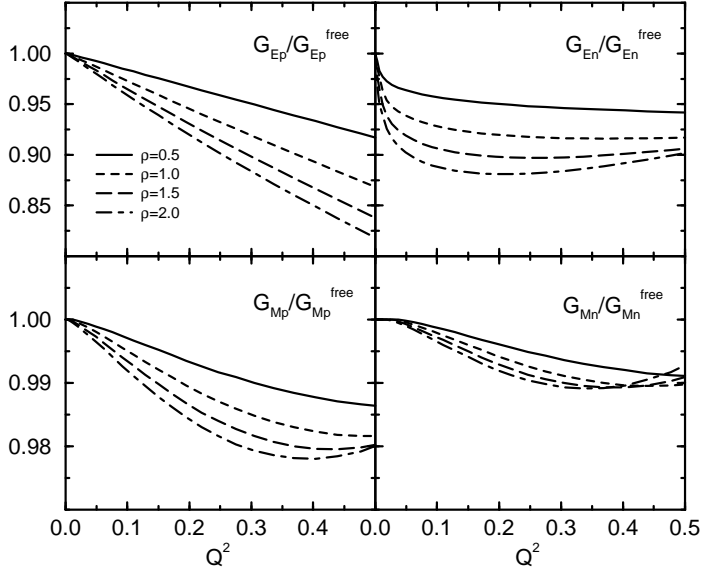,height=9cm}
\end{minipage}
\begin{minipage}[t]{16.5cm}
\caption{The nucleon electromagnetic form factors
in the nuclear medium (relative to those in free space case)~\cite{Lu-eN}.
The free space bag radius is 1 fm and the density is quoted in units of
the saturation density of symmetric nuclear matter, 
$\rho_0 = 0.15$ fm$^{-3}$.}
\label{medfig3}
\end{minipage}
\end{center}
\end{figure}
The charge form factors are much more sensitive to
the nuclear medium density than
the magnetic ones. The latter are nearly one order of magnitude
less sensitive. Increasing density obviously
suppresses the electromagnetic form factors for small $Q^2$.
For a fixed $Q^2$ (less than $0.3 \mbox{ GeV}^2$), the form factors
decrease almost linearly with respect to the nuclear baryon 
density, $\rho$.
At $Q^2 \sim 0.3 \mbox{ GeV}^2$, the proton and neutron  charge form factors
are  reduced by roughly 5\% and 6\% for $\rho = 0.5 \rho_0$, and
8\% for the normal nuclear density, $\rho_0$;
similarly, the proton and neutron magnetic form factors are 1\% and 0.6\%
smaller for $\rho = 0.5 \rho_0$,
and 1.5\% and 0.9\% for the normal nuclear density.

The best experimental constraints on the changes in these form factors
come from the analysis of $y$-scaling data. For example, in 
iron nucleus (Fe) the
nucleon root-mean-square radius cannot vary by more than 3\%~\cite{sick}.
However, in the kinematic range covered by this analysis, 
the $eN$ cross section is predominantly magnetic, 
so this limit applies essentially to $G_{M}$.
(As the electric and magnetic form factors contribute typically in the
ratio 1:3 the corresponding limit on $G_{E}$ would be
nearer 10\%.) For the QMC model considered here, the calculated increase
in the root-mean-square radius of the magnetic form factors is less than
0.8\% at $\rho_0$.
For the electric form factors the best experimental limit seems to
come from the Coulomb sum-rule, where a variation
bigger than 4\% would be excluded~\cite{JOU}.
This is similar in size to the
variations calculated here (e.g., 5.5\% for $G^p_E$ at $\rho_0$)
and not sufficient to reject them.

\subsubsection{\it The QMC predictions and Experimental results 
(EM interactions)
\label{subsec:QMCpredic}}

Now, we study the medium modification of the bound 
proton electromagnetic form factors in nuclei, $^4$He, 
$^{16}$O, and $^{40}$Ca~\cite{LuEMFFA}.
We emphasize that the bound proton electromagnetic 
form factors in $^4$He was predicted before the experimental data from 
JLab were taken.
The quark wave function, as well as the nucleon wave function
(both are Dirac spinors), are determined
once a solution to equations of motion are found self-consistently.
The orbital electromagnetic form factors for a bound proton,
in local density approximation, are given by
\begin{equation}
G_{E,M}^\alpha(Q^2)= \int G_{E,M} (Q^2,\rho_B(\vec{r}))
\rho_{p\alpha}(\vec{r}) \,d\vec{r},
\end{equation}
where $\alpha$ denotes a specified orbit with appropriate quantum numbers,
and $G_{E,M}(Q^2,\rho_B(\vec{r}))$ is the density-dependent form factor
of a ``proton'' immersed in nuclear matter with local baryon density,
$\rho_B(\vec{r})$\footnote{In a more sophisticated treatment,
for example, using a full distorted wave calculation, the weighting may
emphasize the nuclear surface somewhat more~\cite{kelly}.}.
Using the nucleon shell model wave functions (Dirac spinors),
the local baryon density and the local proton density in the specified
orbit, $\alpha$, are easily evaluated as
\begin{eqnarray}
\rho_B(\vec{r}) &=&  \sum_\alpha^{\rm occ} d_\alpha 
\psi^\dagger_\alpha(\vec{r})\psi_\alpha(\vec{r}),
\nonumber \\
\rho_{p\alpha}(\vec{r}) &=& (t_\alpha+{1\over 2})
\psi^\dagger_\alpha(\vec{r})\psi_\alpha(\vec{r}),
\end{eqnarray}
where $d_\alpha= (2j_\alpha+1)$ refers to the degeneracy of nucleons occupying
the orbit $\alpha$ and $t_\alpha$ is the eigenvalue of the isospin operator,
$\tau^N_3/2$.
Notice that the quark wave function only depends on the
surrounding baryon density through the scalar $\sigma$ field in the 
QMC model. Therefore, 
this part of the calculation of $G_{E,M}(Q^2,\rho_B(\vec{r}))$ is
the same as for nuclear matter~\cite{Lu-eN,Lu-He3}, as   
described in section~\ref{subsec:EMff}.
The notable medium modifications of the quark wave function inside the bound
nucleon in QMC include a reduction of its frequency and an enhancement
of the lower component of the Dirac spinor.
As explained in section~\ref{subsec:EMff}, the corrections arising 
from recoil and center of mass motion for the bag are made  
using the Peierls-Thouless projection method, 
combined with  Lorentz contraction of the internal
quark wave function and with the perturbative pion cloud added
afterwards~\cite{Lu97}. Note that
possible off-shell effects~\cite{offshell} and
meson exchange currents~\cite{MEC,KTRiskaE} are ignored 
in the present approach.
The resulting nucleon electromagnetic form factors agree with  experiment
quite well in free space~\cite{Lu97}.
Because of  the limitations of the bag model
the form factors  are expected to be most reliable at low momentum
transfer (say, less than 1 $\mbox{GeV}^2$).
To cut down theoretical uncertainties, we prefer to show the ratios
of the form factors with respect to corresponding free space values.
Throughout this section, we use the renormalized $\pi NN$ coupling constant,
$f^2_{\pi NN} \simeq 0.0771$~\cite{Bugg}.
The bag radius in free space and the current quark mass are 
taken to have the standard values, 0.8 fm and 5 MeV, respectively.

Fig.~\ref{he4.ps} shows the ratio of the electric and magnetic form factors
for $^4$He (which has only one state, $1s_{1/2}$) with respect to the
free space values.
\begin{figure}
\begin{center}
\begin{minipage}[t]{8cm}
\hspace*{-2cm}
\epsfig{file=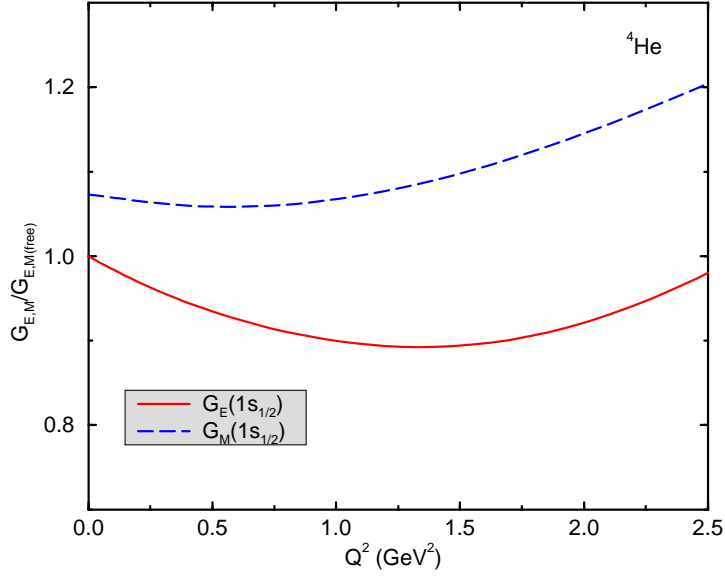,height=9cm}
\end{minipage}
\begin{minipage}[t]{16.5cm}
\caption{Ratio of in-medium to free space electric and magnetic form factors
for the proton in $^4$He.
(The free bag radius and current quark masses were respectively taken to be 
0.8 fm and 5 MeV in all figures.)
(Taken from Ref.~\cite{LuEMFFA}.)}
\label{he4.ps}
\end{minipage}
\end{center}
\end{figure}
As expected, both the electric and magnetic rms  radii become slightly larger,
while the magnetic moment of the proton increases by about 7\%.
Fig.~\ref{co16B.ps} shows the ratio  of the electric and magnetic form
factors for $^{16}$O with respect to the free space values,
which has  one $s$-state, $1s_{1/2}$, and two $p$-states,
$1p_{3/2}$ and $1p_{1/2}$.
For comparison, we also show in Fig.~\ref{co16B.ps}
the corresponding ratio of form factors (those curves with triangle symbols)
using a variant of QMC where the bag constant is allowed to decrease
by 10\%~\cite{LU-1} (see section~\ref{subsec:modfqmc}).
\begin{figure}
\begin{center}
\begin{minipage}[t]{8cm}
\hspace*{-2cm}
\epsfig{file=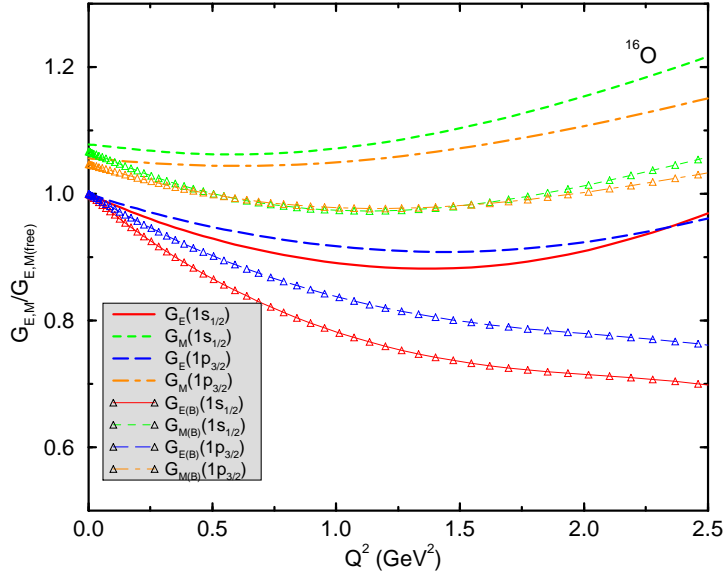,height=9cm}
\end{minipage}
\begin{minipage}[t]{16.5cm}
\caption{Ratio of in-medium to free space electric and magnetic form factors
for the  $s$- and $p$-shells of $^{16}$O.
The curves with triangle symbols represent the corresponding ratio calculated
in a variant of QMC with a 10\% reduction of the bag constant, $B$ 
(from Ref.~\cite{LuEMFFA}).}
\label{co16B.ps}
\end{minipage}
\end{center}
\end{figure}
The momentum dependence of the form factors
for the $s$-wave nucleon is  more suppressed as the inner orbit in 
$^{16}$O experiences a larger average baryon density than in $^4$He.
The magnetic moment for the $s$-orbit nucleon is similar to that in $^4$He,
but it is reduced  by $2 - 3$\% in the $p$-orbit.
Since the difference between the two $p$-orbits is rather small, we do not
plot the results for $1p_{1/2}$. Finally,  
{}Fig.~\ref{2ratio.ps} shows the ratio of the electric and magnetic form
{}factors for $^{16}O$ with respect to the free space values. 
For comparison, we also show in Fig.~\ref{2ratio.ps}
the corresponding ratio of form factors (those curves with triangle symbols)
using the MQMC model~\cite{LU-1}.
\begin{figure}
\begin{center}
\begin{minipage}[t]{8cm}
\hspace*{-2cm}
\epsfig{file=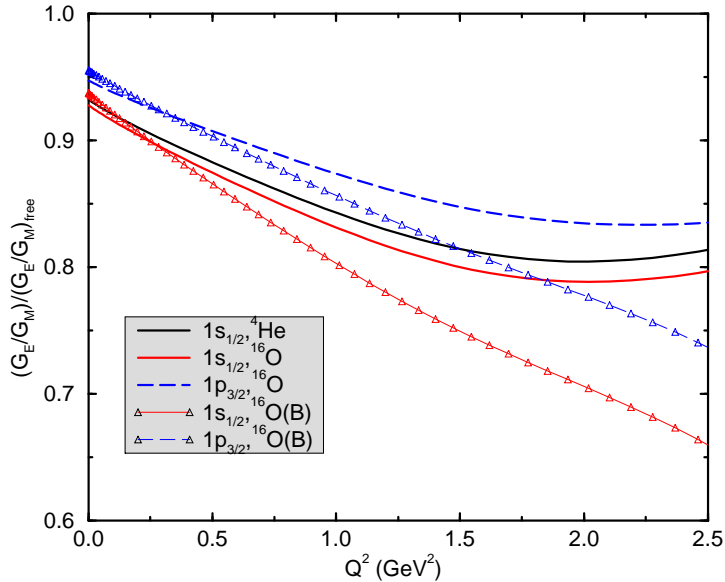,height=9cm}
\end{minipage}
\begin{minipage}[t]{16.5cm}
\caption{Ratio of electric and magnetic form factors in-medium, divided
by the free space ratio. As in the previous figure, curves with triangle symbols
represent the corresponding
values calculated in a variant of QMC  with a 10\% reduction of $B$ 
(from Ref.~\cite{LuEMFFA}).}
\label{2ratio.ps}
\end{minipage}
\end{center}
\end{figure}
It is evident that the effect of a possible reduction in $B$ is quite large
and will severely reduce the electromagnetic form factors for a bound nucleon
since the bag radius is quite sensitive to the value of $B$.

{}From the experimental point of view, it is more reliable to show the ratio,
$G_E/G_M$, since it can be derived directly from the ratio of transverse
to longitudinal polarization of the outgoing proton,
with minimal systematic errors.
We find that $G_E/G_M$ runs from roughly 0.41  at $Q^2 = 0 $ to 0.28 and 0.20
at $Q^2 = 1$ GeV$^2$ and $2$ GeV$^2$, respectively, for a proton
in the $1s$ orbit in $^4$He or $^{16}$O.
The  ratio of  $G_E/G_M$ with respect to the corresponding free
space ratio is presented in Fig.~\ref{2ratio.ps}.
The result for the $1s$-orbit  in $^{16}$O is close to that in  $^4$He
and   2\% lower than that for the $p$-orbits in $^{16}$O.
The effect on this ratio of ratios of a reduction in $B$ by the maximum
permitted from other constraints~\cite{sick} is quite significant,
especially for larger $Q^2$.

For completeness, we have also calculated the orbital electric and magnetic
form factors for heavy nuclei such as $^{40}$Ca and $^{208}$Pb.
The form factors for the proton in selected orbits 
are shown in Fig.~\ref{ca-pb.ps}.
\begin{figure}
\begin{center}
\begin{minipage}[t]{8cm}
\hspace*{-2cm}
\epsfig{file=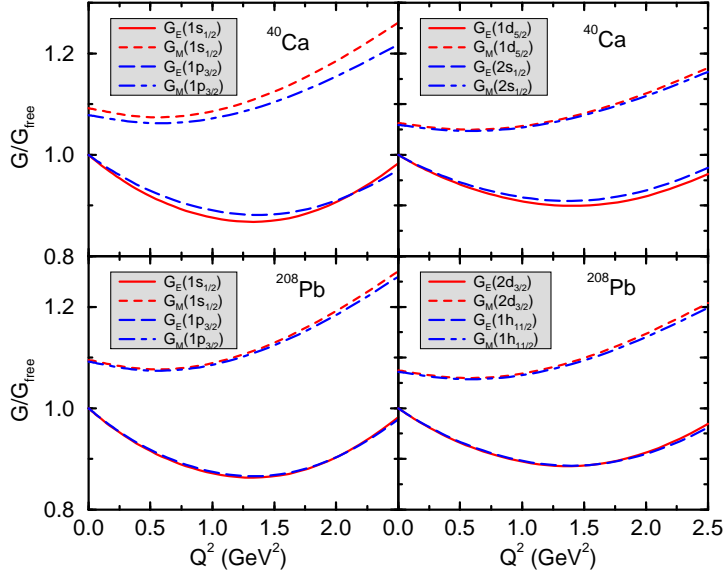,height=9cm}
\end{minipage}
\begin{minipage}[t]{16.5cm}
\caption{Ratio of in-medium to free space electric and magnetic form factors
in specific orbits, for $^{40}$Ca and $^{208}$Pb 
(from Ref.~\cite{LuEMFFA}).}
\label{ca-pb.ps}
\end{minipage}
\end{center}
\end{figure}
Because of the larger central baryon density of heavy nuclei,
the proton electric and magnetic form factors
in the inner orbits ($1s_{1/2}$, $1p_{3/2}$ and $1p_{1/2}$ orbits)
suffer much stronger medium modifications
than those in light nuclei.
That is to say, the $Q^2$ dependence is further suppressed, while
the magnetic moments appear to be larger.
Surprisingly, the nucleons in peripheral orbits ($1d_{5/2}$, $2s_{1/2}$,
and $1d_{3/2}$ for $^{40}$Ca
and $2d_{3/2}$, $1h_{11/2}$, and $3s_{1/2}$ for $^{208}$Pb)
still show significant medium effects,
comparable to those in $^4$He.

Finally, we would like to add some comments on the
magnetic moment in a nucleus.  In the present calculation, we
have only calculated the contribution from the intrinsic
magnetization (or spin) of the nucleon, which is modified
by the scalar field in a nuclear medium~\cite{SAI-q}.
As shown in the figures we have found that the intrinsic
magnetic moment is enhanced in matter because of the change in the quark
structure of the nucleon.  We know, however, that there
are several, additional contributions to the nuclear magnetic
moment, such as  meson exchange currents, higher-order correlations, etc.
As is well known in relativistic nuclear models like QHD, there is a
so-called magnetic moment problem in mean-field approximation~\cite{walecka95}.
To cure this problem, one must
calculate the convection current matrix element within relativistic
random phase approximation (RRPA)~\cite{kurasawa85}.
However, at high momentum transfer we expect
that it should be feasible to detect the enhancement of the intrinsic
spin contribution which we have predicted because the long-range
correlations, like RRPA, should decrease much faster in that region.

So far, we have discussed the QMC model predictions on the bound 
proton electromagnetic form factors.
Experimentally, the electromagnetic form factors of
bound protons were studied in polarized ($\vec e, e' \vec p$) scattering
experiments on $^{16}$O and $^4$He~\cite{POLE} at MAMI and 
Thomas Jefferson Laboratory (JLab), respectively.
We show in Fig.~\ref{JlabQMC} the outcome of the 
``super ratio", $R/R_{{\rm PWIA}}$, which was made for the final analysis 
of the polarization transfer measurements on $^4$He~\cite{POLE} 
performed at JLab.
\begin{figure}
\begin{center}
\begin{minipage}[t]{8cm}
\hspace*{-3cm}
\epsfig{file=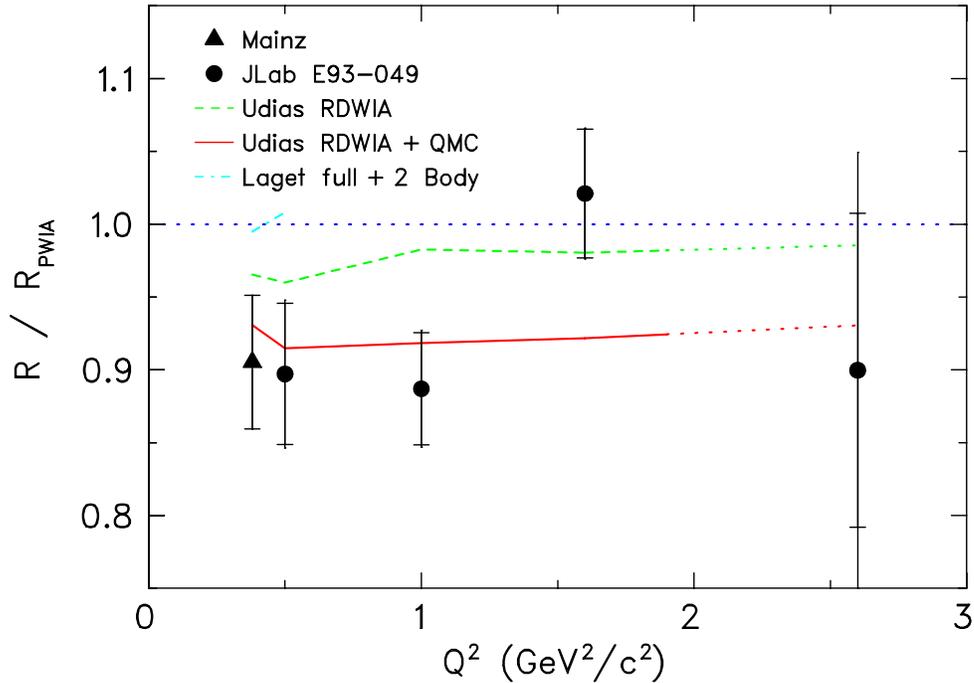,width=9cm,angle=90}
\end{minipage}
\begin{minipage}[t]{16.5cm}
\caption{Super ratio $R/R_{PWIA}$, as a function of $Q^2$, taken from
Ref.~\cite{POLE} (by private communication with S. Strauch). 
See caption of Fig.~1 in Ref.~\cite{POLE} for detailed explanations.
In the figure, "Laget" and "Udias", respectively, correspond to 
the calculations reported in Refs.~\cite{Laget} and~\cite{Udias}.
}
\label{JlabQMC}
\end{minipage}
\end{center}
\end{figure}
In Fig.~\ref{JlabQMC}, $R_{PWIA}$ stands for the prediction
based on the relativistic plane-wave impulse approximation (PWIA),
and the measured ratio $R$ is defined by:
\bge
R = \frac{(P_x'/P'_z)_{^4{\rm He}}}{(P_x'/P'_z)_{^1{\rm H}}}.
\label{sratio}
\ene
The analysis concluded that ratio of the electric ($G^p_E$)
to magnetic ($G^p_M$) Sachs proton form factors differs by
$\sim$10\% in $^4$He from that in $^1$H.
As can be seen from Fig.~\ref{JlabQMC}, conventional models 
employing free proton form factors,
phenomenological optical potentials, and bound state wave
functions, as well as relativistic corrections,
meson exchange currents (MEC), isobar contributions and final state
interactions~\cite{POLE,Laget,Udias,KELLY},
all fail to account for the observed effect in $^4$He.
Indeed, a better agreement with the data is obtained, in addition
to these standard nuclear-structure corrections, the small correction associated with the change in  
the internal structure of the bound proton predicted by the QMC 
model is taken into account.

\subsubsection{\it Nucleon substructure effect on 
the longitudinal response functions
\label{subsubsec:longi}}

There is still considerable interest in the longitudinal response (LR) for
quasielastic electron scattering off a nucleus.  Within the framework of non-relativistic
nuclear models and the impulse approximation,
it may be difficult to reproduce the observed, quenched LR~\cite{LRF}.
In the mid '80s, several groups calculated the LR function using QHD-I and argued that
the contribution of the relativistic random-phase approximation (RRPA) is very vital in reducing the
LR~\cite{HOR-3}.
                                                                                
Since the nucleon has, however, the internal structure, it may affect the LR for electron
scattering~\cite{SAI-LR}.
We first briefly review the calculation of the LR
function for quasielastic electron scattering from (iso-symmetric)
nuclear matter in QHD~\cite{HOR-3}.  The starting point
is the lowest order polarization insertion (PI), $\Pi_{\mu \nu}$, for the
$\omega$ meson.  This describes the coupling of a virtual vector meson
or photon to a particle-hole or nucleon-antinucleon
excitation:
\be
\Pi_{\mu \nu}(q) = -ig_v^2 \int \frac{d^4k}{(2\pi)^4}
\mbox{Tr}[G(k) \gamma_\mu G(k+q) \gamma_\nu] ,
\label{LRFunc}
\ee
where $G(k)$ is the self-consistent nucleon propagator
in relativistic Hartree approximation (RHA).
One can separate the PI into two pieces: one is the density dependent part,
$\Pi_{\mu \nu}^D$, and the other is the
vacuum one, $\Pi_{\mu \nu}^F$.
The former is finite, but the latter is divergent and
must be renormalized~\cite{SAI-LR}.
                                                                                
In the Hartree approximation, where only the lowest one-nucleon loop is
considered, the LR function,
$S_L^H$, is then simply proportional to
$S_L^H(q) \propto G_{pE}^2(q) {\Im} \Pi_L(q)$. Here, $\Pi_L (= \Pi_{33} - \Pi_{00})$ is the longitudinal (L)
component of the PI and $G_{pE}$ is the proton electric form factor,
which is usually parameterized by a dipole form in free space:
$G_{pE}(Q^2) = 1/(1 + Q^2/0.71)^2$
with the space-like four-momentum transfer, $Q^2 = - q^2$, in units of
GeV$^2$.
                                                                                
In RRPA, the L component of the PI, $\Pi_L^{RPA}$, involves the sum of
the ring diagrams to all orders.
It involves the $\sigma$-$\omega$ mixing in the nuclear 
medium~\cite{MORI,HOR-3,SAI-LR}, and
is given by
\be
\Pi_L^{RPA}(q) = \frac{(1 - \Delta_0 \Pi_s) \Pi_L + \Delta_0 \Pi_m^2}{
\epsilon_L},
\label{PILRPA}
\ee
where $\epsilon_L$ is the L dielectric function~\cite{SAI-9,SAI-omega,SAI-LR} 
and $\Delta_0$ is the free $\sigma$-meson propagator.
Here $\Pi_s$ and $\Pi_m$ are respectively the scalar and the time
component of the mixed PIs.  
The vacuum component of the scalar PI is again divergent
and we need to renormalize it~\cite{SAI-9,SAI-omega,SAI-LR}.  
For the mixed PI there is no vacuum polarization and
it vanishes at zero density.
                                                                                
To discuss the effect of changes in the
internal structure of the nucleon, 
we can consider the following modifications to QHD:
\begin{enumerate}
\item Meson-nucleon vertex form factor \\
Since both the mesons and nucleons are composite they have finite
size.  As the simplest example, we take a monopole form factor, $F_N(Q^2)$,
at the vertex
with a cut off parameter, $\Lambda_N = 1.5$ GeV~\cite{SAI-LR}.
\item Modification of the proton electric form factor \\
We have studied the electromagnetic form
factors of the nucleon in nuclear medium, using 
the QMC model (see section~\ref{subsec:EMff}).
The main result of that calculation is 
that the ratio of the electric form factor
of the proton in medium to that in free space
decreases essentially linearly as a function of $Q^2$, and that
it is accurately parameterized as $R_{pE}(\rho_0,Q^2) \equiv
G_{pE}(\rho_0,Q^2)/G_{pE}(Q^2) \simeq
1 - 0.26 \times Q^2$ at $\rho_0$~\cite{Lu-eN,SAI-LR}.
\item Density dependence of the coupling constants \\
In the QMC model, the confined quark in the nucleon couples to the $\sigma$ field
which gives rise to an attractive force.
As a result,
the coupling between the $\sigma$ and nucleon is expected to be
reduced at finite density.
The coupling between the vector meson and nucleon
remains constant because it is related to the baryon number.
\end{enumerate}
                                                                                
To study the LR of nuclear matter, we first have
to solve the nuclear ground state within RHA.
To take into account the modifications 1 and 3, we replace the
$\sigma$- and $\omega$-nucleon coupling constants in QHD by:
$g_s \to g_s(\rho_B) \times F_N(Q^2)$ and
$g_v \to g_v \times F_N(Q^2)$,
where the density dependence of $g_s(\rho_B)$ is given by solving the
nuclear matter problem self-consistently in QMC~\cite{SAI-LR}.
Requiring the usual saturation condition for nuclear matter,
we found the coupling constants:
$g_s^2(0)=61.85$ and $g_v^2=62.61$ (notice that
$g_s$ decreased by about 9\% at $\rho_0$).
In the calculation we fix the quark mass, $m_q$, to be 5 or 300 MeV~\cite{SAI-1,GUI-3},
$m_\sigma=550$ MeV
and $m_\omega=783$ MeV, while the bag parameters are chosen so as
to reproduce the free nucleon mass with the
bag radius $R_N=0.8$ fm (see section~\ref{subsec:matter}).
Within RHA, this yields the effective nucleon mass $M^*/M=0.81$ at $\rho_0$
and the incompressibility $K=281$ MeV.

\begin{figure}
\epsfysize=9.0cm
\begin{center}
\begin{minipage}[t]{8 cm}
\hspace*{-1cm}
\epsfig{file=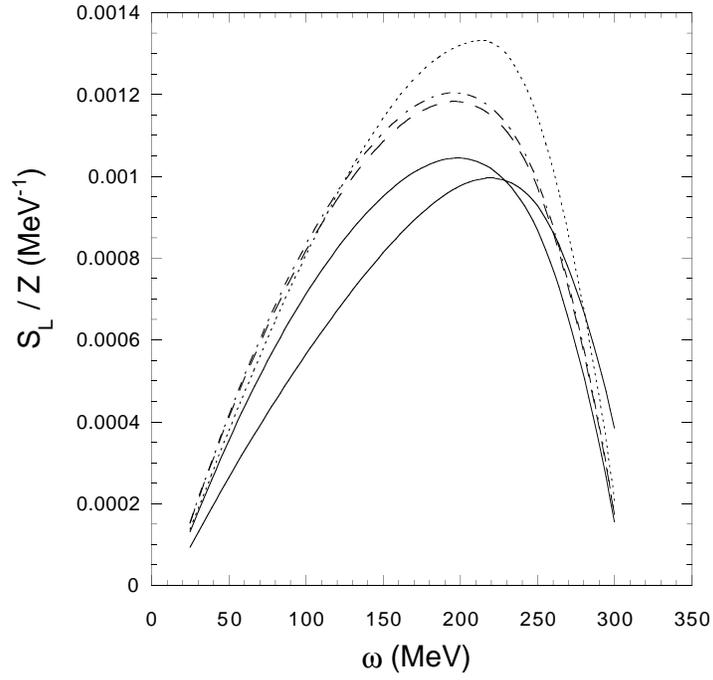,height=9cm}
\end{minipage}
\begin{minipage}[t]{16.5 cm}
\caption{LR functions in the QMC model.
We fix $q$ = 550 MeV and $\rho_B = \rho_0$
(from Ref.~\cite{SAI-LR}) -- see text for details. 
}
\label{fig:resp}
\end{minipage}
\end{center}
\end{figure}
\begin{figure}
\epsfysize=9.0cm
\begin{center}
\begin{minipage}[t]{8 cm}
\hspace*{-1cm}
\epsfig{file=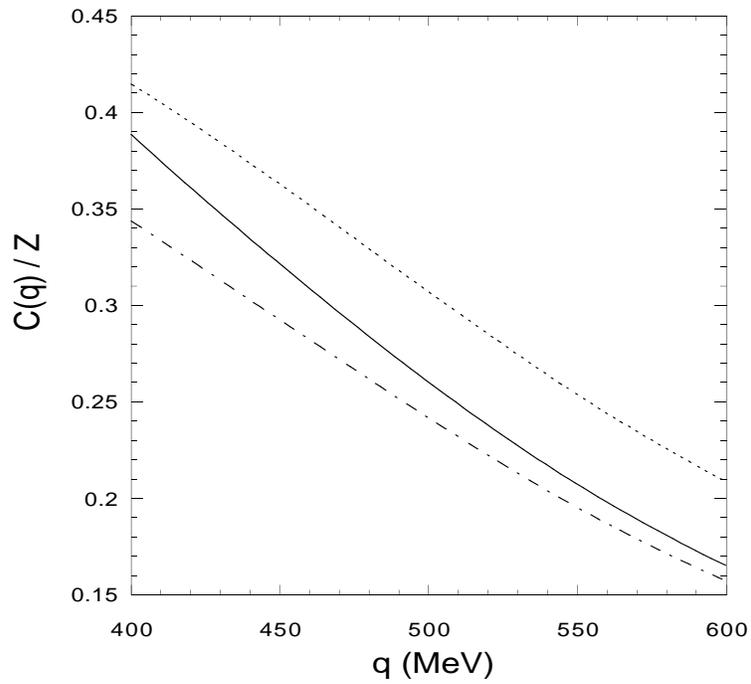,height=9cm,width=10cm}
\end{minipage}
\begin{minipage}[t]{16.5 cm}
\caption{Coulomb sum, $C(q)/Z$, at $\rho_0$ 
(from Ref.~\cite{SAI-LR}) -- see text for details.}
\label{fig:csr}
\end{minipage}
\end{center}
\end{figure}

Our result is shown in Fig.~\ref{fig:resp}.  
The dotted curve is the result of the Hartree approximation, where
the proton electric form factor is the same as in free space.
The dashed curve is the result of the full
RRPA, without the modifications 1 and 2. The dot-dashed curve shows the
result of the full RRPA with the meson-nucleon form factor but $R_{pE}=1$.
The upper (lower) solid curve shows the result of the full RRPA for
$m_q = 5~(300)$ MeV, including all modifications.
Because of the density dependent coupling,
the reduction of the response function from the Hartree
result, caused by the full RRPA, is much smaller than that in QHD.
In contrast, the modification of the proton electric form factor is very
significant, yielding a much bigger reduction in the response.
We can see that the effect of the meson-nucleon form factor
is relatively minor.
                                                                                
It is also interesting to see the quark mass dependence of the LR.
As an example, we consider the case of $m_q =
300$ MeV.  In comparison with the case $m_q$ = 5 MeV,
it is a little smaller and the peak position is shifted to the higher
energy transfer side.  This is related to the smaller effective nucleon
mass in the case $m_q$ = 300 MeV than when $m_q$ = 5 MeV.
                                                                                
The Coulomb sum, $C(q) = \int^q_0 dq_0 \/ S_L(q, q_0)$,
is shown in  Fig.~\ref{fig:csr} as a function of three-momentum transfer,
$q=|\vec{q}|$.  For high $q$, the strength is about 20\% lower in the full
calculation (the solid and dot-dashed curves are
for $m_q$ = 5 and 300 MeV, respectively)
than for the Hartree response with $R_{pE}=1$ (the dotted curve).  For low $q$,
the full calculation with the constituent quark mass remains much lower
than the Hartree result, while in case of the light quark mass it
gradually approaches the Hartree one.  This difference is caused by that
the effective nucleon mass for $m_q$ = 5 MeV being larger in matter than that
for $m_q$ = 300 MeV.
                                                                                
It also seems appropriate to comment on the 
transverse response (TR) from nuclear
matter.  In Ref.~\cite{Lu-eN}, it was found that
the in-medium modification of the nucleon magnetic form factor within QMC
is very small.  Therefore,
one would expect the total change in the TR caused by
RRPA correlations and the effect of the variation of the structure of
the nucleon to be much smaller than in the LR.
                                                                                
In summary, the reduction of the
$\sigma$-nucleon coupling constant with density decreases the
contribution of the RRPA, while the modification of the proton electric form factor
in medium reduces the LR considerably.  The LR or the Coulomb sum is
reduced by about 20\% in total, with RRPA correlations and the variation
of the in-medium nucleon structure contributing about fifty-fifty.
It will be interesting to extend this work to calculate both the
LR and TR functions for finite nuclei, and compare directly with 
the experimental results~\cite{MOR}.

\subsubsection{\it Axial form factors - neutrino-nucleus scattering 
\label{subsec:Axialff}}

We first discuss the axial vector form factor of the 
bound nucleon and its application to neutrino-nucleus scattering. At the end of this section, 
we briefly summarize the scalar and vector form factors 
of the in-medium nucleon as well. 

The extension to the in-medium modification of the bound nucleon
axial form factor $G_A^*(Q^2)$
can be made in a straightforward manner~\cite{Axialff}.
(Hereafter we denote the in-medium quantities by an asterisk $^*$.)
Since the induced pseudoscalar form factor, $G_P(Q^2)$,
is dominated by the pion pole and can be derived using
the PCAC relation~\cite{CBM}, we do not discuss it here.
The relevant axial current operator is then simply given by
\begin{eqnarray}
A^\mu_a(x) &=& \sum_q \overline{\psi}_q(x) \gamma^\mu\gamma_5
{\tau_a\over 2} \psi_q(x) \theta(R_B -r),
\label{Acurrent}
\end{eqnarray}
where $\psi_q(x)$ is the quark field operator for flavor $q$.
Similarly to the case of electromagnetic form factors, in the 
Breit frame the resulting bound nucleon axial form factor is
given by ICBM~\cite{Axialff}:
\begin{eqnarray}
G_A^*(Q^2) &=& \left(\frac{M_N^*}{E_N^*}\right)^2 
G^{\rm sph\, *}_A\left(\left(\frac{M_N^*}{E_N^*}\right)^2 Q^2\right), \label{GAmedium}
\\
G_A^{sph\, *}(Q^2) &=& {5\over 3}
\int\! d^3r
\left\{
\left[j_0^2(\Omega_q r/R_N^*)
-\beta_q^{*2}j_1^2(\Omega_q r/R_N^*)\right]j_0(Qr)\right.
\nn\\
& & \hspace{20ex} \left.
+2\beta_q^{*2}j_1^2(\Omega_q r/R_N^*)[j_1(Qr)/Qr] \right\}
\,\,K(r)/D_{PT}.
\label{PTGA}
\end{eqnarray}
In Fig.~\ref{axial} we show the in-medium axial vector form factor,   
$G^*_A(Q^2)$, divided by that in free space,   
$G_A(Q^2)$, calculated at nuclear densities 
$\rho_B = (0.5,0.7,1.0,1.5) \rho_0$ with $\rho_0 = 0.15$~fm$^{-3}$.
\begin{figure}
\begin{center}
\begin{minipage}[t]{8cm}
\hspace*{-1cm}
\epsfig{file=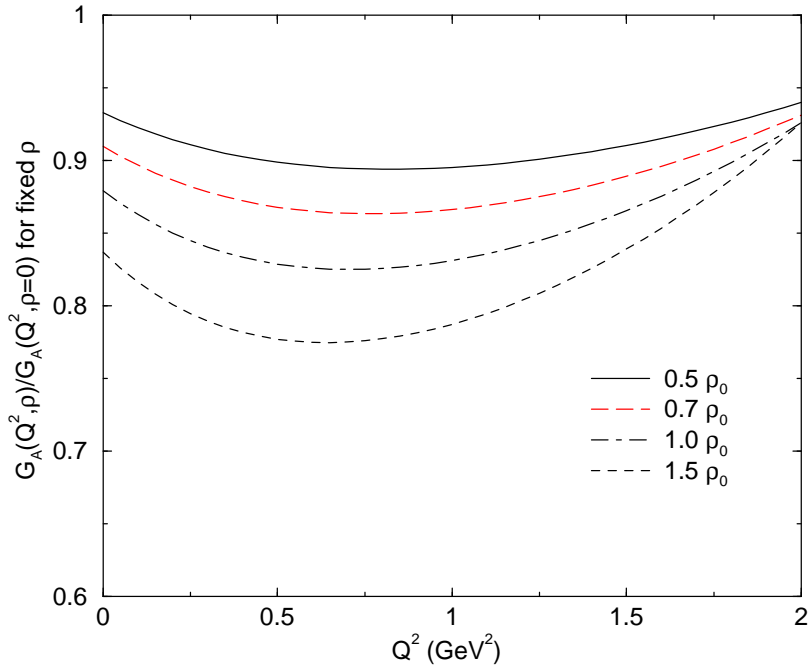,height=9cm}
\end{minipage}
\begin{minipage}[t]{16.5cm}
\caption{The ratio of in-medium to free axial form factors~\cite{Axialff}.
The free space value, $G_A(0) = g_A = 1.14$ is used in the calculation.}
\label{axial}
\end{minipage}
\end{center}
\end{figure}
At $Q^2 = 0$ the space component $G_A^*(Q^2=0) \equiv g_A^*$ is
quenched~\cite{gaspace,KTRiskaA} by about 
10 \% at normal nuclear matter density.
The modification calculated here may correspond to the
``model independent part" in meson exchange language,
where the axial current attaches itself to one of the two nucleon legs,
but not to the exchanged meson~\cite{gaspace,KTRiskaA}. This is because
the axial current operator in Eq.~(\ref{Acurrent})
is a one-body operator which operates on
the quarks and pions belonging to a bound nucleon.
A more detailed discussion will be made later.
The medium modification of the bound nucleon axial form factor
$G_A^*(Q^2)$ may be observed for instance in neutrino-nucleus scattering,
similar to that observed in the ``EMC-type'' experiments 
(see section~\ref{subsubsec:EMC}), or in a similar
experiment to the polarization transfer measurements performed on
$^4$He~\cite{POLE}.
However, at present the experimental uncertainties seem to be too large
to detect such a medium effect directly.
We should also note that the medium modification of the parity-violating
$F_3$ structure functions of a bound nucleon in deep-inelastic
neutrino induced reactions can be extracted using the calculated in-medium
axial form factor $G_A^*(Q^2)$ and quark-hadron duality,
as we will discuss in section~\ref{subsubsec:BGNSF}. 

Below, we study the effect of
the bound nucleon form factors on neutrino-nucleus
scattering~\cite{NeuA}.\footnote{Because the renormalization of axial-vector
form factors are the
same for the time and space components in this study (quenched), 
we will not discuss the time component.}
As an example, we compute the inclusive
$^{12}$C($\nu_\mu,\mu^-$)$X$ cross
sections that have been measured by the LSND
collaboration~\cite{LSND1,LSND2,LSND3}.
It is known that the existing calculations for the 
total cross section based on the nucleon and meson degrees
of freedom overestimate the data by
$\sim$30\% to $\sim$100\%~\cite{LSND1,LSND2,LSND3,Kolbe}.
Because our aim is to focus on the effects due to the 
internal structure change of the bound nucleon, we use 
a relativistic Fermi gas model~\cite{Hungchong1,Hungchong2},
which is simple and transparent for the purpose, 
while implementing the bound nucleon form factors calculated 
in the QMC model. Thus, we do not include the other 
nuclear structure corrections~\cite{Kolbe,Delta}.

Of course, it is difficult to  
separate exactly the effects we consider here from the standard
nuclear-structure corrections, particularly 
from meson exchange current (MEC). 
However, since the relevant current operators in this study 
are one-body quark (pion) operators 
acting on the quarks (pion cloud) 
in the nucleon, double counting with the model-dependent
MEC~\cite{KTRiskaE,KTRiskaA} (the current operators act on the exchanged
mesons) is expected to be avoided.
The same is also true for the model-independent meson pair currents,
because they are based on the {\it anti-nucleon} degrees of
freedom~\cite{KTRiskaE,gaspace,KTRiskaA}.
For the vector current, a double counting with MEC may be practically
avoided because the analyses for the $^4$He($\vec{e},e'\vec{p}$)$^{\,3}$H
experiments~\cite{POLE} have shown. (See section~\ref{subsec:QMCpredic}.)
For the axial-vector current, the quenching of the axial coupling constant 
($g_A = G_A(0)$) due to the model-independent meson pair currents
was estimated~\cite{KTRiskaA} using a Fermi gas model.
The quenching due to the pair currents
amounts to only 2\% at normal 
nuclear matter density, thus contributing negligibly
to the cross section.
Hence, the double counting from the interference
between the axial-vector and vector currents
is also expected to be small, in considering
the analyses for the $^4$He($\vec{e},e'\vec{p}$)$^{\,3}$H 
experiments.
Thus, the effect we consider here, which originates from the change of 
the internal quark wave function, is additional
to the standard nuclear-structure corrections.

Assuming G-parity (no second-class current), the charged-current vector 
and axial form factors for free nucleons with mass $M_N$ are defined by:
\bg
\left< {p' s'}| V^\mu_a(0) |{p s} \right>
&=& \overline{u}_{s'}(p')\left[F_1(Q^2)\gamma^\mu +
i \frac{F_2(Q^2)}{2M_N} \sigma^{\mu \nu} (p'-p)_\nu\right]
\frac{\tau_a}{2} u_s(p),
\label{vff}\\
\left< {p' s'}| A^\mu_a(0) |{p s} \right>
&=& \overline{u}_{s'}(p')\left[G_A(Q^2)\gamma^\mu +
 \frac{G_P(Q^2)}{2M_N} (p'-p)^\mu\right]\gamma_5 \frac{\tau_a}{2} u_s(p),
\label{aff}
\en
where $Q^2 \equiv -(p'-p)^2$, and other notations should be self-explanatory.
The vector form factors, $F_1(Q^2)$ and $F_2(Q^2)$, 
are related to the electric ($G_E(Q^2)$) and 
magnetic ($G_M(Q^2)$) Sachs form factors by 
the conserved vector current hypothesis. 
The induced pseudoscalar form factor, $G_P(Q^2)$,
is dominated by the pion pole and can be calculated using the 
PCAC relation~\cite{CBM}. 
Nevertheless, the contribution from $G_P(Q^2)$ to the cross section 
is proportional to (lepton mass)$^2/M_N^2$, and small
in the present study.
We note that, since there is another vector in nuclear medium,
the nuclear (matter) four-velocity,
various other form factors may arise, 
in addition to those in Eqs.~(\ref{vff}) and~(\ref{aff}).
The modification of the nucleon internal structure
studied here may also be expected to contribute to such form factors.
However, at this stage, information on such form factors
is very limited and not well under control
both in theoretically and experimentally.
Thus, we focus on the in-medium changes of the free form
factors given in Eqs.~(\ref{vff}) and~(\ref{aff}), and
study their effects on neutrino-nucleus scattering.

As already discussed, 
the electromagnetic and axial vector form factors in nuclear medium 
in the Breit frame are calculated by ICBM~\cite{Lu97}:
\bg
G_{E,M,A}^{QMC\, *}(Q^2) &=& \left(\frac{M_N^*}{E_N^*}\right)^2 
 G^{\rm sph\, *}_{E,M,A}\left(\left(\frac{M_N^*}{E_N^*}\right)^2 Q^2\right) .
\label{GEMA}
\en
The explicit expressions for Eq.~(\ref{GEMA})
are given in Eqs.~(\ref{PTE}), (\ref{PTM}) and~(\ref{PTGA}) with proper 
substitutions.
The ICBM includes a Peierls-Thouless projection to account for
center of mass and recoil corrections, and a Lorentz contraction of the
internal quark wave function~\cite{Lu97,LP70}.

Now we calculate the ratios of the bound to free nucleon form factors, 
[$G_{E,M,A}^{QMC\,*}/G_{E,M,A}^{ICBM\,free}$],
to estimate the bound nucleon form factors.
Using the empirical parameterizations 
in free space $G_{E,M,A}^{emp}$~\cite{EMffpara,Axialffpara},
the bound nucleon form factors $G_{E,M,A}^*$ are calculated by 
\bge
G_{E,M,A}^*(Q^2) =
\left[ \frac{G_{E,M,A}^{QMC*}(Q^2)}{G_{E,M,A}^{ICBM\, free}(Q^2)} \right]
G_{E,M,A}^{emp}(Q^2).
\label{boundNff}
\ene
Note that the pion cloud effect is not included in the axial 
vector form factor in the present treatment~\cite{Axialff}. 
However, the normalized $Q^2$ dependence (divided by $g_A = G_A(0) = 1.14$)
relatively well reproduces the empirical
parameterization~\cite{Axialff}. 
Furthermore, the relative modification of $G^*_A(Q^2)$  
due to the pion cloud is expected to be small, 
since the pion cloud contribution
to entire $g_A$ is $\sim$8\%~\cite{CBM} without a specific 
center-of-mass correction.
In the calculation the standard values in the QMC model are used, i.e., 
the current quark mass $m_q (= m_u = m_d) = 5$ MeV assuming SU(2) symmetry,
and the free nucleon bag radius $R_N = 0.8$~fm.

First, in Fig.~\ref{mediumff} we show
ratios of the bound to free nucleon form factors calculated
as a function of $Q^2$
for $\rho_B = \rho_0 = 0.15$ fm$^{-3}$ (the normal nuclear matter density) and
$0.668 \rho_0$ (the Fermi momentum $k_F = 225$ MeV for $^{12}$C).
\begin{figure}
\begin{center}
\begin{minipage}[t]{8cm}
\hspace*{-1cm}
\epsfig{file=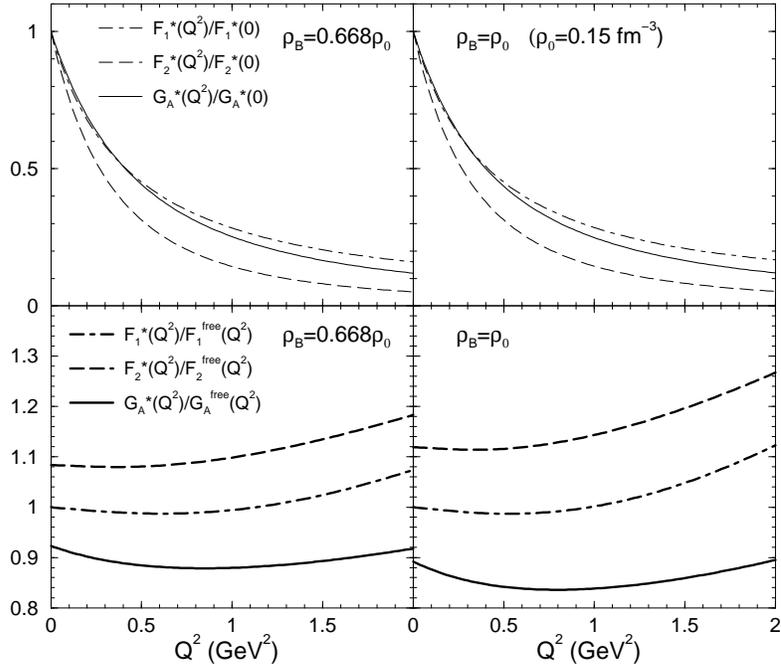,width=9cm,angle=-90}
\end{minipage}
\begin{minipage}[t]{16.5cm}
\caption{Calculated ratios for the bound nucleon form factors
(from Ref.~\cite{NeuA}).}
\label{mediumff}
\end{minipage}
\end{center}
\end{figure}
The lower panels in Fig.~\ref{mediumff} show the enhancement of
momentum dependence of $F_2^*(Q^2)$ and
$G_A^*(Q^2)$, as well as the enhancement of $F^*_2(0)$ and
quenching of $G^*_A(0)$~\cite{SAI-q,Lu-eN,Axialff}.
Although the modification of the $Q^2$ dependence
is small, we emphasize that this effect
originates from change in the nucleon internal structure.
The main origin of this new $Q^2$ dependence is the effect of 
Lorentz contraction on the quark wave function,  
amplified by the reduced effective nucleon
mass. (See also Eq.~(\ref{GEMA}).)
Note that, the relative change of the bound nucleon form factor 
$F_2^*(Q^2) [G_E^{p*}(Q^2)]$ to that of the free nucleon 
is an enhancement [quenching] of
$\sim$8\% [4\%] in $^{12}$C at $Q^2 = 0.15$ GeV$^2$, and we are focusing
on this relative change.
                                                                                
Now, we investigate the effect of the bound nucleon form factors
on charged-current neutrino-nucleus scattering. We compute
the inclusive $^{12}$C($\nu_\mu,\mu^-$)$X$ differential and total
cross sections, which have been measured
by the LSND collaboration~\cite{LSND1,LSND2,LSND3}.
The formalism used in the calculation is described 
in Ref.~\cite{Hungchong1}, and that the 
empirical parameterizations of the electromagnetic~\cite{Hungchong1,EMffpara}
and axial~\cite{Axialff,Axialffpara} form factors
for the free nucleon.
(See Eq.~(\ref{boundNff}).)
A relativistic Fermi gas model is used implementing the bound nucleon
form factors to calculate the differential cross section 
$\left< d\sigma/dE_\mu \right>$, averaged over the LSND muon neutrino 
spectrum $\Phi(E_{\nu_\mu})$~\cite{Hungchong1}
for the full range of the LSND experimental 
spectrum~\cite{LSND1,LSND2,LSND3},
$0 \le E_{\nu_\mu} \le 300$ MeV:
\bge
\left< \frac{d\sigma}{dE_\mu} \right> =
\frac{\int_0^\infty (d\sigma/dE_\mu) \Phi(E_{\nu_\mu}) dE_{\nu_\mu}}
{\int_0^\infty \Phi(E_{\nu_\mu}) dE_{\nu_\mu}}.
\label{dsigma}
\ene

Fig.~\ref{QMCdsigma} shows the result for  
$\left< d\sigma / dE_\mu \right>$ calculated using the nucleon masses,
$M_N$ and $M_N^*$.
\begin{figure}
\begin{center}
\begin{minipage}[t]{8cm}
\hspace*{-1cm}
\epsfig{file=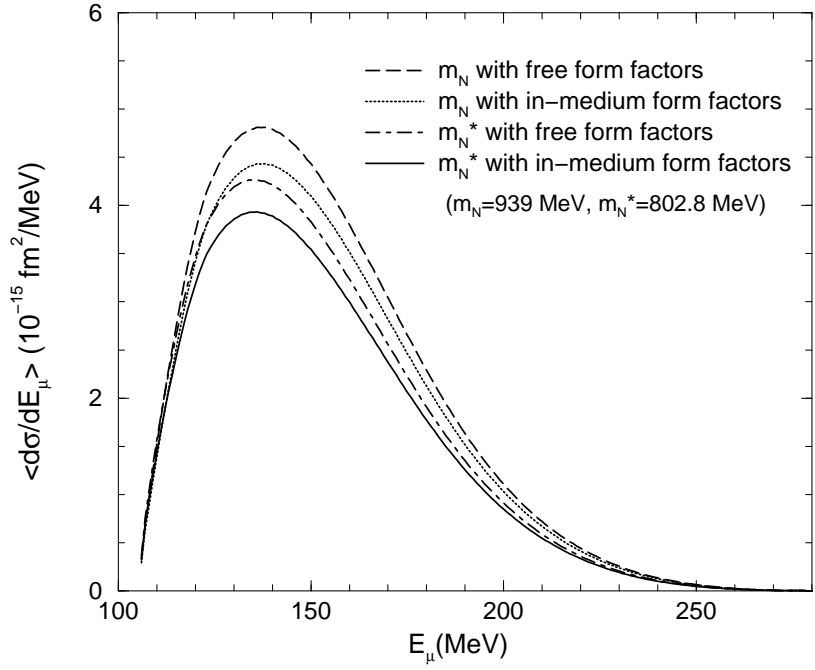,width=9cm,angle=-90}
\end{minipage}
\begin{minipage}[t]{16.5cm}
\caption{Angle-integrated inclusive
$^{12}$C($\nu_\mu,\mu^-$)$X$ differential cross section as
a function of the emitted
muon energy $E_\mu$ using $E_B = 0$ for all cases
(from Ref.~\cite{NeuA}).}
\label{QMCdsigma}
\end{minipage}
\end{center}
\end{figure}
For the Fermi momentum $k_F = 225$ MeV ($\rho_B = 0.668 \rho_0$) for
$^{12}$C, we use the QMC calculated value, $M_N^* = 802.8$ MeV.
A moderate quenching of the cross section can be observed
due to the in-medium form factors for both cases.
Although the effective nucleon mass 
can account for, to some extent, the binding  
effect (the Hugenholtz-van Hove theorem~\cite{binding}), 
there is an alternative to include the binding effect, i.e., 
the ``binding energy'' $E_B$ is introduced and the available reaction energy 
$E$ is replaced by $E - E_B$. 
In this case, we use the free nucleon mass in the calculation.
Since $E_B$ is an effective way of accounting for 
the binding effect~\cite{eb}, we regard $E_B$ as a parameter and 
perform calculations for $E_B = 20, 25$ and $30$ MeV.
(E.g., $E_B = 25-27$ MeV is commonly used
for the $^{16}$O nucleus~\cite{be25}.)
We emphasize that our aim is not to reproduce the LSND
data, but to estimate the corrections due to the
bound nucleon form factors.

In Fig.~\ref{Fermigas} we present the
results of $\left< d\sigma / dE_\mu \right>$ for
$E_B = 20, 25$ and $30$ MeV.
\begin{figure}
\begin{center}
\begin{minipage}[t]{8cm}
\hspace*{-1cm}
\epsfig{file=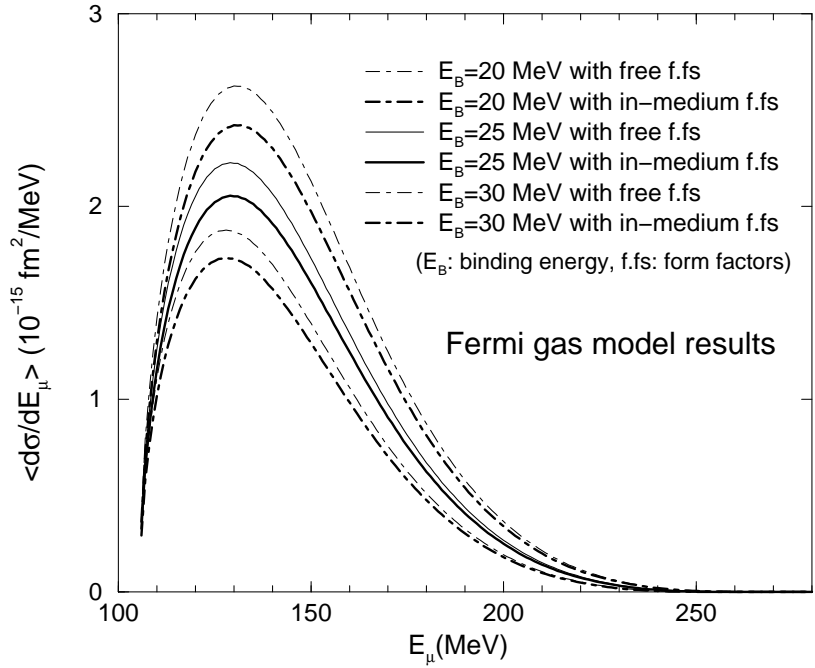,width=9cm,angle=-90}
\end{minipage}
\begin{minipage}[t]{16.5cm}
\caption{Same as Fig.~\ref{QMCdsigma},
but using $M_N = 939$ MeV for all cases
(from Ref.~\cite{NeuA}).}
\label{Fermigas}
\end{minipage}
\end{center}
\end{figure}

In both Figs.~\ref{QMCdsigma} and~\ref{Fermigas},
the bound nucleon form factors reduce
the differential cross section.
In Fig.~\ref{Fermigas}, as the binding energy $E_B$ increases, the peak 
position shifts downward for both cases  
with the free and bound nucleon form factors. 
The similar tendency due to $M^*_N$
is also seen in Fig.~\ref{QMCdsigma}. 
                                                                                
The total cross section is given by integrating Eq.~(\ref{dsigma})
over the muon energy.
We denote the cross section calculated with the free [bound] nucleon
form factors, $F_{1,2}(Q^2)$ and $G_{A,P}(Q^2)$
[$F^*_{1,2}(Q^2)$ and $G^*_{A,P}(Q^2)$],
as $\left<\sigma(F,G)\right>$ [$\left<\sigma(F^*,G^*)\right>$].
Thus, $\left<\sigma(F,G)\right>$ calculated with 
$M_N$ and $E_B = 0$ corresponds to the free Fermi gas model result. 
The results 
with $E_B = 0$ and either $M_N$ or $M_N^*$ are listed in the 
top group rows in Table~\ref{table}.
The LSND experimental data~\cite{LSND1,LSND2,LSND3} 
are also shown in the bottom group rows in Table~\ref{table}.
\begin{table}
\begin{center}
\begin{minipage}[t]{16.5cm}
\caption{Calculated total cross sections for $^{12}$C($\nu_\mu,\mu^-$)$X$.
See the text for notations.}
\label{table}
\end{minipage}
\vspace{3ex}
\begin{tabular}{llcc}
\hline\hline
Notation& Type of calculation 
&$E_B$ (MeV) 
&$\left<\sigma\right>$ in 10$^{-40}$ cm$^2$\\
\hline\hline
$\left<\sigma(F,G)\right>$ &$M_N,F_{1,2}(Q^2),G_{A,P}(Q^2)$ &0 &32.5\\
$\left<\sigma(F^*,G^*)\right>$ &$M_N,F_{1,2}^*(Q^2),G_{A,P}^*(Q^2)$ &0 &30.0\\
$\left<\sigma(F,G)\right>$ &$M_N^*, F_{1,2}(Q^2),G_{A,P}(Q^2)$ &0 &28.4\\ 
$\left<\sigma(F^*,G^*)\right>$ &$M_N^*,F_{1,2}^*(Q^2),G_{A,P}^*(Q^2)$&0&26.2\\
\hline\hline
$\left<\sigma(F^*,G)\right>$& $M_N, F_{1,2}^*(Q^2), G_{A,P}(Q^2)$ &0   &33.5\\
$\left<\sigma(F,G^*)\right>$& $M_N, F_{1,2}(Q^2), G_{A,P}^*(Q^2)$ &0   &29.1\\
\hline\hline
$\left<\sigma(F,G)\right>$& $M_N, F_{1,2}(Q^2), G_{A,P}(Q^2)$      &20  &16.1\\

$\left<\sigma(F^*,G^*)\right>$& $M_N,F_{1,2}^*(Q^2),G_{A,P}^*(Q^2)$&20  &14.8\\

$\left<\sigma(F,G)\right>$& $M_N,F_{1,2}(Q^2),G_{A,P}(Q^2)$        &25  &13.2\\

$\left<\sigma(F^*,G^*)\right>$& $M_N,F_{1,2}^*(Q^2),G_{A,P}^*(Q^2)$&25  &12.2\\

$\left<\sigma(F,G)\right>$& $M_N,F_{1,2}(Q^2),G_{A,P}(Q^2)$        &30  &10.7\\

$\left<\sigma(F^*,G^*)\right>$& $M_N,F_{1,2}^*(Q^2),G_{A,P}^*(Q^2)$&30  & 9.9\\
\\
 &Experiment~\cite{LSND1} (2002)& &$10.6 \pm 0.3 \pm 1.8$\\
 &Experiment~\cite{LSND2} (1997)& &$11.2 \pm 0.3 \pm 1.8$\\
 &Experiment~\cite{LSND3} (1995)& &$ 8.3 \pm 0.7 \pm 1.6$\\
\hline\hline
\end{tabular}
\end{center}
\end{table}
As expected~\cite{LSND1,LSND2,LSND3}, the free Fermi gas result
overestimates the data 
by a factor of three. The results obtained using the bound nucleon 
form factors,
with either $M_N$ or $M_N^*$, similarly overestimate the LSND data.
In order to make discussions more quantitative, we define a ratio:
\bge
R(\delta\sigma) \equiv \left[ \left<\sigma(F,G)\right> -
\left<\sigma(F^*,G^*)\right> \right] / \left<\sigma(F,G)\right>. 
\label{delsigma}
\ene
For the total cross sections calculated with $(M_N, M_N^*)$ and $E_B = 0$,
we get $R(\delta\sigma) = (7.7, 7.7)$\%, respectively.
Thus, the correction due to the bound nucleon form factors
to the total cross section
is not sensitive to $M_N$ or $M_N^*$ in the case of $E_B = 0$.
                                                                                
Next, we investigate which bound nucleon form factor
gives dominant corrections to the total cross section.
We calculate the total cross section with $M_N$,
using the free and bound form factors for two cases,
[$F^*_{1,2}(Q^2)$ and $G_{A,P}(Q^2)$] and
[$F_{1,2}(Q^2)$ and $G^*_{A,P}(Q^2)$].
They
are denoted by
$\left<\sigma(F^*,G)\right>$ and $\left<\sigma(F,G^*)\right>$, respectively.
The results are given in the middle group rows in Table~\ref{table}. 
Together with the results in the upper group rows in 
Table~\ref{table}, we obtain inequalities for the total cross 
sections calculated with $M_N$ and $E_B = 0$:
\bge
\left<\sigma(F,G^*)\right> < \left<\sigma(F^*,G^*)\right> 
< \left<\sigma(F,G)\right> < \left<\sigma(F^*,G)\right>.
\label{inequality}
\ene
This shows that the most dominant reduction
is driven by the axial form factor,
$G_A^*(Q^2)$. 
(The induced pseudoscalar form factor 
$G_P(Q^2)$  
gives only a few percent contribution
when calculated using all free form factors.)
Furthermore, 
$F^*_{1,2}(Q^2)$ enhance the total cross section 
(mostly due to $F^*_2(Q^2)$),  
as can be 
seen from the lower panel in Fig.~\ref{mediumff}. 

The total cross sections 
for $E_B = 20, 25$ and $30$ MeV  
are listed in the bottom group rows in Table~\ref{table}. 
The bound nucleon form factors for these cases also
reduce the total cross section relative to those calculated 
with the free form factors.
In addition, the results are rather sensitive
to the values for $E_B$.
However, we find
$R(\delta\sigma) = (8.1,7.6,7.5)\%$ for $E_B = (20,25,30)$MeV,
respectively. Thus, the effect of the bound nucleon form factors
to the reduction rate is again not sensitive to $E_B$.
To draw a more definite conclusion, it is essential to perform a more
precise, elaborate calculation within the framework of
RPA~\cite{Hungchong1}
including the effect of bound nucleon form factors.
However, even at the present stage,
it is important to point out that
the correction due to the in-medium form factors
arising from the bound nucleon internal structure change, 
could be significant for a precise estimate
of the charged-current neutrino-nucleus cross section.

{}Finally, we here comment on the scalar and vector form factors of the in-medium nucleon. 
Using the QMC model, it is possible to calculate the form factors at 
$\sigma$- and $\omega$-nucleon strong-interaction vertices in nuclear matter. 
In Ref.~\cite{svform}, the Peierls-Yoccoz projection technique is used to take account of center 
of mass and recoil corrections, and the Lorentz contraction is also applied to 
the internal quark wave function. 
The form factors are reduced by the nuclear 
medium relative to those in vacuum. At normal nuclear matter density and 
$Q^2 = 1$ GeV$^2$, the reduction rate in the scalar form factor is about 15\%, 
which is almost identical to that in the vector one. One can parameterize the ratios 
of the form factors in symmetric nuclear matter to those in vacuum as a 
function of nuclear density and momentum transfer~\cite{svform}. 

\subsection{\it Quark-hadron duality, the nuclear EMC effect 
and nuclear structure functions
\label{subsec:duality}}
\subsubsection{\it The nuclear EMC effect
\label{subsubsec:EMC}}

The European Muon Collaboration (EMC) effect tells us the fact that a nuclear structure function measured in
deep inelastic scattering in the valence quark regime (Bjorken variable $x\ge 0.3$) is considerably
reduced compared with that of a free nucleon~\cite{EMCEXP}.  Despite much experimental and
theoretical progress~\cite{EMC}, no unique and universally accepted explanation
of the EMC effect has yet emerged.
Because the measurements in deep inelastic scattering are performed at very high momentum transfers, we can expect
that the observed results really reflect quark degrees of freedom in a nucleus.
The simple parton model interpretation is that a quark in the bound nucleon carries less momentum than in free
space seems uncontested, but its underlying origin is still elusive.
                                                                                
The most popular explanation may be the conventional nuclear binding effect that is responsible for the depletion
of the quark momentum in matter. It then suggests that the nuclear structure function,
$F_{2}^A(x)$, can be expressed by the following convolution form:
\begin{equation}
{F_{2}^A(x)\over A}=\int dy \ f_{N/A}(y) F_2^N(x/y), \label{deep}
\end{equation}
where $F_2^N$ is the free nucleon structure function, and
the light front distribution function $f_{N/A}(y)$, which gives the probability
that a nucleon carries a fractional momentum $y$ of the nucleus $A$. 
If $k^\mu$ is the momentum of a nucleon and $P^\mu$ is the momentum of the target nucleus,
$y=(k^0+k^3)A/P^+=(k^0+k^3)A/M_A=(k^0+k^3)/{\bar M}_N$, in which
the nucleus is taken to be at rest with $P^+=M_A$ ($M_A$ the nuclear mass).  
One can easily
use conventional nuclear physics to obtain the probability that a
nucleon carries a three momentum ${\vec k}$ but if one uses only
naive considerations, one faces a puzzle when deciding how to
choose the value of $k^0$. Should one use the average separation
energy or the average nucleon mass ${\bar M}_N$, or possibly
the effective mass in the chosen many-body theory~?
                                                                                
In the binding explanation, $k^0$ is usually given by the free nucleon mass
$M_N$ minus the average separation energy $\epsilon$. Then, $f_{N/A}(y)$
is narrowly peaked at $y=1-\epsilon/M_N$ 
($\epsilon$ becomes as much as 70 MeV for 
infinite nuclear matter~\cite{BEN,MUT}),
which thus leads to a significant reduction in the value of 
the nuclear structure function (see also
Ref.~\cite{BOF}). However, in any case, it is necessary to supply a derivation to avoid the need to
arbitrarily choose a prescription for $k^0$.

To do this, one needs to develop the light front dynamics for 
nuclear system~\cite{MIL-LF} because,
in the parton model, $x$ is the ratio of the plus component of the momentum of the struck
quark to that of the target and it is the plus component of
the momentum which was observed to be depleted by the EMC.
In Ref.~\cite{MILL-EMC,MIL-LF}, the function $f_{N/A}(y)$ is shown to be the one which  maintains the covariance of
the formalism, and in which the nucleons carry the entire
plus-momentum $P^+$ of the  nucleus. This result
is obtained independently of the specific relativistic mean field
theory used, so no such theory contains
the binding effect discussed above. The only binding effect arises from
the average binding energy of the nucleus (16 MeV for infinite nuclear matter and 8 MeV for a finite nucleus),
and is far too small to explain the observed depletion of the structure
function. This conclusion has also been obtained by Birse~\cite{BIRS}.
                                                                                
The generality of this result may be supported by
the Hugenholtz-van Hove theorem~\cite{binding}, which states that
the  energy of the level at the Fermi surface, 
$E_F$, is equal to the nuclear mass divided by $A$:
\be
E_F=M_A/A\equiv {\bar M}_N.
\label{hvh}
\ee
This theorem is the consequence  of using the
condition that the total pressure of the nucleus vanishes at equilibrium, and
the assumption that nucleons are the only degrees of freedom contributing to
the nuclear energy. Thus this theorem is a signal that $P^+=P_N^+$ or that
nucleons account for the entire plus momentum of the nucleus.
                                                                                
Thus, the relativistic mean-field approach may lead to results in severe
disagreement with experiment, i.e., the depth
of the minimum in the EMC ratio is not monotonically decreasing with $A$, and it has a
smaller magnitude than experiment~\cite{MILL-EMC}. Furthermore, the plus momentum distributions give
$\langle y\rangle\simeq 1$ which indicates that nearly all of the
plus momentum is carried by the nucleons. In order to reproduce
the data, the nucleon plus momentum must be decreased by some
mechanism that becomes more important at larger $A$.
Nucleon-nucleon correlations cannot take plus momentum from
nucleons, and explicit mesonic components in the nuclear Fock
state wave function carrying plus momentum are limited by
Drell-Yan experiments~\cite{ALD}.
Therefore, it appears that the nuclear EMC effect 
must be due to something outside of conventional
nucleon-meson dynamics.
                                                                                
It is of great interest to apply the QMC model to explain the nuclear EMC effect, because the quarks in
the bound nucleon respond to the nuclear environment 
and the quark eigenvalue, $x_0$, decreases
in matter (see section~\ref{subsec:matter}).  
That should be responsible for the
depletion of the quark
momentum in a nucleus.  In Ref.~\cite{SAI-EMC,ST}, 
the quark momentum distribution
function, $q_A^{(2)}(x)$,
in nucleus $A$ was calculated using the QMC model
\be
q_A^{(2)}(x) = \int d^3r \ \rho_A(r) q_{N/A}^{(2)}(x,k_F(r)) , \label{qmdist}
\ee
where, for simplicity, the local-density approximation is assumed and $\rho_A$ is the nuclear density
distribution ($\rho_A(r) = 2/(3\pi^2)k_F(r)^3$, $k_F$ the local Fermi momentum). The distribution
of valence quarks in nuclear matter is then given by
\be
q_{N/A}^{(2)}(x,k_F(r)) = \int_x^A dy \ f_{N/A}(y,k_F(r)) q_{N}^{(2)}(x/y,k_F(r),\mu^2) , \label{qmdist2}
\ee
where the Fermi motion is incorporated as a convolution of $f_{N/A}$ with the quark distribution function of the
bound nucleon, $q_{N/A}^{(2)}$, which is calculated in terms of the nucleon bag
model~\cite{TON-BAG} with
the scalar and vector mean-fields generated by QMC. For a free nucleon, the dominant intermediate state is a
two-quark bag, while for nuclei it is the two-quark bag state bound to the residual ($A-1$) nuclear debris.
                                                                                
Using this framework, we have found that the usual impulse approximation, in which the interaction
of the intermediate two-quark bag state with the rest of the nucleus is ignored, gives an overestimate of the
nuclear EMC effect by a factor of two or three~\cite{SAI-EMC,ST}.  
In contrast, the inclusion of the interaction
between the two-quark bag and the debris can provide a 
reasonable amount of the
nuclear EMC effect~\cite{SAI-EMC,ST}.
This outcome is physically very sensible. 
The nuclear binding is the result of the attractive scalar mean-field
experienced by the three constituent quarks. 
By ignoring the binding of the pair of spectator quarks, that is, by
treating the quarks as though they were free in the nucleus, 
it is assumed that
the kinematics of the hard
scattering of the quark struck by the virtual photon 
carries the binding of all
three quarks. This is clearly
not correct.
                                                                                
In Refs.~\cite{SAI-EMC,ST}, the calculation is somewhat crude 
and needs to be made more realistic one. Nevertheless,  
the physical insight gained is much more general. Recently, Miller {\it et al.}
have performed similar
calculations using the QMC model supplemented by the 
chiral quark-soliton picture~\cite{SMITH}.
They concluded that their model can simultaneously 
describe the nuclear EMC
effect and the related Drell-Yan experiments.
Finally, we mention that Cloet {\it et al.}~\cite{Cloet:2005rt} 
have taken the covariant 
generalization the QMC model developed by 
Bentz and Thomas~\cite{bentz} and 
shown that not only does it reproduce the EMC effect in nuclear matter 
but that it predicts a nuclear effect on the proton spin structure function 
which is twice as large as in the unpolarized case. It is clearly 
very important to pursue this prediction experimentally 
as soon as possible.

\subsubsection{\it Bloom-Gilman duality and the nuclear EMC effect
\label{subsubsec:BGEMC}}

The relationship between form factors and structure functions, or more
generally between inclusive and exclusive processes, has been studied
in a number of contexts over the years.
Drell \& Yan~\cite{DY} and West~\cite{WEST} pointed out long ago that,
simply on the basis of scaling arguments, the asymptotic behavior of
elastic electromagnetic form factors as $Q^2 \to \infty$ can be related
to the $x \to 1$ behavior of deep-inelastic structure functions.

Furthermore, the relationship between resonance (transition) form
factors and the deep-inelastic continuum has been studied in the framework
of quark-hadron, or Bloom-Gilman~\cite{BG}, 
duality: the equivalence of the averaged structure function 
in the resonance region and the scaling
function which describes high $W$ data.
The high precision Jefferson Lab data~\cite{JLABF2} on the $F_2$
structure function suggests that the resonance--scaling duality also
exists locally, for each of the low-lying resonances, including
surprisingly the elastic~\cite{JLABPAR}, to rather low values of $Q^2$.
(For a recent extensive review on quark-hadron duality, see Ref.~\cite{MEK}.)

In the context of QCD, Bloom-Gilman duality can be understood within
an operator product expansion of moments 
of structure functions~\cite{RUJ,JI}: the weak $Q^2$ dependence 
of the low $F_2$ moments can
be interpreted as indicating that higher twist ($1/Q^2$ suppressed)
contributions are either small or cancel.
However, while allowing the duality violations to be identified and
classified according to operators of a certain twist, it does not explain
why some higher twist matrix elements are intrinsically small.

Whatever the ultimate microscopic origin of Bloom-Gilman duality, for our
purposes it will be sufficient to note the {\em empirical fact} that local
duality is realized in lepton-proton scattering down to
$Q^2 \sim 0.5$~GeV$^2$ at the 10-20\% level for the lowest moments of the
structure function.
In other words, here we are not concerned about {\em why} duality works,
but rather {\em that} it works.

Motivated by the experimental verification of local duality, one can use
measured structure functions in the resonance region to directly extract
elastic form factors~\cite{RUJ}.
Conversely, empirical electromagnetic form factors at large $Q^2$ can
be used to predict the $x \to 1$ behavior of deep-inelastic structure
functions~\cite{BG,CM,ELDUAL,QNP}.
The assumption of local duality for the elastic case implies that the area
under the elastic peak at a given $Q^2$ is equivalent to the area under
the scaling function, at much larger $Q^2$, when integrated from the pion
threshold to the elastic point~\cite{BG}.
Using the local duality hypothesis, de R\'ujula et al.~\cite{RUJ}, and
more recently Ent et al.~\cite{JLABPAR}, extracted the proton's magnetic
form factor from resonance data on the $F_2$ structure function at large
$x$, finding agreement to better than 30\% over a large range of $Q^2$
($0.5 \alt Q^2 \alt 5$~GeV$^2$).
In the region $Q^2 \sim 1$--2~GeV$^2$ the agreement was at the $\sim 10\%$
level. An alternative parameterization of $F_2$ was suggested in
Ref.~\cite{SIMULA}, which because of a different behavior in the unmeasured
region $\xi \agt 0.86$, where $\xi = 2 x / (1 + \sqrt{1 + x^2/\tau})$ is
the Nachtmann variable, with $\tau = Q^2/4M_N^2$, led to larger differences
at $Q^2 \agt 4$~GeV$^2$.
However, at $Q^2 \sim 1$~GeV$^2$ the agreement with the form factor data
was even better.
As pointed out in Ref.~\cite{REPLY}, data at larger $\xi$ are needed to
constrain the structure function parameterization and reliably extract
the form factor at larger $Q^2$.
Furthermore, since we will be interested in {\em ratios} of form factors
and structure functions only, what is more relevant for our analysis is
not the degree to which local duality holds for the {\em absolute}
structure functions, but rather the {\em relative} change in the duality
approximation between free and bound protons.

Applying the argument in reverse, one can formally differentiate the local
elastic duality relation~\cite{BG} with respect to $Q^2$ to express the
scaling functions, evaluated at threshold,
$x = x_{\rm th} = Q^2 / (W^2_{\rm th} - M_N^2 + Q^2)$, with
$W_{\rm th} = M_N + m_\pi$, in terms of $Q^2$ derivatives of elastic form
factors. Explicit explanations and derivations will be given 
in section~\ref{subsubsec:BGNSF} for the case of the bound nucleon.
In Refs.~\cite{BG,ELDUAL} the $x \to 1$ behavior of the neutron to proton
structure function ratio was extracted from data on the elastic
electromagnetic form factors.
(Nucleon structure functions in the $x \sim 1$ region are important as
they reflect mechanisms for the breaking of spin-flavor SU(6) symmetry in
the nucleon~\cite{MT}.)
Extending this to the case of bound nucleons~\cite{WKT}, one finds that as
$Q^2 \to \infty$ the ratio of bound to free proton structure functions is:
\begin{eqnarray}
\label{SFdual}
{ F_2^{p\, *} \over F_2^p }
&\to& { dG_M^{p\, *\, 2} / dQ^2 \over
        dG_M^{p\, 2 } / dQ^2 }\ .
\end{eqnarray}
At finite $Q^2$ there are corrections to Eq.~(\ref{SFdual}) arising from
$G_E^p$ and its derivatives, as discussed in Ref.~\cite{ELDUAL}.
(In this analysis we use the full, $Q^2$ dependent 
expressions~\cite{ELDUAL,QNP}.)
Note that in the nuclear medium, the value of $x$ at which the pion
threshold arises is shifted:
\begin{eqnarray}
x_{th} &\to& x^*_{th}\
=\ \left( { m_\pi ( 2 M_N + m_\pi ) + Q^2 \over
            m_\pi (2 (M_N^* + 3V^q_\omega) + m_\pi ) + Q^2 }
   \right) x_{th}\ ,
\end{eqnarray}
where $V^q_\omega = g^q_\omega \omega$ is the vector potential felt by
the nucleon and (consistent with chiral expectations and phenomenological
constraints) we have set $m_\pi^* = m_\pi$.
However, the difference between $x_{th}$ and $x^*_{th}$ has a
negligible effect on the results for most values of $x$ considered.

Using the duality relations between electromagnetic form factors and
structure functions, in Fig.~\ref{fig:F2p} 
we plot the ratio $F_2^{p *}/F_2^p$ as
a function of $x$, with $x$ evaluated at threshold, $x = x_{th}$
(solid lines).
\begin{figure}
\begin{center}
\begin{minipage}[t]{8cm}
\hspace*{-2cm}
\epsfig{figure=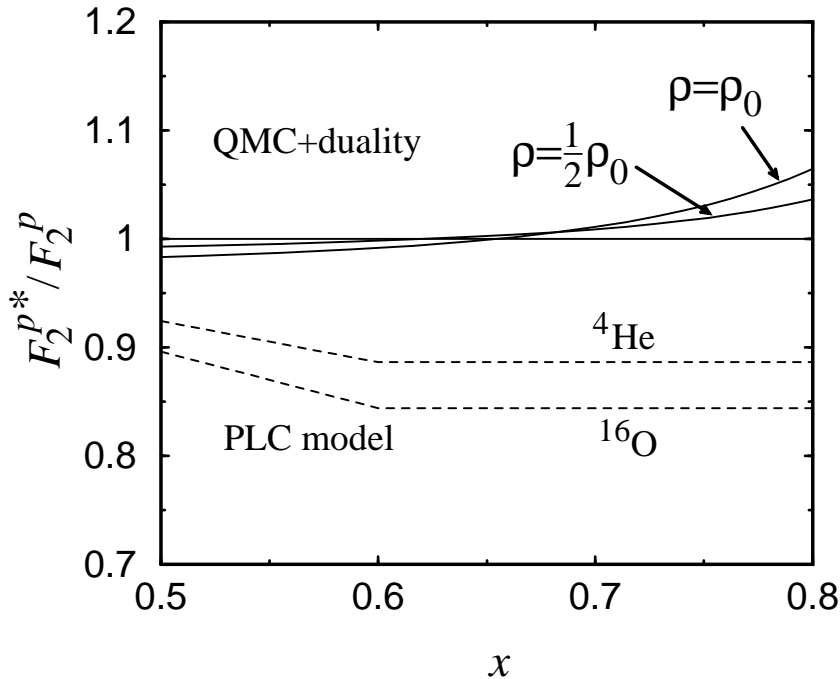,height=9cm}
\end{minipage}
\begin{minipage}[t]{16.5cm}
\caption{In-medium to free proton $F_2$ structure function ratio as a
function of $x$ at threshold, $x=x_{th}$, extracted from the
polarization transfer data~\cite{POLE} within the QMC model
and local duality, at nuclear matter density, $\rho = \rho_0$,
and at $\rho = {1 \over 2}\rho_0$ (solid) with $\rho_0=0.15$ fm$^{-3}$.
For comparison the results of the PLC suppression model~\cite{FS} 
are shown for $^4$He and $^{16}$O (dashed)
(from Ref.~\cite{WKT}).}
\label{fig:F2p}
\end{minipage}
\end{center}
\end{figure}
Note that at threshold the range of $Q^2$ spanned between $x=0.5$ and
$x=0.8$ is $Q^2 \approx 0.3$--1.1~GeV$^2$.
Over the range $0.5 \alt x \alt 0.75$ the effect is almost negligible,
with the deviation of the ratio from unity being $\alt 1\%$ for
$\rho={1\over 2}\rho_0$ and $\alt 2\%$ for $\rho=\rho_0$ 
($\rho_0=0.15$ fm$^{-3}$).
For $x \agt 0.8$ the effect increases to $\sim 5\%$, where,
larger $x$ corresponds to larger $Q^2$, and the analysis in terms 
of the QMC model becomes unreliable for $x \agt 0.9$.
However, in the region where the analysis can be considered reliable, the
results based on the bound nucleon form factors inferred from the
polarization transfer data~\cite{POLE} and local duality imply that the
nucleon structure function undergoes small modification in medium.

It is instructive to contrast this result with models of the EMC effect
in which there is a large medium modification of nucleon structure.
For example, let us consider the model of Ref.~\cite{FS}, where it is
assumed that for large $x$ the dominant contribution to the structure
function is given by the point-like configurations (PLC) of partons which
interact weakly with the other nucleons.
The suppression of this component in a bound nucleon is assumed to be the
main source of the EMC effect.
This model represents one of the extreme possibilities that the EMC
effect is solely the result of deformation of the wave function of bound
nucleons, without attributing any contribution to nuclear pions or other
effects associated with nuclear binding~\cite{MSS}.
Given that this model has been so successfully applied to describe the
nuclear EMC effect, it is clearly important to examine its consequences
elsewhere.

The deformation of the bound nucleon structure function in the PLC
suppression model is governed by the function~\cite{FS}:
\begin{eqnarray}
\label{deltaPLC}
\delta(k) &=& 1 - 2 (k^2/2M_N + \epsilon_A)/\Delta E_A ,
\end{eqnarray}
where $k$ is the bound nucleon momentum, $\epsilon_A$ is the nuclear
binding energy, and $\Delta E_A \sim 0.3$--0.6~GeV is a nucleon
excitation energy in the nucleus.
For $x \agt 0.6$ the ratio of bound to free nucleon structure functions
is then given by~\cite{FS}:
\begin{eqnarray}
\label{plc}
{ F_2^{N\, *}(k, x) \over F_2^N(x) }
&=& \delta(k) .
\end{eqnarray}
The $x$ dependence of the suppression effect is based on the assumption
that the point-like configuration contribution in the nucleon wave
function is negligible at $x \alt 0.3$ ($F_2^{N\, *}/F_2^N = 1$), and for
$0.3 \alt x \alt 0.6$ one linearly interpolates between 
these values~\cite{FS}.
The results for $^4$He and $^{16}$O are shown 
in Fig.~\ref{fig:F2p} (dashed lines)
for the average values of nucleon momentum, $\langle k^2 \rangle$, in
each nucleus.
The effect is a suppression of order 20\% in the ratio
$F_2^{N\, *}/F_2^N$ for $x \sim 0.6$--0.7.
In contrast, the ratios extracted on the basis of duality, using the QMC
model constrained by the $^4$He polarization transfer data~\cite{POLE},
show almost no suppression ($\alt 1$--2\%) in this region.
(See section~\ref{subsec:QMCpredic}.)
Thus, for $^4$He, the effect in the PLC suppression model is an order
of magnitude too large at $x \sim 0.6$, and has the opposite sign for
$x \agt 0.65$.

Although the results extracted from the polarization transfer
measurements~\cite{POLE} rely on the assumption of local duality, we
stress that the corrections to duality have been found to be typically
less than 20\% for $0.5 \alt Q^2 \alt 2$~GeV$^2$~\cite{JLABF2,SIMULA}.
The results therefore appear to rule out large bound structure function
modifications, such as those assumed in the point-like configuration
suppression model~\cite{FS}, and instead point to a small medium
modification of the intrinsic nucleon structure, complemented
by standard binding effects.

As a consistency check on the analysis, one can also examine the change
in the form factor of a bound nucleon that would be implied by the
corresponding change in the structure function in medium.
Namely, from the local duality relation~\cite{RUJ,QNP}:
\begin{eqnarray}
\label{elint}
\left[ G_M^p(Q^2) \right]^2
&\approx& { 2 - \xi_0 \over \xi_0^2 }
          { (1 + \tau) \over (1/\mu_p^2 + \tau) }
          \int_{\xi_{\rm th}}^1 d\xi\ F_2^p(\xi)\ ,
\end{eqnarray}
one can extract the magnetic form factor by integrating the $F_2^p$
structure function over $\xi$
between threshold, $\xi = \xi_{\rm th}$, and $\xi=1$.
Here $\xi_0 = \xi(x=1)$, and $\mu_p$ is the proton magnetic moment.
In Fig.~\ref{fig:PLC} we show the PLC model predictions for 
the ratio of the magnetic
form factor of a proton bound in $^4$He to that in vacuum, derived from
Eqs.~(\ref{plc}) and~(\ref{elint}), using the parameterization for
$F_2^p(\xi)$ from Ref.~\cite{JLABPAR}, and an estimate for the in-medium
value of $\mu_p^*$ shown in Fig.~\ref{he4.ps}.
\begin{figure}
\begin{center}
\begin{minipage}[t]{8cm}
\hspace*{-2cm}
\epsfig{figure=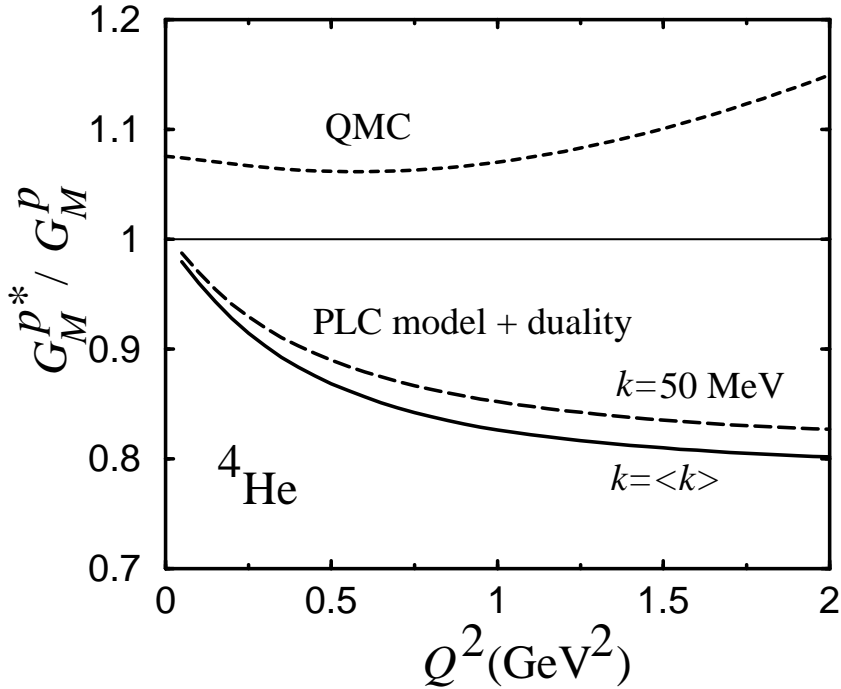,height=9cm}
\end{minipage}
\begin{minipage}[t]{16.5cm}
\caption{Ratio of in-medium to free proton magnetic form factors,
extracted from the PLC suppression model~\cite{FS}
for the EMC ratio in $^4$He, using the $F_2^p$ data from
Refs.~\cite{JLABF2,JLABPAR} and local duality, for
$k=\langle k \rangle$ (solid) and $k = 50$~MeV (long-dashed).
The QMC model prediction (short-dashed) is shown for comparison
(from Ref.~\cite{WKT}).}
\label{fig:PLC}
\end{minipage}
\end{center}
\end{figure}
Taking the average nucleon momentum in the $^4$He nucleus,
$k = \langle k \rangle$, the result is a suppression of about 20\% in
the ratio $G^{p *}_M/G^p_M$ at $Q^2 \sim 1$--2~GeV$^2$ (solid curve).
Since the structure function suppression in the PLC model depends on the
nucleon momentum (Eq.~(\ref{deltaPLC})), we also show the resulting form
factor ratio for a momentum typical in the $(\vec e, e' \vec p)$
experiment, $k = 50$~MeV (long dashed).
As expected, the effect is reduced, however, it is still of the order
15\% since the suppression also depends on the binding energy, as well
as on the nucleon mass, which changes with density rather than with momentum.
In contrast, the QMC calculation, which is consistent with the 
MAMI and JLab $^4$He quasi-elastic data~\cite{POLE} 
(see section~\ref{subsec:QMCpredic}), 
produces a ratio which is typically 5--10\%
{\it larger} than unity (dashed).
Without a very large compensating change in the in-medium electric form
factor of the proton (which seems to be excluded by $y$-scaling
constraints), the behavior of the magnetic form factor implied by the
PLC model $+$ duality would produce a large {\em enhancement} of the
polarization transfer ratio, rather than the observed 
small suppression~\cite{POLE}.

In the context of the QMC model, the change in nucleon form factors
allowed by the data imply a modification of the in-medium structure
function of $\alt 1$--2\% at $0.5 \alt x \alt 0.75$ for all nuclear
densities between nuclear matter density, $\rho=\rho_0$, and
$\rho={1\over 2}\rho_0$.
While the results rely on the validity of quark-hadron duality, the
empirical evidence suggests that for low moments of the proton's $F_2$
structure function the duality violations due to higher twist corrections
are $\alt 20\%$ for $Q^2 \agt 0.5$~GeV$^2$~\cite{JLABF2}, and decrease
with increasing $Q^2$.

The results place rather strong constraints on models of the nuclear EMC
effect, especially on models which assume that the EMC effect arises from
a large deformation of the nucleon structure in medium.
For example, we find that the PLC suppression model~\cite{FS} predicts
an effect which is about an order of magnitude larger than that allowed
by the data~\cite{POLE}, and has a different sign.
The findings therefore appear to disfavor models with large medium
modifications of structure functions as viable explanations for the
nuclear EMC effect, although it would be desirable to have more data
on a variety of nuclei and in different kinematic regions.

These results have other important practical ramifications.
For instance, the PLC suppression model was used~\cite{SSS} to
argue that the EMC effects in $^3$He and $^3$H differ significantly at
large $x$, in contrast to calculations~\cite{AFNAN,PACE} based on
conventional nuclear physics using well-established bound state wave
functions which show only small differences.
Based on the findings presented here, one would conclude that the
conventional nuclear physics description of the $^3$He/$^3$H system
should indeed be a starting point for nuclear structure function
calculations, as the available evidence suggests little room for large
off-shell corrections. On top of this, we expect a small 
correction due to the change of the 
internal structure of the bound nucleon,  
as discussed in section~\ref{subsec:QMCpredic},  
would indeed reflect the real explanation.
Finally, let us stress that quark-hadron duality is a powerful tool with
which to simultaneously study the medium dependence of both exclusive
and inclusive observables, and thus provides an extremely valuable guide
toward a consistent picture of the effects of the nuclear environment on
nucleon substructure.

\subsubsection{\it Bloom-Gilman duality and the nuclear structure functions
\label{subsubsec:BGNSF}}

As we discussed in section~\ref{subsubsec:BGEMC}, we can study 
the nuclear structure functions at large Bjorken-$x$ based on 
the Bloom-Gilman duality~\cite{KSSdual}.
Since, our main concern is to see the effect of the
bound nucleon internal structure change included entirely in the
bound nucleon form factors and the pion threshold in a nuclear
medium, the elastic contribution to the bound nucleon structure
functions for charged-lepton scattering may be given
by~\cite{CM,ELDUAL,QNP}:
\begin{eqnarray}
F_1^{BN*} &=& \frac{1}{2} (G^*_M)^2
\delta(x-1),
\label{F_1BN} \\
F_2^{BN*} &=&
\frac{1}{1+\tau} \left[(G^*_E)^2 + \tau (G^*_M)^2\right]
\delta(x-1),
\label{F_2BN} \\
g_1^{BN*} &=&
\frac{1}{2(1+\tau)} G^*_M \left(G^*_E + \tau G^*_M\right)
\delta(x-1),
\label{g_1BN} \\
g_2^{BN*} &=&
\frac{\tau}{2(1+\tau)} G^*_M \left(G^*_E - G^*_M\right)
\delta(x-1),
\label{g_2BN}
\end{eqnarray}
while those for (anti)neutrino scattering for an isoscalar nucleon,
$N \equiv \frac{1}{2}(p + n)$, may be given by~\cite{FK}:
\begin{eqnarray}
F_1^{WNBN*}
&=& \frac{1}{4} \left[(G^{V*}_M)^2 + (1 + 1/\tau)(G^*_A)^2 \right]
\delta(x-1),
\label{F_1WNBN} \\
F_2^{WNBN*}
&=& \frac{1}{2} \left[\frac{(G^{V*}_E)^2 + \tau (G^{V*}_M)^2}{1 + \tau}
+ (G^*_A)^2 \right]
\delta(x-1),
\label{F_2WNBN} \\
F_3^{WNBN*}
&=& G^{V*}_MG^*_A
\delta(x-1),
\label{F_3WNBN}
\end{eqnarray}
where $\tau = Q^2/4M_N^2$ ($M_N$, the free nucleon mass) and $x$ the
Bjorken variable. $G^*_E$ $[G^*_M]$ is the bound nucleon electric
[magnetic] Sachs form factor, $G^{V*}_{E,M} =
G^{p*}_{E,M} - G^{n*}_{E,M}$ the corresponding isovector
electromagnetic form factors, and $G^*_A$ is the (isovector) axial
vector form factor~\cite{NeuA,FK}.

Using the Nachtmann variable, $\xi = 2x/(1+\sqrt{1+x^2/\tau})$,
local duality equates the scaling bound nucleon structure function
$F_2^*$ and the contribution from $F_2^{BN*}$ of
Eq.~(\ref{F_2BN}):
\begin{equation}
\int_{\xi^*_{th}}^1 F_2^*(\xi) d\xi
=
\int_{\xi^*_{th}}^1 F_2^{BN*}(\xi,Q^2) d\xi,
\label{LQHD}
\end{equation}
where $\xi^*_{th}$ is the value at the pion threshold in a nuclear
medium given below. Below, we consider symmetric nuclear
matter, and we can assume that the pion mass in-medium ($m^*_\pi$) is
nearly equal to that in free space ($m_\pi$), and $\xi^*_{th}$ may 
be given by~\cite{WKT}:
\begin{equation}
\xi^*_{th} = \xi (x^*_{th}), \label{xi_th_medium}
\end{equation}
with
\begin{equation}
x^*_{th} =
x_{th} 
\frac{m_\pi (2M_N + m_\pi) + Q^2}{m_\pi [2(M_N^* + 3 V^q_\omega) + m_\pi]
+ Q^2},
\qquad
x_{th} = \frac{Q^2}{m_\pi (2M_N + m_\pi) + Q^2},
\label{x_th_medium}
\end{equation}
where $x^*_{th}$ [$x_{th}$] is the Bjorken-$x$ at the
pion threshold in medium [free space], and
$M_N^*$ and $3 V^q_\omega$ are respectively
the effective mass and the vector potential of the bound nucleon.
Inserting Eq.~(\ref{F_2BN}) into the r.h.s. of Eq.~(\ref{LQHD}),
we get~\cite{ELDUAL,QNP,Rujula}:
\begin{equation}
\int_{\xi^*_{th}}^1 F_2^*(\xi) d\xi
=
\frac{\xi_0^2}{4-2\xi_0}
\left[ \frac{(G^*_E)^2 + \tau (G^*_M)^2}{1 + \tau} \right],
\label{F_2_LQHD}
\end{equation}
where $\xi_0 = \xi (x = 1)$.
The derivative in terms of $\xi^*_{th}$ in both sides of Eq.~(\ref{F_2_LQHD})
with $\xi_0$ fixed gives~\cite{ELDUAL,QNP,FK}:
\begin{equation}
F_2^*(\xi^*_{th}) \equiv F_2^*(x^*_{th})
=
-2\beta^* \left[\frac{(G^*_M)^2 - (G^*_E)^2}{4M_N^2(1 + \tau )^2}
+ \frac{1}{1 + \tau}
\left(\frac{d(G^*_E)^2}{dQ^2} + \tau\frac{d(G^*_M)^2}{dQ^2}\right)\right],
\label{F_2_x_th}
\end{equation}
where $\beta^* =
(Q^4/M_N^2)(\xi^2_0/\xi^{*3}_{th})[(2-\xi^*_{th}/x^*_{th})/(4-2\xi_0)]$.
Similarly, we get expressions for other bound nucleon structure
functions at $x = x^*_{th}$:
\begin{eqnarray}
F_1^*(x^*_{th}) &=& -\beta^*\frac{d(G^*_M)^2}{dQ^2},
\label{F_1_x_th}\\
g_1^*(x^*_{th})
&=&
-\beta^* \left[\frac{G^*_M(G^*_M - G^*_E)}{4M_N^2(1 + \tau )^2}
+ \frac{1}{1 + \tau}
\left(\frac{d(G^*_E G^*_M)}{dQ^2} + \tau\frac{d(G^*_M)^2}{dQ^2}\right)\right],
\label{g_1_x_th}\\
g_2^*(x^*_{th})
&=&
-\beta^* \left[\frac{G^*_M(G^*_E - G^*_M)}{4M_N^2(1 + \tau )^2}
+ \frac{\tau}{1 + \tau}
\left(\frac{d(G^*_E G^*_M)}{dQ^2} - \frac{d(G^*_M)^2}{dQ^2}\right)\right],
\label{g_2_x_th}\\
F_1^{WN*}(x^*_{th})
&=&
-\frac{\beta^*}{2} \left[\frac{-(G^*_A)^2}{4M_N^2\tau^2}
+ \frac{d(G^{V*}_M)^2}{dQ^2}
+ \frac{1 + \tau}{\tau}\frac{d(G^*_A)^2}{dQ^2} \right],
\label{F_1WN_x_th}\\
F_2^{WN*}(x^*_{th})
&=&
-\beta^*\left[\frac{(G^{V*}_M)^2 - (G^{V*}_E)^2}{4M_N^2(1 + \tau )^2}
+ \frac{1}{1 + \tau}
\left(\frac{d(G^{V*}_E)^2}{dQ^2} + \tau\frac{d(G^{V*}_M)^2}{dQ^2}\right)
+ \frac{d(G^*_A)^2}{dQ^2} \right],
\label{F_2WN_x_th}\\
F_3^{WN*}(x^*_{th})
&=&
-\beta^* \frac{(2G^{V*}_MG^*_A)}{dQ^2}.
\label{F_3WN_x_th}
\end{eqnarray}
First, we show
in Figs.~\ref{fig_F12} and~\ref{fig_g12} ratios of
the bound to free nucleon structure functions without the effect
of Fermi motion (dash-doted lines), calculated for the
charged-lepton scattering for baryon densities $\rho_B =
\rho_0$ with $\rho_0 = 0.15$ fm$^{-3}$. 
\begin{figure}
\begin{center}
\begin{minipage}[t]{8cm}
\epsfig{file=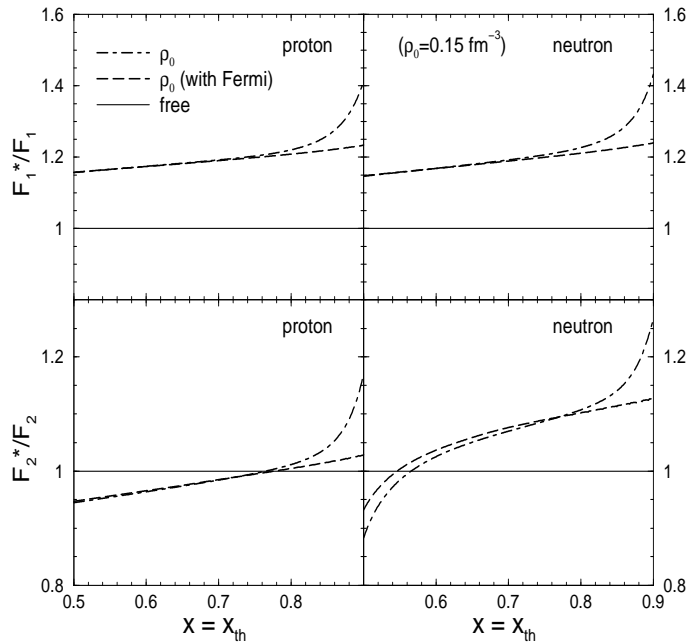,width=9cm,height=9cm,angle=-90}
\end{minipage}
\begin{minipage}[t]{16.5cm}
\caption{Ratios for the charged-lepton scattering structure functions
$F^*_{1,2}/F_{1,2}$, those extracted using the
bound and free nucleon form factors.
Effect of Fermi motion
is included by the convolution with the
nucleon momentum distribution~\cite{MILL-EMC} for {\it both} the
structure functions extracted using the free and bound nucleon form
factors, and then ratios are calculated
(the dashed lines denoted by "with Fermi")
(from Ref.~\cite{KSSdual}).
}
\label{fig_F12}
\end{minipage}
\end{center}
\end{figure}
\begin{figure}
\begin{center}
\begin{minipage}[t]{8cm}
\epsfig{file=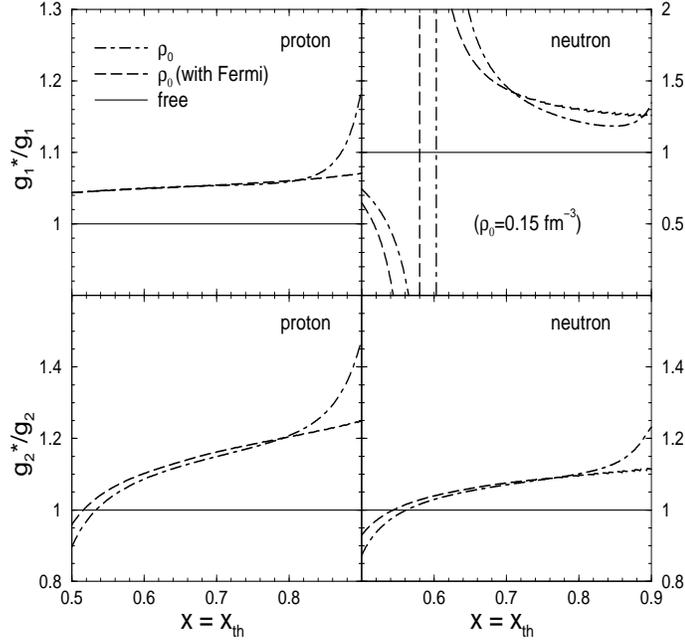,width=9cm,height=9cm,angle=-90}
\end{minipage}
\begin{minipage}[t]{16.5cm}
\caption{Same as Fig.~\ref{fig_F12}, but for $g^*_{1,2}/g_{1,2}$
(from Ref.~\cite{KSSdual}).
}
\label{fig_g12}
\end{minipage}
\end{center}
\end{figure}

Because $x_{th} =
(0.7,0.9)$ in free space (see also Eq.~(\ref{x_th_medium}))
correspond to $Q^2 \simeq (0.65,2.5)$ GeV$^2$, we regard the
results shown in the region, $0.7 \alt x \alt 0.9$, as the present
local duality predictions. The corresponding $Q^2$ range is also
more or less within the reliability of the bound nucleon form
factors calculated~\cite{POLE,Lu-eN,NeuA}.
In the region, $0.8
\alt x \alt 0.9$, all the bound nucleon structure functions
calculated, $F^*_{1,2}$ and $g*_{1,2}$, are enhanced relative to
those in a free nucleon.
(We have checked that also the enhancement becomes larger as the
baryon density increases.)
However, the depletion observed in the
EMC effect, occurring just before the enhancement as $x$
increases, is absent for all of them. Probably, the conventional
binding effect, which is not entirely included in the present study, may
produce some depletion~\cite{ST}. (We may note also that 
the conclusion drawn by Smith and Miller~\cite{MILL-EMC}, that the
depletion of the deep inelastic nuclear structure functions
observed in the valence quark regime, is due to some effect beyond
the conventional nucleon-meson treatment of nuclear physics.)
Thus, only the effect of
the bound nucleon internal structure change introduced via the
bound nucleon form factors and the pion threshold shift in the
present local duality framework, cannot explain the observed
depletion in the EMC effect for the relevant Bjorken-$x$
range $0.7 \alt x \alt 0.85$.
However, it can explain a part of the enhancement
at large Bjorken-$x$ ($x \agt 0.85$).

In order to see whether or not the conclusions drawn above are affected
by Fermi motion, we also calculate the ratios by convoluting
the nucleon momentum distribution
obtained in Ref.~\cite{MILL-EMC} with the value $\bar{M}_N = 931$ MeV.
(See Eq.~(\ref{hvh}).)
Namely, we convolute the nucleon momentum distribution with {\it both} the
structure functions extracted using the free and bound nucleon
form factors first, and then calculate ratios.
The corresponding results are shown
in Figs.~\ref{fig_F12} and~\ref{fig_g12} (dashed lines,
denote by "with Fermi").
Note that, because the upper value of
$x_{th} (x_{th}^{max} \sim 0.91)$ is limited for a reliable extraction
of the structure functions
by the reliable $Q^2$ range for the nucleon elastic
form factors in the present case, we had to cut the contribution 
from the region,
$x_{th} \ge x_{th}^{max}$, in the convolution integral.
This would effectively suppress the enhanced part of the
bound nucleon structure functions.
The obtained results show a similar feature, except the region
$x_{th} \agt 0.8$.
However, even the region $x_{th} \agt 0.8$,
the enhancement feature remains the same.
Thus, we conclude that, the obtained enhancement
of the bound nucleon structure functions $F^*_{1,2}$ and $g^*_{1,2}$ in the
charged lepton scattering,
is intrinsic and not smeared by the effect of Fermi motion.
This is more obvious in the region $0.7 \alt x_{th} \alt 0.8$.
In that region, the effect of Fermi motion is nearly
canceled out in the ratios, as one could expect.

Next, we show in Fig.~\ref{fig_Fjwmed} the bound nucleon structure
functions calculated from a charged current, $F^{WN*}_{1,2,3}$,
together with those in vacuum for the (anti)neutrino scattering.
For a reference, we show also $\frac{18}{5}F_2^{\gamma N} \equiv
\frac{18}{5} \frac{1}{2}\left(F^p_2+F^n_2\right)$ in vacuum for
the charged-lepton scattering.
\begin{figure}
\begin{center}
\begin{minipage}[t]{8cm}
\hspace*{-1cm}
\epsfig{file=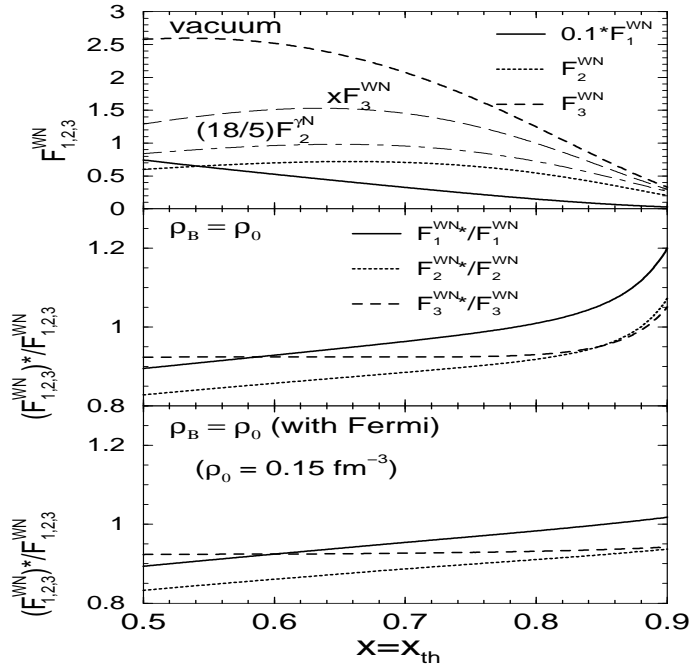,width=9cm,height=9cm,angle=-90}
\end{minipage}
\begin{minipage}[t]{16.5cm}
\caption{Structure functions calculated for (anti)neutrino
scattering (for an isoscalar nucleon), for baryon density
$\rho_B = 0$ and $\rho_0$. For a reference,
$\frac{18}{5}F_2^{\gamma N} \equiv \frac{18}{5}\frac{1}{2}(F_2^p+F_2^n)$
for the charged-lepton scattering in vacuum is also shown (top panel).
Ratios are shown for both with and without the effect of Fermi motion.
(See also caption of Fig.~\ref{fig_F12} for
the effect of Fermi motion.)
(Taken from Ref.~\cite{KSSdual}.)
}
\label{fig_Fjwmed}
\end{minipage}
\end{center}
\end{figure}
The effect of Fermi motion is included in the same way as that
was included in the charged-lepton scattering case.
Similarly to the charged-lepton
scattering case, $F^{WN*}_{1,2,3}$ in symmetric nuclear matter are
enhanced at large $x$ without the effect of Fermi motion,
but only in the region, $0.85 \alt x \alt
0.9$. This is due to the contribution from the in-medium axial
vector form factor $G^*_A$. Although $G^*_A$ falls off faster than
the free space $G_A$ in the range $Q^2 \alt 1$ GeV$^2$, the $Q^2$ dependence
turns out to be slightly enhanced in the range $Q^2 \agt 1$
GeV$^2$, due to the Lorentz contraction of the internal quark wave
function of the bound nucleon~\cite{NeuA}. Then, the contribution
from the $Q^2 \agt 1$ GeV$^2$ region gives a suppression. (See
Eqs.~(\ref{F_1WN_x_th})-(\ref{F_3WN_x_th}), but neglecting small
contributions from the non-derivative terms with respect to $Q^2$,
which are suppressed by $\sim 1/\tau^2$ as $Q^2$ increases.)
With the effect of Fermi motion, the enhancement and quenching features
at $x_{th} \agt 0.8$ are less pronounced because of
the convolution procedure
applied in the present treatment.
                                                                                
After having calculated $F_2^*$ for both the charged-lepton and
(anti)neutrino scattering, we can study also charge symmetry
breaking in parton distributions focusing on the effect of the
bound nucleon internal structure change.
In free space, it was
studied in Ref.~\cite{FK} based on the local duality. A measure of
charge symmetry breaking in parton distributions at $x = x^*_{th}$
may be given by~\cite{FK}:
\begin{eqnarray}
\left[\frac{5}{6}F_2^{WN*}(x^*_{th})\right.
&-& 3\left.F_2^{\gamma N*}(x^*_{th})\right]
\nonumber\\
&=&
3\left\{\frac{13}{18}\beta^*\left[\frac{d(G^{p*}_M)^2}{dQ^2}
+ \frac{d(G^{n*}_M)^2}{dQ^2}\right]
+ \frac{5}{9}\beta^*\frac{d(G^{p*}_MG^{n*}_M)}{dQ^2}
- \frac{5}{18}\beta^*\frac{d(G^*_A)^2}{dQ^2}\right\}.
\label{CSB}
\end{eqnarray}
In Fig.~\ref{fig_F2CSBmed} we show normalized ratios, divided by
$\frac{1}{2}\left[\frac{5}{6}F_2^{WN*}+3F_2^{\gamma N*}\right]$,
for baryon densities $\rho_B = 0$ and $\rho_0$ with and without the
effect of Fermi motion. 
\begin{figure}
\begin{center}
\begin{minipage}[t]{8cm}
\hspace*{-1cm}
\epsfig{file=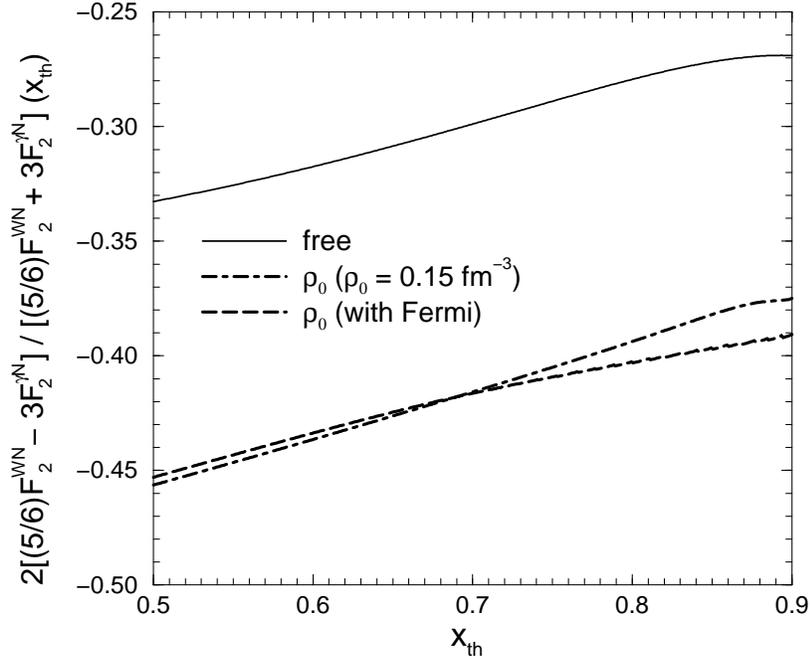,width=9cm,angle=-90}
\end{minipage}
\begin{minipage}[t]{16.5cm}
\caption{Normalized ratios, divided by
$\frac{1}{2}\left[\frac{5}{6}F_2^{WN*}+3F_2^{\gamma N*}\right]$,
for charge symmetry breaking in parton
distributions, for baryon densities $\rho_B = 0$ and
$\rho_0$, with and without the effect of Fermi motion.
(See also caption of Fig.~\ref{fig_F12} for the effect of
Fermi motion.)
(Taken from Ref.~\cite{KSSdual}.)
}
\label{fig_F2CSBmed}
\end{minipage}
\end{center}
\end{figure}
The results show that the charge symmetry breaking in symmetric
nuclear matter is enhanced due to the effect of the bound nucleon
internal structure change.
(We have checked that the breaking becomes larger
as the baryon density increases.)
Note that, because the quantity is the ratio
by definition, it is very insensitive to the effect of Fermi motion
in entire region of $x_{th}$ considered.
Thus, charge symmetry breaking looks to
be more appreciable in nuclei than in the case of a free nucleon,
and/or could affect fragmentation in heavy ion collisions. In
particular, the results imply that the NuTeV anomaly~\cite{NuTeV},
which was observed in the measurements using iron target, would be
enhanced even more than the analysis made~\cite{FK} in free space.
(See e.g., Ref.~\cite{Londergan} for detailed discussions.)
However, the present status of experimental accuracies would not
allow to detect the effect distinctly.

We summarize the results and conclusions of the present study
based on the local quark-hadron duality:
\begin{enumerate}
\item The effect of the change of the bound nucleon 
internal structure in
the nuclear medium is to enhance the structure
functions at large Bjorken-$x$ ($x \agt 0.85$) for 
charged-lepton scattering and especially $F_1^{WN}$ in
(anti)neutrino scattering.
\item The $x$ dependence of the bound nucleon structure functions
obtained for the charged-lepton scattering and $F_{2,3}^{WN}$ in
(anti)neutrino scattering
is different. Namely, the former [latter] is enhanced [quenched] in the
region $0.8 \alt x \alt 0.9$ [$0.7 \alt x \alt 0.85$].
\item The effect of the change in the 
bound nucleon internal structure change cannot
explain the depletion observed in the EMC effect (for the charged-lepton
scattering) for the relevant Bjorken-$x$ range in this study,
$0.7 \alt x \alt 0.85$, but it can explain a part of the enhancement
occurring in the larger region of $x$.
\item Charge symmetry breaking in parton distributions in nuclei,
or higher baryon densities, would be enhanced relative to that in free space
by the internal structure change of a bound nucleon.
\item The conclusions obtained above are insensitive to the effect of
Fermi motion in the treatment of the present study.
                                                                                
\end{enumerate}
Finally, we note again that the present study has not included
in a rigorous manner the
conventional nuclear effects, such as binding and Fermi motion.
However, even solely from the present results,
we can conclude that,
in addition to the conventional nuclear effects,
the effect of a change in the internal structure of a bound nucleon 
may be appreciable in various nuclear structure functions
at large Bjorken-$x$.

\section{Hadronic reactions in nuclear medium
\label{sec:reactions}}

We now turn to the study of how the internal structure changes of 
hadrons can have an impact on hadronic reactions. 
As examples, we discuss subthreshold kaon production, 
$D$ and $\overline{D}$ meson production in nucleus, and $J/\psi$ suppression in heavy ion collisions 
in sections,~\ref{subsec:kaon},~\ref{subsec:Dmeson} 
and~\ref{subsec:jpsi}, respectively. 

\subsection{\it Subthreshold kaon production and in-medium effects
\label{subsec:kaon}}

The properties of kaons in nuclear matter have recently  attracted
enormous interest because of their capacity to signal chiral symmetry
restoration or give information on the possibility of kaon
condensation in neutron
stars~\cite{BR,Kaplan,Ko1,Lee,Maruyama}. Studies with a variety of
models~\cite{Tsushima_k,Waas1,Waas2,SibirtsevaK} indicate that the
antikaon potential is attractive while the kaons feel a repulsive
potential in nuclear matter. The results from kaonic
atoms~\cite{Friedman},  as well as an
analysis~\cite{Li1,Cassing2,Li2,Bratkovskaya,Cassing3} of the $K^-$
production from heavy ion
collisions~\cite{Schroter,Senger1,Barth,Laue,Senger2},
are in reasonable agreement with the former expectation for
antikaons. However, the analysis of available data on
$K^+$ production from heavy ion collisions at SIS
energies~\cite{Barth,Laue,Senger2,Senger3} contradicted 
the predictions that the kaon potential is repulsive.
The comparison between the heavy ion calculations and the
data~\cite{Li2,Bratkovskaya,Cassing3,Senger3,Li3} indicated 
that the $K^+$-meson spectra were best described by neglecting
any in-medium modification of the kaon properties. 
Furthermore, the introduction of even a weakly repulsive
$K^+$-nucleus potential usually resulted in a
substantial underestimate of the experimental data on kaon
production.
However, subsequent studies including the in-medium modification 
of kaon property~\cite{Fuchs_K,Hartnack_K,Chen_K,Zheng,Mishra_K} 
all suggest that the repulsive $K^+$-nucleus potential can be 
consistently incorporated in kaon production simulations.
The main uncertainties lie in the elementary kaon production 
cross sections which can not be constrained by experiments, those 
involving the $\Delta(1232)$ resonance. 
They serve as dominant kaon production mechanisms.
Keeping this point in mind, however, we would like to study the 
impact of property changes of kaon, nucleon, hyperons in nuclear medium 
on subthreshold kaon production in heavy ion collisions.

Since in heavy ion collisions at
SIS energies~\cite{Schroter,Senger1,Barth,Laue,Senger2}
a lot of $K^+$-mesons are expected to be produced by secondary pions,
we investigate the kaon production reactions,
$\pi{+}N{\to}Y{+}K$ ($Y = \Lambda, \Sigma$ hyperons),
in nuclear matter~\cite{TST_strange}. To do this, we combine 
the in-medium property changes of hadrons predicted by the QMC model 
with earlier studies of kaon production in free space.
All parameters in the QMC model are fixed,   
as explained in section~\ref{sec:properties}, 
and the effects of the medium on the reaction cross
sections and {\em amplitudes} are calculated. The result is impressive in
that {\em the medium effects explain the nuclear production data, without any
adjustment of the parameters determined elsewhere, including the
standard repulsive kaon-nucleus interaction}.

First, in Fig.~\ref{laka8}, we show the total potential,
$U_{TOT}=V_s+V_v$ of Eqs.~(\ref{spot}) and~(\ref{vpot}) for nucleon ($N$) 
and $\Lambda$ hyperon in symmetric nuclear matter. 
This indicates that both nucleon and hyperon
potentials approach minima around normal nuclear matter density,
which reflects the fact that around $\rho_0$ the energy density
of nuclear matter is minimized.
\begin{figure}
\epsfysize=9.0cm
\begin{center}
\begin{minipage}[t]{8 cm}
\psfig{file=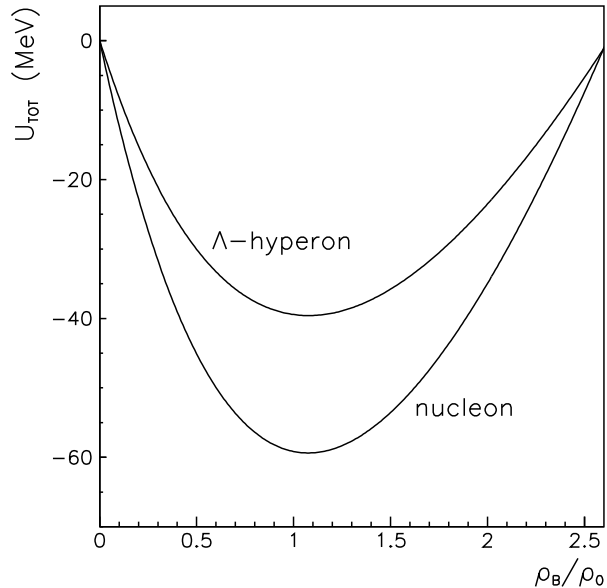,height=9cm}
\end{minipage}
\begin{minipage}[t]{16.5 cm}
\caption{Total potential $U_{tot}$ for nucleon and
$\Lambda$-hyperon shown as a function of the baryon density, $\rho_B$,
in units of the nuclear matter saturation density, $\rho_0$=0.15
fm$^{-3}$ (from Ref.~\cite{TST_strange}).
}
\label{laka8}
\end{minipage}
\end{center}
\end{figure}

Next, in Fig.~\ref{laka9} we show the density dependence of the total
$K$ and $K^*$-meson potentials at zero momenta.
\begin{figure}
\begin{center}
\begin{minipage}[t]{8cm}
\psfig{file=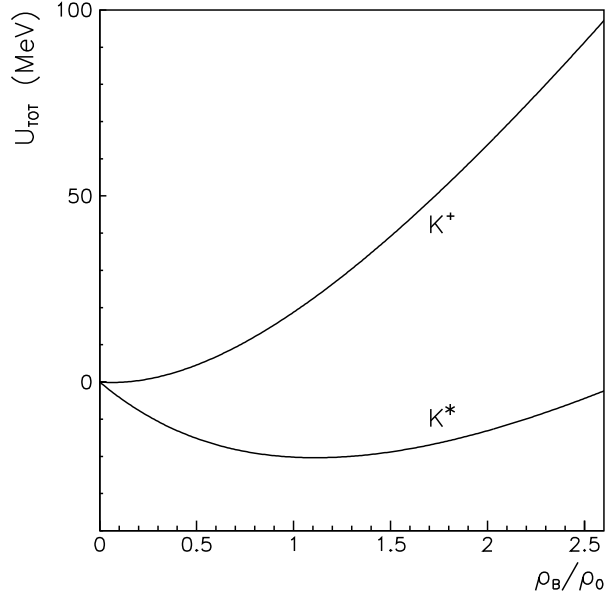,height=9cm}
\end{minipage}
\begin{minipage}[t]{16.5cm}
\caption{Total potential $U_{tot}$ for $K^+$  and
$K^\ast(892)^+$ mesons plotted as function of the baryon density, $\rho_B$,
in units of  saturation density, $\rho_0$=0.15
fm$^{-3}$ of the nuclear matter (from Ref.~\cite{TST_strange}).
}
\label{laka9}
\end{minipage}
\end{center}
\end{figure}
The total kaon potential is repulsive as explained before,
and depends substantially on the baryon density.
(See Eq.~(\ref{vpot}).) 
The $K^*$-meson total potential is attractive
at baryon densities below $\simeq 2.7 \rho_0$.
As for the pion, the Goldstone boson nature of it and the chiral symmetry 
suggest us that the pion mass modification in nuclear medium is 
marginal, as already mentioned before. Thus, we do not consider 
possible medium effects on the pion. 

{}For the energy dependence of the cross section $\pi N \to Y K$
($Y=\Lambda,\Sigma$) we use the
resonance model~\cite{TsushimaRes,SibirtsevRes},
which could describe the energy dependence of the total cross sections,
$\pi N \to Y K$, quite well, and has been used widely in
kaon production simulation
codes~\cite{Cassing2,Li2,Bratkovskaya,Li3,likpot,kcode}, and
extend the model by including medium modification
of the hadron properties, not only in the kinematic factors such as
the flux and the phase space, but also in the reaction amplitudes.
Kaon and hyperon production processes in $\pi N$ collisions
in the resonance model~\cite{TsushimaRes,SibirtsevRes} 
are shown in Figs.~\ref{pbyklafig} and~\ref{pbyksifig}.
Because different intermediate states and final states
contribute to the $\pi N \to \Lambda K$ and $\pi N \to \Sigma K$ reactions,
the in-medium modification of the reaction amplitudes for 
these reactions should also be different.
\begin{figure}
\begin{center}
\begin{minipage}[t]{8cm}
\psfig{file=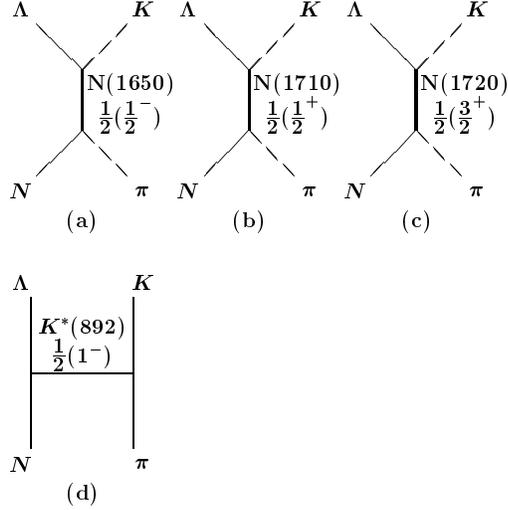,height=12cm}
\vspace{-3cm}
\caption{Processes included for the $\pi N \to \Lambda K$ reactions.}
\label{pbyklafig}
\end{minipage}
\end{center}
\end{figure}
\begin{figure}
\begin{center}
\begin{minipage}[t]{8cm}
\psfig{file=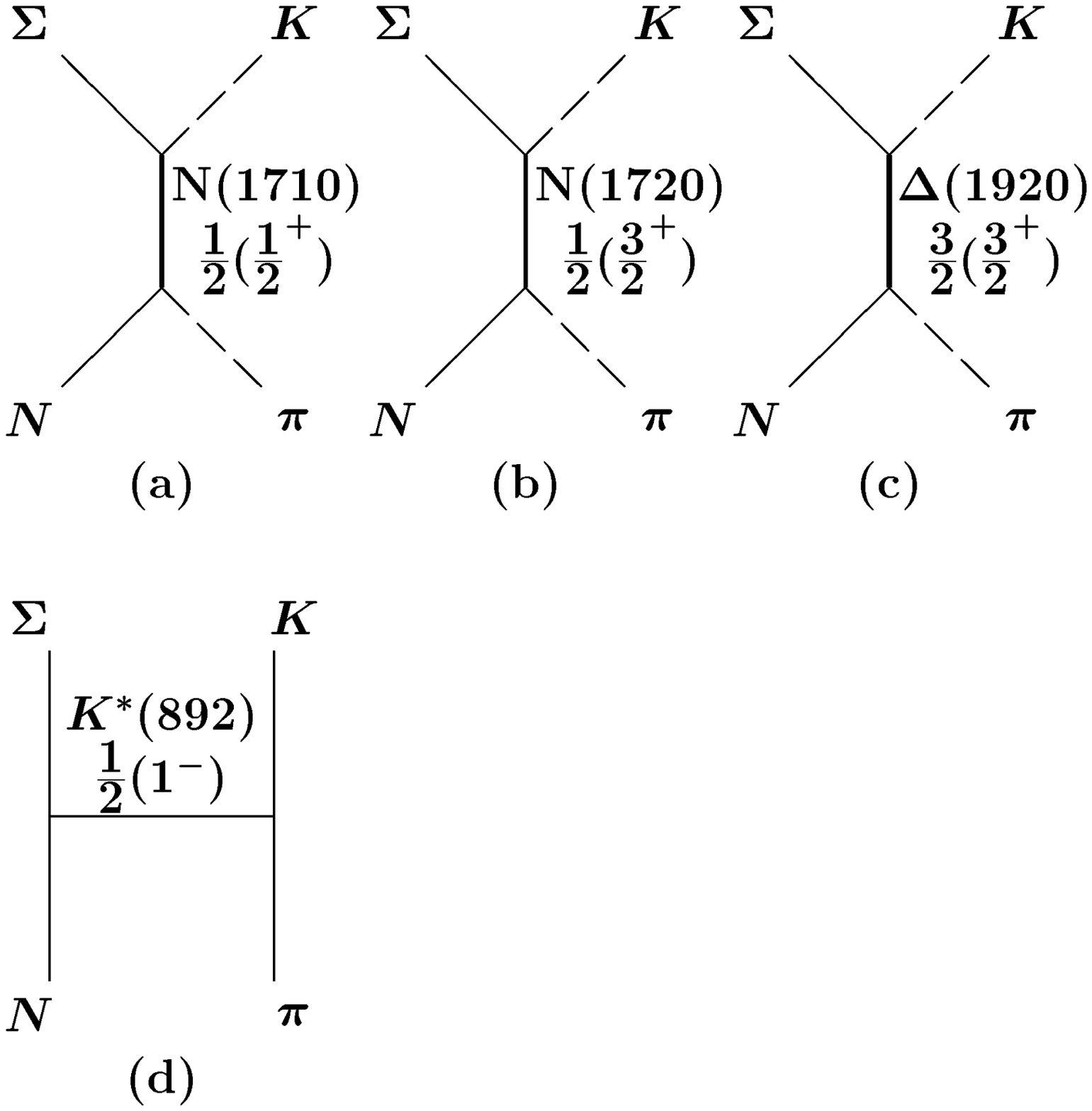,height=12cm}
\vspace{-3cm}
\caption{Processes included for the $\pi N \to \Sigma K$ reactions.}
\label{pbyksifig}
\end{minipage}
\end{center}
\end{figure}

Now, we need to discuss the in-medium modification of the resonance masses, 
which enter the resonance propagators and modify the reaction amplitudes.
At present it seems that there is no reliable estimate for the in-medium
modification of masses for the higher mass baryon resonances.

Based on the QMC model, we assume that the light quarks in 
the baryon resonances are
responsible for the mass modification in nuclear medium,
as explained in section~\ref{subsubsec:scbmatter}.
However, there is a possibility that the excited state light quarks
may couple differently to the scalar $\sigma$ field from those in the
ground states, although we expect that the difference should be small.
Thus, we estimate the range for the in-medium baryon resonance masses
by the following two extreme cases, i.e., (i) all light quarks
including those in the excited states play the same role for the
mass modification as those in the ground states,
(ii) only the ground state light quarks play the role
as in the usual QMC model.
These two cases are expected to give the maximum
and minimum limits for the mass modifications
of the baryon resonances.
Specifically, the range for the effective masses
of the baryon resonance in medium is given:
\begin{eqnarray}
M_{N(1650)} - \delta M^*_N &\leq& M^*_{N(1650)} \leq
M_{N(1650)} - \frac{2}{3} \delta M^*_N, \label{mr1} \\
M_{N(1710)} - \delta M^*_N &\leq& M^*_{N(1710)} \leq
M_{N(1710)} - \frac{1}{3} \delta M^*_N, \label{mr2} \\
M_{N(1720)} - \delta M^*_N &\leq& M^*_{N(1720)} \leq
M_{N(1720)} - \frac{1}{3} \delta M^*_N, \label{mr3} \\
M_{\Delta(1920)} - \delta M^*_N &\leq& M^*_{\Delta(1920)} \leq
M_{\Delta(1920)} - \frac{1}{3} \delta M^*_N, \label{mr4} \\
{\rm with} \hspace{2em} \delta M^*_N &=& M_N - M^*_N. \label{dmn}
\label{mresonance}
\end{eqnarray}
These in-medium resonance masses may be expected to modify the
resonance propagators in the reaction amplitudes.
To avoid introducing extra unknown parameters, 
we approximate the in-medium resonance widths appearing
in the propagator by the free space ones. From Eqs.~(\ref{mr1})-(\ref{dmn}), 
we will show results for the cross section calculated using 
the lower limit for the resonance masses.
However, we have also performed the calculation using the upper limit
for the resonance masses and confirmed that our conclusion
remains the same.

The dispersion relation in nuclear matter relating the total energy
$E$ and the momentum $\vec{p}$ of the particle $h$ 
with free space mass $m_h$ is written as
\begin{equation}
E=\sqrt{\vec{p}^{\,\, 2} + (m_h + V^h_s)^2 } + V^h_v,
\label{dispa}
\end{equation}
where $V^h_s$, $V^h_v$ denote the scalar and
vector potentials in nuclear matter.
(See Eqs.~(\ref{spot}) and~(\ref{vpot}).) 
The threshold,
$\sqrt{s_{th}}$, for the  reaction $\pi{+}N{\to}Y{+}K^+$ is given
as the sum of the total energies of the final $K^+$-meson and
$Y$-hyperon, taking their momenta to be zero and hence
\begin{equation}
\sqrt{s_{th}}=m_K+m_Y+V_s^K+V_s^Y+V_v^K+V_v^Y,
\label{th1}
\end{equation}
where now the upper indices denote kaons and hyperons.
The solid lines in Fig.~\ref{laka10}
show the $K^+\Lambda$ and $K^+\Sigma$ reaction thresholds,
$\sqrt{s_{th}}$, as a function of the baryon density.
Obviously, in free space the scalar and vector potentials vanish
and the reaction threshold equals to the sum of
the bare masses of the produced particles, which is shown by the
dashed lines in Fig.~\ref{laka10} for the $K^+\Lambda$ and $K^+\Sigma$
final states.
\begin{figure}
\begin{center}
\begin{minipage}[t]{8cm}
\psfig{file=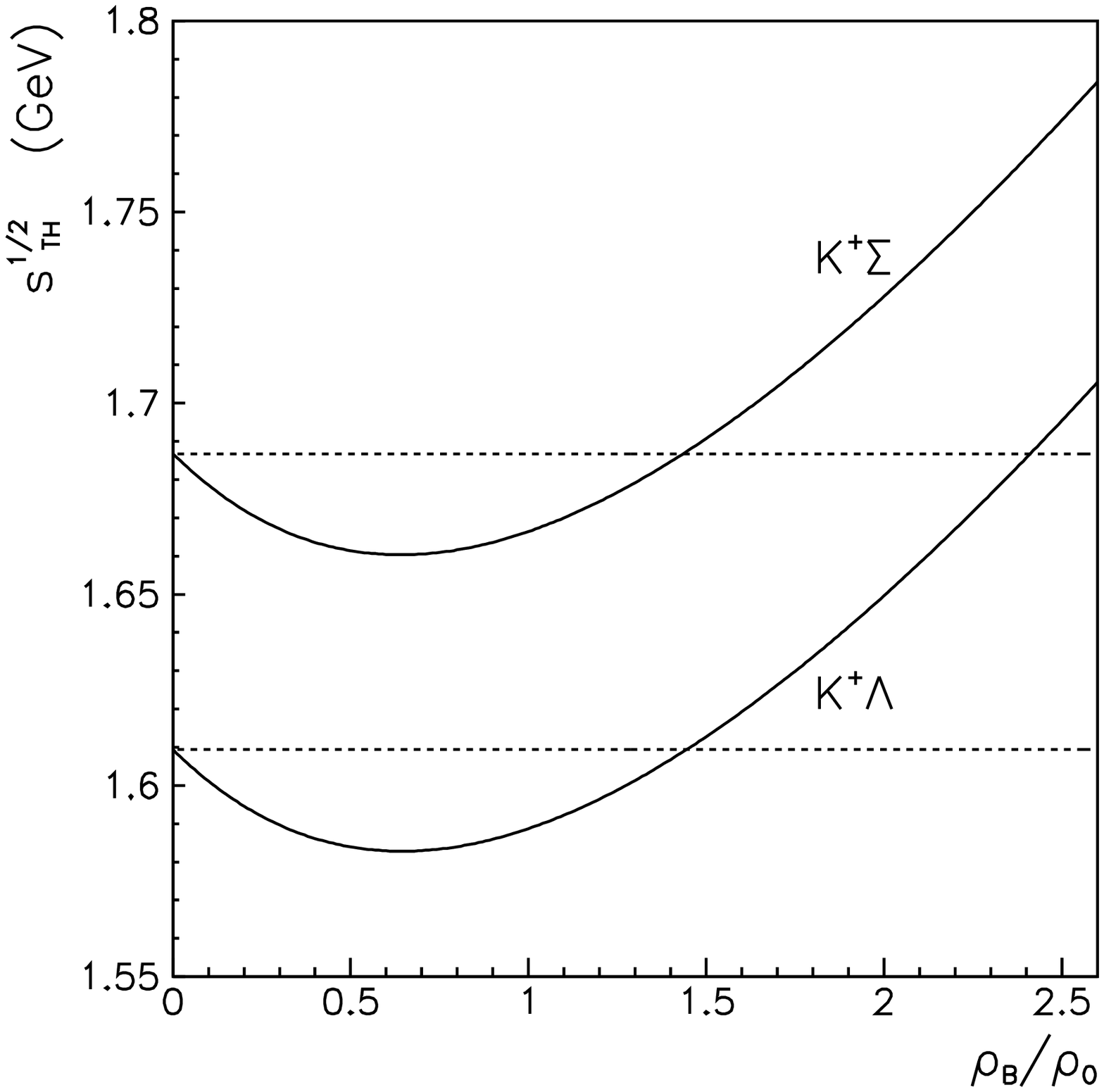,height=9cm}
\end{minipage}
\begin{minipage}[t]{16.5cm}
\caption{The threshold energy, $\sqrt{s_{th}}$,
for $K^+\Lambda$
and $K^+\Sigma$ production given by their total in-medium energy
at zero momentum, as a function of the baryon density $\rho_B$
in units of saturation density of nuclear
matter, $\rho_0$=0.15 fm$^{-3}$. The solid lines indicate results 
calculated, while the dashed lines show the threshold in free space
(from Ref.~\cite{TST_strange}).}
\label{laka10}
\end{minipage}
\end{center}
\end{figure}

It is important, that while the $K^+$-meson energy at zero
momentum increases with the baryon density
(see Fig.~\ref{laka9}), because of the negative  $\Lambda$ and $\Sigma$
potentials the reaction threshold in nuclear matter at
$\rho_B{<}$1.4$\rho_0$ ($\rho_0=0.15$ fm$^{-3}$) 
is shifted below than that in free space.
The maximal downward shift of the
reaction threshold in nuclear matter occurs at
baryon densities around $\rho_B{\simeq}0.6\rho_0$.
This value is the result of  competition between the simple,
linear dependence on density of the vector potentials and the more
complicated, non-linear behavior of the scalar potentials. (A similar
competition leads to the saturation of the binding energy of normal
nuclear matter in the QMC model.)
Furthermore, the maximum of the downward shift of the
$\pi{+}N{\to}Y{+}K^+$ reaction threshold amounts to roughly
30~MeV. We also found that at  baryon densities
$\rho_B{>}$0.2~fm$^{-3}$ the strangeness production threshold
in ${\pi}N$ collisions is higher than the free space case.

Now we apply the resonance model to calculate the in-medium
amplitudes, while keeping the coupling constant as well as the form
factors at the values found in free
space. While this assumption certainly cannot be completely correct
in nuclear matter, there are no presently  established
ways to improve  this
part of our calculation. In principle, since we started from the
reaction amplitude itself, it is possible to include
in-medium modifications of the coupling constants as well as the
form factors when reliable calculations of the changes of these
quantities in nuclear matter  become available.
In the following calculations we include the vector and scalar
potentials for the interacting (initial) nucleons and final kaons and hyperons,
as well as for the intermediate baryonic resonances and $K^*$-meson.
                                                                                
Fig.~\ref{laka1} shows the results calculated for the differential
cross section for the 
$\pi^-{+}p{\to}\Lambda{+}K^0$ reaction 
at the invariant collision energy $\sqrt{s}$=1683 MeV. It is  calculated
both in  free space (the solid line) and in  nuclear
matter, at baryon densities $\rho_B$=$\rho_0$ (the dashed line) and
$\rho_B$=2$\rho_0$ (the dotted line). For  comparison, we also
show in Fig.~\ref{laka1} the experimental data collected
in free space~\cite{Knasel,Baker}.
The important finding is that not only the absolute magnitude, but
also the shape (the dependence on the $\cos \theta$)
of the $\pi^-{+}p{\to}\Lambda{+}K^0$ differential cross section,
depends on the baryon density.
\begin{figure}
\begin{center}
\begin{minipage}[t]{8cm}
\psfig{file=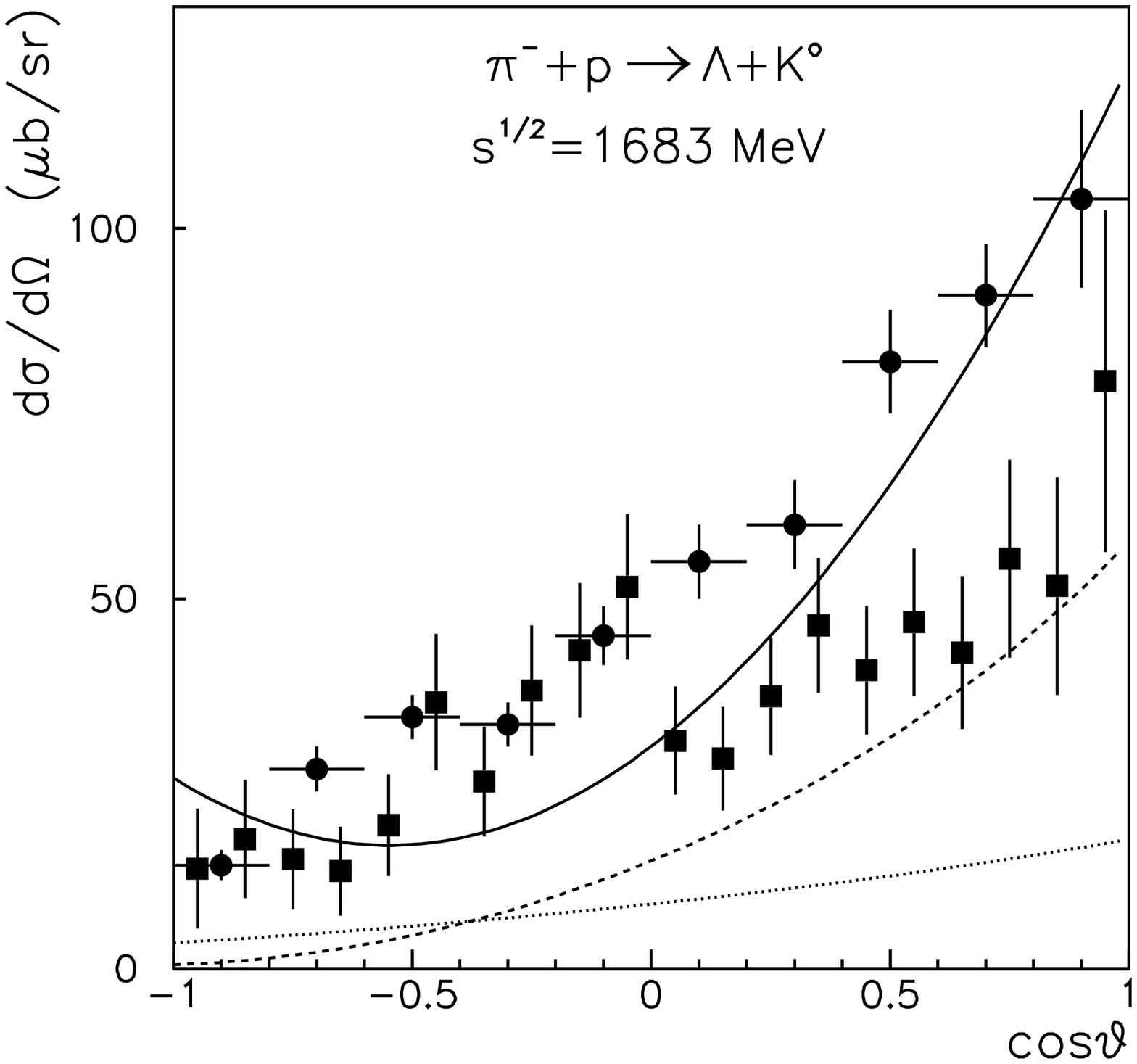,height=9cm}
\end{minipage}
\begin{minipage}[t]{16.5cm}
\caption{The $\pi^-{+}p{\to}\Lambda{+}K^0$ differential
cross section at invariant collision energy $\sqrt{s}$=1683 MeV as
a function of the $\cos\theta$ of the  kaon emission angle in
the center of mass system.
The experimental data are from Ref.~\cite{Knasel} (the squares) and
from Ref.~\cite{Baker} (the circles). The lines show our calculations
in free space (solid) and in nuclear  matter at baryon densities 
$\rho_B$=$\rho_0$ (dashed) and $\rho_B$=2$\rho_0$ (dotted),
with $\rho_0 = 0.15$ fm$^{-3}$
(from Ref.~\cite{TST_strange}).}
\label{laka1}
\end{minipage}
\end{center}
\end{figure}

One of the simplest ways to construct the in-medium reaction
cross section is to  take into account  only the in-medium
modification of the flux and phase space factors while leaving amplitude
in matter the same as that in free space,
without including any medium effect~\cite{Pandharipande}.
To shed more light on the problem of how  the reaction amplitude itself
is modified in nuclear matter, we show in
Fig.~\ref{laka3} the reaction amplitudes squared
in arbitrary units for
the $\pi^-{+}p{\to}\Lambda{+}K^0$ reaction,
calculated at $\sqrt{s}$=1.7 GeV and 1.9 GeV, in
free space (the solid lines), $\rho_B = \rho_0$ (the dashed lines) and
$\rho_B = 2 \rho_0$ (the dotted lines).
Our calculation clearly indicates that the $\pi^-{+}p{\to}\Lambda{+}K^0$
reaction amplitude in nuclear matter differs substantially
from that in free space at these energies,
and that the amplitudes depend on the baryon density.
\begin{figure}
\begin{center}
\begin{minipage}[t]{8cm}
\psfig{file=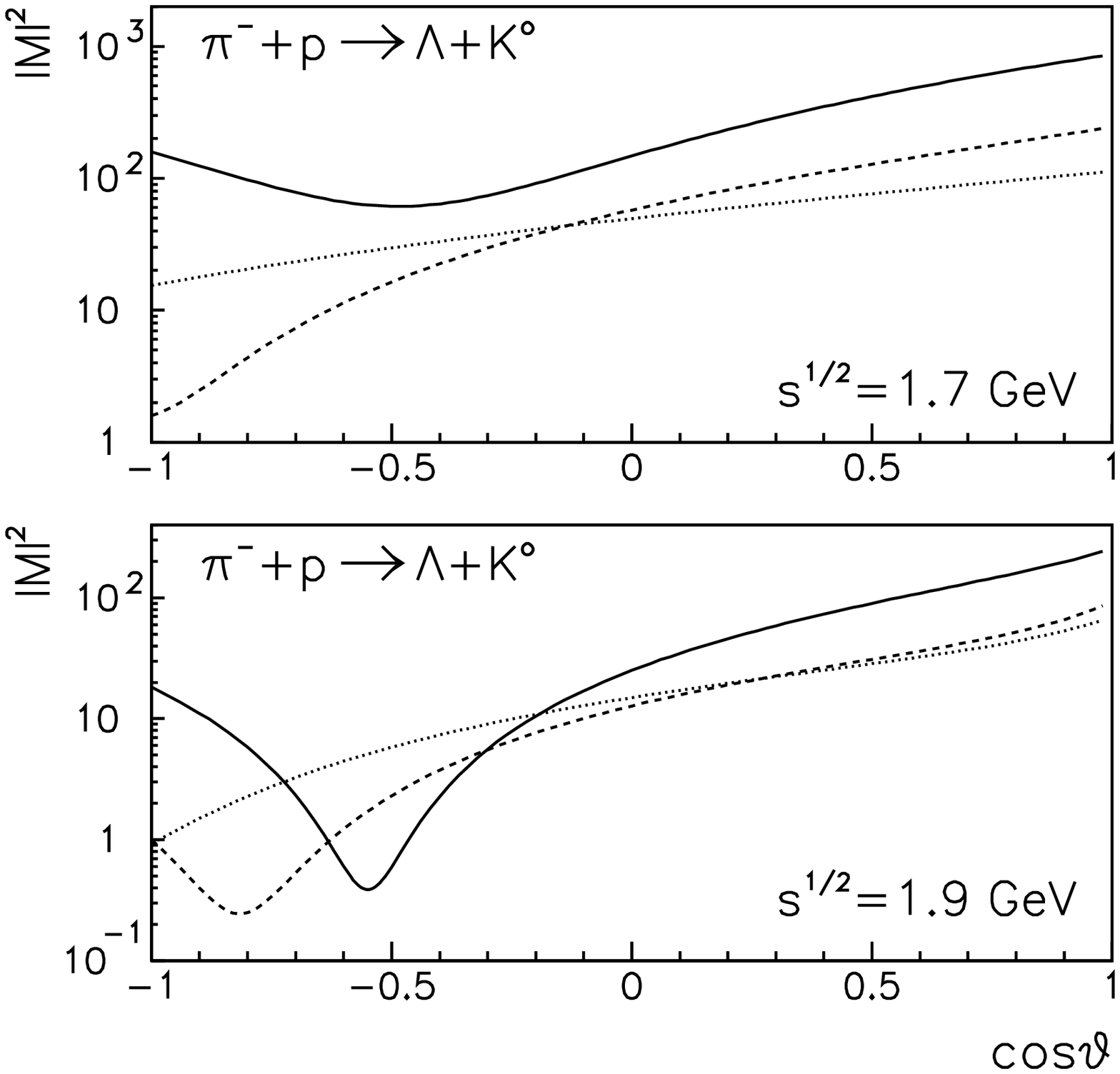,height=9cm}
\end{minipage}
\begin{minipage}[t]{16.5cm}
\caption{The (dimensionless) invariant amplitude squared for
the $\pi^-{+}p{\to}\Lambda{+}K^0$ reaction,
as a function of the $\cos \theta$ (the  $K^+$-meson
emission angle in the center of mass system), calculated for the
invariant collision energies $\sqrt{s}$=1.7 GeV (upper) and
1.9 GeV (lower). The lines show the result for free space (solid)
and for nuclear matter at baryon densities $\rho_B$=$\rho_0$ (dashed)
and $\rho_B$=2$\rho_0$ (dotted)
(from Ref.~\cite{TST_strange}).}
\label{laka3}
\end{minipage}
\end{center}
\end{figure}
                                                                                
Finally, the energy dependence
of the total $\pi^-{+}p{\to}\Lambda{+}K^0$
cross section is shown in Fig.~\ref{laka4}, as a function of the
invariant collision energy, $\sqrt{s}$. The experimental data
in free space are taken from Ref.~\cite{LB}. The calculations for free space
are in reasonable agreement with the data, as shown by the solid line.
The dashed line in Fig.~\ref{laka4} shows the results obtained for
nuclear matter at $\rho_B$=$\rho_0$, while the
dotted line is the calculation at $\rho_B$=2$\rho_0$.
\begin{figure}
\begin{center}
\begin{minipage}[t]{8cm}
\psfig{file=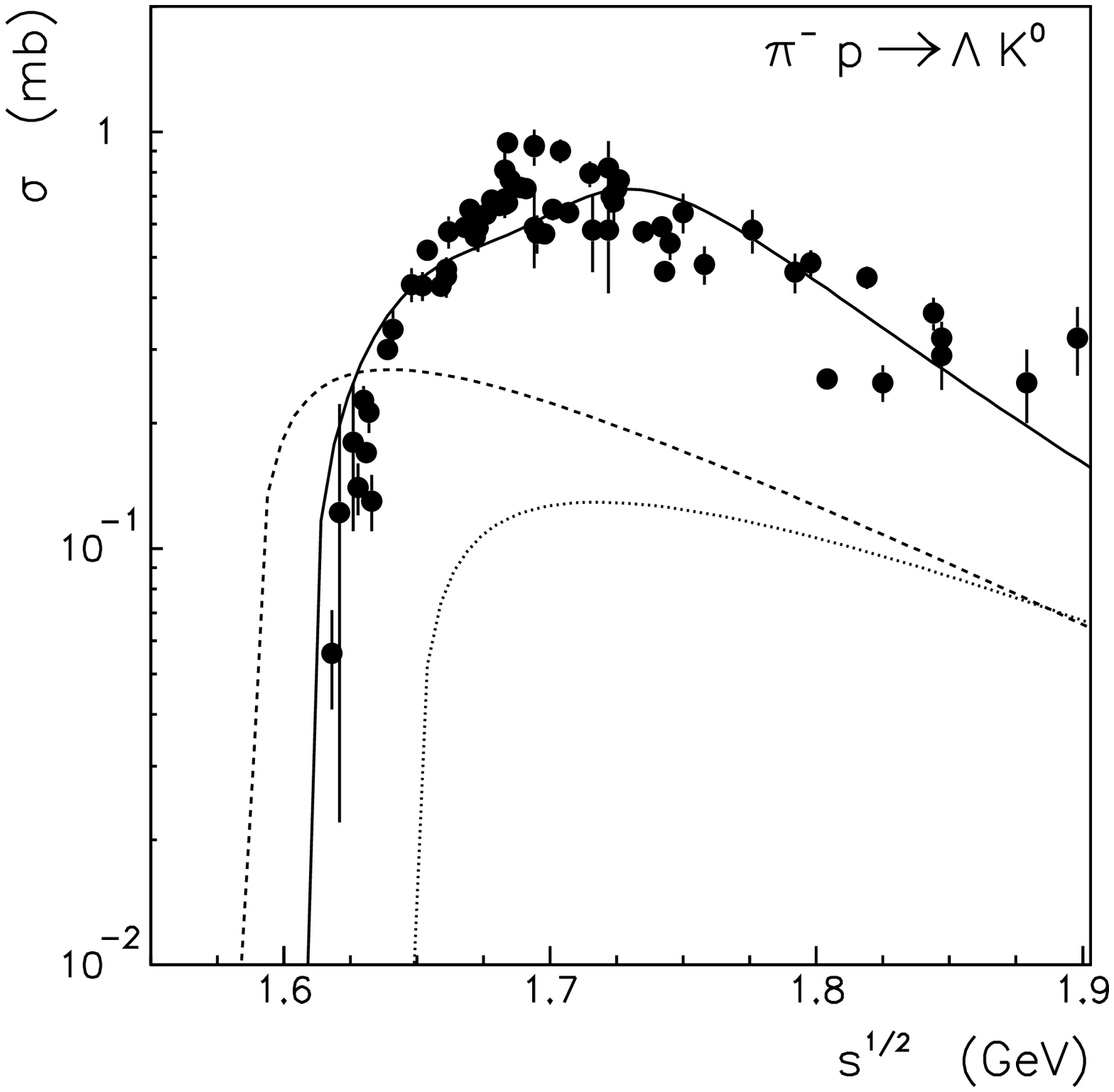,height=9cm}
\end{minipage}
\begin{minipage}[t]{16.5cm}
\caption{Energy dependence of the
total cross section, $\pi^-{+}p{\to}\Lambda{+}K^0$,
as a function of the invariant collision energy, $\sqrt{s}$,
calculated for different baryon densities.
The data in free space are taken from Ref.~\cite{LB}.
The lines correspond to the results calculated for
free space (solid) and for nuclear matter at baryon densities
$\rho_B$=$\rho_0$ (dashed) and $\rho_B$=2$\rho_0$ (dotted). (Only the
solid curve should be compared directly with the data.)
(Taken from Ref.~\cite{TST_strange}.)}
\label{laka4}
\end{minipage}
\end{center}
\end{figure}
Clearly the total $\pi^-{+}p{\to}\Lambda{+}K^0$ reaction cross section
depends substantially  on the baryon density.
Furthermore, as already discussed previously,
the reaction threshold at
baryon density $\rho_B$=$\rho_0$ is shifted downward as compared to
that in free space, while at $\rho_B$=2$\rho_0$
it is shifted upward.

It is obvious that heavy ion collisions probe a range of baryon densities 
from $\rho_B$=0 up to several times normal nuclear matter density. 
The calculation of the time
and spatial dependence of the baryon density distribution is
a vital aspect of  dynamical heavy
ion simulations. However, a first estimate
of the density averaged total $\pi^-{+}p{\to}\Lambda{+}K^0$
cross section can be gained from Fig.~\ref{laka4}.
Of course, the data is only available in free space and
should only be directly compared with the solid curve. Nevertheless, it
is suggestive for the problem of in-medium production to note that a
crude average of the in-medium cross sections over
the range 0${<}\rho_B{<}2\rho_0$ would be quite close to the free space
cross section at energies around the free space
threshold. This seems to provide a reasonable
explanation of why the heavy
ion calculations including~\cite{Cassing3,Senger3,Li3}  the
$\pi{+}p{\to}\Lambda{+}K$  cross section in
free space, that is without a repulsive kaon potential, can
reproduce the data~\cite{Schroter,Senger1,Barth,Laue,Senger2}.
A more quantitative calculation and discussion of this effect will be given
later.

Next, we consider the $\pi{+}N{\to}\Sigma{+}K$ reaction in nuclear matter.
The $\pi{+}N{\to}\Sigma{+}K$ reaction involves different dynamics
in comparison with the $\pi{+}p{\to}\Lambda{+}K$ reaction, because
the reaction involves different intermediate baryonic
resonances. For instance, although the $N(1650)$ resonance couples
to ${\Lambda}N$ channel, it does not couple to the ${\Sigma}N$ state.
Moreover, the $\Delta(1920)$ resonance couples to
${\Sigma}N$ channel, but does not couple to the ${\Lambda}N$ channel
in the resonance model~\cite{TsushimaRes,SibirtsevRes}.
For this reason, the dependence on the baryon density of the
reaction in nuclear matter, $\pi{+}N{\to}\Sigma{+}K$,
is quite different from that of
$\pi{+}N{\to}\Lambda{+}K$.

\begin{figure}
\begin{center}
\begin{minipage}[t]{8cm}
\psfig{file=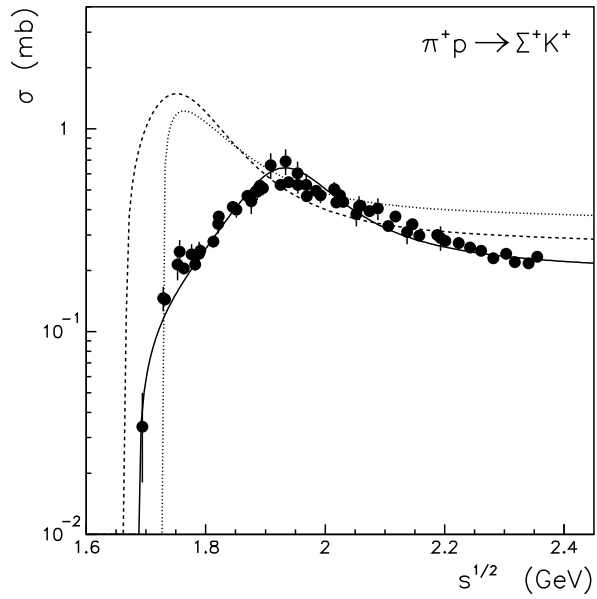,width=9cm,height=9cm}
\end{minipage}
\begin{minipage}[t]{16.5cm}
\caption{Energy dependence of the
total cross section, $\pi^+{+}p{\to}\Sigma^+{+}K^+$,
as a function of the invariant collision energy, $\sqrt{s}$,
calculated for different baryon densities.
The data in free space are taken from Ref.~\cite{LB}. The
lines indicate calculations for
free space (solid) and for nuclear matter at baryon densities
$\rho_B$=$\rho_0$ (dashed) and $\rho_B$=2$\rho_0$ (dotted).
(Taken from Ref.~\cite{TST_strange}.)}
\label{laka5}
\end{minipage}
\end{center}
\end{figure}
                                                                                
In Fig.~\ref{laka5} we show the energy dependence of the
total cross section, $\pi^+{+}p{\to}\Sigma^+{+}K^+$,
as a function of the invariant collision energy,
$\sqrt{s}$. The experimental data in free space are taken from
Ref.~\cite{LB}. The free space data are well
reproduced by the calculations in free space, as shown in
Fig.~\ref{laka5} by the solid line. The dashed line indicates the
results obtained for nuclear matter at baryon density
$\rho_B$=$\rho_0$, while the dotted line shows the result at
$\rho_B$=2$\rho_0$.
                                                                                
Again, as already discussed previously, the
density dependence of the hadron masses and the vector
potentials leads to a shift of the reaction thresholds
in nuclear matter. Because of the density dependence of the
$\Sigma$-hyperon potential, the threshold at normal nuclear matter
density ($\rho_B$=$\rho_0$) is shifted downward
compared with that in free space.
At $\rho_B$=2$\rho_0$ the ${\Sigma}K$ reaction threshold
is shifted upward relative to the threshold in free space.
Moreover, the magnitude of the $\pi^+{+}p{\to}\Sigma^+{+}K^+$ cross section
depends much more strongly on the density than the
$\pi^-{+}p{\to}\Lambda{+}K^0$ reaction.
                                                                                
Figs.~\ref{laka6} and~\ref{laka7} show the energy dependence of the
total cross sections
for the $\pi^-{+}p{\to}\Sigma^0{+}K^0$ and $\pi^-{+}p{\to}\Sigma^-{+}K^+$
reactions, respectively. The data in free space~\cite{LB}
are well reproduced
with the calculations in free space, which are indicated by the solid lines.
The cross sections calculated for nuclear matter,
except for $\pi^-{+}p{\to}\Sigma^0{+}K^0$ at $\rho_B = 2 \rho_0$, are
substantially enhanced in comparison with those in free space,
at energies just above the in-medium reaction
thresholds.
\begin{figure}
\begin{center}
\begin{minipage}[t]{8cm}
\psfig{file=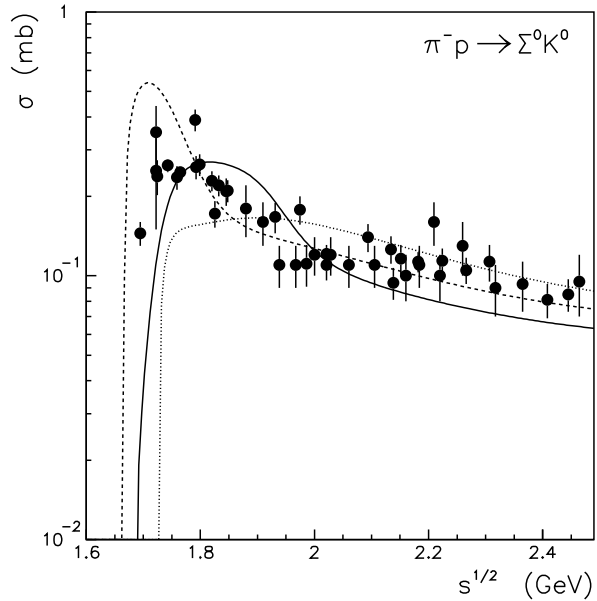,height=9cm}
\end{minipage}
\begin{minipage}[t]{16.5cm}
\caption{Energy dependence of the
total cross section, $\pi^-{+}p{\to}\Sigma^0{+}K^0$,
as a function of the invariant collision energy, $\sqrt{s}$,
calculated for different baryon densities.
The data in free space are taken from Ref.~\cite{LB}. The
lines indicate the calculations for
free space (solid) and for nuclear matter at baryon densities
$\rho_B$=$\rho_0$ (dashed) and $\rho_B$=2$\rho_0$ (dotted)
(from Ref.~\cite{TST_strange}).}
\label{laka6}
\end{minipage}
\end{center}
\end{figure}
\begin{figure}
\begin{center}
\begin{minipage}[t]{8cm}
\psfig{file=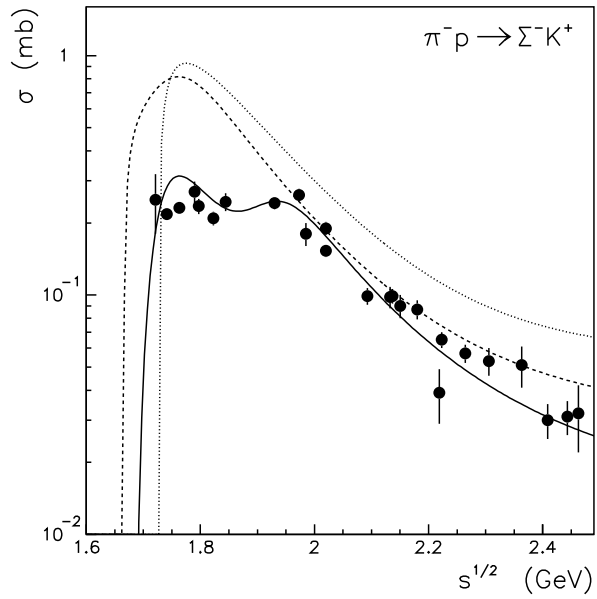,height=9cm}
\end{minipage}
\begin{minipage}[t]{16.5cm}
\caption{Energy dependence of the
total cross section, $\pi^-{+}p{\to}\Sigma^-{+}K^+$,
as a function of the invariant collision energy, $\sqrt{s}$,
calculated for different baryon densities.
The data in free space are taken from Ref.~\cite{LB}. The
lines indicate our calculations for
free space (solid) and for nuclear matter at baryon densities
$\rho_B$=$\rho_0$ (dashed) and $\rho_B$=2$\rho_0$ (dotted)
(from Ref.~\cite{TST_strange}).}
\label{laka7}
\end{minipage}
\end{center}
\end{figure}

It is expected that in relativistic heavy ion
collisions at SIS energies  nuclear matter can be compressed up to
baryonic densities of order $\rho_B{\simeq}{3}\rho_0$~\cite{Senger3}.
The baryon density
$\rho_B$ available in heavy ion collisions evolves with the interaction
time, $t$, and is given by the dynamics of the heavy ion collision.
Moreover, the density is large in the very center of the
collision. In the following estimates we investigate
the density dependence of the production cross section for central
central heavy ion collisions. However, it should be remembered
that at the edges,
where most particles are expected to be located,
the density dependence of the strangeness production
mechanism is not strong compared to that of the central
zone of the collision.
To calculate the $K^+$-meson production cross section averaged
over the available density distribution we adopt the
density profile function obtained in dynamical
simulations~\cite{Hombach} of $Au{+}Au$ collisions at 2 AGeV and at impact
parameter $b{=}0$. This can be parametrized as
\begin{equation}
\label{time1}
\rho_B (t) = \rho_{max} \, \exp
\left( \frac{[\, t\,- \,{\bar t} \, \, ]^2}{{\Delta t}^2}
\right),
\label{dprofile}
\end{equation}
where the parameters, $\rho_{max}$=3$\rho_0$,
${\bar t}$=13~fm and ${\Delta t}$=6.7~fm, were fitted to the heavy
ion calculations~\cite{Hombach}.

\begin{figure}
\begin{center}
\begin{minipage}[t]{8cm}
\psfig{file=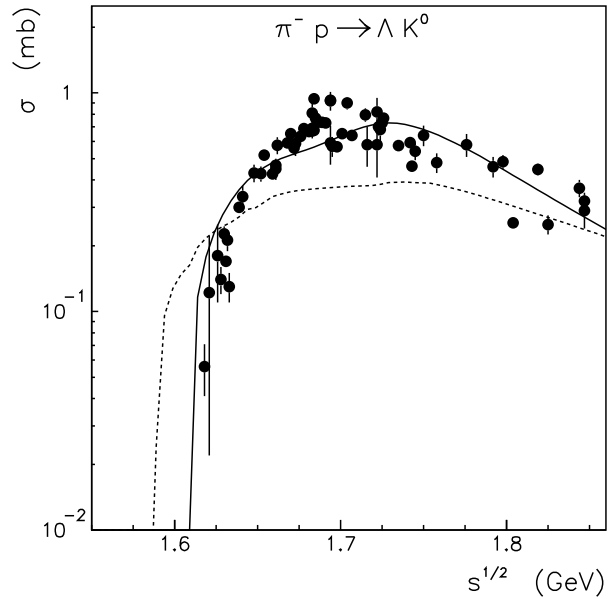,height=9cm}
\end{minipage}
\begin{minipage}[t]{16.5cm}
\caption{Energy dependence of the
total cross section, $\pi^-{+}p{\to}\Lambda{+}K^0$,
as a function of the invariant collision energy, $\sqrt{s}$.
The data in free space are taken from Ref.~\cite{LB}. The
solid line indicates our calculation for
free space. The dashed line shows the cross section calculated
by averaging over the density function profile~\cite{Hombach}
given by the time evolution obtained for
$Au{+}Au$ collisions at 2 AGeV (see Eq.~(\ref{dprofile}))
(from Ref.~\cite{TST_strange}).}
\label{laka4a}
\end{minipage}
\end{center}
\end{figure}
                                                                                
The total cross section for the $\pi^-{+}p{\to}\Lambda{+}K^0$ reaction
integrated over the time range $5{\le}t{\le}23$~fm and weighted by
the time dependent density profile given in Eq.~(\ref{dprofile}),
is shown by the dashed line in Fig.~\ref{laka4a}. The
limits of the $t$ integration were taken from the simulations of the
$Au{+}Au$ collision time evolution in Ref.~\cite{Hombach}. The circles and
solid line in Fig.~\ref{laka4a} show the experimental data
in free space~\cite{LB}
and the calculations in free space, respectively.

One can see that the total cross section averaged over
the collision time (time dependent density profile) for
the $\pi^-{+}p{\to}\Lambda{+}K^0$ reaction
is quite close to the result given in
free space integrated up to energies above the production threshold,
(i.e., $\sqrt{s} \simeq$ 1.7 GeV).
That the results shown in Fig.~\ref{laka4a} actually explain why the
heavy ion calculations with the free space kaon production cross
section might quite reasonably reproduce the experimental data,
will be discussed more quantitatively below.
                                                                                
As a matter of fact, the total cross section averaged over
the time dependent density profile,
shown by the dashed line in Fig.~\ref{laka4a},
should additionally be averaged over the
invariant collision energy distribution available in heavy
ion reactions. The number of meson-baryon collisions, $N_{mB}$,
for the central $Au{+}Au$ collisions at 2 AGeV is given
in Ref.~\cite{CBJ} as a function of the invariant collision energy,
$\sqrt{s}$. It can be parametrized for $\sqrt{s}{>}$1~GeV as
\begin{equation}
\frac{dN_{mB}}{d\sqrt{s}} = N_0 \, \exp{\left( \frac{[\,
\sqrt{s} \,- \,{\sqrt{s_0}} \, \, ]^2}{[{\Delta \sqrt{s}}]^2}
\right)},
\label{donsi}
\end{equation}
where the normalization factor $N_0$=6$\times$10$^4$ GeV$^{-1}$,
while $\sqrt{s_0}$=1 GeV
and $\Delta \sqrt{s}$=0.63~GeV. Note that, at SIS
energies $N_{mB}$ is almost entirely given by the pion-nucleon
interactions, and heavy meson and baryon collisions contribute
only to the high energy tail of the
distribution in Eq.~(\ref{donsi}) -- with quite small densities~\cite{CBJ}.
Finally, if we also average the
calculated, in-medium, total cross section for
$\pi^-{+}p{\to}\Lambda{+}K^0$, shown by the dashed line
in Fig.~\ref{laka4a}, over the available energy distribution given
in Eq.~(\ref{donsi}), we obtain an average total
kaon production cross section of
${<}K{>}$=65~$\mu$b for
central $Au{+}Au$ collisions at 2 AGeV.
This result is  indeed
compatible with the calculations using the free space total cross section
of the $\pi^-{+}p{\to}\Lambda{+}K^0$ reaction,
which provide an average total kaon production
cross section of ${<}K{>}$=71~$\mu$b
for central $Au{+}Au$ collisions at 2 AGeV.
Note that the inclusion of even a slight modification of the
$K^+$ mass because of the nuclear medium
(without the corresponding changes introduced here)
leads to a substantial reduction of the inclusive $K^+$ spectra
(by as much as a factor of 2 or 3),
compared to that calculated using the free space properties
for the relevant hadrons~\cite{Bratkovskaya}.
                                                                                
We stress that at SIS energies  reaction channels with
a $\Sigma$-hyperon in the final state play a minor role,
because of the upper limit of the energy available in the
collisions.  As was illustrated by Fig.~\ref{laka10},
the downwardly shifted $\pi{+}N{\to}\Sigma{+}K$ reaction threshold
at small densities is still quite high
compared to that for the reaction with
a $\Lambda$-hyperon in the final state.

In summary, we averaged the kaon production cross section
over the baryon density profile, which depends on the evolution
time of the heavy ion collision, in order to evaluate the impact of
our microscopic calculations on the heavy ion results.
Furthermore, in order to compare
with the experimental data more quantitatively, we calculated
the effective total kaon production cross section
by averaging over the invariant collision energy
distributions available in heavy ion reactions. We found that
at low collision energies,
the density or time averaged $K^+$-meson production total
cross section, calculated using
the in-medium properties for the $K^+$ meson, hyperons and relevant
hadrons, is very close to that calculated using the total cross section
given in free space.
                                                                                
Thus, the present results provide an explanation of why the
analyses~\cite{Li2,Cassing3,Senger3,Li3} of
available data on $K^+$ production from heavy ion collisions at SIS
energies~\cite{Barth,Laue,Senger2,Senger3} for the
$K^+$ spectra, were able to be reasonably described when one neglected 
any in-medium modification of the kaon and hadronic properties
-- i.e., adopting
the $K^+$-meson production cross section given in free space.
                                                                                
The conclusion of this section is that, if one accounts for
the in-medium modification of the production amplitude
(i.e., the in-medium properties of the $K^+$-meson and hadrons) correctly,
it is possible to understand $K^+$ production data in heavy ion collisions
at SIS energies consistently as it should be, even if the $K^+$-meson 
feels the theoretically expected, repulsive mean field potential. 
Thus, it is very important to understand the $K^+$ production data 
microscopically based on the in-medium dynamics, 
by including the in-medium changes of the $K^+$-meson and hadron 
properties in the reaction amplitudes. 
To include the in-medium properties of hadrons only in the purely 
kinematic effects, is not enough for the better understanding of   
the subthreshold kaon production in heavy ion collisions.

\subsection{\it $D$ and $\Dbar$ meson production in nuclei
\label{subsec:Dmeson}}

Antiproton annihilation on nuclei provides new possibilities
for studying  open charm production, exploring the properties
of charmed particles in nuclear matter and  measuring the
interaction of charmed hadrons.
                                                                                
Hatsuda and Kunihiro~\cite{Hatsuda} proposed that the light
quark condensates may be substantially reduced in hot and
dense matter and that as a result  hadron masses would be
modified. At low density the ratio of the hadron mass
in medium to that in vacuum can be directly linked to the ratio
of the quark condensates~\cite{SAI-q,Hatsuda1,Nelson,Brown,Lutz}.
Even if the change in the ratio of the quark condensates is small,
the absolute difference between the in-medium and vacuum masses of
the hadron is expected to be larger for the heavy hadrons.
In practice, any detection of the modification of the mass of a
hadron in matter deals with the measurement of effect associated with
this absolute difference.
                                                                                
It was found in Refs.~\cite{Hayashigaki1,Klingl2,Hayashigaki2} 
that the in medium change of quark condensates is smaller 
for heavier quarks, $s$ and
$c$, than those for the light quarks, $u$ and $d$.  Thus the
in-medium modification of the properties of heavy hadrons  may be
regarded as being  controlled mainly by the light quark condensates. \,
As a consequence we \, expect that charmed mesons,
which consist of a light
quark and heavy $c$ quark, should serve as suitable probes of
the in-medium modification of hadron properties.
                                                                                
As for the $\bar{K}$ and $K$-mesons, with their quark contents
$\bar{q}s$  and $q\bar{s}$ ($q{=}u,d$ light quarks), respectively, the
$D$ ($\bar{q}c$) and $\Dbar$ ($q\bar{c}$) mesons will satisfy different
dispersion relations in nuclear matter because of the different sign of
the $q$ and $\bar{q}$ vector coupling. Some experimental confirmation
of this effect has come from
measurements~\cite{Li1,Cassing2,Cassing3,Schroter,Barth,Laue,LiCB,Ritman,Shin}
of $K^-$ and $K^+$-meson production from  heavy
ion collisions. The $D^+$ and $D^-$ production from $\bar{p}A$
annihilation might yield an even cleaner signal for the in-medium
modification of the $D$ and $\Dbar$ properties.

Because of charm conservation, $D$ and $\Dbar$ mesons are produced
pairwise and can be detected by their semileptonic decay channels.
The threshold for the $\bar{p}N{\to}D\Dbar$ reaction in
vacuum opens at an antiproton energy around 5.57~GeV,
but it is lowered in the $\bar{p}A$ annihilation
by the in-medium modification of the $D$ and $\Dbar$ masses,  
as well as by  Fermi motion.
                                                                                
The interaction of the $D$-mesons with nuclear matter
is of special interest~\cite{Kharzeev}.
Note that the $DN$ interaction should be very
different from that of charmonia ($J{/}\Psi{N}$), since the
interaction between the nucleons and the heavy mesons which do not
contain $u$ and $d$ quarks is expected to proceed
predominantly through gluon
exchange. On the other hand, as for $\bar{K}$-mesons, the $D$-mesons
might be strongly absorbed in matter because of the charm
exchange reaction $DN{\to}\Lambda_c\pi$, while the $\Dbar$-mesons
should not be absorbed. As will be shown later, the specific
conditions of the $D\Dbar$ pair production in  $\bar{p}A$
annihilation provide a very clean and almost model independent
opportunity for the experimental reconstruction of the charm-exchange
mechanism.

As before, we calculate effective masses, 
$m^*_D(r)$, and mean field potentials, $V^q_{\sigma,\omega,\rho}(r)$,
at position $\vec{r}$ ($r=|\vec{r}|$) in the nucleus.
The scalar and vector potentials (neglecting the Coulomb force)
felt by the $D$ and $\Dbar$ mesons, which depend only on
the distance from the center of the nucleus, $r$, 
are given by~\cite{AlexD}(see Eqs.~(\ref{spot}) and~(\ref{vpot})):
\begin{eqnarray}
V^{D^\pm}_s(r)
&\equiv& U_s(r) = m^*_D(r) - m_D,
\label{spotD}\\
V^{D^\pm}_v(r) &=&
 \mp  (\tilde{V}^q_\omega(r) - \frac{1}{2}V^q_\rho(r)),
\label{vpotD2}
\end{eqnarray}
The $\rho$-meson mean field potential, $V^q_\rho(r)$,
(and the Coulomb potential) which are small and expected to give
a minor effect, will be neglected below.
Note that $\tilde{V}^q_\omega=1.4^2V^q_\omega$, and also 
$V^q_\rho$ is negative in a nucleus with a neutron excess.

For the following calculations we define the total potential as
\begin{equation}
U^{D^\pm}(r) = V_s(r) + V^{D^\pm}_v(r),
\end{equation}
where $V_s$ and $V_v$ denote the scalar and vector
potentials, respectively.
                                                                                
In Fig.~\ref{comic6} we show the total potentials  
for the $D^-$ and $D^+$-mesons
as a function of the nuclear radius calculated for
$^{12}C$ and $^{197}Au$. 
\begin{figure}
\begin{center}
\begin{minipage}[t]{8 cm}
\epsfig{file=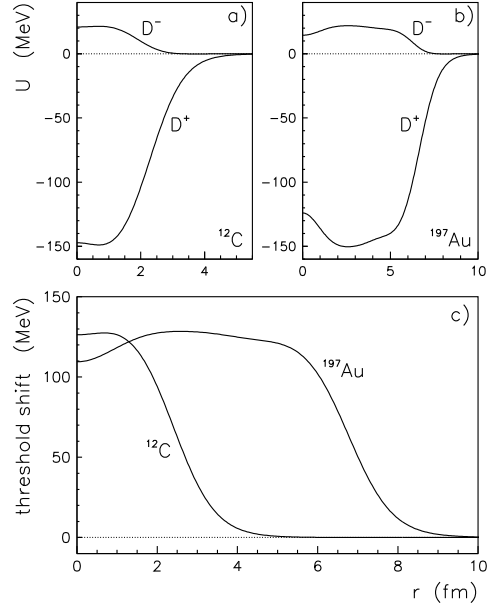,height=9cm}
\end{minipage}
\begin{minipage}[t]{16.5 cm}
\caption{The $D^-$ and $D^+$ potentials calculated for
$^{12}C$ (a) and $^{197}Au$ (b) as a function of the nuclear radius.
We also show the downward shift in the threshold for $D^+D^-$ production
for $^{12}C$ and $^{197}Au$, in (c)
(from Ref.~\cite{AlexD}).}
\label{comic6}
\end{minipage}
\end{center}
\end{figure}
%
The in-medium dispersion relation,
for the total energy $E_{D^\pm}$ and the momentum
$p = |\vec{p}|$ of the $D^\pm$-meson is given by
\begin{equation}
E_{D^\pm}(r) =
\sqrt{p^2 + (m_D + V_s(r))^2 } + V^{D^\pm}_v(r),
\label{totale}
\end{equation}
where the bare $D \equiv D^{\pm}$-meson mass is 
$m_D{=}1.8693$~GeV~\cite{PDG00,PDG98,PDG02}.
Note that the total $D^-$-meson potential is repulsive, while the
$D^+$ potential is attractive, which is analogous to the
case for the $K^+$ and $K^-$ mesons, respectively~\cite{Tsushima_k}.
                                                                                
The amount of downward shift of the $\bar{p}N{\to}D^+D^-$ reaction
threshold in nuclei, associated with the in-medium modification of the
$D$ and $\Dbar$ scalar potentials and the vector potentials,
is simply $2V_s$, and is shown in Fig.~\ref{comic6}c) for $^{12}C$ and
$^{197}Au$ as a function of the nuclear radius. The
threshold reduction is quite large in the central region of these
nuclei and should be detected as an enhanced production of the
$D^+D^-$ pairs. Note that a similar situation holds for the
$K^+$ and $K^-$ production and, indeed, enhanced $K^-$-meson production
in heavy ion collisions, associated with the reduction of the
production threshold, has been partially confirmed
experimentally~\cite{Li1,Cassing2,Cassing3,Schroter,
Barth,Laue,LiCB,Ritman,Shin}.

The $D\Dbar$ production in antiproton-nucleus
annihilation was calculated using the cascade model~\cite{Sibirtsev1}
adopted for $\bar{p}A$ simulations. The detailed
description of the initialization procedure as well as the interaction
algorithm are given in Ref.~\cite{Sibirtsev1}.
The reaction zone was initialized  with the use of the
momentum dependent $\bar{p}N$ total cross section, given
as~\cite{PDG94}:
\begin{equation}
\sigma_{\bar{p}N}=38.4 + 77.6p^{-0.64}+0.26(\ln{p})^2-1.2\ln{p},
\end{equation}
where $p$ denotes the antiproton laboratory momentum and
the cross section was taken to be the same for the proton and the
neutron target (in good agreement with the
data~\cite{PDG94}).
                                                                                
The $\bar{p}N{\to}D\Dbar$ cross section was calculated with
quark-gluon string model proposed in Ref.~\cite{Kaidalov3}.
In the following we will concentrate on
the production of $D^+$ and $D^-$-mesons and thus take into account
only two possible reactions, namely $\bar{p}p{\to}D^+D^-$ and
$\bar{p}n{\to}D^0D^-$. Note that the relation,
\begin{equation}
4\sigma (\bar{p}p\to D^+D^-)= \sigma (\bar{p}n\to D^0D^-)
\end{equation}
is due to the difference in the number of the quark planar
diagrams~\cite{Kaidalov3}.
                                                                                
Furthermore, to account for the $D^-$ and $D^+$-meson propagation
in nuclear matter one needs to introduce the relevant cross sections for
elastic and inelastic $DN$ scattering. Since no data
for the $DN$ interaction are available we use a diagrammatic approach
illustrated by Fig.~\ref{comic14}a,b). Let us compare the $D^-N{\to}D^-N$
and the $K^+N{\to}K^+N$ reactions in terms of the
quark lines. Apart from the difference between the
$c$ and $s$ quarks, both reactions are very similar
and can be understood in terms of rearrangement of the $u$ or $d$ quarks.
Thus, in the following calculations we assume that
$\sigma_{D^-N{\to}D^-N}{=}\sigma_{K^+N{\to}K^+N}$.

\begin{figure}
\begin{center}
\begin{minipage}[t]{8cm}
\epsfig{file=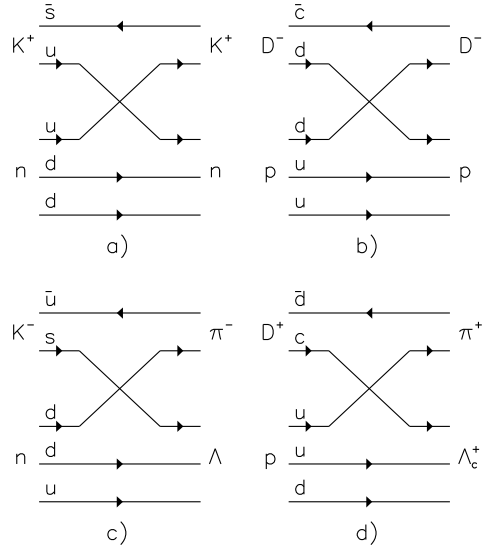,height=9cm}
\end{minipage}
\begin{minipage}[t]{16.5cm}
\caption{Quark diagrams for  $K^+n{\to}K^+n$ (a) and
$D^-p{\to}D^-p$ (b) elastic scattering and for
$K^-n{\to}\Lambda\pi^-$ (c), $D^+p{\to}\Lambda_c^+\pi^+$ (d)
inelastic scattering.}
\label{comic14}
\end{minipage}
\end{center}
\end{figure}

The $K^+N$ cross section was taken from Ref.~\cite{SibirtsevaK}, which
gives a parametrization of the available experimental data.
The total $K^+N$ cross section, averaged over neutron and proton targets,
is shown in Fig.~\ref{comic3}a) by the dashed line - as a
function of the kaon momentum in the laboratory system.
Note, that within a wide range of
kaon momentum $\sigma_{K^+N}$ is  almost constant and approaches a
value of ${\simeq}20$~mb. We adopt the value  $\sigma_{D^-N}{=}20$~mb,
noting that it is entirely due to the elastic scattering channel.
                                                                                
Now, Fig.~\ref{comic14}c,d) shows both the $K^-N{\to}\Lambda\pi$
and $D^+N \to \Lambda_c\pi$ processes, which are again quite similar
in terms of the rearrangement
of the $s$ and $c$ quarks, respectively. Thus we assume
that $\sigma_{D^+N{\to}\Lambda_c\pi}{\simeq}$
$\sigma_{K^-N{\to}\Lambda\pi}$.

\begin{figure}
\begin{center}
\begin{minipage}[t]{8cm}
\epsfig{file=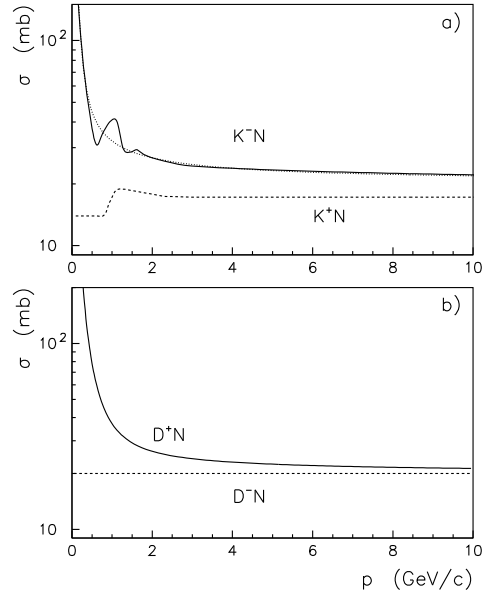,height=9cm}
\end{minipage}
\begin{minipage}[t]{16.5cm}
\caption{a) The total $K^-N$ (solid) and $K^+N$ (dashed line)
cross sections obtained~\cite{SibirtsevaK} as the best
fit to the available experimental data~\cite{PDG98}
and shown as function of the kaon momentum. The dotted line
show the result as explained in the text. b) The $D^-N$ (dashed)
and $D^+N$ total cross sections used in the calculations
(from Ref.~\cite{AlexD}).}
\label{comic3}
\end{minipage}
\end{center}
\end{figure}

The total $K^-N$ cross section is shown by the solid line in
Fig.~\ref{comic3}. Again it is averaged over proton and the
neutron  and taken as a parametrization~\cite{SibirtsevaK}
of the experimental data. At low momenta the $K^-N$
cross section shows  resonance structures due to the
strange baryonic resonances~\cite{PDG98}, while at high momenta
it approaches a constant value. Apart from the contribution from
these intermediate baryonic resonances the inelastic
$K^-N{\to}\Lambda\pi$ cross section can be written  as
\begin{eqnarray}
\label{matr}
\sigma_{K^-N{\to}\Lambda\pi}=\frac{|M|^2}{16\ \pi \ s} 
\left\lbrack \frac{(s-m_\Lambda^2-m_\pi^2)^2-4m_\Lambda^2m_\pi^2}
{(s-m_K^2-M_N^2)^2-4M_N^2m_K^2}\right\rbrack^{1/2},
\end{eqnarray}
where $s$ is the square of the invariant collision energy
and $m_K$, $M_N$,
$m_\Lambda$, $m_\pi$ are the masses of kaon, nucleon,
$\Lambda$-hyperon and pion, respectively. In Eq.~(\ref{matr})
the $|M|$ denotes the matrix element of the $K^-N{\to}\Lambda\pi$
transition, which was taken as a constant. Now the total $K^-N$
cross section is given as a sum of the cross section for the inelastic
channel Eq.~(\ref{matr}) and for the elastic one, where the latter was
taken to be  20.5~mb. The dotted line in Fig.~\ref{comic3}
shows our result for the total $K^-N$ cross section obtained with
$|M|{=}$11.64~GeV$\cdot$fm, which reproduces the
trend of the data reasonably well.
                                                                                
A similar approach was used to construct the $D^+N$
total cross section. It was assumed that at high momenta
the $D^+N$ elastic cross section equals  the $D^-N$ cross section,
while the $D^+N{\to}\Lambda_c^+\pi$ cross section was calculated from
Eq.~(\ref{matr}), with appropriate replacements of the particle masses.
The final results are shown in Fig.~\ref{comic3} and were
adopted for the following calculations.
\begin{figure}
\begin{center}
\begin{minipage}[t]{8cm}
\epsfig{file=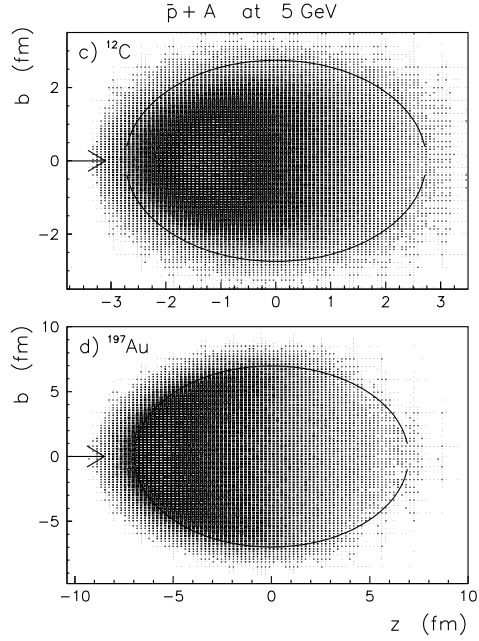,height=9cm}
\end{minipage}
\begin{minipage}[t]{16.5cm}
\caption{The plot of the annihilation zone for $\bar{p}{+}^{12}C$ (a)
and $\bar{p}{+}^{197}Au$ (b) reactions at a beam energy of 5 GeV.
The solid
line indicates the rms radius of the target nucleus.
The arrows show the direction of the antiproton beam
(from Ref.~\cite{AlexD}).}
\label{comic5}
\end{minipage}
\end{center}
\end{figure}
%
We wish to emphasize that the status of the $D$-meson-nucleon
interactions is still unknown and itself one of the
important goals of the $\bar{p}A{\to}D\Dbar X$ studies.
Our approach is necessary in order to estimate the expected
sensitivity of the experimental measurements to the
$DN$ interaction and to study the possibility
of evaluating the $D^+N$ and $D^-N$ cross sections.
                                                                                
In comparison to  low energy antiprotons that  annihilate
at the periphery of the nucleus, because of the large $\bar{p}N$
annihilation probability,  antiprotons with
energies above 3~GeV should penetrate the nuclear interior. They
can therefore  probe the nuclear medium at normal baryon density
$\rho_0$ and hence yield information about the
in-medium properties of the particles. Indeed, as is illustrated
by Fig.~\ref{comic6}, the $D$-meson potential  deviates
strongly from zero in the  interior of the nuclei considered. 
{}Figure~\ref{comic5} illustrates the reaction zone for the $\bar{p}C$ and
$\bar{p}Au$ annihilations at an antiproton beam energy of 5~GeV.
The plots are given as a functions of the impact parameter, $b$,  
and the $z$-coordinate, assuming that the beam is oriented along
the $z$-axis, which is shown by arrows in Fig.~\ref{comic5}.
The annihilation zone is concentrated in the front
hemisphere of the target nuclei. Actually the antiprotons
penetrate sufficiently deeply to test densities near that of  normal
nuclear matter and hence the shift in the $D^+D^-$
production threshold should be manifest.
\begin{figure}
\begin{center}
\begin{minipage}[t]{8cm}
\epsfig{file=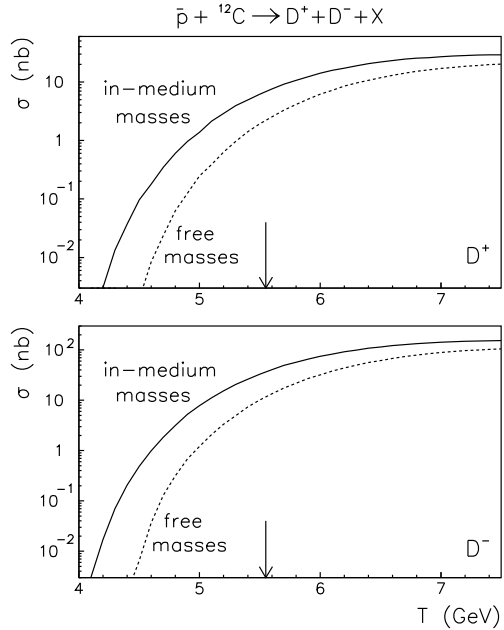,height=9cm}
\end{minipage}
\begin{minipage}[t]{16.5cm}
\caption{The total cross section for $D^+$  and
$D^-$-meson  production in $\bar{p}C$ annihilation
as a function of the antiproton energy. The results are shown
for calculations with free (dashed lines) and in-medium masses (solid
lines) for the $D$-mesons. The arrow indicates the reaction
threshold on a free nucleon
(from Ref.~\cite{AlexD}).}
\label{comic10}
\end{minipage}
\end{center}
\end{figure}
                                                                                
Now we calculate the total $\bar{p}A{\to}D^+D^-X$ production
cross section as function of the antiproton beam energy and
show the results in Fig.~\ref{comic10} for a carbon target and
in Fig.~\ref{comic1} for gold. The vacuum
$\bar{p}N{\to}D^+D^-$ cross section is also shown in
Fig.~\ref{comic1}. Note that the difference
between the $D^+$ and $D^-$-meson production rates is caused
by the $D^+$-absorption in nuclear matter.

Obviously the production threshold is substantially reduced
as compared to the antiproton annihilation on a free nucleon.
Apparently,  part of this reduction is due to the Fermi
motion~\cite{Sibirtsev3,Debowski,Sibirtsev4}, however
the results calculated with in-medium $D$-meson masses indicate a
much stronger threshold reduction than those using 
free masses for the final $D$-mesons.
                                                                                
Note that, because of their relatively long mean life, the $D$-mesons
decay outside the nucleus and their in-medium masses cannot be
detected through a  shift of  the invariant  mass of the
decay products (unlike  the leptonic decay of the vector mesons).
Thus it seems that the modification of
the $D^+$ and $D^-$-meson masses in nuclear matter can best be detected
experimentally as for the shift of the in-medium
$K^+$ and $K^-$-meson masses, namely as an enhanced $D$-meson
production rate at energies below the threshold for
the $\bar{p}N{\to}D^+D^-$ reaction in free space.

\begin{figure}
\begin{center}
\begin{minipage}[t]{8cm}
\epsfig{file=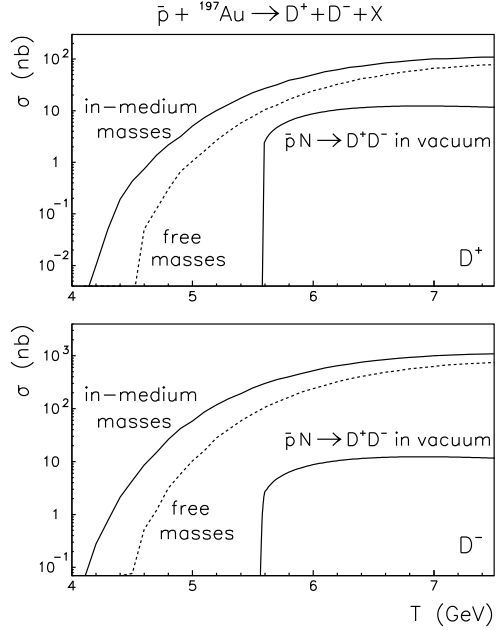,height=9cm}
\end{minipage}
\begin{minipage}[t]{16.5cm}
\caption{The total cross section for $D^+$  and
$D^-$-meson production in $\bar{p}Au$ annihilation
versus the antiproton energy. The results are shown
for calculations with free (dashed lines) and in-medium masses (solid
lines) for the $D$-mesons. For a comparison the 
$\bar{p}N{\to}D^+D^-$ cross section in free space is also indicated
(from Ref.~\cite{AlexD}).}
\label{comic1}
\end{minipage}
\end{center}
\end{figure}

We should further note that experimentally it may be difficult to distinguish
whether such an enhancement is due to  the modification of
the  $D^+$ and $D^-$-meson masses in nuclear matter, or to
the Fermi motion, or to other processes that are not yet included
in our study. In principle, the high momentum component of the
nuclear spectral function can provide sufficient energy for  particle
production far below the reaction threshold in free
space~\cite{Sibirtsev3}. However, the results in
Refs.~\cite{Debowski,Sibirtsev4} calculated with realistic
spectral functions~\cite{BEN,Sick,Atti} indicate that such effects
are actually negligible, while a more important contribution
comes from multistep production mechanisms. For instance, the dominant
contribution for $K^+$ production in $pA$ collisions
comes from the secondary ${\pi}N{\to}YK^+$ process, which
prevails over the direct $pN{\to}NYK^+$ reaction~\cite{Debowski,Badala}.
Thus, the interpretation of the data
depends substantially on the reliable determination of the production
mechanism.

We emphasize that an additional advantage of the $D^+D^-$
production in $\bar{p}A$ annihilation is the possibility to reconstruct
the production mechanism directly.  Let us denote $M_X$ as 
the missing mass of the target nucleon in the 
$\bar{p}N{\to}D^+D^-$ reaction. Obviously, in vacuum  $M_X$ is equal
to the free nucleon mass and can be reconstructed for
antiproton energies above the $D^+D^-$ production threshold. 
When analyzing $M_X$ in $\bar{p}A$ annihilations
one expects the distribution $d\sigma{/}dM_X$ to be
centered close to the mass of the bound nucleon
- below the free nucleon mass. The shape of the
distribution $d\sigma{/}dM_X$ reflects information of the spectral
function of the nucleus~\cite{BEN,Sick,Atti}.
                                                                                
The preceding discussion is based on the hypothesis that
the reaction $\bar{p}N{\to}D^+D^-$
is the dominant mechanism for  $D^+D^-$ pair production.
By  measuring both the $D^+$ and $D^-$ mesons one can directly
check this hypothesis.
%
\begin{figure}
\begin{center}
\begin{minipage}[t]{8cm}
\epsfig{file=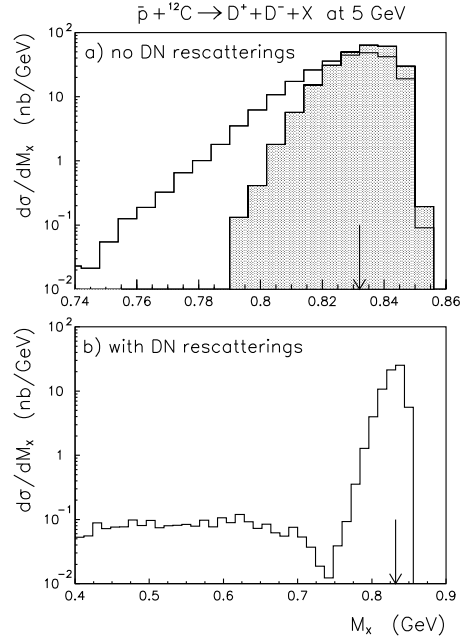,height=9cm}
\end{minipage}
\begin{minipage}[t]{16.5cm}
\caption{The missing mass distribution calculated for
$\bar{p}C$ annihilation at 5~GeV. The upper part shows the results
obtained without (hatched histogram) and with
account of the in-medium potentials (open histogram), but neglecting
the $D$-meson interactions in the nucleus. The hatched histogram
is normalized to the open histogram. The lower part shows the
calculations with $D^+$ and $D^-$ potentials and with $DN$
interactions
(from Ref.~\cite{AlexD}).}
\label{comic13}
\end{minipage}
\end{center}
\end{figure}
Let us first neglect the $D$-meson interactions in the nuclear
environment and analyze the $M_X$ spectrum for
$\bar{p}C$ annihilation at 5~GeV. Fig.~\ref{comic13}(a) shows
the missing mass distribution calculated without (hatched histogram)
and with the inclusion of the $D^+$ and $D^-$-meson potentials in 
carbon nucleus $^{12}$C.
Recall that the results calculated with free
masses provide much smaller $\bar{p}C{\to}D^+D^-X$ production
cross sections (see Fig.~\ref{comic10}). Thus, for the purpose of the
comparison, the result obtained without potentials is 
renormalized to those with in-medium masses in Fig.~\ref{comic13}a).
The arrow in Fig.~\ref{comic13}a) indicates the density averaged
mass of the bound nucleon in the carbon target~\cite{SAI-7}.
Indeed, both histograms are centered around the expected value.
However, the results calculated with the potentials shows a substantially
wider distribution. This effect can be understood easily in terms
of the downward shift of the $D$-meson production threshold 
in $^{12}$C.
 
Fig.~\ref{comic13}(b) shows the $M_X$ distribution calculated with
in-medium masses, taking into account both $D^+$ and
$D^-$-meson interactions in the nuclear environment. Note that the
distribution below $M_X{\simeq}0.75$~GeV results from
secondary $DN$ elastic rescattering and its strength is
proportional to the $DN$ elastic cross section.
A deviation of the actual experimental missing mass distribution from
those shown in Fig.~\ref{comic13}(b) may directly imply a 
contribution from $D^+D^-$ reaction mechanisms, other than
direct production.
                                                                                
In principle, the missing mass $M_X$ reconstruction appears  
to be a very promising tool for the detection of the in-medium
mass modification of the $D^\pm$ mesons, although 
it requires a detailed knowledge of the nuclear
spectral function~\cite{BEN,Sick,Atti} as well as an accurate
calculation of the $M_X$ distribution.

In summarizing this section, we have shown that $\bar{p}A$ annihilation
at energies below the $\bar{p}N{\to}D^+D^-$ reaction threshold in 
free space offer reasonable conditions for detecting  
the changes in $D$-meson properties in nuclear environment. 
In-medium modification of the
$D$-meson mass can be observed as an enhanced $D^+D^-$
production at antiproton energies below ${\simeq}5.5$~GeV.
The advantage of the $\bar{p}A{\to}D^+D^-X$ reaction is the
possibility to reconstruct directly the primary production mechanism
and hence to avoid a misinterpretation that such an enhancement
is due to the contribution from multistep production processes.
The missing mass reconstruction allows one to 
separate the effect due to the high momentum component
of the nuclear spectral function.
Thus, the study of the in-medium modification of the $D$-meson mass is very
promising, even with a target as light as carbon, where the total
$D^+D^-$ mass reduction is sizable and the nuclear spectral function
is under control~\cite{BEN,Sick,Atti}.
Furthermore, we have shown that the $\bar{p}A$ annihilation provides
favorable conditions for studying the $D$-meson interaction
in nucleus. The difference in the $D^+$ and $D^-$-meson
momentum spectra from antiproton annihilation on heavy nuclei
provides a very clean signature for the charm exchange 
reaction $D^+N{\to}\Lambda_c^+\pi$, 
and thus can be used to determine the $DN$
rescattering and absorption cross sections.

\subsection{\it $\JPsi$ suppression
\label{subsec:jpsi}}

There is a great deal of interest in possible
signals of Quark-Gluon Plasma (QGP)
formation (or precursors to its formation), and $\JPsi$ suppression
has been considered to be one of the most promising 
candidates. Indeed, an anomalous result has been reported 
there~\cite{RHI,Qm97,Abreu}.
On the other hand, there may be other mechanisms which produce an
increase in $\JPsi$ absorption in a hot, dense medium.
(For recent reviews on QGP, see e.g., Refs.~\cite{STAR,Jacobs}.) 
We are particularly interested in the rather exciting suggestion, based on the
QMC model~\cite{GUI-1,SAI-1}, that the charmed mesons,
$D$, $\Dbar$, $D^*$ and $\overline{D^*}$, should suffer substantial
changes in their properties in a nuclear 
medium~\cite{Tsushima_d}. 
(See sections~\ref{subsubsec:scbmatter},~\ref{subsec:mesonA} 
and~\ref{subsec:Dmeson}.) 
This might be expected to have a considerable impact on charm 
production in heavy ion collisions.

In Ref.~\cite{Tsushima_d} it was found, for example, that at a density
$\rho_B{=}3\rho_0$ ($\rho_0{=}0.15$~fm$^{-3}$)
the $D$-meson would feel an attractive scalar
potential of about $120$~MeV
and an attractive vector potential
of about $250$~MeV.
(See sections~\ref{subsec:mesonA} and~\ref{subsec:Dmeson}.)
These potentials are comparable to
those felt by a $K^-$-meson~\cite{Tsushima_k,Waas2,SibirtsevaK,Kminus}, 
while the total potential
felt by the $D$ is much larger than that for the vector mesons,
$\rho$, $\omega$ and $\phi$~\cite{SAI-3,SAI-q,Tsushima_etao}.
Within QMC it is expected that the mass of
the $\JPsi$ should only be changed by a tiny amount in nuclear
matter~\cite{SAI-3,SAI-q,Tsushima_k,Tsushima_etao}.
A similar result has also been obtained using QCD
sum rules~\cite{Hayashigaki1,Klingl2}.

In the light of these results, it seems that the charmed mesons,
together with the $K^-$, are probably the best candidates to
provide us information on the partial restoration of
chiral symmetry.  Both open charm production and the
dissociation of charmonia in matter may therefore be used as new
ways of detecting the modification of particle properties
in a nuclear medium.
                                                                                
The suppression of $\JPsi$ production observed in relativistic
heavy ion collisions, from $p{+}A$ up to central $S{+}U$ collisions,
has been well understood in terms of charmonium absorption in
the nuclear medium. However, recent data from
$Pb{+}Pb$ collisions show a considerably stronger $\JPsi$
suppression~\cite{Qm97}. In an attempt to explain this
``anomalous'' suppression of
$\JPsi$ production, many authors have studied one of two possible
mechanisms, namely
hadronic processes~\cite{Brodsky,Capella,Martins,Mueller,Capella1}
and QGP formation~\cite{Matsui}
(see Ref.~\cite{RHI} for a review).

In the hadronic dissociation scenario~\cite{Brodsky} it is well
known that the $\JPsi$ interacts with pions and $\rho$-mesons
in matter, forming charmed mesons through the reactions,
$\pi{+}\JPsi {\to} D^*+\overline{D}, \overline{D^*}+D$ and
$\rho{+}J/\Psi {\to} D+\overline{D}$.
The absorption of $J/\Psi$ mesons on pions and $\rho$-mesons has been
found to be important (see Refs.~\cite{Cassing3,Capella,Gavin} and
references therein) in general and absolutely
necessary in order to fit the data on $J/\Psi$ production.
Furthermore, the absorption on comovers should certainly play
a more important role in $S+U$ and $Pb+Pb$ experiments, where hot,
high density mesonic matter is expected to be achieved.
However, recent calculations for the processes in free space~\cite{Mueller},
indicate a much lower cross sections than that necessary
to explain the data for $J/\Psi$ suppression.

$J/\Psi$ dissociation on comovers, combined with the absorption on
nucleons, is the main mechanism proposed as an alternative to
that of Matsui and Satz~\cite{Matsui} -- namely the dissociation of the
$J/\Psi$ in a QGP.
Note that both the hadronic and QGP scenarios predict $J/\Psi$
suppression but no mechanism has yet been found to separate them
experimentally.
Within the hadronic scenario the crucial point is the required
dissociation strength. In particular, one needs a total cross section for
the $\pi,\rho{+}J/\Psi$ interaction of around $1.5{\div}3$~mb
in order to explain
the data in heavy ion simulations~\cite{Cassing3}.
Recent calculations~\cite{Mueller} of the reactions
$\pi{+}J/\Psi {\to} D+\overline{D^*}, \overline{D}+D^*$ and
$\rho{+}J/\Psi {\to} D+\overline{D}$, based on $D$ exchange,
indicate a  much lower cross section than this.
                                                                                
The main uncertainty in the discussion of the
$J/\Psi$ dissociation on a meson gas is given by the estimates
of the $\pi,\rho{+}J/\Psi$ cross section~\cite{Mueller}.
The predictions available for the  $\pi{+}J/\Psi$ cross section are given
in Refs.~\cite{Martins,Mueller}.
Following the meson exchange model
of Ref.~\cite{Mueller}, we show by the dotted line in 
Fig.~\ref{char1} the $\pi{+}J/\Psi$ (a) and $\rho{+}J/\Psi$ (b)
dissociation cross sections calculated in free space.
\begin{figure}
\begin{center}
\begin{minipage}[t]{8cm}
\psfig{file=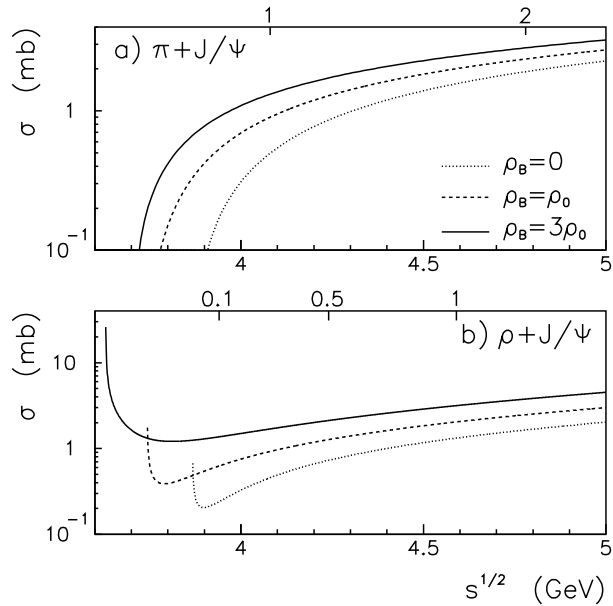,height=9cm}
\end{minipage}
\begin{minipage}[t]{16.5cm}
\caption{$\pi{+}J/\Psi$ (a) and $\rho{+}J/\Psi$ (b) dissociation
cross sections as functions of the invariant collision energy, $s^{1/2}$.
Results are shown for vacuum (the dotted line), $\rho_0$ (the dashed line)
and $3\rho_0$ (the solid line)
(from Ref.~\cite{jpsi}).}
\label{char1}
\end{minipage}
\end{center}
\end{figure}
Moreover, the upper axis of Fig.~\ref{char1}(a) indicates the $\pi$-meson
kinetic energy, $T_\pi$, given in the $J/\Psi$ rest frame,
which indicates that the pions should be sufficiently
hot to be above the $DD^\ast$ production threshold and to
dissociate the  $J/\Psi$-meson. By taking a thermal
pion gas with average $T_\pi{\simeq}$150~MeV, one might conclude
that independent of the $\pi{+}J/\Psi$ dissociation
model used~\cite{Martins,Mueller}, the rate of this
process is small. However, this situation changes drastically when
the in-medium potentials of the charmed mesons are taken into account,
because they lower the $\pi{+}J/\Psi{\to}\overline{D}{+}D^\ast$
threshold as will be shown later. The upper axis of Fig.~\ref{char1}(b)
shows the $\rho$-meson kinetic energy $T_\rho$ in the
$J/\Psi$ rest frame. As was discussed in
Refs.~\cite{RHI,Cassing3,Mueller}, the $J/\Psi$ dissociation
might proceed on thermal $\rho$-mesons, because of the low
$\rho{+}J/\Psi{\to}\overline{D}{+}D^\ast$ reaction threshold.

As far as the meson properties in free space are concerned,
the Bethe-Salpeter (BS) and Dyson-Schwinger (DS) approaches
have been widely used~\cite{mir}. The application of BS approach
to the description of heavy-light quark systems
allows one to describe the D and B meson properties in free
space quite well~\cite{Weiss}. The DS approach at finite baryon density
was used~\cite{Blaschke} for the calculation of the in-medium
properties of $\rho$, $\omega$ and $\phi$ mesons.
The modification of the $\rho$ and $\omega$ meson masses
resulting from the DS equation is close to the calculations
with the QMC model~\cite{GUI-1,SAI-1,SAI-3,SAI-q,Tsushima_etao},
while the $\phi$-meson mass reduction from Ref.~\cite{Blaschke}
is larger than the QMC result.
(See section~\ref{subsubsec:scbmatter}.)

Based on the QMC model, the scalar and vector potentials
felt by the $D (D^+,D^0)$ and $\overline{D} (D^-,\overline{D^0})$ mesons in 
symmetric nuclear matter 
are given by~\cite{Tsushima_d}:
\begin{eqnarray}
V^{D^\pm}_s  
&\equiv& U_s = m^*_D - m_D,
\label{spotd}\\
V^{D^\pm}_v &\equiv& \pm U_v  
= \mp  (\tilde{V}^q_\omega - \frac{1}{2}V^q_\rho),
\label{vpotd}
\end{eqnarray}
where, $\tilde{V}^q_\omega = 1.4^2 V^q_\omega$, which is assumed to
be the same as that for the $K^+$ and $K^-$
mesons~\cite{Tsushima_k}. (See Eq.~(\ref{vpot}).)
The isovector meson mean field potential, $V^q_\rho$,
is zero in symmetric nuclear matter.

We show the $D$, $D^*$ and $\rho$-meson potentials used in further
calculations in Fig.~\ref{char3} as a function of the
baryon density, in units of $\rho_0$=0.15 fm$^{-3}$.
\begin{figure}
\begin{center}
\begin{minipage}[t]{8 cm}
\epsfig{file=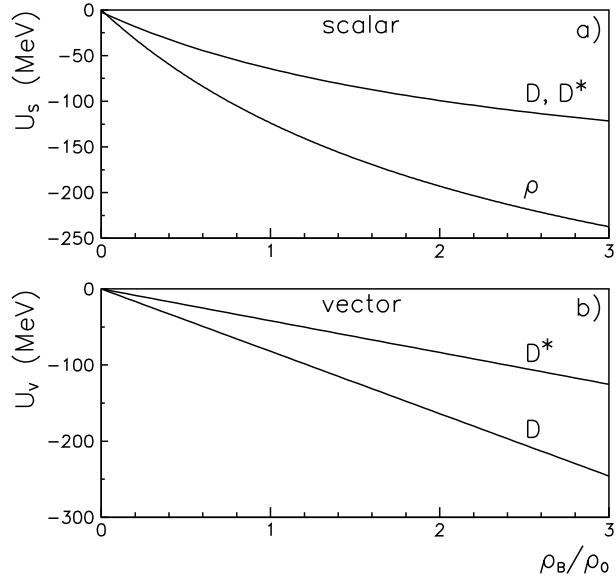,height=9cm}
\end{minipage}
\begin{minipage}[t]{16.5 cm}
\caption{The scalar (a) and vector (b) potentials for the
$D$, $D^*$ and $\rho$ mesons, calculated for nuclear matter
as functions of the baryon density, in units of the
saturation density of nuclear matter, $\rho_0$=0.15 fm$^{-3}$.
Scalar potentials for $D$ and $D^*$ are indistinguishable
(from Ref.~\cite{jpsi}).}
\label{char3}
\end{minipage}
\end{center}
\end{figure}
Note that these potentials enter not only in the final state
phase space (which becomes larger since the scalar masses
are reduced in matter), but also in the reaction amplitude and
the initial $\rho$-meson mass. 
(Recall the discussions in section~\ref{subsec:kaon} 
for subthreshold kaon production.)
Furthermore, as observed earlier,
the properties of the $J/\Psi$ meson are not
significantly altered in medium within QMC.
                                                                                
Note that the total $D^-$-meson potential is repulsive, while the
$D^+$ potential is attractive, which is analogous to the
case for the $K^+$ and $K^-$ mesons, 
respectively~\cite{Tsushima_k,Waas2,SibirtsevaK,Kminus}.
The threshold reduction is quite large when the nuclear density becomes
large for the $D^+D^-$ pairs. Note that a similar situation holds for the
$K^+$ and $K^-$ production and, indeed, enhanced $K^-$-meson production
in heavy ion collisions, associated with the reduction of the
production threshold, has been partially confirmed
experimentally~\cite{Cassing3,Laue}.

We first discuss the thermally averaged
cross sections, ${\langle}{\sigma}v{\rangle}$,  for ${\pi{+}J/\Psi}$
and ${\rho{+}J/\Psi}$ dissociation in Figs.~\ref{char2} and~\ref{char4},
when the free masses are used
for the charmed mesons and the in-medium potentials are set to zero.
They are shown by the dotted lines.
These results are needed for comparison with the calculations
including the potentials.
%
\begin{figure}
\begin{center}
\begin{minipage}[t]{8cm}
\epsfig{file=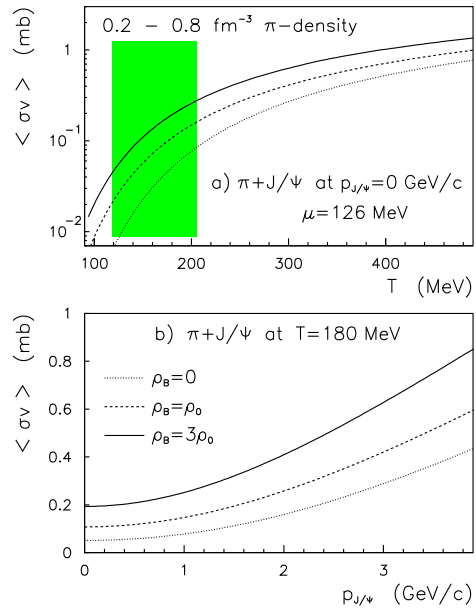,height=9cm}
\end{minipage}
\begin{minipage}[t]{16.5cm}
\caption{Thermally averaged $\pi{+}J/\Psi$ absorption
cross section as a function of the pion gas temperature, $T$,
and chemical potential, $\mu$, at $p_{J/\Psi}{=}0$
(a) and as a function of the
$J/\Psi$-momentum at $T{=}180$~MeV (b). The results are shown
using the same notation as in Fig.~\ref{char1}. 
The shadowed area indicates the
temperatures expected to be achieved in heavy ion collisions
(from Ref.~\cite{jpsi}).}
\label{char2}
\end{minipage}
\end{center}
\end{figure}
%
Since the pions are almost
in thermal equilibrium, their energy spectrum is given by a Bose
distribution with temperature, $T$, and chemical potential,
$\mu$, where we have used the value, $\mu{=}126$~MeV~\cite{Kataja}.
The thermally averaged ${\pi{+}J/\Psi}$
cross section can be obtained by averaging over the
$\pi$-spectrum at fixed $J/\Psi$-momentum. The dotted line
in Fig.~\ref{char2}a) shows ${\langle}{\sigma}{v}\rangle$ as a
function of the pion gas temperature, $T$, which was calculated with
zero $J/\Psi$ momentum relative to the pion gas.
                                                                                
The shadowed area in Fig.~\ref{char2}a) indicates the temperature range
corresponding to the pion densities $0.2 - 0.8$~fm$^{-3}$,
which are expected to be achieved in the heavy ion collisions presently
under consideration.
In vacuum the $\pi{+}J/\Psi$ dissociation cross section
is less than about $0.3$~mb.
The thermally averaged absorption cross section for temperature,
$T{=}180$~MeV,
is shown in Fig.~\ref{char2}(b) (the dotted line) as a
function of the $J/\Psi$ momentum. 
The thermally averaged cross section which we find,
${\langle}{\sigma}v{\rangle}$, would be very difficult to detect
with the present experimental capabilities.

\begin{figure}
\begin{center}
\begin{minipage}[t]{8cm}
\epsfig{file=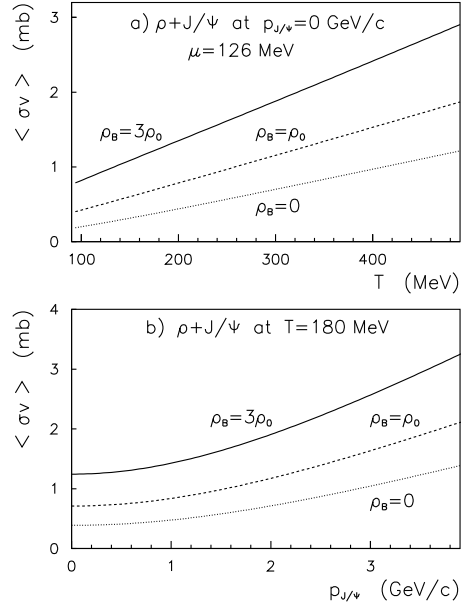,height=9cm}
\end{minipage}
\begin{minipage}[t]{16.5cm}
\caption{Thermally averaged $\rho{+}J/\Psi$ absorption
cross section as a function of the $\rho$-meson gas temperature $T$
with $p_{J/\Psi}{=}0$ (a) and as a function of the
$J/\Psi$-momentum at $T{=}180$~MeV (b). Notations are
similar to Fig.~\ref{char2}
(from Ref.~\cite{jpsi}).}
\label{char4}
\end{minipage}
\end{center}
\end{figure}

A similar situation holds for the $\rho{+}J/\Psi$ dissociation,
as illustrated by the dotted line in Fig.~\ref{char4}.
Indeed, the $J/\Psi$
absorption on comovers seems to be negligible~\cite{Mueller}
in comparison with that needed to explain the $J/\Psi$ suppression,
provided that we use the vacuum properties of the charmed mesons.
On the other hand, this situation changes dramatically when we
consider the effect of the vector and
scalar potentials felt by the charmed $D$, $D^\ast$ mesons and also  
$\rho$ mesons (the vector potential is zero), 
as calculated using Eqs.~(\ref{spotD}),~(\ref{vpotD2}) and~(\ref{spot}).
The cross sections calculated for $\pi, \rho{+}J/\Psi$ collisions
with the in-medium potentials are shown in Fig.~\ref{char1},
for densities, $\rho_0$ (the dashed line) and $3\rho_0$
(the solid line). The dotted line in Fig.~\ref{char1}
indicates  the free space cross sections.
                                                                                
Clearly the $J/\Psi$ absorption cross sections are substantially enhanced
for both the $\pi{+}J/\Psi$ and $\rho{+}J/\Psi$ reactions, not only
because of the downward shift of the reaction threshold, but also
because of the in-medium effect on the reaction amplitude.
Moreover, now the $J/\Psi$ absorption
on comovers becomes  density dependent -- a
crucial finding given the situation in actual heavy ion collisions.
These effects have never been considered before.
That is, this was the first calculation of the comover 
absorption cross section as
a function of baryon density.
                                                                                
The thermally averaged, in-medium $\pi{+}J/\Psi$ and $\rho{+}J/\Psi$
absorption cross sections, ${\langle}{\sigma}v{\rangle}$,
are shown by the dashed and solid lines in Figs.~\ref{char2} 
and~\ref{char4}, respectively. We find that
${\langle}{\sigma}v{\rangle}$ depends
very strongly on the nuclear density. Even for
$p_{J/\Psi}{=}0$, with a pion gas
temperature of 120~MeV, which is close to the saturation pion density,
the thermally averaged $J/\Psi$ absorption cross section on the
pion, at $\rho_B{=}3\rho_0$, is about a factor of 7 larger than that
at $\rho_B{=}0$ (i.e., with no effect of the in-medium modification
-- see Fig.~\ref{char2}(a)).

The thermally averaged $\rho{+}J/\Psi$ dissociation cross section
at $\rho_B{=}3\rho_0$
becomes larger than 1~mb. Thus, the $J/\Psi$ absorption on $\rho$-mesons
should be appreciable, even though the $\rho$-meson density
is expected to be small in heavy ion collisions.
We note that dynamical calculations~\cite{Cassing3}
suggest that the $\rho$-meson density should be around half of 
the pion density in $Pb{+}Pb$ collisions.
                                                                                
In order to compare our results with the NA38/NA50
data~\cite{Qm97,Abreu} on $J/\Psi$ suppression in $Pb{+}Pb$
collisions, we have adopted the heavy ion model proposed in
Ref.~\cite{Capella} with the $E_T$ model from
Ref.~\cite{Capella1}. We  introduce the absorption
cross section on comovers as function of the density of comovers,
while the nuclear absorption cross section is taken
as 4.5~mb~\cite{Capella1}.
The final results are shown in Fig.~\ref{char7}.
The dashed line in Fig.~\ref{char7} shows the calculations
with the phenomenological constant cross section for $J/\Psi$
absorption on comovers ${\langle}{\sigma}v{\rangle}$$\simeq$1~mb
and is identical to the results given in Ref.~\cite{Capella1}.
The solid line in
Fig.~\ref{char7} shows the calculations with the
density dependent cross section ${\langle}{\sigma}v{\rangle}$
for $J/\Psi$ absorption on comovers calculated 
based on the QMC model. Both curves clearly reproduce
the data~\cite{Qm97} quite well,
including most recent results from NA50 on the ratio
of $J/\Psi$ over Drell-Yan cross sections, as a function
of the transverse energy up to $E_T$=100~GeV.
It is important to note that if one neglected the in-medium modification of
the $J/\Psi$ absorption cross section the large cross section
${\langle}{\sigma}v{\rangle}$$\simeq$1~mb could not be justified by
microscopic theoretical calculations and hence 
the NA50 data~\cite{Qm97,Abreu} could not be described.

\begin{figure}
\begin{center}
\begin{minipage}[t]{8cm}
\epsfig{file=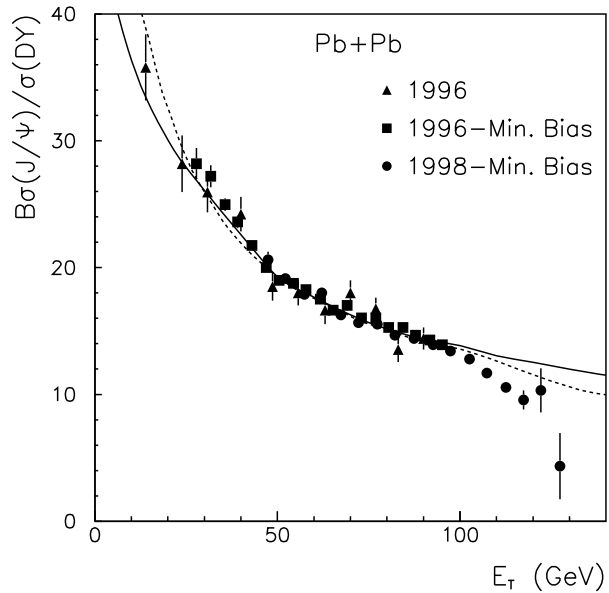,height=9cm}
\end{minipage}
\begin{minipage}[t]{16.5cm}
\caption{The ratio of the $J/\Psi$ over Drell-Yan cross sections from
$Pb{+}Pb$ collisions as function of the transverse energy $E_T$.
Data are from Ref.~\cite{Qm97,Abreu}. The solid line shows our calculations
with the density dependent cross section for
$J/\Psi$ absorption on comovers. The dashed line indicates the
calculations with phenomenological cross section
${\langle}{\sigma}v{\rangle}$$\simeq$1~mb~\cite{Capella1}.
For both calculations the nuclear absorption cross section was
taken as 4.5~mb
(from Ref.~\cite{jpsi}).}
\label{char7}
\end{minipage}
\end{center}
\end{figure}

Furthermore, we notice
that our calculations with in-medium modified absorption
provide a significant improvement in the understanding of the
data~\cite{Qm97} compared to the models quoted by NA50~\cite{Abreu}.
The basic difference between our results and those quoted by
NA50~\cite{Abreu} is that in previous
heavy ion calculations~\cite{RHI,Cassing3,Capella} the cross section for
$J/\Psi$ absorption on comovers was taken as a free parameter to
be adjusted to the data~\cite{Qm97,Abreu} and was never
motivated theoretically.
                                                                                
To summarize this section, we have studied $J/\Psi$ dissociation 
in a gas of $\pi$ and $\rho$ mesons taking into account the density
dependence of the scalar and vector potentials which
the mesons feel in nuclear matter.
We have shown a substantial density dependence of the  $J/\Psi$ 
absorption rate as a result of the changes in the properties 
of the charmed mesons in-medium. This aspect has
never been considered before when analyzing $J/\Psi$ suppression in
heavy ion collisions.
Moreover, when we introduce density dependent cross sections on comovers, 
based on the QMC model, into a heavy ion 
calculation, the result achieves a very good agreement with 
the NA50 data~\cite{Qm97,Abreu} up to transverse energy of about 100~GeV.

\section{Summary and outlook
\label{sec:outlook}}

We have studied various nuclear phenomena, starting at the quark level, 
using a self-consistent model for nuclear physics, namely the QMC model. 
Although there are many kinds of relativistic mean-field 
theory for nuclear physics,  
only the QMC model incorporates explicit quark degrees of freedom 
into nuclear many-body 
systems. As discussed in section~\ref{subsec:QMCQHD}, 
at the {\em hadronic} level, 
it is certainly possible to cast the QMC model into 
a form similar to that of a QHD-type mean-field 
model by re-defining the scalar field.  
However, at the same time, the QMC model can describe 
how the internal structure of hadrons changes in 
a nuclear medium. That is the greatest advantage of the present model and
it opens a tremendous number of 
opportunities for future work. 

Since the discovery of QCD as the fundamental theory of the strong interaction, numerous attempts have 
been made to derive the nuclear force within quark models. 
The QMC model stands between the traditional 
meson-exchange picture and the hard core 
quark models, namely, it is a mean-field model in the sense of 
QHD but with the couplings of $\sigma$ and $\omega$ mesons 
to confined quarks, rather than to the point-like 
nucleon.  After a considerable amount of work, 
one finds that the effect of the internal, quark structure 
of the nucleon is absorbed into the scalar polarizability 
in the effective nucleon mass in matter. 
It is the dependence of the scalar polarizability on the 
scalar field in matter (or it is numerically equivalent 
to the dependence on nuclear density) that is the heart of the 
QMC model and leads to the novel, saturation 
mechanism of the binding energy of nuclear matter as a function of density. 

Because the scalar polarizability plays the important 
role in the QMC model, it is of great interest 
to study whether the dependence of the scalar 
polarizability on the scalar field can be extracted from the 
fundamental theory, i.e., QCD.  In Ref.~\cite{TONY}, it is shown that 
the remarkable progress in resolving the problem of 
chiral extrapolation of lattice QCD 
data gives one confidence that the pion loop contributions are under control. 
In the case of the nucleon, one can then use this 
control to estimate the effect of applying 
a chiral invariant scalar field to the nucleon, i.e., to estimate the 
scalar polarizability of the nucleon. The resulting value is in excellent 
agreement with the range found in the QMC model, which is vital to 
describe many phenomena in nuclear physics. 
Thus, in a very real sense, the results presented 
in Ref.~\cite{TONY} provide a direct connection 
between the growing power to compute hadron 
properties from QCD itself and fundamental 
properties of atomic nuclei. Further work   
in this direction is very important and will be 
a focus for our field in the coming years.

It is also of great interest to perform more 
realistic calculations for nuclear systems. 
Feldmeier~\cite{feldm} proposed a molecular 
dynamics approach to solve the many-body 
problem of interacting identical fermions with 
spin $1/2$ approximately. The interacting system is represented by an 
antisymmetrized many-body wave function consisting of 
single-particle states which are localized in phase 
space. The equations of motion for the parameters characterizing 
the many-body state (e.g. positions, 
momenta and spin of the particles) are derived from a 
quantum variational principle. 
This is called the fermionic molecular dynamics (FMD) model. 
Later, based on FMD, 
a powerful technique to treat a fermionic system was  
developed by Horiuchi {\it et al.}~\cite{AMD} 
and applied not only to nuclear structure but also nuclear 
reactions. This is sometimes called 
the antisymmetrized molecular dynamics (AMD). 
In AMD, basis wave functions of the system are given by 
Slater determinants where the spatial part of each 
single-particle wave function is a Gaussian wave packet.  
The model has a remarkable feature that 
the wave function can represent various clustering 
structures as well as shell-model-like structures, although 
no inert cores and no clusters are assumed. 
Therefore, it is possible to calculate nuclear systems 
in a model-independent way. Although the formulation of AMD 
is non-relativistic, if we choose a proper nuclear force and 
adopt the non-relativistic version of the QMC 
model (see section~\ref{subsec:nonrelativistic}), 
it would be possible to 
construct a more realistic nuclear model which 
covers a wide range of nuclei from stable to unstable ones. 
This may be a new challenge and provide a further connection 
between quark degrees of freedom (QCD) and  
conventional nuclear physics. 

As seen through this article, many nuclear phenomena 
now seem to indicate that the traditional approach 
may have its limitations and 
suggest a need for subnucleonic degrees of freedom.  
There is no doubt that hadrons consist of quarks, antiquarks 
and gluons and that they can respond to the 
environment and change their characters in matter.  
We would like to emphasize here that 
{\it quarks play an important role in nuclei and nuclear matter}. 

\vspace{2em}
\noindent{\bf Acknowledgment}\\
We would like to thank P.A.M. Guichon, J. Haidenbauer, Hungchong Kim, 
G. Krein, D.H. Lu, W. Melnitchouk, A. Sibirtsev, and F.M. Steffens 
for valuable discussions and collaborations. This work was partly 
supported by Academic Frontier Project (Holcs, Tokyo University of Science, 
2005) of MEXT. It was also supported in part by DOE contract 
DE-AC05-84ER40150, under which SURA operates Jefferson Laboratory.


\end{document}